\documentclass{jfm}

\usepackage[english]{babel}
\usepackage[usenames, dvipsnames]{color}
\usepackage{graphicx}
\usepackage{caption}
\usepackage{subcaption}
\usepackage{natbib}
\usepackage{fixltx2e}
\usepackage{textgreek}
\usepackage{textcomp}
\usepackage{hyperref}
\usepackage{multirow}
\usepackage{array}
\usepackage{pbox}

\usepackage{xcolor}
\definecolor{cinnamon}{rgb}{0.82, 0.41, 0.12}
\definecolor{pink}{rgb}{0.858, 0.188, 0.478}
\definecolor{black}{rgb}{0.0, 0.0, 0.0}

\def\sz#1{{\textcolor{black}{#1}}}

\graphicspath{{./pdf_figures/}}

\begin{document}

\newtheorem{lemma}{Lemma}
\newtheorem{corollary}{Corollary}

\shorttitle{Turbulent flow in a square duct with spherical particles} 
\shortauthor{S. Zade, P. Costa, W. Fornari,  F. Lundell and L. Brandt} 

\title{Experimental investigation of turbulent suspensions of spherical particles in a square duct}

\author
 {
 Sagar Zade\aff{1},
 Pedro Costa\aff{1},
 Walter Fornari\aff{1},\\
 Fredrik Lundell\aff{1},
 \and
 Luca Brandt\aff{1}
 \corresp{\email{luca@mech.kth.se}}
  }

\affiliation
{
\aff{1}
Linn\'e Flow Centre and SeRC (Swedish e-Science Research Centre), \\ KTH Mechanics, SE 100 44 Stockholm, Sweden
}

\maketitle

\begin{abstract}
We report experimental observations of turbulent flow with spherical particles in a square duct. Three particle sizes namely: $2H/d_{p}$ = 40, 16 and 9 ($2H$ being the duct full height and $d_{p}$ being the particle diameter) are investigated. The particles are nearly neutrally-buoyant with a density ratio of 1.0035 and 1.01 with respect to the suspending fluid. 
Refractive Index Matched - Particle Image Velocimetry (RIM-PIV) is used for fluid velocity measurement even at the highest particle volume fraction (20\%) 
and Particle Tracking Velocimetry (PTV) for the particle velocity statistics for the flows seeded with particles of the two largest sizes, whereas only pressure measurements are reported for the smallest particles.  
Settling effects are seen
at the lowest bulk Reynolds number $Re_{2H}\approx$ 10000 whereas, at the highest $Re_{2H}\approx$ 27000, particles are in almost full suspension. 
The friction factor of the suspensions is found to be significantly larger than that of single-phase duct flow at the lower $Re_{2H}$ investigated; however, the difference decreases when increasing the flow rate and the total drag approaches the values  of the single phase flow at the higher Reynolds number considered, $Re_{2H}=27000$.
The pressure drop is found to decrease with the particle diameter for volume fractions lower than $\phi$ = 10\% for nearly all $Re_{2H}$ investigated. However, at the highest volume fraction $\phi$ = 20\%, we report a peculiar non-monotonic behavior:
the pressure drop first decreases and then increases with increasing particle size. 
The decrease of the turbulent drag with particle size
at the lowest volume fractions is related to an attenuation of the turbulence. 
The drag increase \sz{for the two largest particle sizes} at $\phi$ = 20\%, however, occurs despite this large reduction of the turbulent stresses, and it is therefore due to significant particle-induced stresses.
At the lowest Reynolds number, the particles reside mostly in the bottom half of the duct where the mean velocity significantly decreases; the flow is similar to that in a moving porous bed near the bottom wall and to turbulent duct flow with low particle concentration near the top wall.

\end{abstract}
 
\section{Introduction}\label{sec:Introduction}
The study of the flow of suspensions is a highly challenging endeavor. However, its sheer prevalence in natural flows and process industry has drawn the attention of researchers since a long time. \citet{einstein1906neue,einstein1911berichtigung} first derived a formula for the effective viscosity of a suspension of spheres under dilute concentration. He was followed by many well-known scientists who contributed in different measure to unravel the mysteries of particle-laden flows. However, as ubiquitous suspensions are, equally varied is their diversity in terms of particle size, density, concentration, shape, stiffness, etc. The analysis of suspensions is further complicated by the chaotic nature and multi-scale dynamics of turbulent flows.

In the industrial use of technology, study of particle-laden flows are relevant in areas like transport of crushed coal, slurries, use of particle dispersions in paper coating and paints, foodstuffs, etc. In nature, flows include transport of sediments on and above the river-bed, red blood cells immersed in plasma, etc. \sz{All this justifies the need to gain fundamental understanding of the physical phenomena occurring in suspensions.}

\subsection{Flow in a square duct}\label{sec:Flow in a square duct}
\begin{figure}
\centering
 \includegraphics[height=0.4\linewidth]{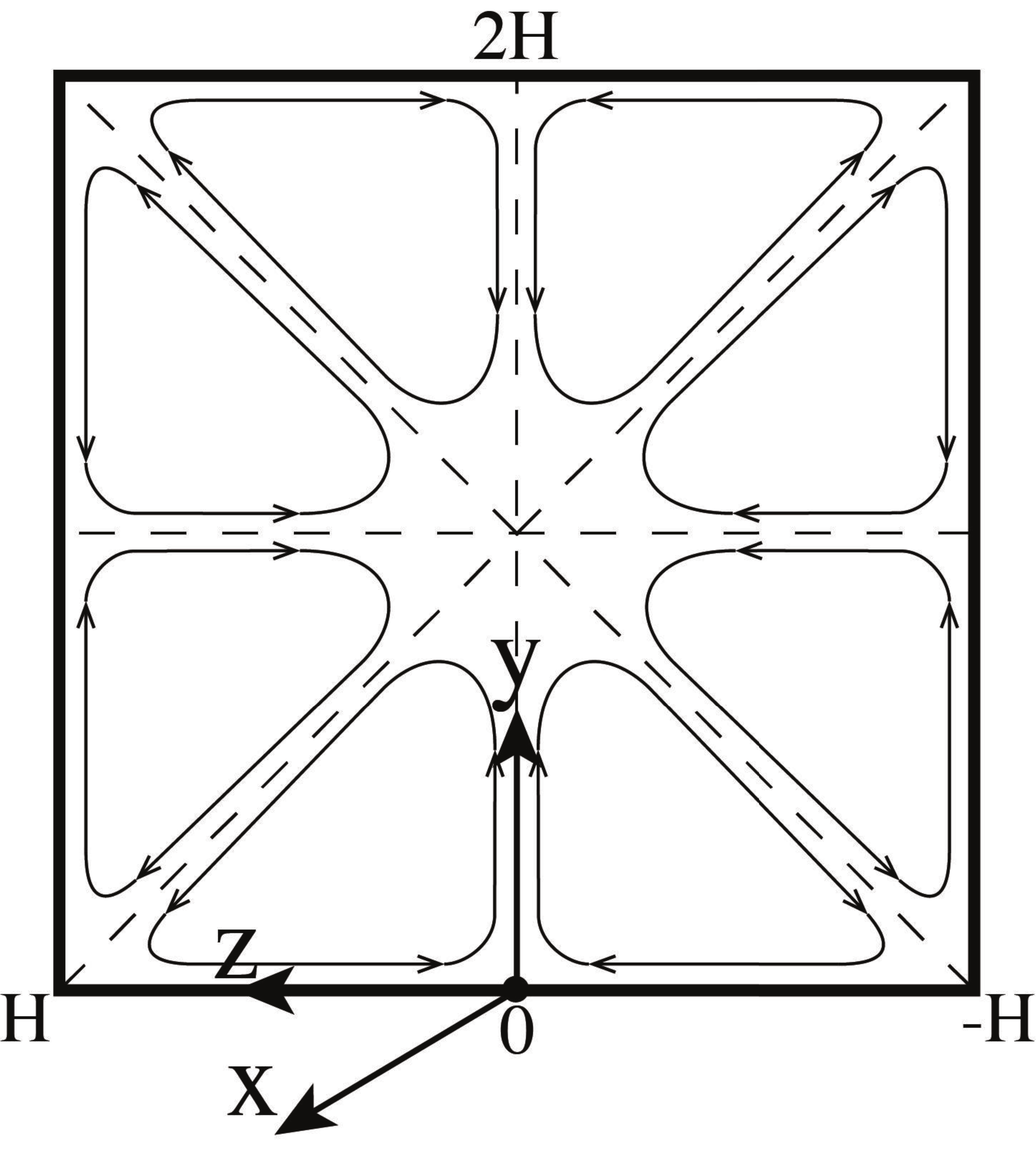}
\caption{Schematic of secondary flow of Prandtl\textquotesingle s second kind and coordinate system used in this study.}
\label{fig:Schematic secondary flow}
\end{figure}

This article focuses on the interaction of rigid spherical particles with a fully developed turbulent flow in a square duct. A square duct poses interesting differences as compared to a plane channel flow, the turbulence in which is more often studied. In a square duct, the flow is inhomogeneous in the two \sz{cross}-stream directions, whereas a plane channel as well as a round pipe has only one direction of inhomogeneity. However, compared to other duct geometries (e.g. a rectangular duct), these two inhomogeneous directions are symmetric. This inhomogeneity and wall-boundedness gives rise to a net cross-stream secondary flow of Prandtl\textquotesingle s second kind \citep{prandtl1926uber}, which is superimposed on the primary axial flow. To note, secondary motion of Prandtl's first kind is much higher in magnitude (generally 10 to 40\% of the bulk velocity). This is driven by pressure gradients and is encountered in curved passages, where centrifugal forces act at an angle to the streamwise direction. On the other hand, the secondary flow of Prandtl\textquotesingle s second kind is driven by gradients in the turbulence-stresses and its magnitude is smaller than the root-mean-square (r.m.s.) velocity of turbulent fluctuations and generally between 1 to 4\% of the bulk velocity in most straight ducts with non-circular cross-section. Nevertheless, these secondary motions, in the form of four pairs of counter-rotating vortices located at the duct corners (see figure \ref{fig:Schematic secondary flow} for a schematic), act to transfer fluid momentum from the centre of the duct to its corners, thereby causing a bulging of the streamwise mean velocity contours toward the corners. Also, their effect on the distribution of wall shear stress and heat transfer rates are quite significant \citep{demuren1991calculation}. Therefore, the simple geometry of a square duct provides an excellent case to test and develop turbulence models that explore secondary flows of the second kind.

Probably, the first qualitative data on turbulence-driven secondary motion in non-circular ducts was presented by \citet{nikuradse1930untersuchungen} using flow visualization techniques. \citet{brundrett1964production} found that the Reynolds normal-stress gradients were responsible for the generation of streamwise vorticity, manifesting in the form of secondary flow. \citet{gessner1973origin} showed that the secondary flow was due to the gradient of the Reynolds shear-stress normal to the corner-bisector. Later on, \citet{gavrilakis1992numerical}, \citet{huser1993direct} and \citet{uhlmann2007marginally} investigated the mechanism behind the generation of secondary flow using numerical simulations.

\subsection{Particle-laden flows}\label{sec:Effect of particles}

The presence of the cross-stream secondary motion, explained in the above section, can have significant effects on the motion of a suspended particle phase. Even though there are limited studies on particle-laden turbulent flows in a square duct, phenomena observed in other canonical turbulent flow cases (e.g. channel or pipe flow) can help us in understanding the fundamental mechanisms of particle-fluid interactions. 

\subsubsection{Numerical simulations using the point particle assumption}\label{subsec:Numerical simulations using the point particle assumption}

Starting with the relatively simpler assumption of point particles (where particle size is smaller than the Kolmogorov length scale of flow, $\eta$), various interesting phenomena have been observed in numerical simulations. In the one-way coupling regime (where the particle concentration is so small that it has negligible effect on the fluid phase), preferential concentration or clustering of particles within specific regions of the instantaneous turbulence field is observed \citep{eaton1994preferential}. In wall-bounded flows, particle inertia also induces a mean particle drift towards the walls, so-called turbophoresis \citep{reeks1983transport}, where the particles migrate from regions of high to low turbulence intensities. This migration is most pronounced when the particle inertial time-scale nearly matches the turbulent near-wall characteristic time-scale \citep{soldati2009physics} and, close to the walls, particles tend to form streaky patterns \citep{sardina2012wall}. 
In the two way coupling regime (where the particle to fluid mass-ratio is high enough to affect the fluid phase but, the volume-ratio is still small so that inter-particle interactions and excluded-volume effects are negligible), \citet{squires1990particle} and \citet{elghobashi1993two} observed that particles alter the spectral distribution of the fluid turbulent kinetic energy in isotropic homogeneous flows. In wall-bounded flows, \citet{kulick1994particle} showed that particles reduce the turbulent near-wall fluctuations and \citet{zhao2010turbulence} showed that this phenomenon can lead to drag reduction. In the case of a square duct, small inertial 
particles may preferentially migrate to the wall-center or to the corner regions \citep{winkler2004preferential,sharma2006turbulent, yao2014numerical,noorani2016aspect}.

\subsubsection{Simulations of finite-size particles and experiments}\label{subsec:Simulations of finite-size particles and experiments}

In industrial and environmental flows, the particles often \sz{have sizes comparable to or larger than the Kolmogorov length scale,} along with varying shape and stiffness, and the point particle approximation is, generally, not applicable. Amongst such finite-sized particles, the most commonly studied cases involve mono-dispersed, rigid, spherical, neutrally-buoyant particles suspended in an incompressible Newtonian fluid \citep{picano2015turbulent,lashgari2016channel,fornari2016rheology,ardekani2017drag}.

Dilute laminar flow of finite-size particles is known to exhibit an increase in the effective viscosity \citep{guazzelli2011physical}. \citet{bagnold1954experiments} showed how inter-particle collisions increase the effective viscosity in the highly inertial regime. Such inertial effects at the particle scale can induce other rheological effects like shear-thickening \citep{picano2013shear} or normal-stress differences \citep{kulkarni2008suspension, yeo2013dynamics}. These macroscopic changes could be explained by changes in the particle micro-structure \citep{morris2009review}, as also shown by the recent direct experimental measurement of the suspension micro-structure in \citet{blanc2013microstructure}. Among other dependencies,  \citet{fornari2016effect} showed how the effective viscosity depends, in the special case of microfluidic application, on the system confinement.

The suspended phase alters the critical $Re$ for the flow to transition from the laminar to the turbulent state (see experiments by \citet{matas2003transition} in a pipe flow and numerical simulations by \citet{loisel2013effect} in a channel flow). \citet{lashgari2014laminar} showed how the distribution of viscous, turbulent and particle stresses \citep{batchelor1970stress} varies when changing the particle volume fraction and the $Re$ in a plane channel flow.

Revisiting preferential concentration, now in the case of finite-sized particles, migration is also observed in wall-bounded laminar flows. In the viscous Stokes regime, particles migrate from regions of high shear to low shear due to irreversible interactions, e.g. towards the centerline in a Poiseuille flow \citep{guazzelli2011physical}. With an increase in particle $Re$, inertial effects become important and particles tend to move away from the centerline and equilibrate at an intermediate position due to the repulsive forces from the wall, see the tubular pinch effect in \citet{segre1962behaviour} and refer to \citet{ho1974inertial} for an explanation of its mechanism. The final equilibrium position is found to depend on the ratio of the particle to pipe diameter, the particle $Re$ and the bulk $Re$ \citep{matas2004inertial}. Particles also tend to form long trains in the flow \citep{matas2004trains}. Returning to the case of a square duct, \citet{chun2006inertial}, in their Lattice-Boltzmann simulations, observed the formation of particle trains that become unstable and break up into small clusters that tend to migrate towards the center of the duct. However, in contrast to the formation of a continuous annular ring in a round pipe, particles in a square duct migrate to one of a small number of equilibrium positions: at the center of the wall or near a corner (depending on $Re$). 
 \citet{choi2011lateral,abbas2014migration,miura2014inertial} experimentally measured the migration of particles in laminar flow in a square duct at low volume fractions. 
 \citet{Hamid2017Laminar} studied numerically the laminar flow in a square duct at different $Re$, up to a particle volume fraction $\phi$ = 20\% and for different duct to particle size ratios.  According to their study, particles largely move to the corners at lower volume fractions and at higher $Re$. 
Recently \citet{fornari2017suspensions} numerically investigated turbulent flows of a suspension of spherical rigid particles in a square duct up to a volume fraction of 20\% and found that at the highest volume fraction, particles preferentially accumulate in the core region and the intensity of the secondary flows reduces below that of the unladen case.

Most of the above observations pertaining to higher volume fractions are made using numerical tools, due to the difficulties in tackling such problems experimentally. 
External bulk measurements such as the overall pressure drop in the presence of particles (mostly sedimenting slurries in pipe flows) have been reported in \cite{doron1987slurry} and \cite{kaushal2002solids} amongst others. However, detailed velocity and concentration measurement in the interior of the suspension at high concentrations is rare, especially in turbulent flows. A few studies in this direction are highlighted below.

\cite{hampton1997migration} 
made use of Magnetic Resonance Imaging Velocimetry (MRI-V) to study the evolution of the axial velocity and concentration distribution of neutrally buoyant particles in a Poiseuille flow up to a particle volume fraction of 50\%. These authors found that the entrance lengths for the development of the concentration and velocity fields rapidly decrease as volume fraction and particle to pipe ratio was increased. \citet{han1999particle} extended MRI-V measurements to particles with non-negligible inertia ($Re_{p}\approx$ 0.2) so that inertial migration towards regions between the pipe-axis and the wall is competing with particle-particle interaction that pushes the particles towards the center. The parabolic velocity profile becomes blunted due to migration of particles towards the center. Another experimental technique used for suspensions is Laser Doppler Velocimetry (LDV). \citet{koh1994experimental} used LDV in a rectangular channel flow and also observed that the particle concentration distribution possessed a maximum near the channel centerline and a minimum at the channel walls \cite[see also][]{lyon1998experimental}. \citet{frank2003particle} studied the particle migration in Brownian suspensions (colloidal particles) in Poiseuille flow  using confocal microscopy, owing to its higher spatial resolution. \sz{Using PIV in a matched refractive index medium, \cite{abbas2017pipe} studied recently the laminar flow of concentrated non-colloidal particles ($\phi$ = 70\%) and observed a Newtonian like variation of pressure drop with bulk velocity even though the mean velocity profile clearly exhibited a shear-thinning behavior. They used the notion of effective viscosity, based on the local particle concentration, in the framework of the suspension balance model \citep{zarraga2000characterization}, to explain this peculiar behavior.} All the above experimental studies focused on low Reynolds number laminar flows. Recently, \citet{gurung2016measurement} used Ultrasound Imaging Velocimetry (UIV) to show that usable data can be obtained even in turbulent particle-laden flows at moderately high volume fractions. However, it was not possible to differentiate between the statistics of the fluid and the particle phase. MRI-V, on the other hand, suffers from low spatial and temporal resolutions when studying complex suspensions in turbulence. Furthermore, they also involve heavy procedures and expensive tools that make them somewhat difficult to use.

\subsubsection{Refractive Index Matched - Particle Image Velocimetry (RIM-PIV)}\label{subsec:Simulations of finite-size particles and experiments}

The de facto workhorse in the field of experimental flow measurement is PIV, which is an optical measurement technique requiring the use of an optically transparent solution as the working fluid. Measuring velocity in suspensions using PIV relies on the use of particles that are transparent i.e. the refractive indices of the fluid and particles are nearly the same with respect to the wavelength of the light used for illumination. However, the particles usually available are composed of materials like plastic, metal, glass etc. For these dispersed particles, volume fraction of 0.5\% in a domain of 5 to 10 cm has been indicated as the limit \citep{poelma2006turbulence}. There are, of course, some limited options available for refractive-index-matching (RIM) fluids to enable the use of PIV \cite[see][]{wiederseiner2011refractive}, \sz{but they are often difficult to scale-up due to issues related to long-time properties of the suspending solution, thermal stability, handling and cost.}

Therefore, there is a need for particles that display long-term optical transparency independent of moderate changes in temperature and solution properties, and that can be used in a fluid that is readily available and easy to handle (e.g. water). Super-absorbent hydrogel particles offer such an advantage. These particles have been successfully used to study the trajectories of large spheres (with size $\approx$ 100 times the Kolmogorov length scale $\eta$) in three dimensions in \cite{klein2012simultaneous}, where the particle surface was meticulously grafted with fluorescent particles to track their complete motion. \citet{bellani2012shape} studied the effect of Taylor-microscale-sized particles on homogeneous isotropic turbulence. In order to track particle rotation, small PIV tracer particles were embedded inside the macroparticle, made via injection molding. However, the particle volume concentration ($\phi$) used in their experiments was only 0.14\%. The simultaneous fluid-particle PIV method requires a large number of tracers that are all internal to a single macroparticle with unambiguous borders, which may be an issue for small macroparticles and large concentrations \citep{byron2013refractive}.

Finite-size particle modulations of the flow cannot be captured by simple modeling techniques such as those using an effective viscosity or other more complex constitutive laws for the stresses, \sz{as shown for e.g., in turbulent channel flows in} \citet{costa2017finite}. \sz{Even for laminar flows, finite-size particle effects, e.g. particle migration, results in large deviation of flow characteristics from the predictions based on an effective fluid model and exhibit a complex dependence on volume fraction and Reynolds number as shown in a recent study for a laminar square duct by \cite{manoorkar2018suspension}.} Therefore, in this study, we perform detailed experiments in a square-duct, for a range of bulk Reynolds number ($Re_{2H}$), particle concentration ($\phi$) and particle size ratio ($2H/d_p$), so as to relate the bulk observables to the local micro-scale dynamics.    

\subsection{Outline}\label{sec:Outline}
The structure of this article is as follows: Firstly, we describe the experimental facility and details about the data acquisition. The following result section includes validation studies with single-phase Newtonian fluid, pressure drop measurements and velocity profiles for different flow cases with particles. 
Finally, we discuss the observations with a focus on the highest (full suspension regime) and the lowest (sedimentation regime) Reynolds number, also referring to the numerical simulations in \citet{fornari2017suspensions} for neutrally-buoyant particles.

\section{Experimental setup}\label{sec:Experimental setup}
\subsection{Particles}\label{sec:Particles}
\begin{figure}
\centering
\begin{subfigure}{.30\textwidth}
  \centering
  \includegraphics[height=1\linewidth]{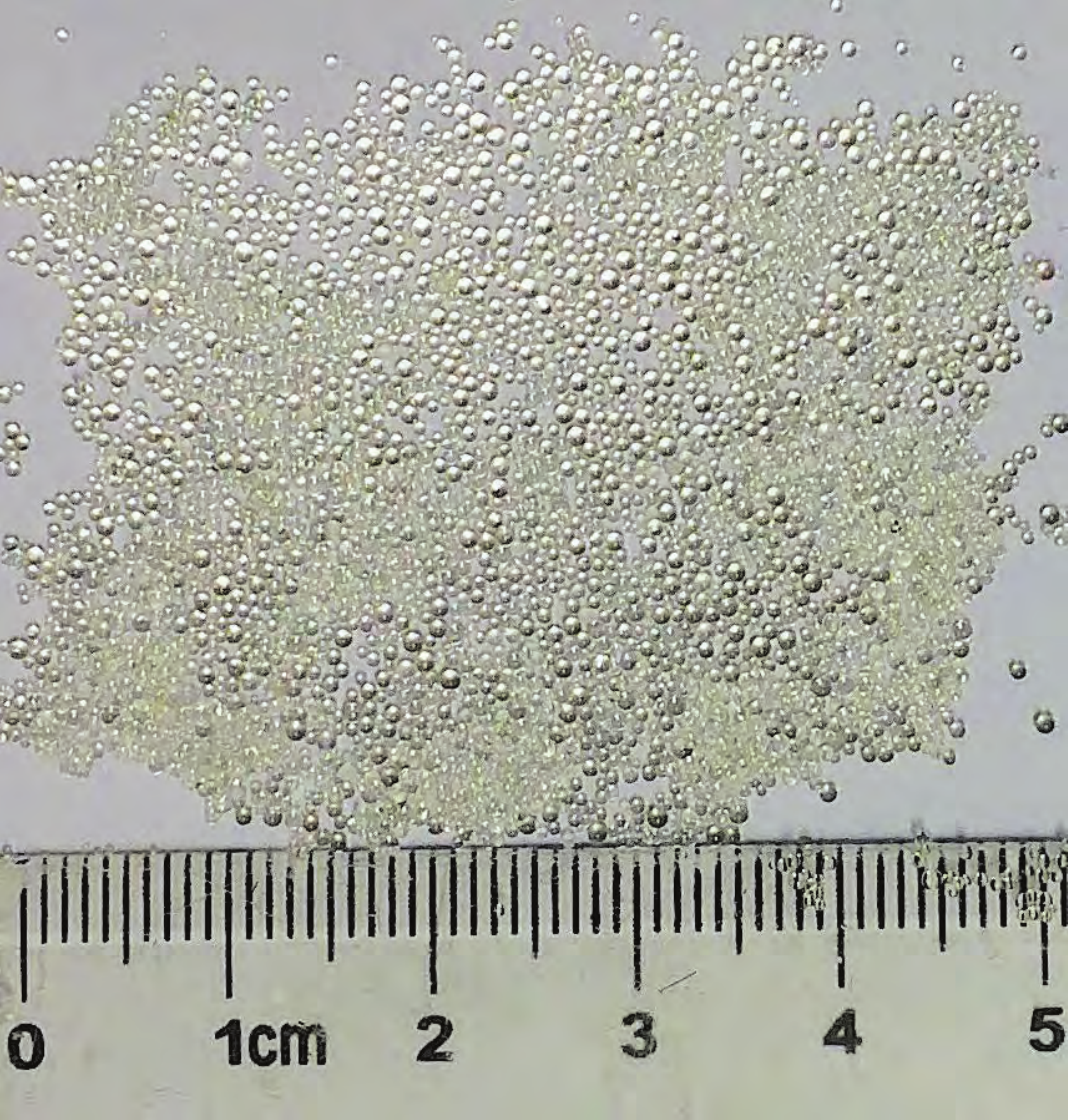}
  \caption{}
  \label{fig:Particles before expansion}
\end{subfigure}%
\begin{subfigure}{.30\textwidth}
  \centering
  \includegraphics[height=1\linewidth]{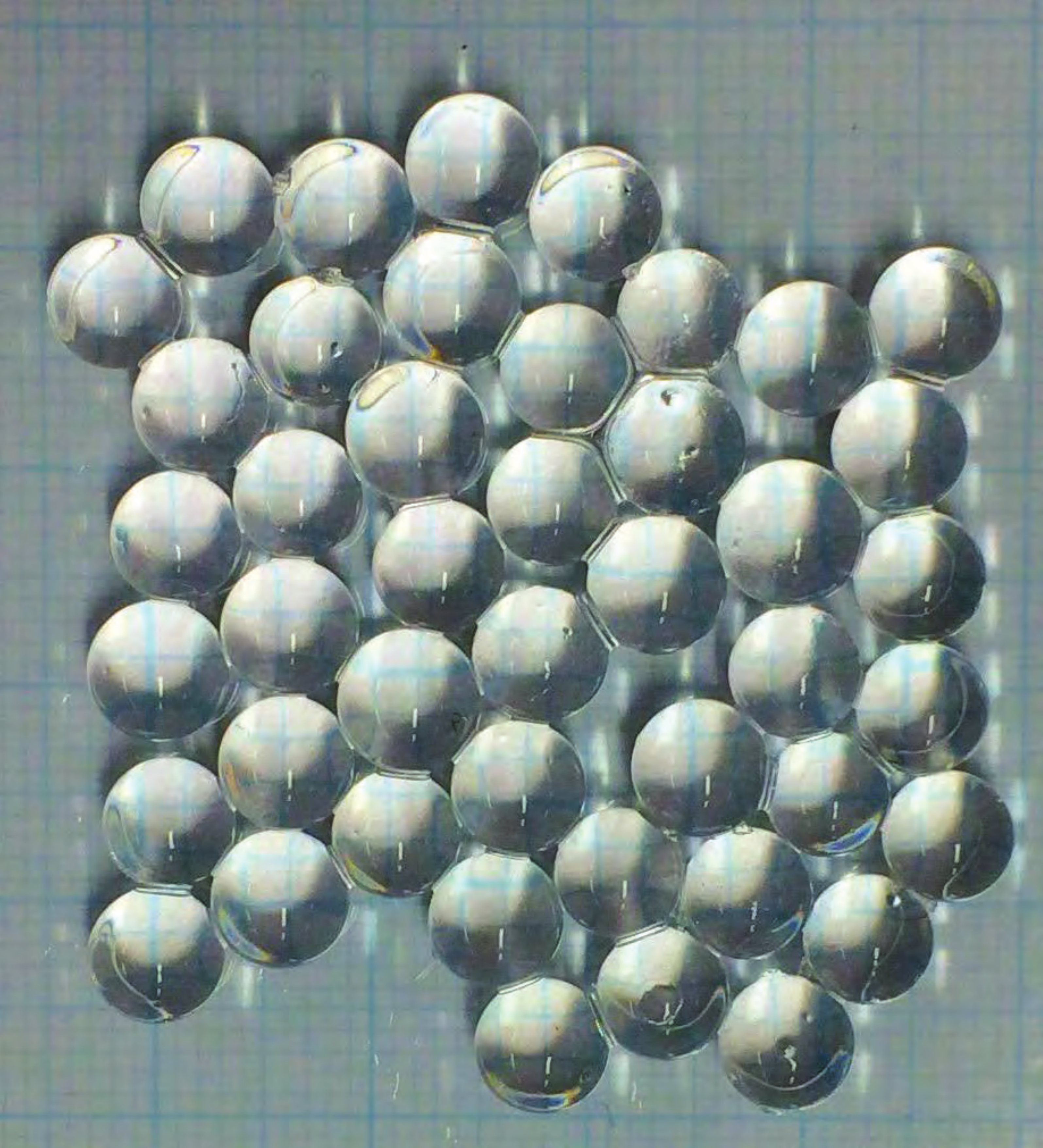}
  \caption{}
  \label{fig:Particles after expansion}
\end{subfigure}
\begin{subfigure}{.30\textwidth}
  \centering
  \includegraphics[height=1\linewidth]{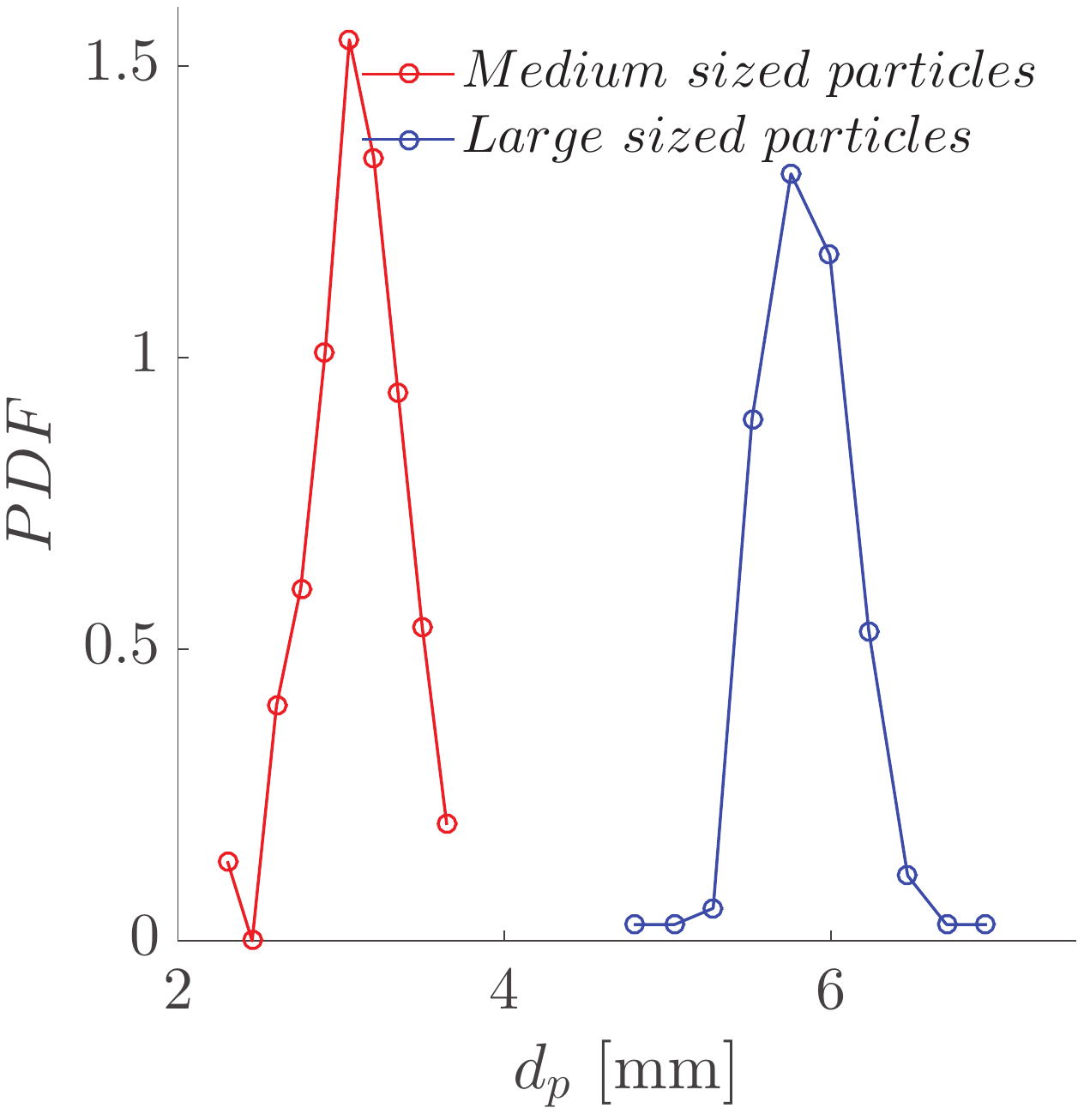}
  \caption{}
  \label{fig:Particles size}
\end{subfigure}
\caption{Image of the particles (a) before expansion (all sizes, in dry state), (b) after expansion (only large-sized particles) and (c) size distribution of medium and large-sized particles (see\ref{tab:Particle property} for details about particle nomenclature).}
\label{fig:Particles before and after expansion}
\end{figure}

The particles are commercially procured super-absorbent (polyacrylamide based) hydrogel. They are delivered with a range in diameter from 0.5 mm -- 1.1 mm (see figure \ref{fig:Particles before expansion}). Once mixed with tap water and left submerged for around 1 day, they grow to an equilibrium size between 3 to 6 mm. Particles are graded into different sizes using a range of sieves, prior to expansion in water (as they become softer and would break while sieving after expansion), into two batches: 0.55 $\pm$ 0.05 mm and 1.0 $\pm$ 0.05 mm which expanded to a size of 3.05 $\pm$ 0.8 mm (3 times standard deviation) and 5.60 $\pm$ 0.9 mm (see figure \ref{fig:Particles after expansion}). The particle size was determined by a digital imaging system and from the PIV images of particles in flow. Figure \ref{fig:Particles size} shows the Probability Distribution Function (PDF) of the particle size for the medium and large-sized particles. The fact that a Gaussian-like particle size distribution, with small size variance, has small effect on the flow statistics has been shown in \cite{FORNARI201854}. Hereafter, the 3.05 mm particles ($2H/d_{p}\approx$ 16) will be denoted as `Medium-sized Particles' or `MP' whereas the 5.60 mm particles ($2H/d_{p}\approx$ 9) will be addressed as `Large-sized Particles' or LP'.

We also use smaller particles, denoted `Small-sized Particles' or `SP', which are 1.25 mm in diameter ($2H/d_{p}\approx$ 40). These were also procured commercially. 
The SP have slightly different density and optical properties as compared to the larger ones. 
In this case, a small quantity of tracer particles ($\approx$ 4\% w/w in the dry state) was added during their manufacturing process to facilitate the tracking of particle rotation \cite[see][]{bellani2012shape}. At the particle concentrations used in this study, it was difficult to visualize the fluid flow beyond a few particle diameters due to the attenuation of light in such a turbid medium. Under such conditions, it was hard to differentiate the border of the particle from the fluid. Thus, PIV was not possible with SP and only pressure drop measurements are reported. 
On the other hand, the MP and LP, without tracers inside, were relatively transparent, so as to enable visualization up to a bulk volume fraction of 20\%. (Note that, at a bulk volume fraction of 20\%, there are areas where the local volume fraction exceeds more than twice this nominal  value due to settling or flow-induced migration).  
To enhance the contrast of the MP and LP, a small amount of Rhodamine was added to the water in which the particles expand. Due to the absence of tracers inside MP and LP, we could only measure translational motion.

The density of the particles was determined using 2 methods: (i) by measuring the volume displacement by a known mass of particles and (ii) by determining the terminal settling velocity in a long liquid settling column. In the first method, a known mass of fully expanded particles was put in a water-filled container of uniform diameter. The rise in the level of water due to the particles was measured using a very precise laser distance meter (optoNTDC 1710, Micro-Epsilon Messtechnik GmbH, resolution = 0.5 \textmugreek m). In the second method, a single particle with a known diameter was gently dropped in a long vertical pipe, filled with water, wide enough  
so as to minimize the wall effects. The settling velocity is determined (after it has reached steady state) by noting the time to pass-by regularly placed distance markers. 
The relation for drag force $F$ on a settling particle
in \cite{crowe2011multiphase},
\begin{equation}
  \frac{F}{\rho_{f}U_{T}^2A}=\frac{12}{Re_{p}}(1+0.15Re_{p}^{0.687})
  \label{eqn:Particle density}
\end{equation} applicable in the transitional regime: 1$<$Re$<$750, 
was used to relate the particle diameter $d_p$ and terminal velocity $U_T$ to the unknown particle density $\rho_{p}$ . $A$ is the projected area of the particle. 
In the above, $Re_p$ is the particle Reynolds number given by $\frac{\rho_{p}U_{T}d_{p}}{\mu_f}$ where, ${\mu_f}$ is the dynamic viscosity of the liquid. Both the above tests were performed at a room temperature of around 20$^\circ$C and yielded nearly the same value of density. 
The latter method was preferred as it was less sensitive to measurement uncertainties.
Table \ref{tab:Particle property} lists the properties of the particles used in the experiments.

\begin{table}
 \begin{center}
  \begin{tabular}{ccccc}
    $2H/d_{p}(\approx)$ & $\frac{\rho_{p}}{\rho_{f}}$ & Name & Appearance & Measurements performed\\[3pt]
    \hline
       40 & 1.01$\pm$0.003 & Small Particles (SP) & `Milky' & Pressure drop only\\
       16 & 1.0035$\pm$0.0003 & Medium Particles (MP) & `Clear' & Pressure drop and velocity\\
       9 & 1.0035$\pm$0.0003 & Large Particles (LP) & `Clear' & Pressure drop and velocity\\
  \end{tabular}
  \caption{Particle properties and the type of measurements performed.}{\label{tab:Particle property}}
 \end{center}
\end{table}

Our solid-gel-like particles are elastically deformable, but the forces (shear and pressure) applied by the fluid were not sufficient to deform them. This was ascertained by observing the particle shape in the PIV movies. Therefore, these particles can, for the purpose of this study, be considered as rigid.

\sz{Since the gels consist mostly of fluid, there exists a layer of fluid between the gel and the contacting surface, making them slippery, so they are able to slide with friction coefficients 
that may be smaller than those observed in solid materials. Also in contrast to hard solid-solid contact, the coefficient of sliding friction has been observed to decrease with increasing normal force \citep{gong1997friction} and increase with sliding velocity along with being a function of contact area for the gels. But, for our spherical particles, in the limit of negligible deformation, we can assume a single point-of-contact between two spheres or a sphere and the wall. 
A recent study \citep{Emorybachelorthesis2017} has measured the friction coefficient between a sphere and flat surface, both made from a Polyacrylamide based gel (which is the same material that our particles are made up of). They observed values around 10\textsuperscript{-2} at sliding velocities between 0.001 to 0.1 $m/s$.}

\subsection{Experimental rig}\label{sec:Experimental rig}
\begin{figure}
\centering
\begin{subfigure}{.45\textwidth}
  \centering
  \includegraphics[height=0.75\linewidth]{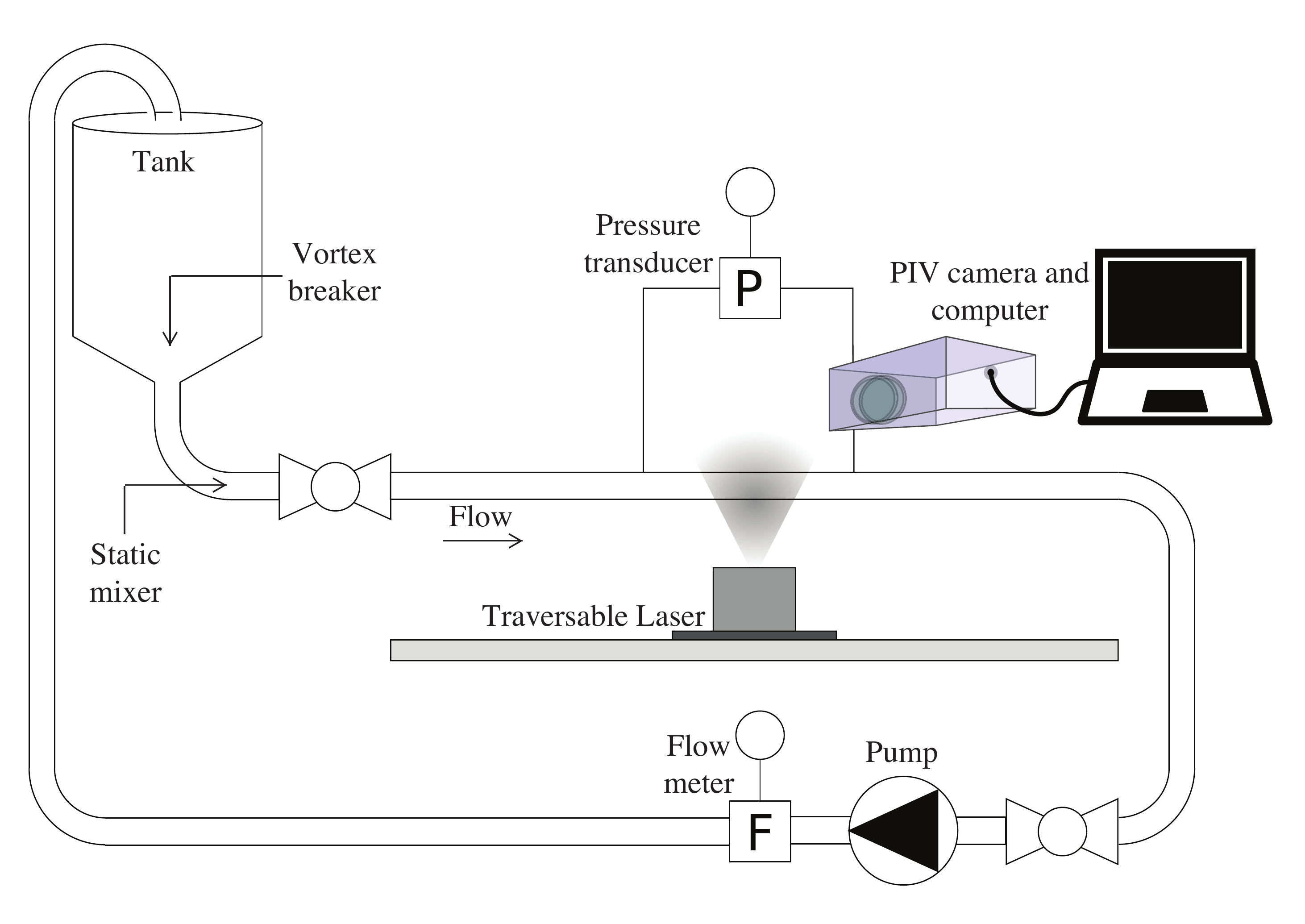}
  \caption{}
  \label{fig:Set-up schematic}
\end{subfigure}%
\begin{subfigure}{.45\textwidth}
  \centering
  \includegraphics[height=0.75\linewidth]{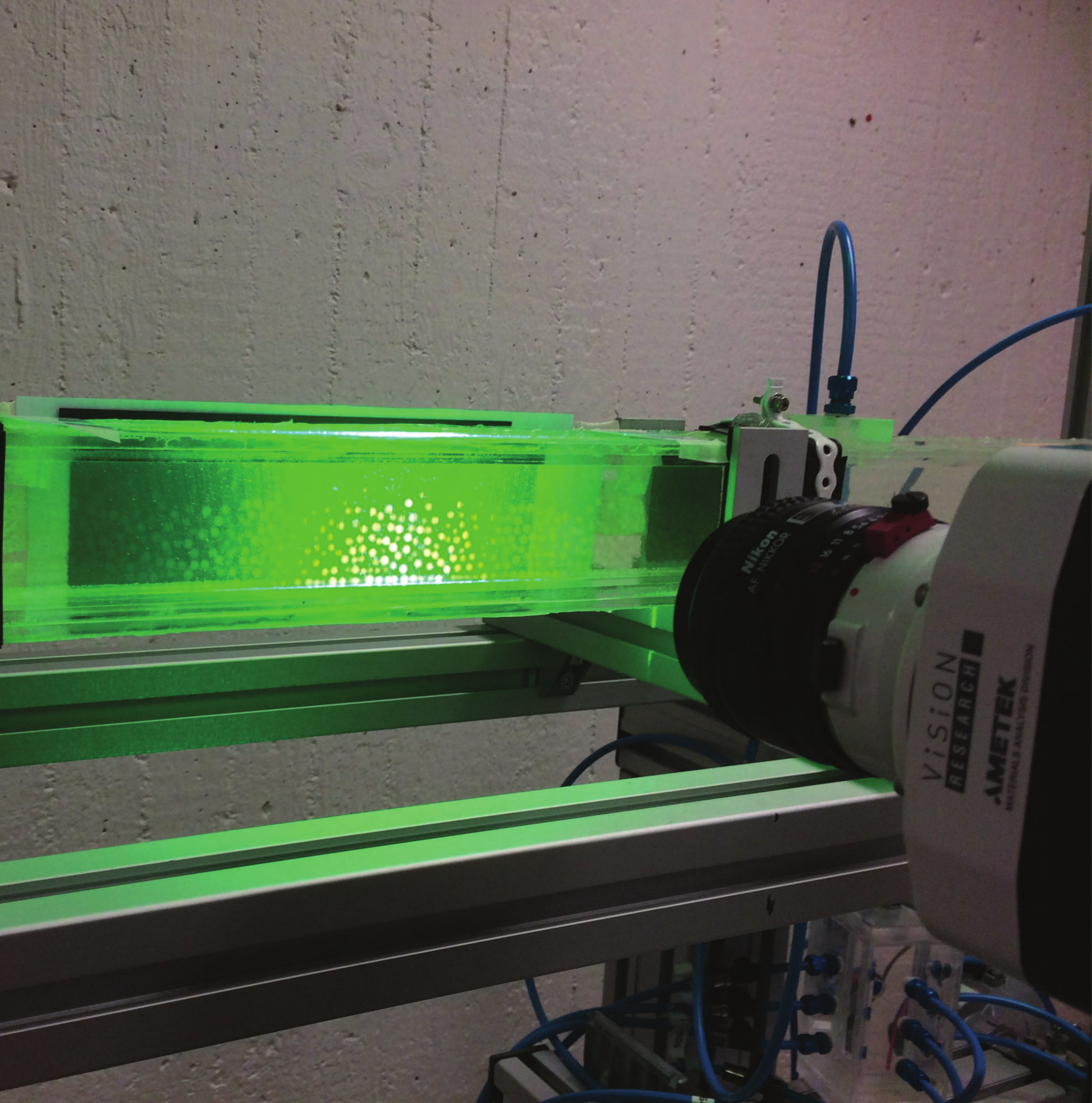}
  \caption{}
  \label{fig:PIV set-up}
\end{subfigure}
\caption{(a) Schematic of the flow-loop \sz{(b) Photo of the section where PIV is performed.}}
\label{fig:Set-up}
\end{figure}

The experiments were performed in a square duct that is 50 mm x 50 mm in cross section and 5 m in length. The entire duct is made of transparent Plexiglas permitting visualization throughout its length. Figure \ref{fig:Set-up schematic} shows a schematic of the flow loop. The fluid is recirculated through a conical tank, open to the atmosphere, where the particle-fluid mixture can be introduced. A vortex breaker has been installed at the conical end of the tank to prevent entrainment of air due to suction at higher flow rates. A static mixer is mounted close to the inlet of the duct with the intention of neutralizing any swirling motions that may arise from the gradual 90$^\circ$ bend at the exit of the tank. It is followed by a section providing a smooth transition from a circular to a square cross-section. Finally, to trigger turbulence at the inlet, a friction tape is lined on the inner walls of the Plexiglas duct at its entrance. The temperature of the solution is maintained at nearly 20$^\circ$C by means of an external heat-exchanger in the tank. Flexible corrugated piping is used to minimize the induction of vibrations from the pump to the duct. As explained in section \ref{sec:Particles}, the particles are soft and can break under high stresses. In order to minimize mechanical breakage of the particles, a very gentle disc pump (Discflo Corporations, CA, USA) has been chosen. It can pump rather large particles at reasonably high volume concentration without pulsations. Such a pump has been previously used to study the laminar-turbulent transition in flow of polymer solution where the degradation of the polymer-chains should be minimized to achieve high drag-reduction for longer time \citep{draad1998laminar}. 

It was observed that the particles marginally shrink in size, while retaining their spherical shape, over many flow cycles. This may due to the mechanical degradation of the particles reducing the water retention capability of the polymer matrix. To minimize the influence of such long term deviations in size and density, the time of experiments was kept short enough to have reasonably constant properties of the particles during the course of the data acquisition but, long enough to get statistically stationary results. The maximum measurement time for any given case did not exceed 20 minutes.

\subsection{Volume flow and pressure measurements}\label{sec:Flow and pressure measurement technique}
An electromagnetic flowmeter (Krohne Optiflux 1000 with IFC 300 signal converter, Krohne Messtechnik GmbH, Germany) is used to measure the flow rate. 
Since the conductivity properties of the hydrogel particles are similar to those of  water, 
the flow rate measured by the electromagnetic flow meter is equal to the flow rate of the mixture. To reduce the risk of faulty readings caused by electrical disturbances in the flow meter, another flow meter with a ceramic electrode was also tested. The results  compared at different particle concentrations and flow rates indicated a deviation less than 1\%. The Reynolds number $Re_{2H}$, used hereafter, is based on the flow rate of the mixture of solid and liquid phase, the viscosity of the liquid and full height of the duct $2H$.

\begin{figure}
\centering
  \includegraphics[width=0.7\linewidth]{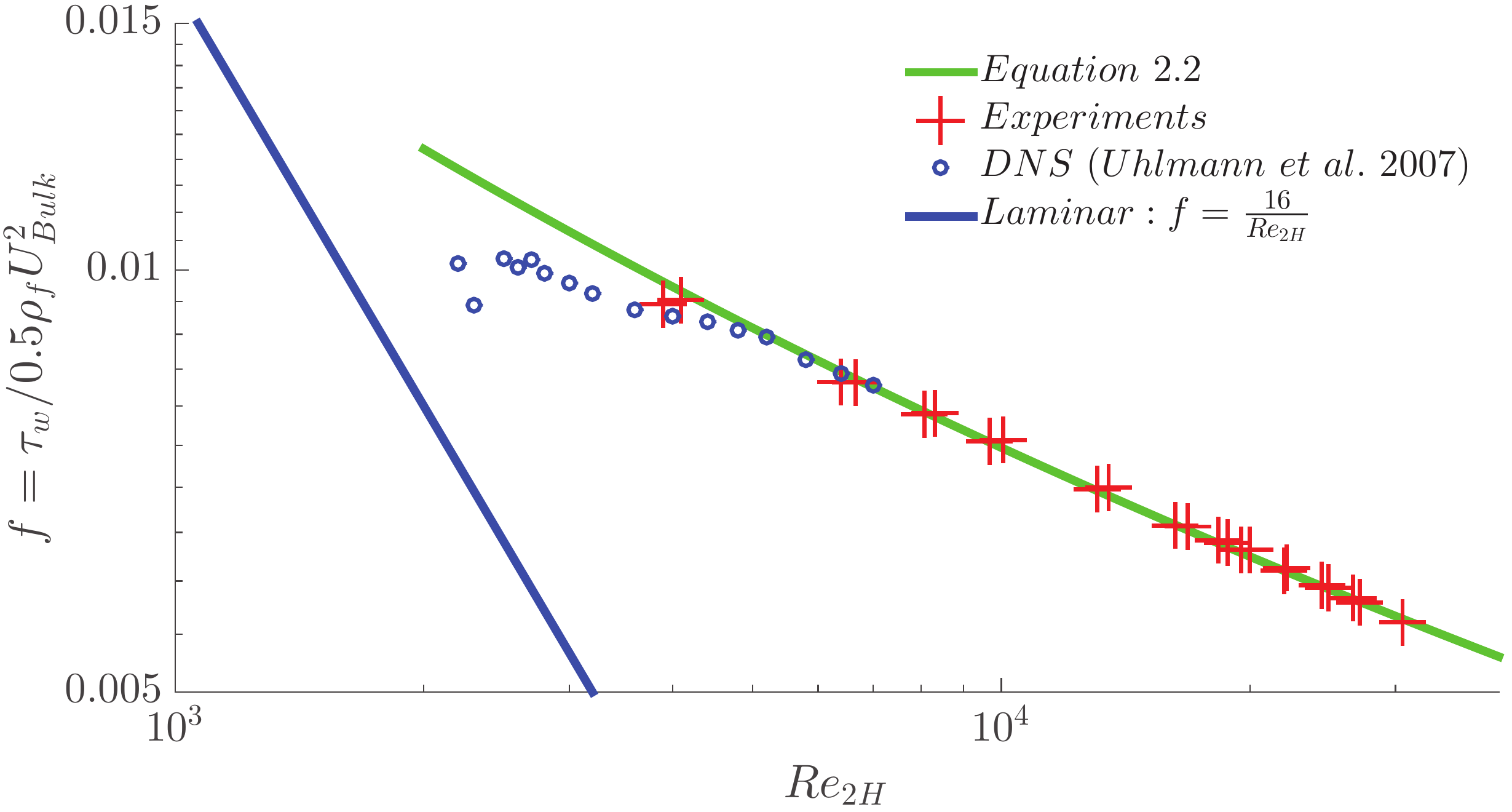}
\caption{Comparison between experiments, DNS simulations and empirical correlation for the Fanning friction factor $f$ as a function of bulk Reynolds number $Re_{2H}$ for single-phase Newtonian fluid. }
\label{fig:Friction factor plot}
\end{figure}

The pressure drop is measured across a length of $54H$ in a region of the duct that is nearly $140H$ from the inlet (the turbulent flow was seen to be fully developed at this entry length) using a differential pressure transducer (0 - 1 kPa, Model: FKC11, Fuji Electric France, S.A.S.). At a given streamwise location, two pressure tappings (one at the top wall and other at the bottom wall) were precisely drilled in the wall normal direction at the center plane of the duct. The pressure tubes emerging from these two upstream holes were joined together into one single tube, then connected to the high-pressure side of the pressure transducer. A similar connection was made at the downstream position of the duct to the low-pressure side of the pressure transducer. Additional pressure tappings were also made to study the evolution of pressure drop over the streamwise length. Figure \ref{fig:Friction factor plot} shows a reasonable agreement between the friction-factor $f={\tau_w}/{(\rho_fU_{Bulk}^2/2)}$ ($U_{Bulk}$ is the bulk velocity given by the ratio of the flow rate to the cross sectional area) for the single phase flow measured in our square duct and the empirical correlation given in \cite{duan2012pressure}, 
\begin{equation}
  {f}={\Big(3.6\log_{10}(\frac{6.115}{Re_{\sqrt{A}}})\Big)}^{-2}.
  \label{eqn:Friction factor}
\end{equation}
The Reynolds number $Re_{\sqrt{A}}$ is based on the characteristic length given by the square root of the cross section area $A$. The friction velocity is given by $u_{\tau} = \sqrt{\tau_w/\rho_f}$. Data acquisition from the camera, flow meter and pressure transducer is performed using a NI-6215 DAQ card using Labview\textsuperscript{TM} software.

\subsection{Velocity measurement technique}\label{sec:Velocity measurement technique}
The coordinate system used in this study is indicated in figure \ref{fig:Schematic secondary flow}. The velocity field is measured using 2D Particle Image Velocimetry (2D-PIV) in 3 cross-stream planes: $z/H$ = 0, 0.4 and 0.8, where $H$ is the half-width of the duct. These measurements are performed at a streamwise distance of $x/H\approx$ 150 from the entrance of the duct. A traversable continuous wave laser (wavelength = 532 nm, power = 2 W) and a high-speed camera (Phantom Miro 120, Vision Research, NJ, USA) are used to capture successive image pairs. Figure \ref{fig:PIV set-up} shows the PIV set-up. The laser light travels from the bottom to the top of the duct.

For imaging the full height of the duct, a resolution of approximately 60 mm/1024 pixels was chosen. The frame rate (acquisition frequency) was chosen so that the maximum pixel displacement did not exceed a quarter of the size of the final interrogation window IW  \citep{raffel2013particle}. Images were processed using an in-house, three-step, FFT-based, cross-correlation algorithm  \cite[used in][]{kawata2014velocity}. The first step consisted of basic PIV with a large IW size (48 x 48 pixel), followed by the discrete-window-shift PIV at the same IW size, and finally, the central-difference-image-correction method \citep{wereley2003correlation} with the final IW size (32 x 32 pixel). The degree of overlap can be estimated from the fact that the final resolution is 1 mm x 1 mm. Usually 1000 image pairs have been observed to be sufficient to ensure statistically converged results.

\begin{figure}
\centering

\begin{subfigure}{.3\textwidth}
  \centering
  \includegraphics[height=1\linewidth]{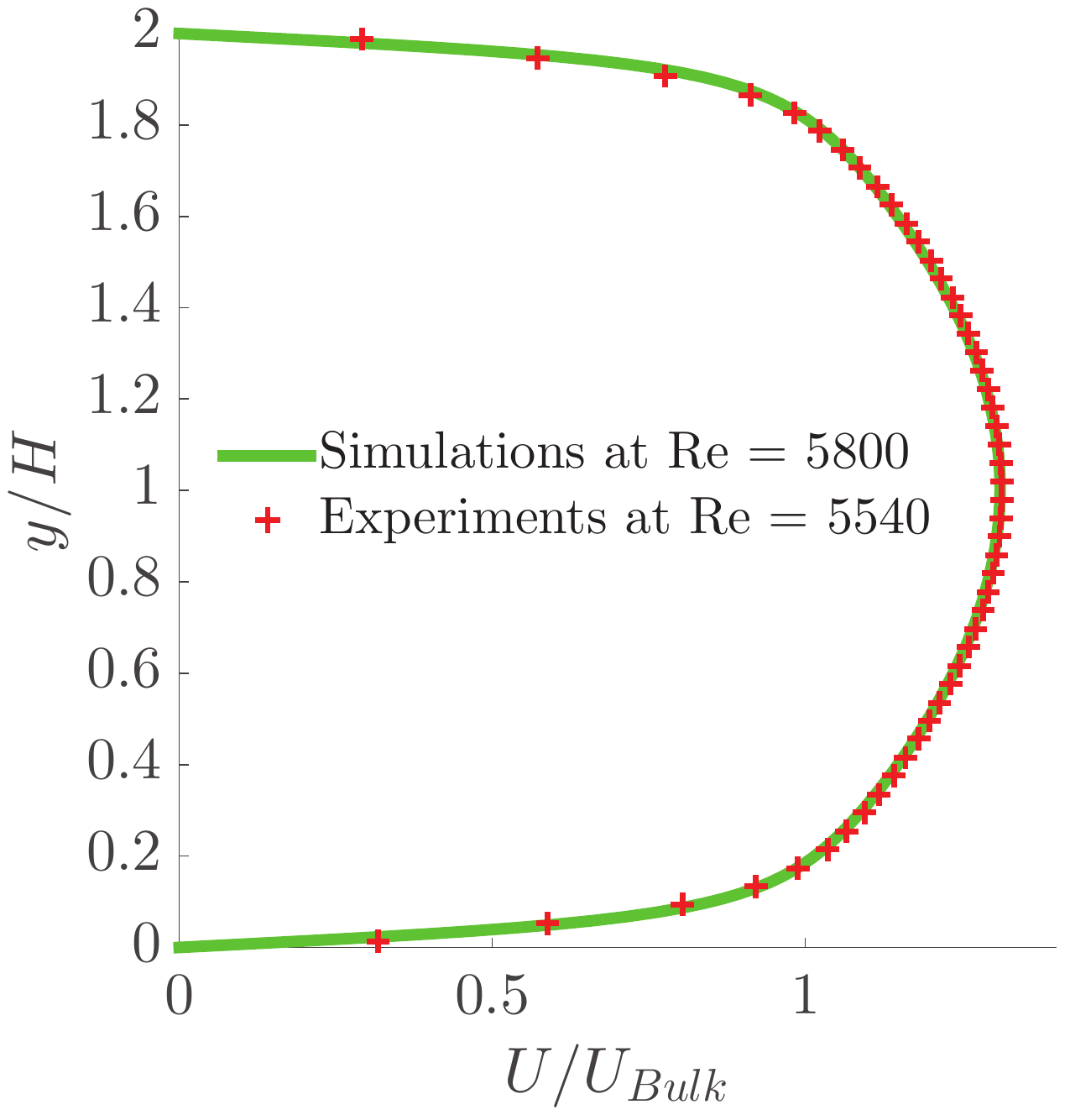}
  \caption{}
  \label{fig:Umean water 0_0H}
\end{subfigure}%
\begin{subfigure}{.3\textwidth}
  \centering
  \includegraphics[height=1\linewidth]{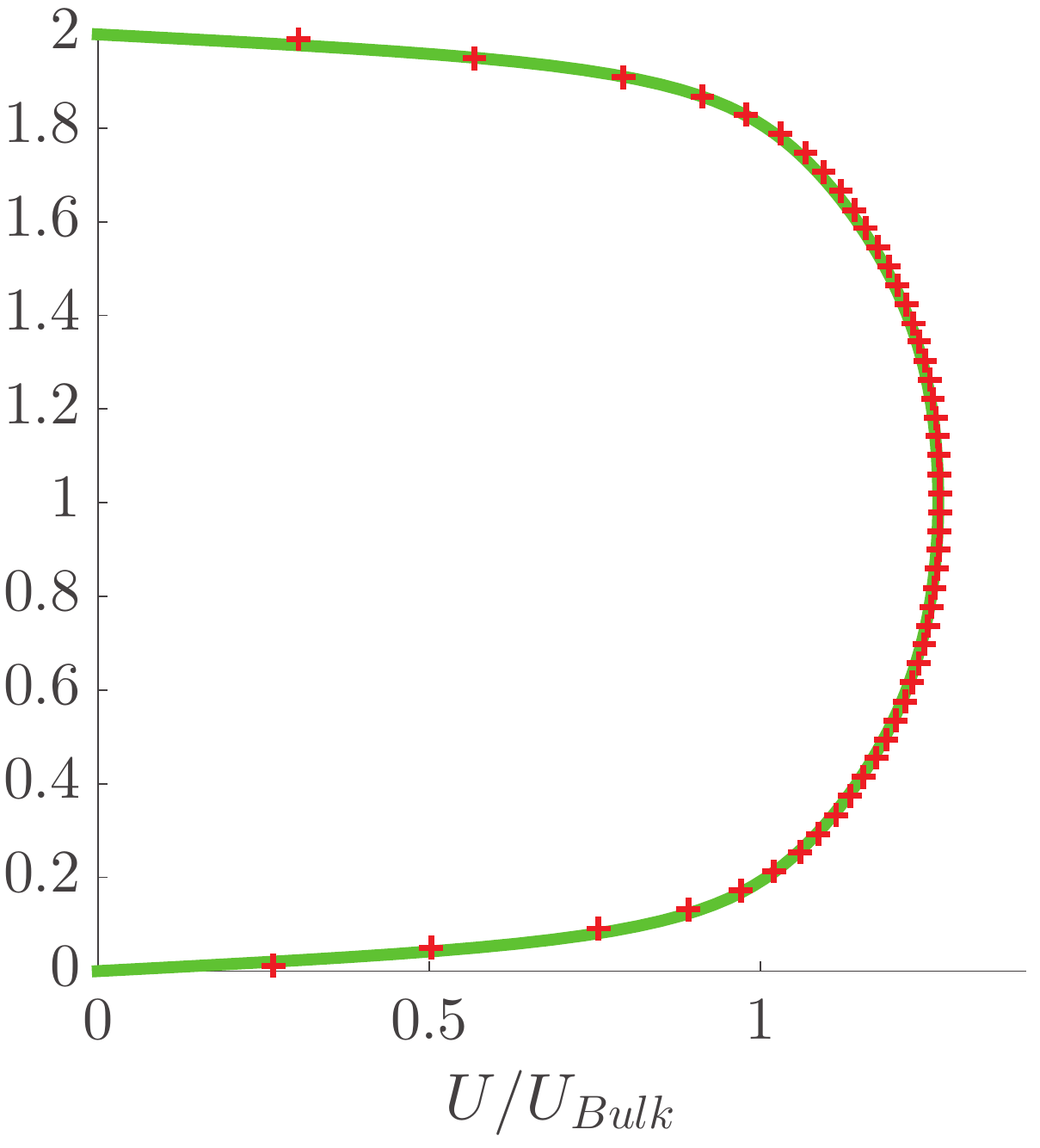}
  \caption{}
  \label{fig:Umean water 0_3H}
\end{subfigure}
\begin{subfigure}{.3\textwidth}
  \centering
  \includegraphics[height=1\linewidth]{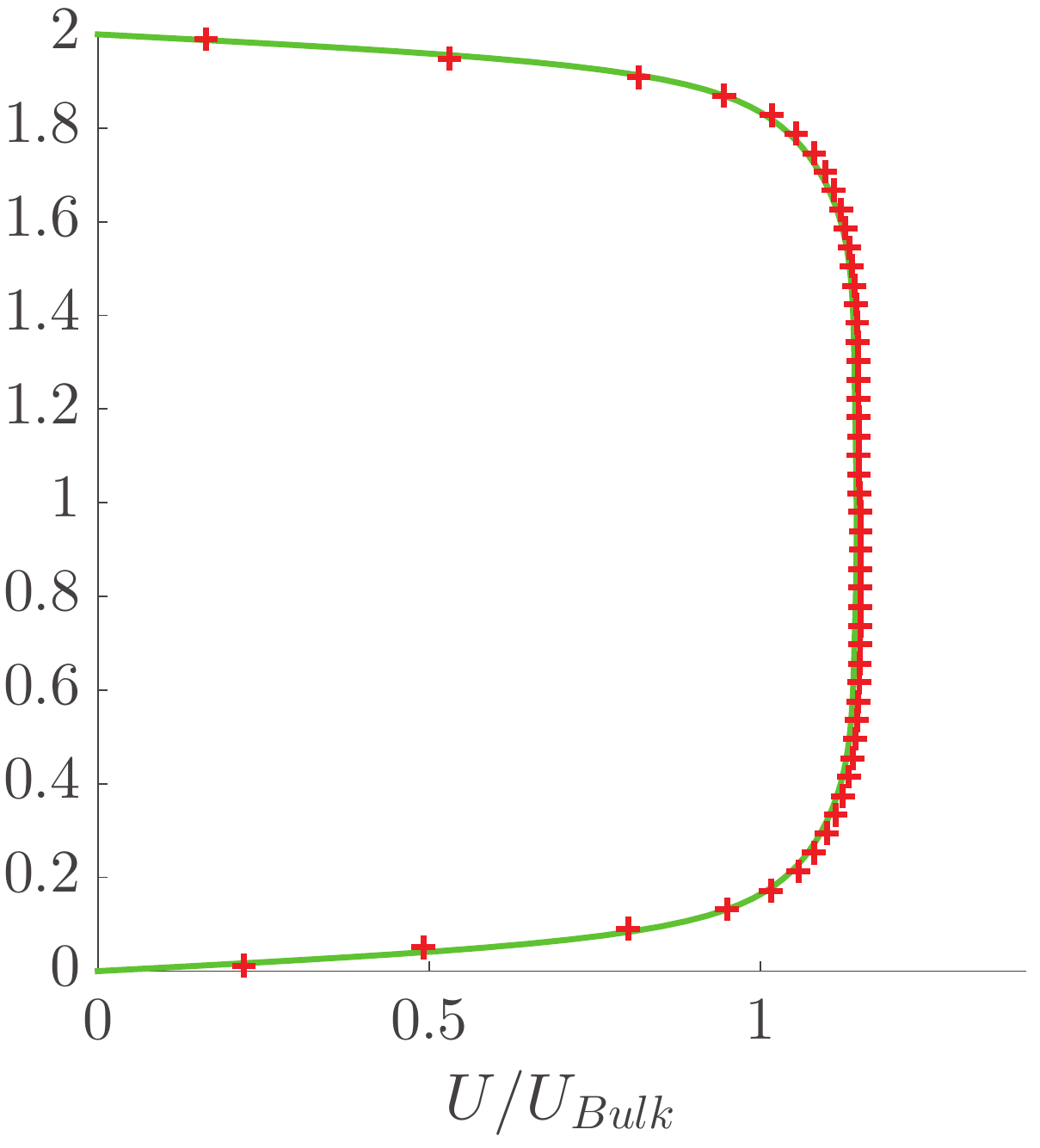}
  \caption{}
  \label{fig:Umean water 0_6H}
\end{subfigure}

\begin{subfigure}{.3\textwidth}
  \centering
  \includegraphics[height=1\linewidth]{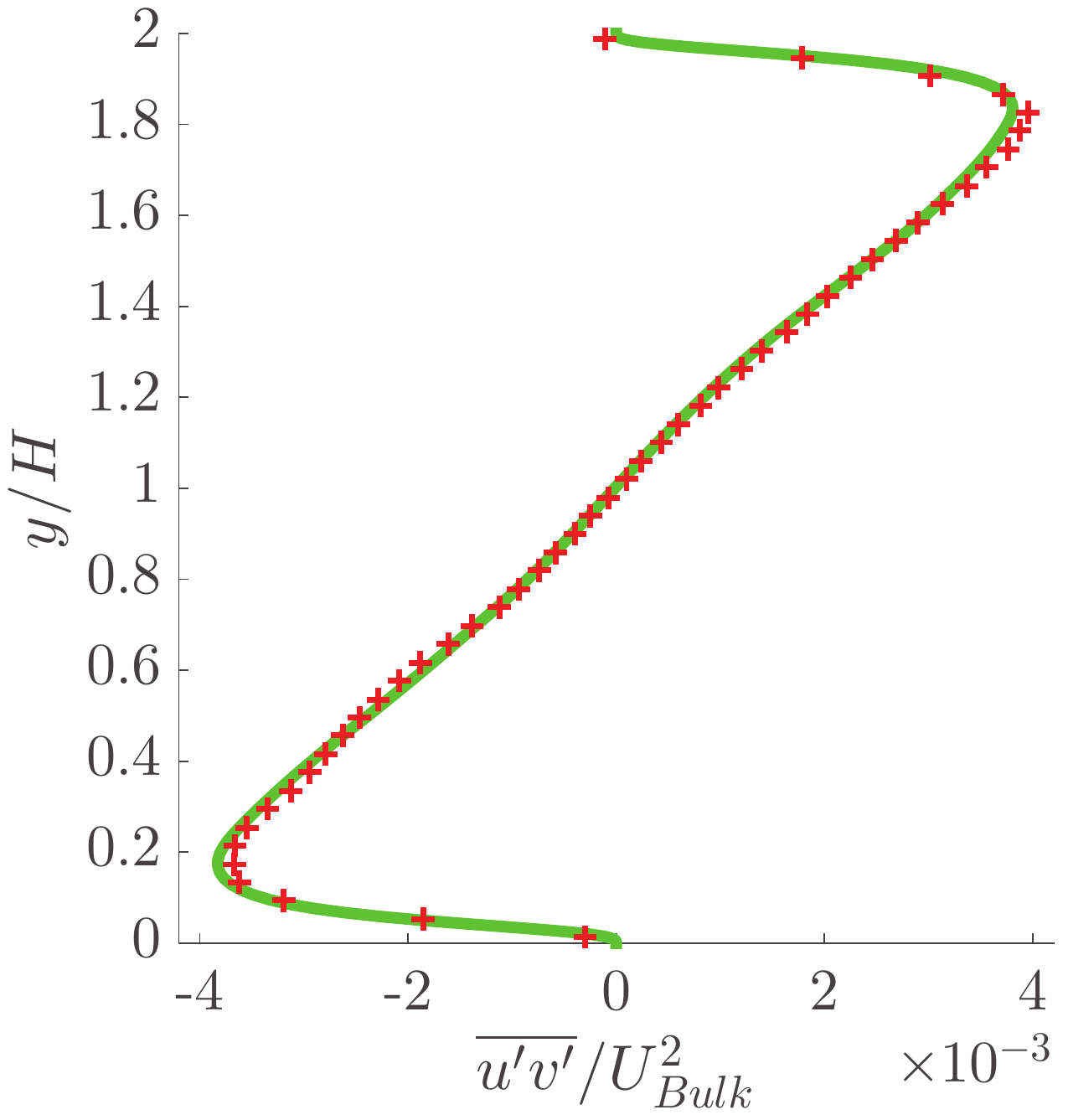}
  \caption{}
  \label{fig:uv water 0_0H}
\end{subfigure}%
\begin{subfigure}{.3\textwidth}
  \centering
  \includegraphics[height=1\linewidth]{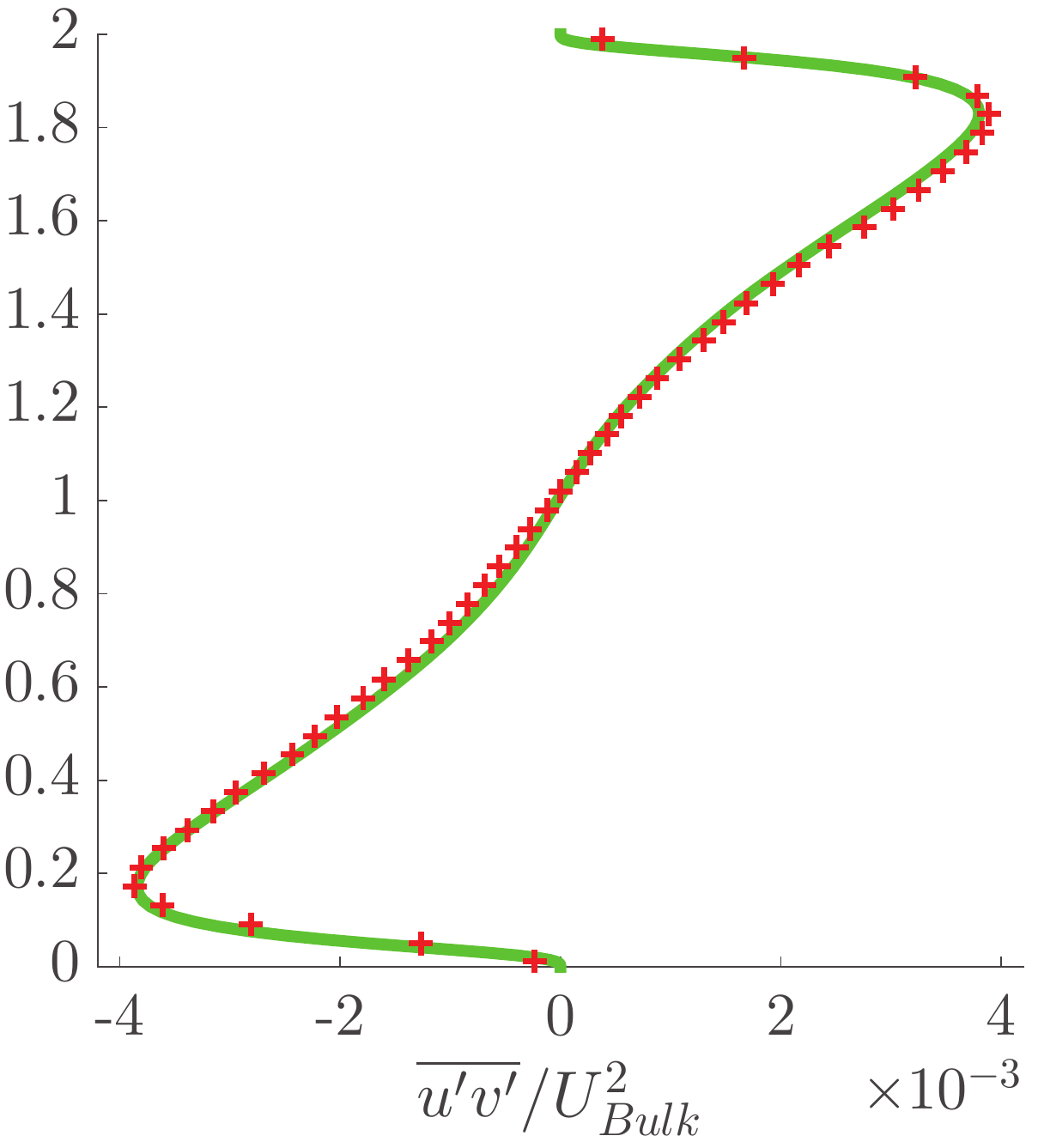}
  \caption{}
  \label{fig:uv water 0_3H}
\end{subfigure}
\begin{subfigure}{.3\textwidth}
  \centering
  \includegraphics[height=1\linewidth]{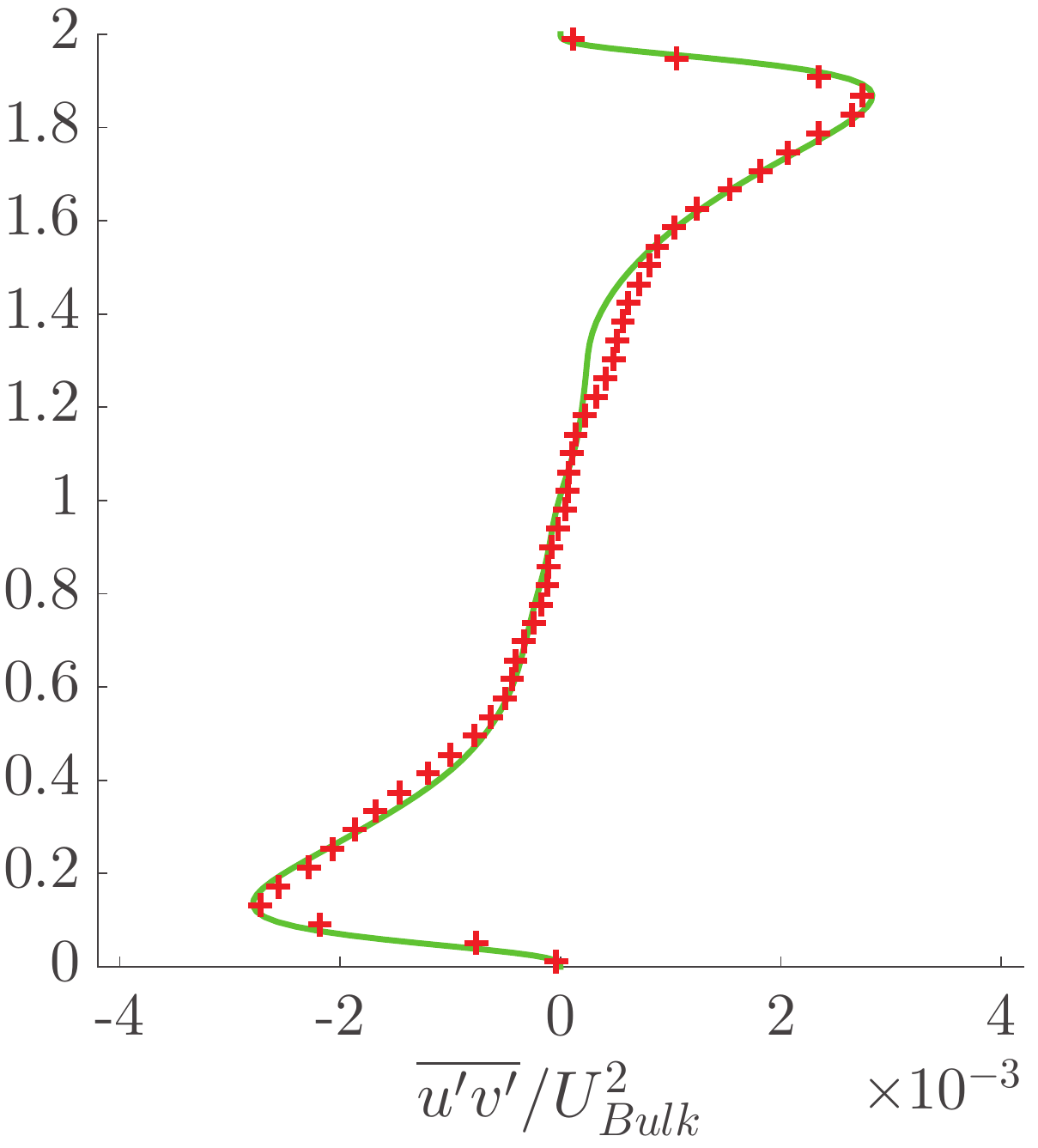}
  \caption{}
  \label{fig:uv water 0_6H}
\end{subfigure}

\begin{subfigure}{.32\textwidth}
  \centering
  \includegraphics[height=1\linewidth]{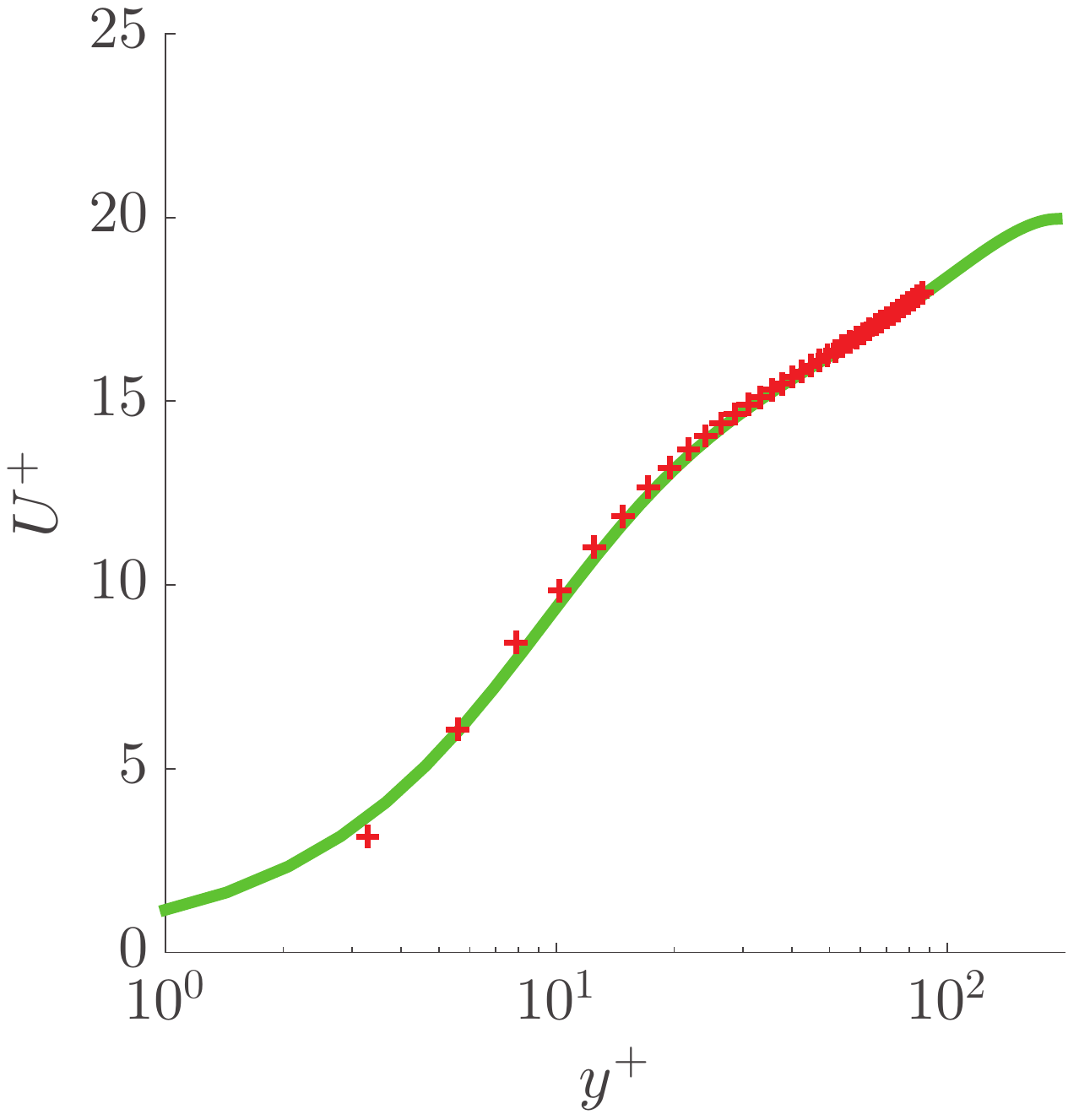}
  \caption{}
  \label{fig:U+ water 0_0H}
\end{subfigure}%
\begin{subfigure}{.32\textwidth}
  \centering
  \includegraphics[height=1\linewidth]{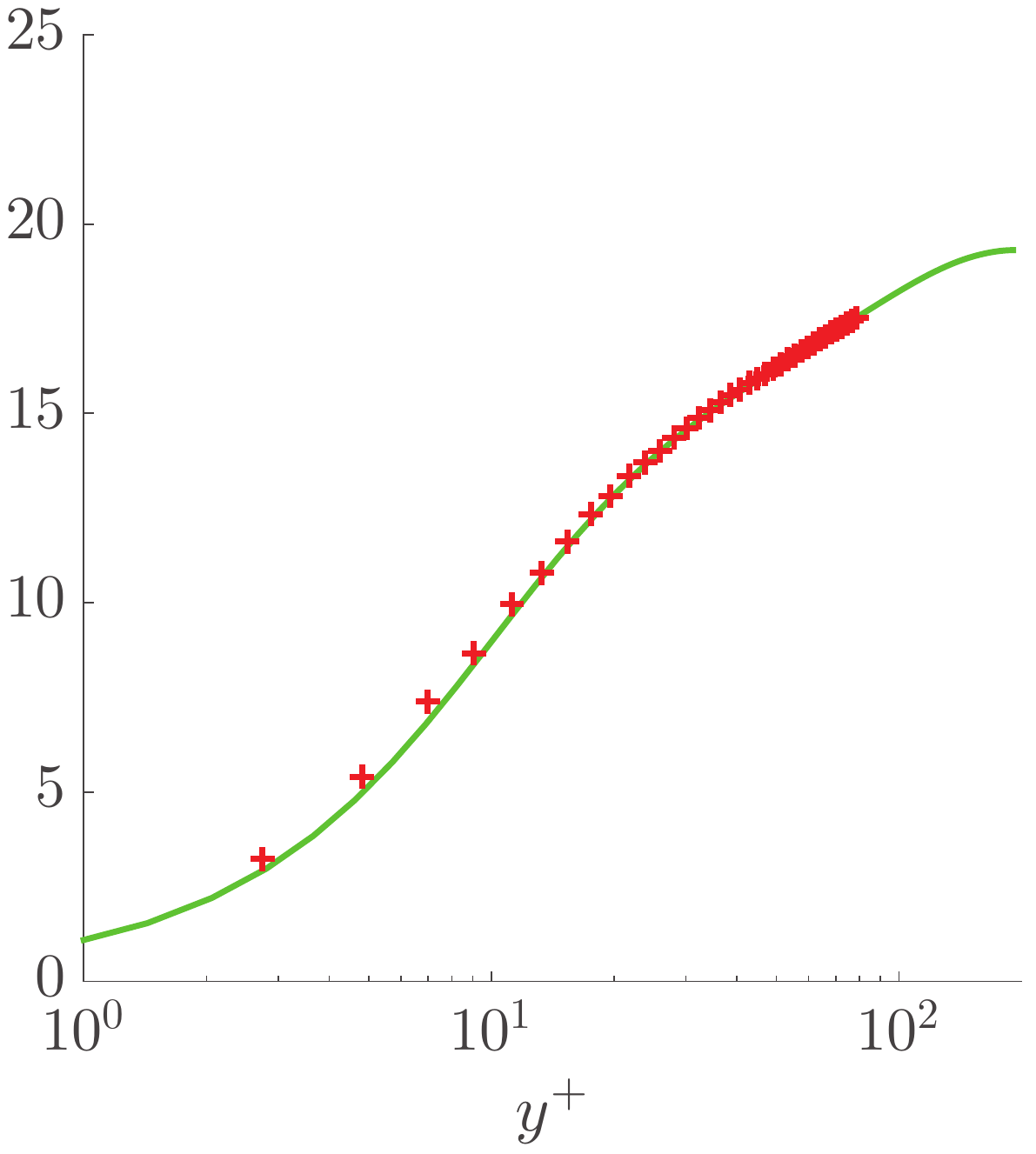}
  \caption{}
  \label{fig:U+ water 0_3H}
\end{subfigure}%
\begin{subfigure}{.32\textwidth}
  \centering
  \includegraphics[height=1\linewidth]{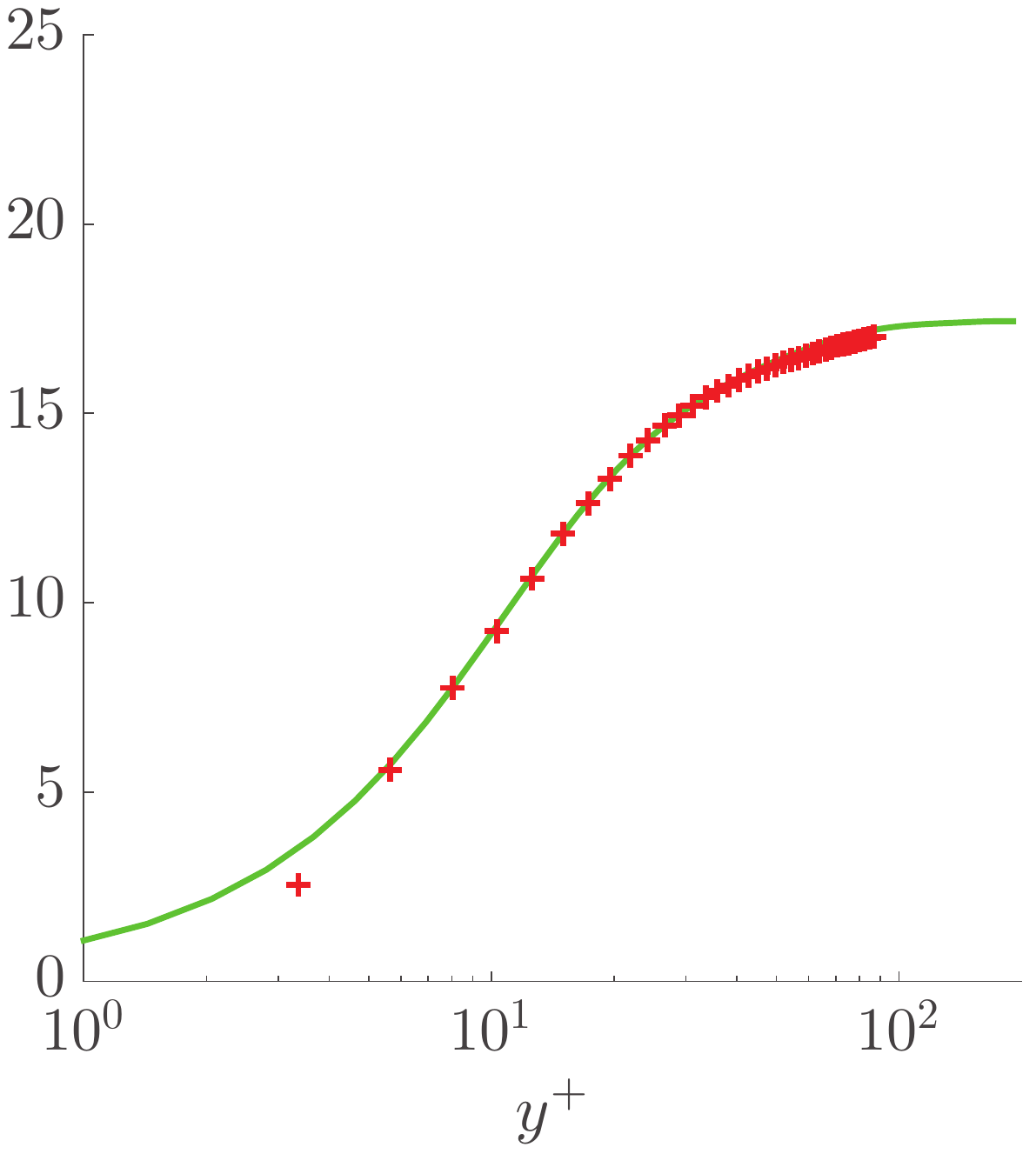}
  \caption{}
  \label{fig:U+ water 0_6H}
\end{subfigure}%

\caption{Comparison of experimental mean streamwise velocity profiles with DNS  of single-phase flow at 3 spanwise planes: (a) z/H=0, (b) z/H=0.3 and (c) z/H=0.6. Panels (d)-(f) show the corresponding Reynolds stress and (g)-(i) the mean streamwise velocity in wall units. The Reynolds number $Re_{2H}$ for DNS and experiments are 5800 and 5540 respectively. }
\label{fig:Single phase comparison}
\end{figure}

Figures \ref{fig:Umean water 0_0H}--\ref{fig:Umean water 0_6H} show the streamwise velocity of single-phase Newtonian flow measured at three different sections: $z/H$ = 0, 0.3 and 0.6, compared with previously published DNS simulations  at $Re_{2H}$=5800 \citep{uhlmann2007marginally}. The experimental Reynolds number is slightly lower at $Re_{2H}\approx$ 5540. The corresponding Reynolds stresses are reported in figures \ref{fig:uv water 0_0H}--\ref{fig:uv water 0_6H} to confirm that the mean velocity and turbulence statistics are in good agreement with the simulations. 
The resolution is around  1mm x 1mm in figures \ref{fig:Umean water 0_0H}--\ref{fig:uv water 0_6H} where the measurements cover the full height of the duct. This resolution is suitable for velocity statistics in the bulk of the flow, away from the walls, but it is inadequate in the near-wall region, characterised by a large variation in velocity over a very small distance. 
Therefore, an separate measurement is conducted by zooming the camera on a small region close to the wall with a resolution of $\approx$ 0.35 mm $\times$ 0.35 mm. Figures \ref{fig:U+ water 0_0H}--\ref{fig:U+ water 0_6H} display the streamwise velocity profile (in wall units) close to the wall. These measurements close to the wall show an even better agreement with the simulations for the location and magnitude of the peak in the variance of the velocity fluctuations (not shown).

The experiments with particles are performed at a higher $Re_{2H}$, ranging from 10000 to 27000. At these high $Re_{2H}$, the magnitude of the secondary flow was observed to be less than 2\% of the bulk velocity in the 3 spanwise planes where the velocities were measured. The measured wall-normal velocity profiles were extremely sensitive to the PIV arrangement (parallelism of the camera, concentration of PIV tracers, etc.) and  exhibited variation (related to e.g. point of zero-crossing, top-down symmetry) between similar experiments. 
Hence, to prevent observations beyond our measurement capabilities, we do not report the secondary flow velocity. 
However, the effect of such secondary motions on the streamwise velocity distribution are of a relatively larger magnitude and hence, such effects are properly captured by the PIV. 

\begin{figure}
\centering
\begin{subfigure}{.3\textwidth}
  \centering
  \includegraphics[height=0.95\linewidth]{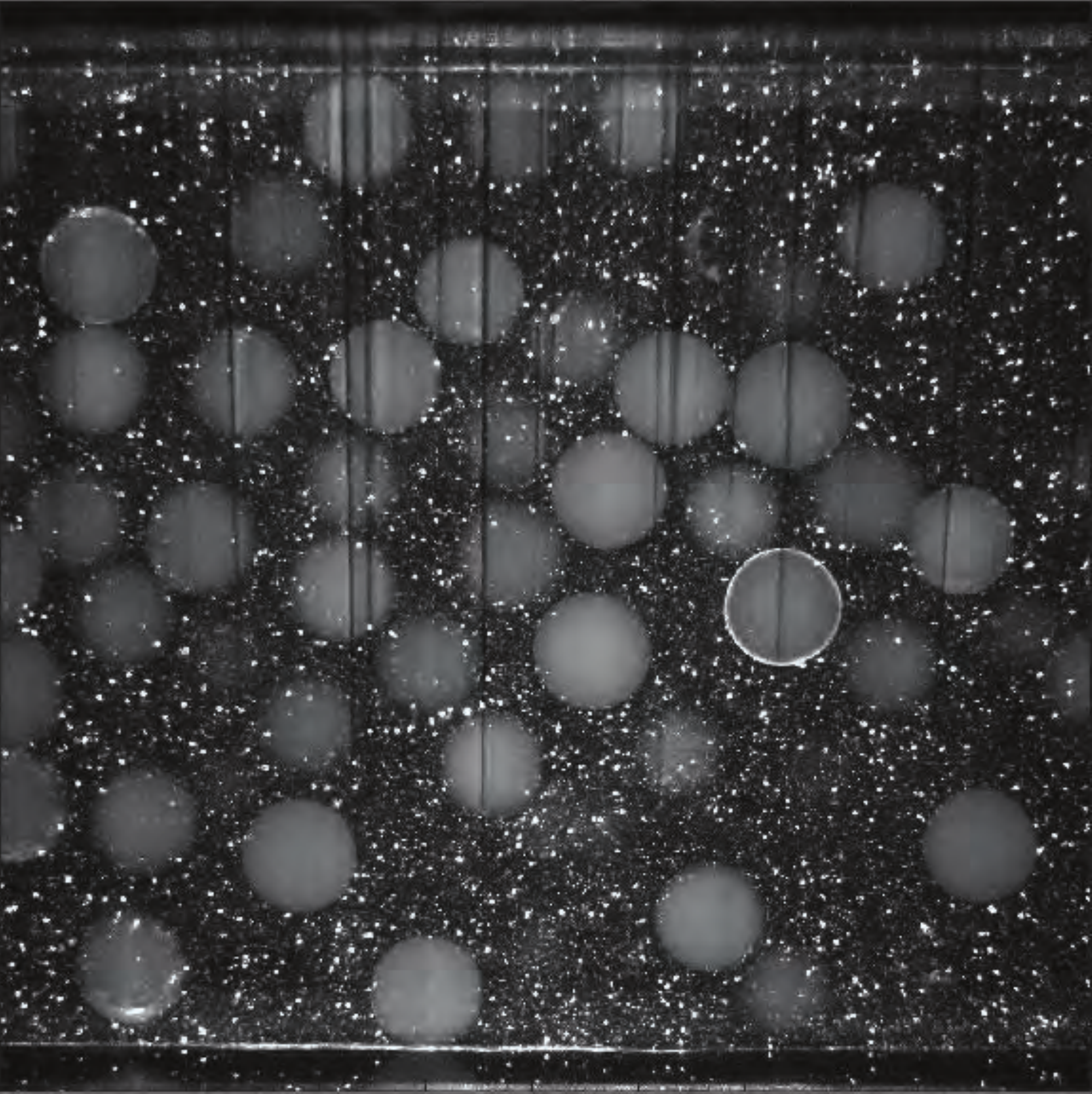}
  \caption{}
  \label{fig:PIV image}
\end{subfigure}%
\begin{subfigure}{.3\textwidth}
  \centering
  \includegraphics[height=0.95\linewidth]{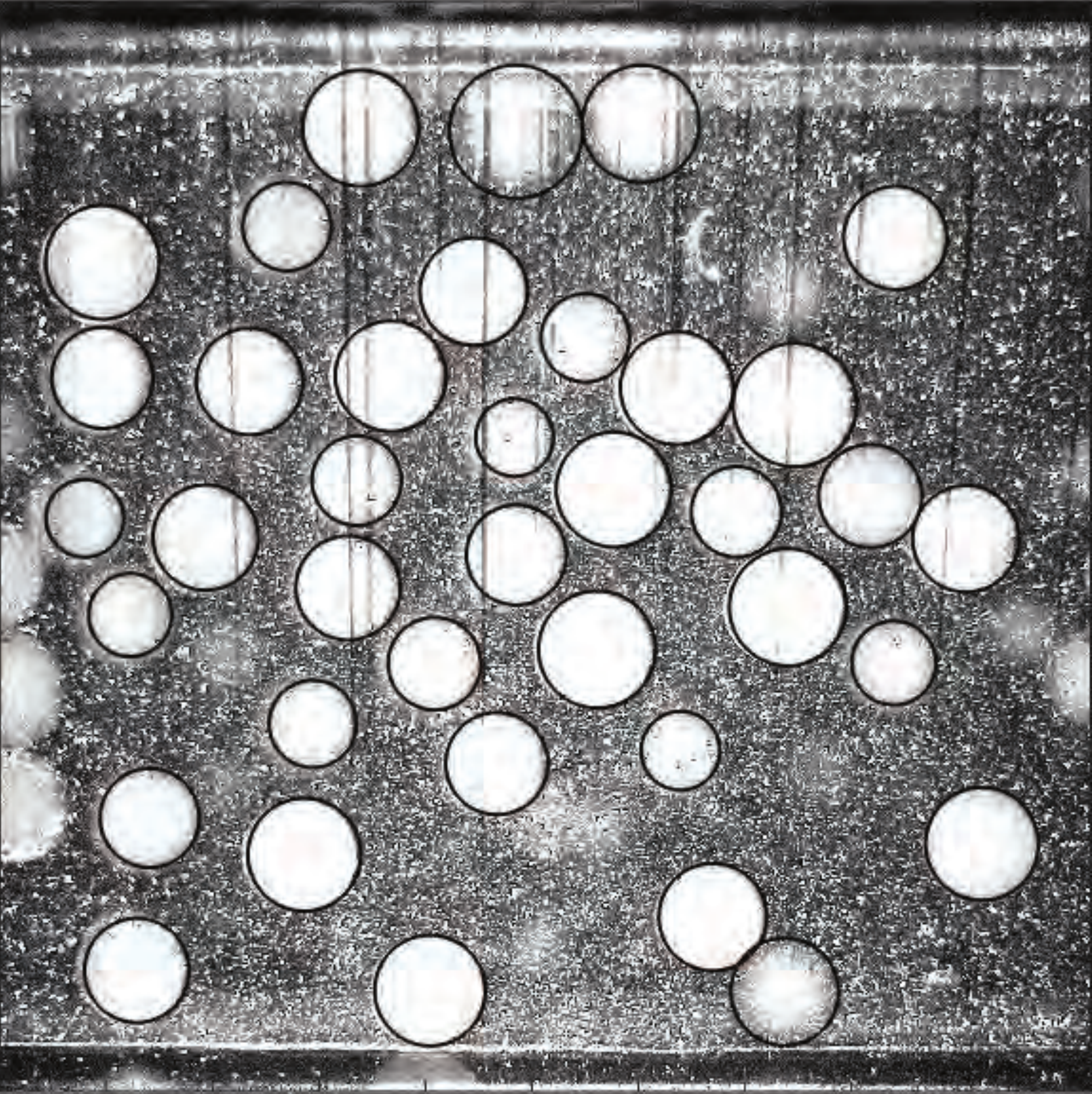}
  \caption{}
  \label{fig:PTV image}
\end{subfigure}
\begin{subfigure}{.3\textwidth}
  \centering
  \includegraphics[height=0.95\linewidth]{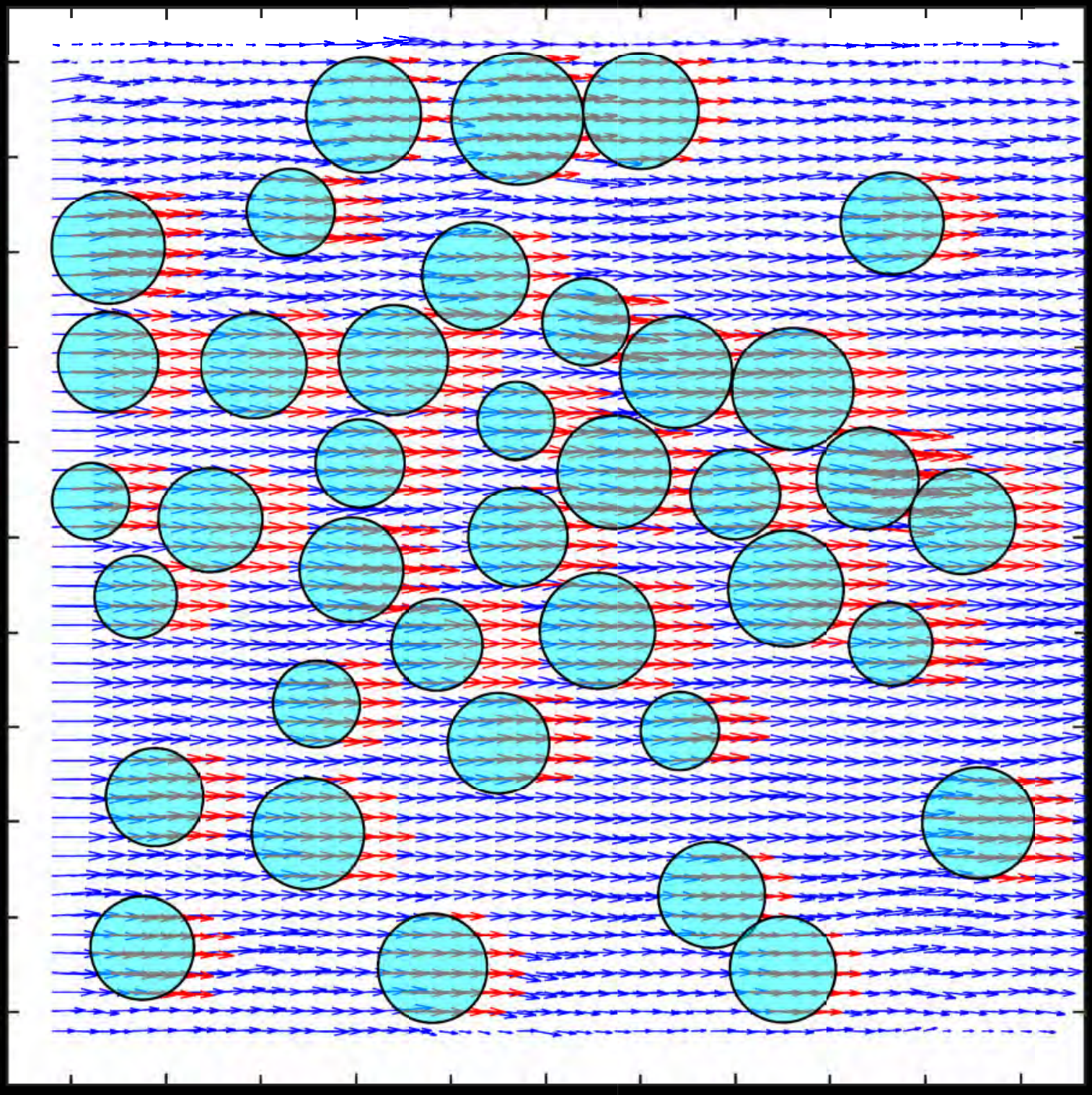}
  \caption{}
  \label{fig:PIV+PTV}
\end{subfigure}
\caption{(a) Image for PIV analysis, (b) image for particle detection and PTV analysis and (c) combined fluid PIV - particle PTV velocity vectors for $\phi$ = 20\% of LP at $Re_{2H}\approx27000$.}
\label{fig:PIV+PTV image and velocity vectors}
\end{figure}

\begin{figure}
\centering
\begin{subfigure}{.36\textwidth}
  \centering
  \includegraphics[height=0.95\linewidth]{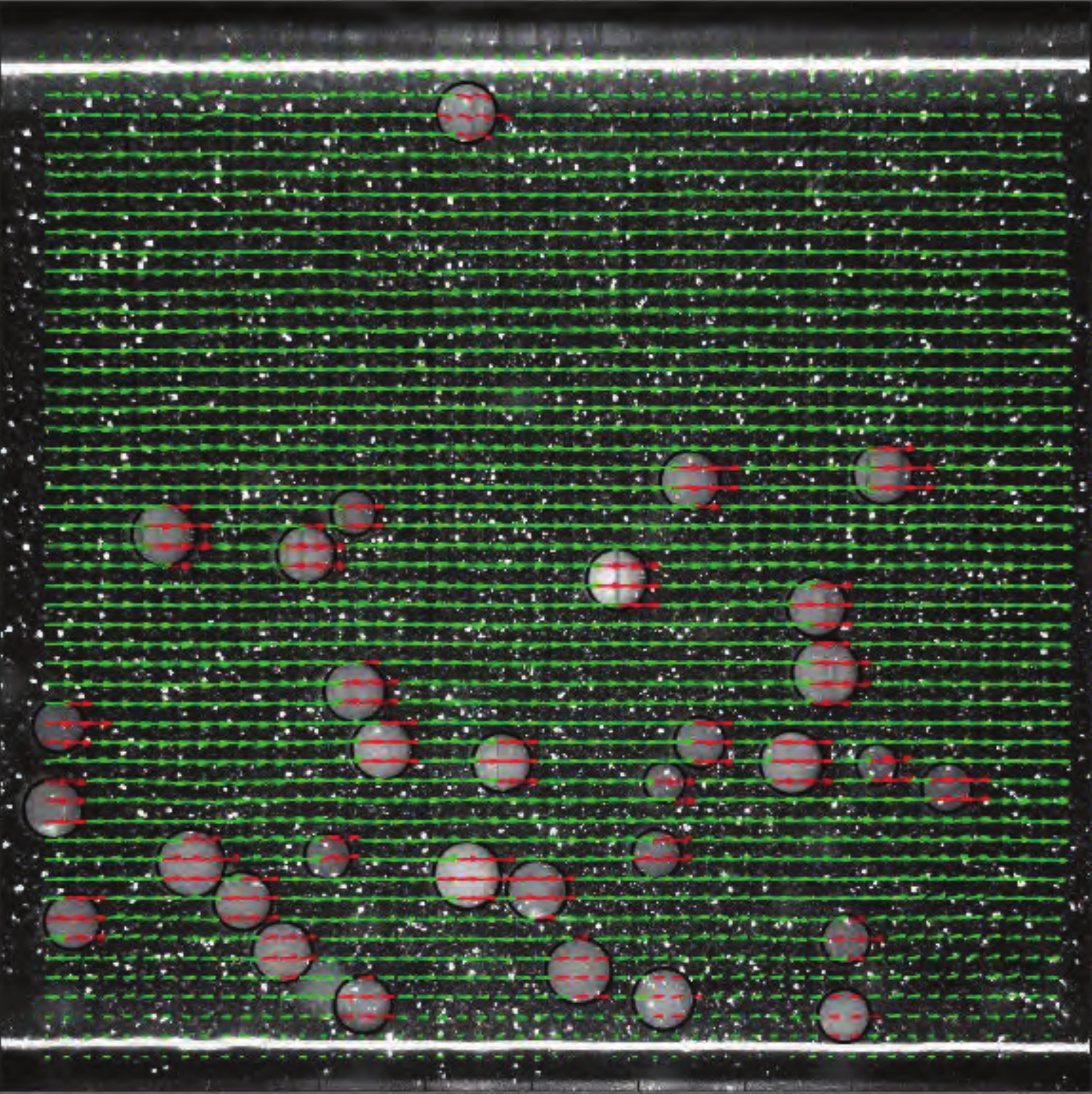}
  \caption{}
  \label{fig:3mm_5p_Re10000}
\end{subfigure}%
\begin{subfigure}{.36\textwidth}
  \centering
  \includegraphics[height=0.95\linewidth]{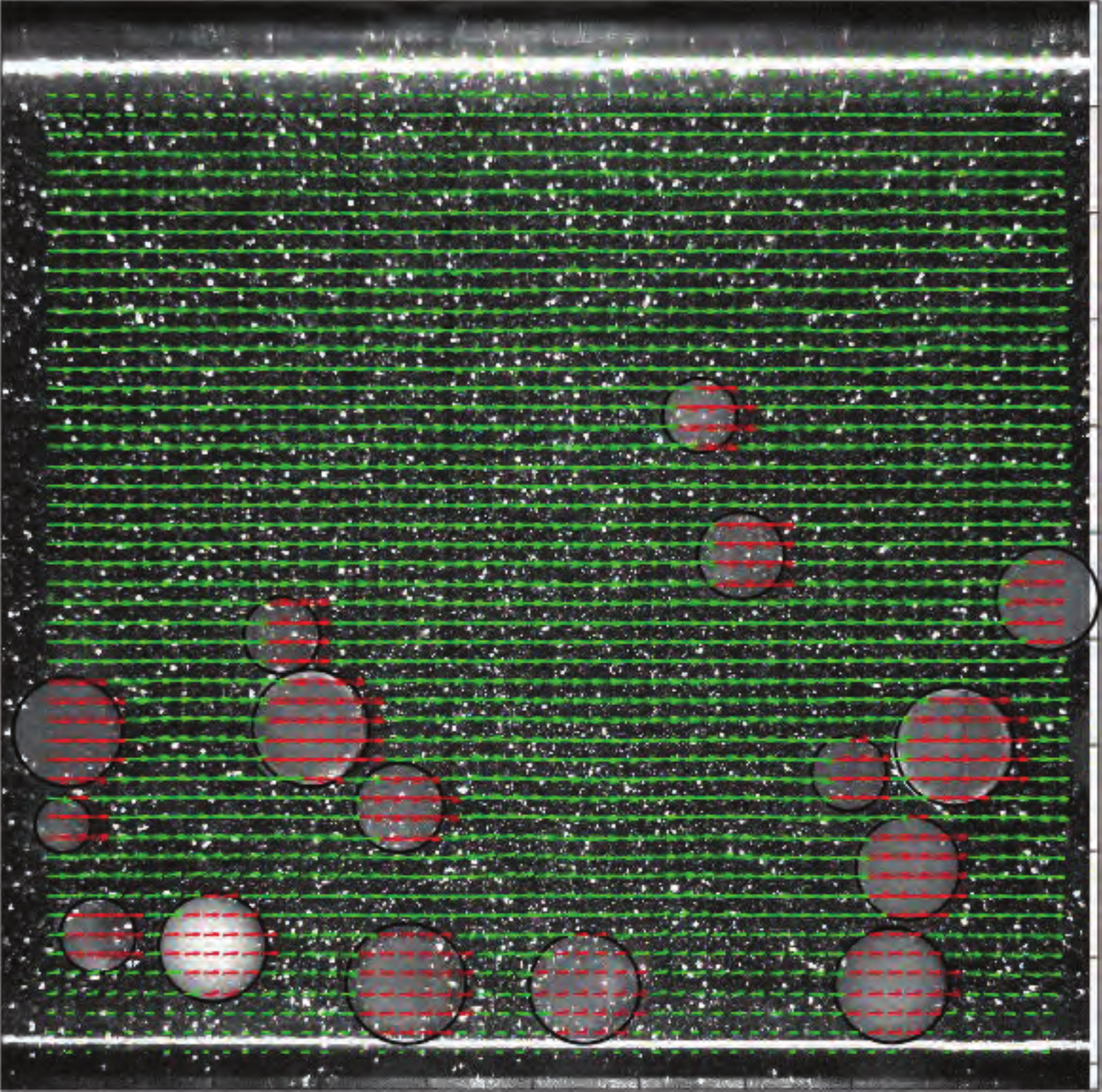}
  \caption{}
  \label{fig:6mm_5p_Re10000}
\end{subfigure}
\begin{subfigure}{.36\textwidth}
  \centering
  \includegraphics[height=0.95\linewidth]{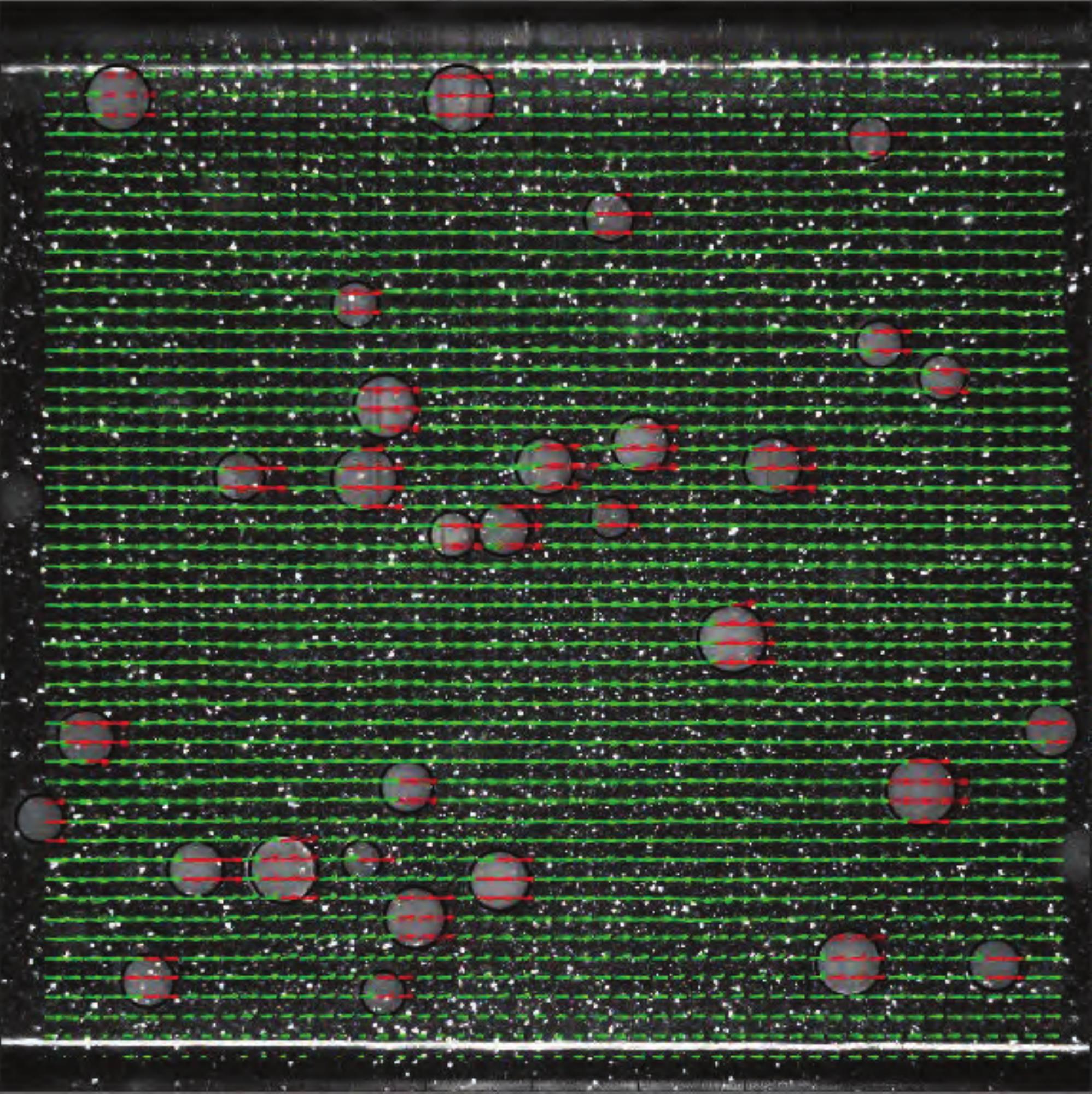}
  \caption{}
  \label{fig:3mm_5p_Re27000}
\end{subfigure}%
\begin{subfigure}{.36\textwidth}
  \centering
  \includegraphics[height=0.95\linewidth]{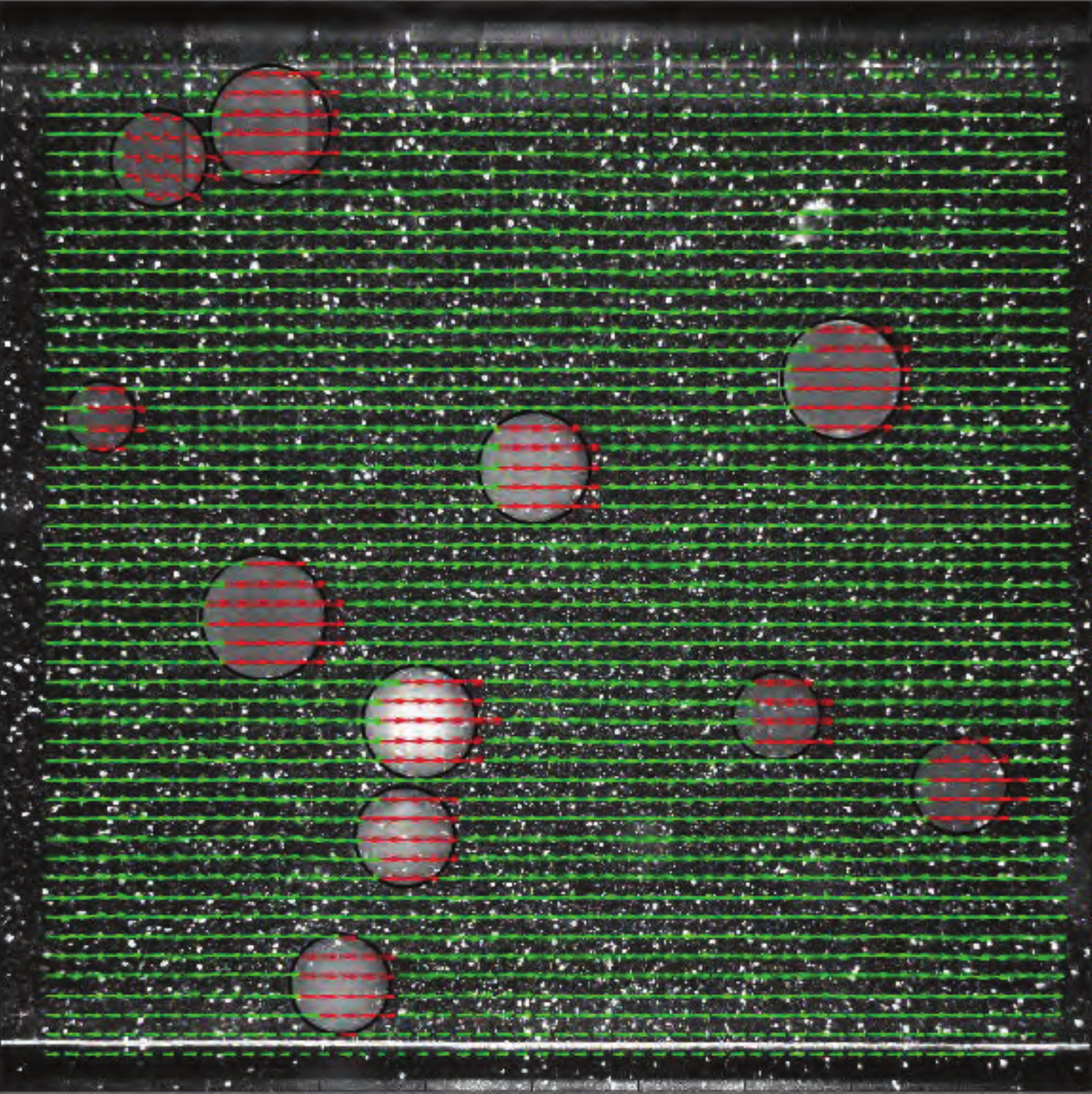}
  \caption{}
  \label{fig:6mm_5p_Re27000}
\end{subfigure}
\caption{Effect of density difference on the particle distribution: non-negligible particle sedimentation in (a) $2H/d_p$ = 16, $Re_{2H}\approx$ 10000 and (b) $2H/d_p$ = 9, $Re_{2H}\approx$10000. Particles in nearly full suspension in (c) $2H/d_p$ = 16, $Re_{2H}\approx$ 27000 and (d) $2H/d_p$ = 9, $Re_{2H}\approx$ 27000. The volume fraction in all the above figures is $\phi$ = 5\%.}
\label{fig:Velocity vectors at 2 Re}
\end{figure}

Figure \ref{fig:PIV+PTV image and velocity vectors} depicts one image from a typical PIV sequence for particle-laden flow. As mentioned earlier, the contrast between particle and fluid was enhanced by using a small quantity of Rhodamine.
Raw images captured during the experiment were saved in groups of 2 different intensity levels. The first group of images (an example being figure \ref{fig:PIV image}) was used for regular PIV processing according to the algorithm mentioned in Sec. \ref{sec:Velocity measurement technique}. The second group of images (cf. figure \ref{fig:PTV image}) was used for detecting the particles only. The images were sharpened and their intensity adjusted so that particles could be easily detected using a circular Hough transform \citep{yuen1990comparative}. From the detected particles in image A and B of the PIV sequence, a nearest neighbour approach was used to determine their translational motion. Particles that were detected only in one image of the pair were, thus, eliminated by the PTV algorithm. 
For the Eulerian PIV velocity field, we define a mask, which assumes the value 1 if the point lies inside the particle and 0, if it lies outside. 
The fluid phase velocity is thus determined on a fixed mesh.
The particle velocity is determined using PTV at its center, which is assigned to the grid points inside the particle (mask equal to 1). 
The velocity field of the particle-phase is, now, available at the same grid points as that of the fluid and the ensemble averaging, reported later, are phase averaged statistics. Figure \ref{fig:PIV+PTV} shows the combined fluid (PIV) and particle (PTV) velocity field.

\begin{figure}
\centering

\begin{subfigure}{.3\textwidth}
  \centering
  \includegraphics[height=1\linewidth]{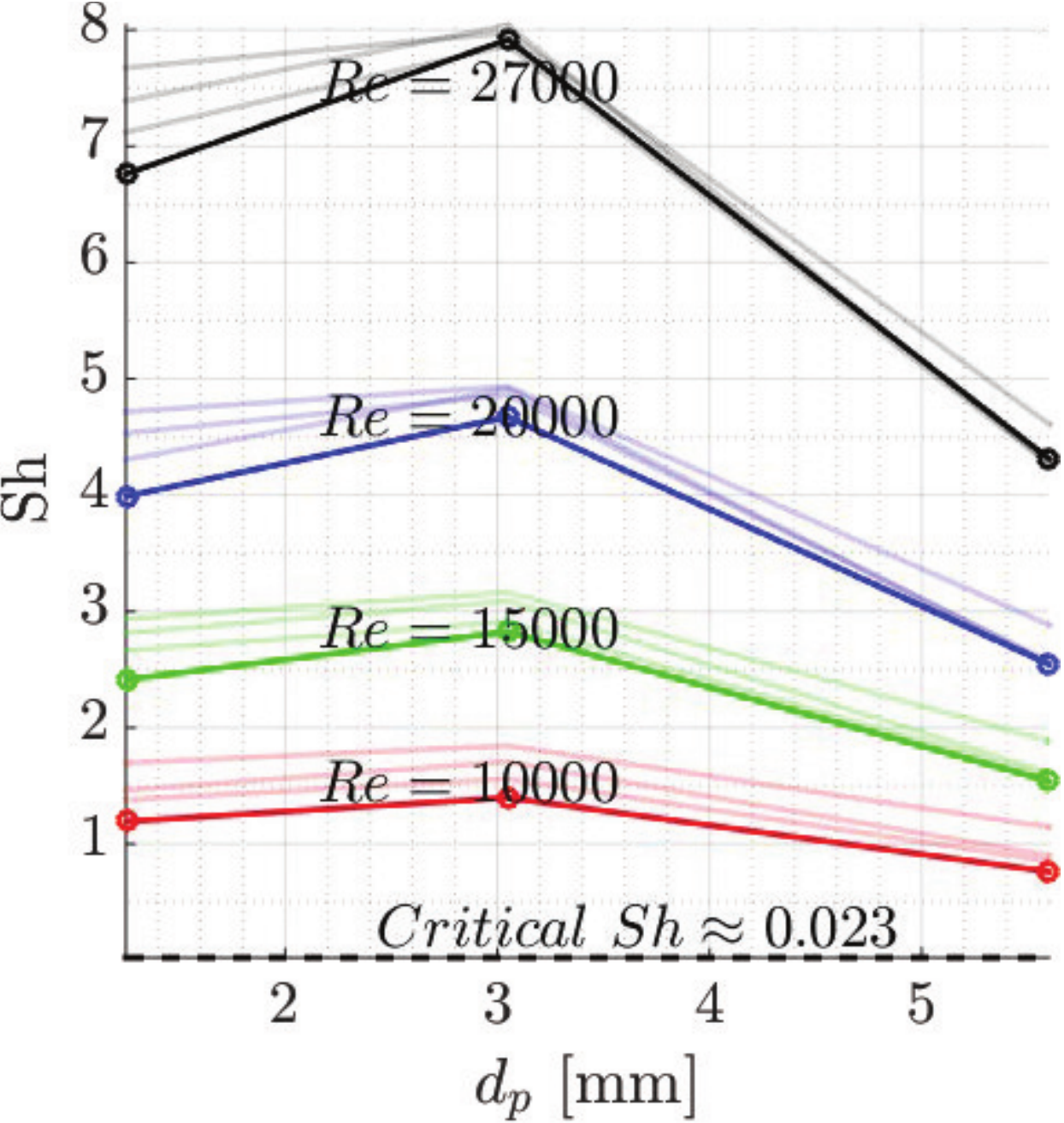}
  \caption{}
  \label{fig:Shields number}
\end{subfigure}
\begin{subfigure}{.3\textwidth}
  \centering
  \includegraphics[height=1\linewidth]{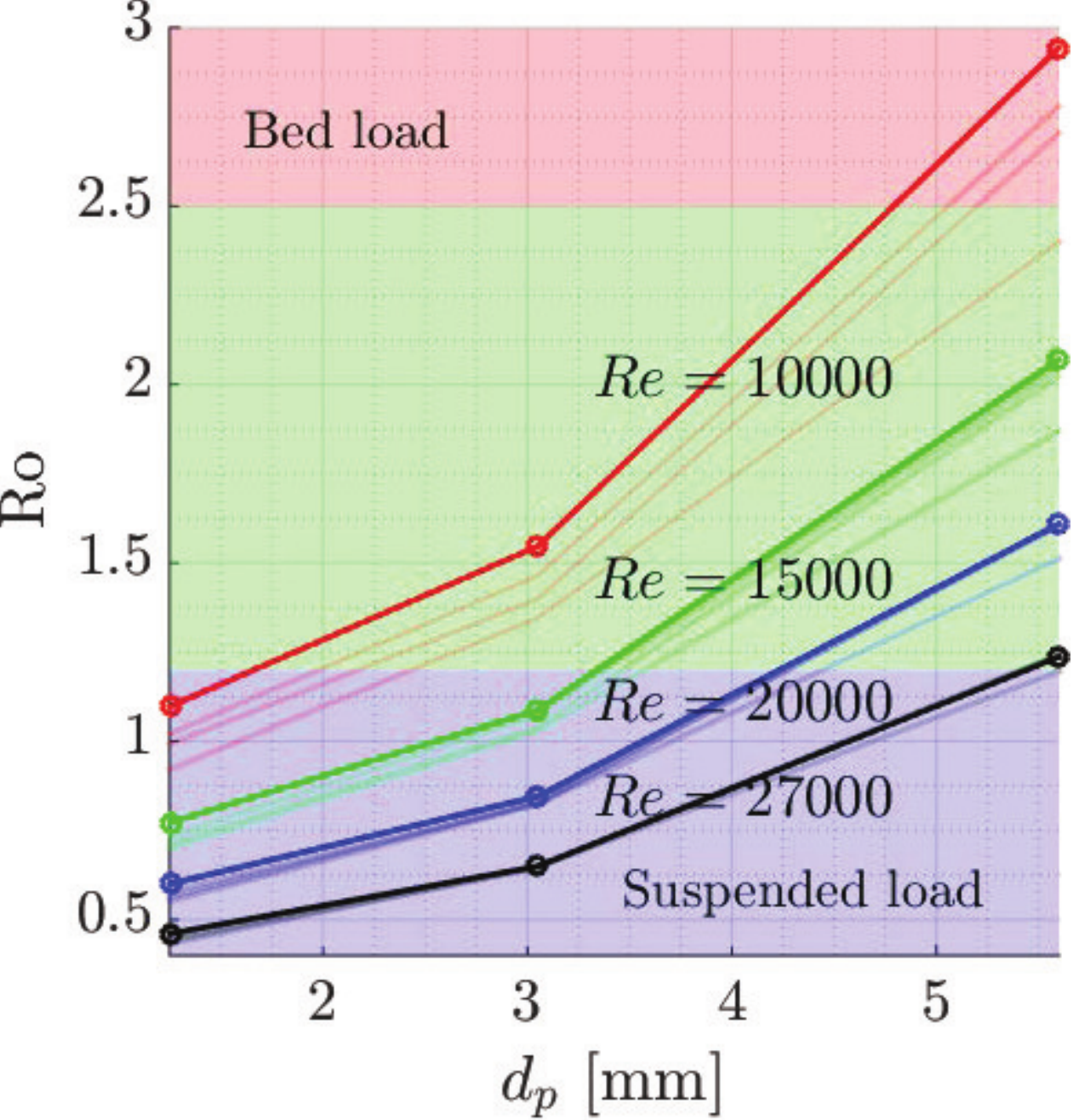}
  \caption{}
  \label{fig:Rouse number}
\end{subfigure}%
\begin{subfigure}{.3\textwidth}
  \centering
  \includegraphics[height=1\linewidth]{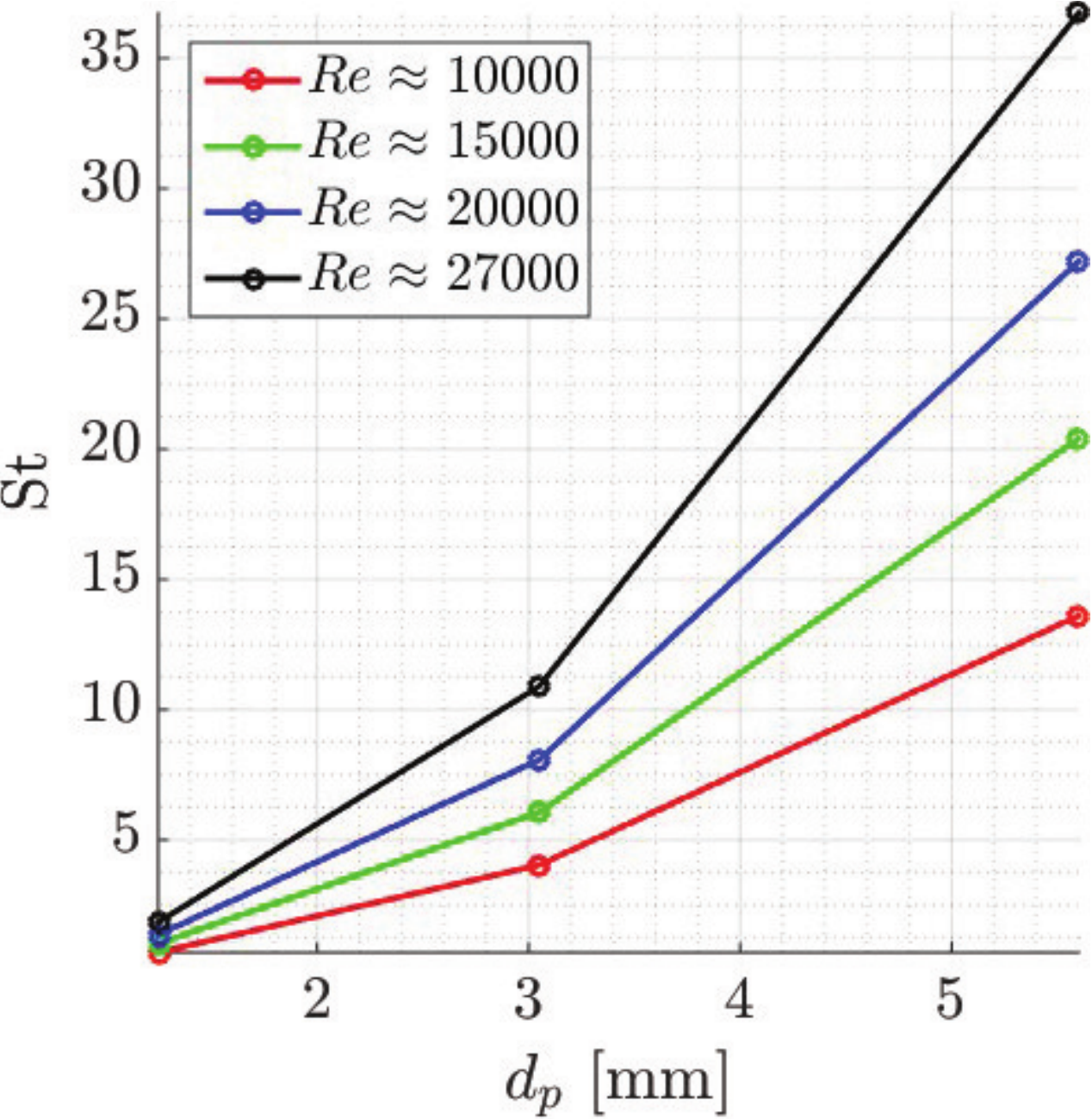}
  \caption{}
  \label{fig:Stokes number}
\end{subfigure}

\caption{Particle Shields number $Sh$ (a), Rouse number $Ro$, based on the friction velocity $u_{\tau}$ (b) and Stokes number $St$ (c), as a function of particle size $d_p$ and Reynolds number $Re_{2H}$. The solid lines in all figures use $u_{\tau}$ and $\tau_{w}$ from single-phase flow (i.e. $\phi=$0\%) as scaling factors. The thin lines in figures (a) and (b) are obtained by dividing with the values pertaining to particle laden flow at $\phi=$ 5, 10 and 20\%.}
\label{fig:Ro_Sh_St}
\end{figure}

As will be shown later in the results section, the area concentration $\phi_{Area}$ in some regions of the flow was as high as 50\%. At such a high concentration, the inter-particle distance is small and particles could not be detected at all times. 
In addition, the detection was slightly inefficient close to the top wall at the highest concentration considered because the intensity of the laser light reduces. Also, particles on the bottom wall cast shadows on the particles above them, making them appear as bright circles cut by dark lines. Detecting such faint `striped' circles was an issue for the particle detection algorithm. These are the reasons why we decided to limit the particle bulk concentration to $\phi$ = 20\% in our study.

\section{Results}\label{sec:Results}

We measure the pressure drop at 3 particle volume fractions ($\phi$ = 5, 10 and 20\%), 3 particle sizes ($2H/d_p\approx$ 40, 16 and 9) at 4 different Reynolds numbers ($Re_{2H}\approx$ 10000, 15000, 20000 and 27000). Particle and fluid velocity field\sz{s} have been measured for all particle sizes except for the smallest particles i.e. $2H/d_p\approx$ 40 as they were not transparent enough (see previous section). 

Even though the particle to fluid density ratio is very close to 1 (see table \ref{tab:Particle property}), sedimentation effects are visible, especially at low $Re_{2H}$. To illustrate this, figure \ref{fig:Velocity vectors at 2 Re} shows a typical PIV image at $\phi$ = 5\% for MP and LP at the lowest and highest $Re_{2H}$. For the lowest $Re_{2H}$, most of the particles occupy the lower half of the duct (see \ref{fig:3mm_5p_Re10000} and \ref{fig:6mm_5p_Re10000}). As the $Re_{2H}$ is increased, more particles are drawn into suspension (see \ref{fig:3mm_5p_Re27000} and \ref{fig:6mm_5p_Re27000}). 

To quantify the role of gravity, we consider two non-dimensional numbers: the Shields number $Sh$ and the Rouse number $Ro$. The former, $Sh={\tau_{w}}/{(\rho_p-\rho_f)gd_p}$, quantifies the relative strength of shear forces and gravitational forces. Here, $\tau_{w}$ is the wall shear stress and $g$ is the acceleration due to gravity. A critical threshold value of $Sh$ is generally used to determine the onset of particle motion on a bed \citep{shields1936application}. Figure \ref{fig:Shields number} depicts the $Sh$ for the different cases in our experiment, which are well above the critical value. With increasing $Sh$, more particles are suspended and the transport of sedimenting particles depends on the ratio of the particle sedimenting velocity $U_T$ to the characteristic turbulent velocity $u_\tau$ i.e. the Rouse number $Ro={U_T}/{\kappa u_\tau}$ \citep{rouse1937}. Here $\kappa$ is the von K\'arm\'an constant. In our experiments, $Ro$ ranges from values for bed load (where particles are transported along the bed: large particles LP at lowest $Re_{2H}$) to full suspension (all particles are suspended by the fluid turbulence: small particles SP at highest $Re_{2H}$) as shown in figure \ref{fig:Rouse number}. The different regimes of particle transport given in \citet{fredsoe1992mechanics} are indicated by the different background colors in figure \ref{fig:Rouse number}. 

To give an estimate of the difference in the time scales of the flow and particles, figure \ref{fig:Stokes number} depicts the particle Stokes number $St={(\rho_p{d_p}^2/{18\rho_f\nu_f}})/(H/U_{Bulk})$ based on the fluid bulk time scale. To estimate the particle size compared to the inner length scales of the flow, we note that the small particle (SP) diameter is $\approx$ 15$\delta_\nu$ (where $\delta_\nu = \frac{\nu}{u_{\tau}}$ is the viscous length scale) at the lowest $Re_{2H}\approx$10000. So, the SP are already around 3 times larger than the thickness of the viscous sub-layer. On the other limit, the size of LP is $\approx$ 165$\delta_\nu$ at the highest $Re_{2H}\approx$27000.

\subsection{Pressure drop}\label{sec:Pressure drop}
\begin{figure}
\centering
 \includegraphics[width=.8\linewidth]{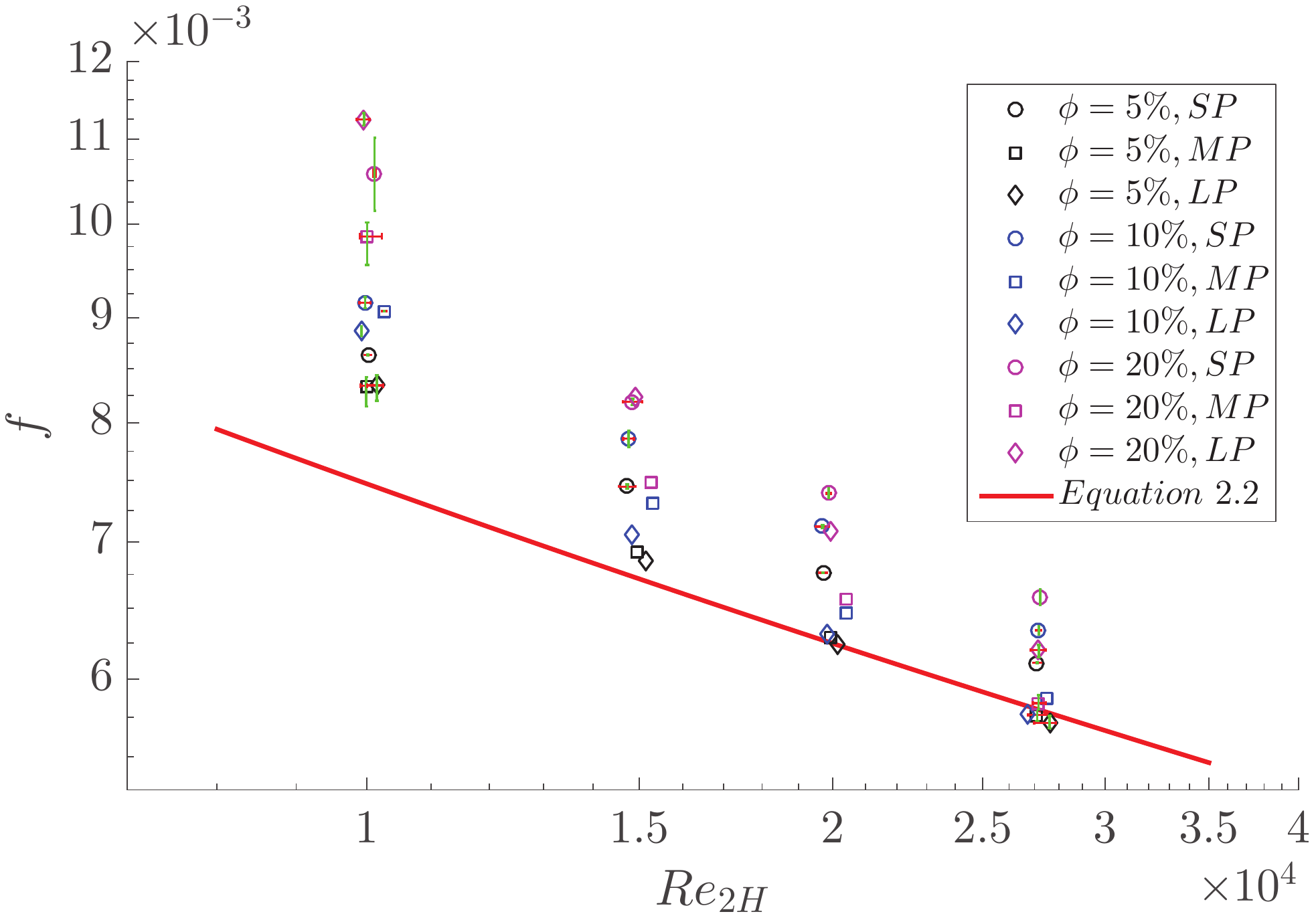}
\caption{Friction factor $f$ as a function of Reynolds number, $Re_{2H}$, particle size $d_p$ and volume fraction $\phi$. Symbols with the same color denote the same concentration. Symbols with same shape denote the same particle size. The solid red line corresponds to single-phase flow.}
\label{fig:f_vs_Re plot}
\end{figure}

We first report the overall pressure drop, in terms of the friction factor $f$, as a function of the Reynolds number $Re_{2H}$ for all the particle sizes $d_p$ and volume fractions $\phi$ considered in this study. This is shown in figure \ref{fig:f_vs_Re plot}. Error-bars are reported for those experiments that were repeated and correspond to the maximum and minimum values observed for the specific case: their small values suggests that the uncertainty in the measurements is relatively small. 

\begin{figure}
\centering
\begin{subfigure}{.44\textwidth}
  \centering
  \includegraphics[height=0.95\linewidth]{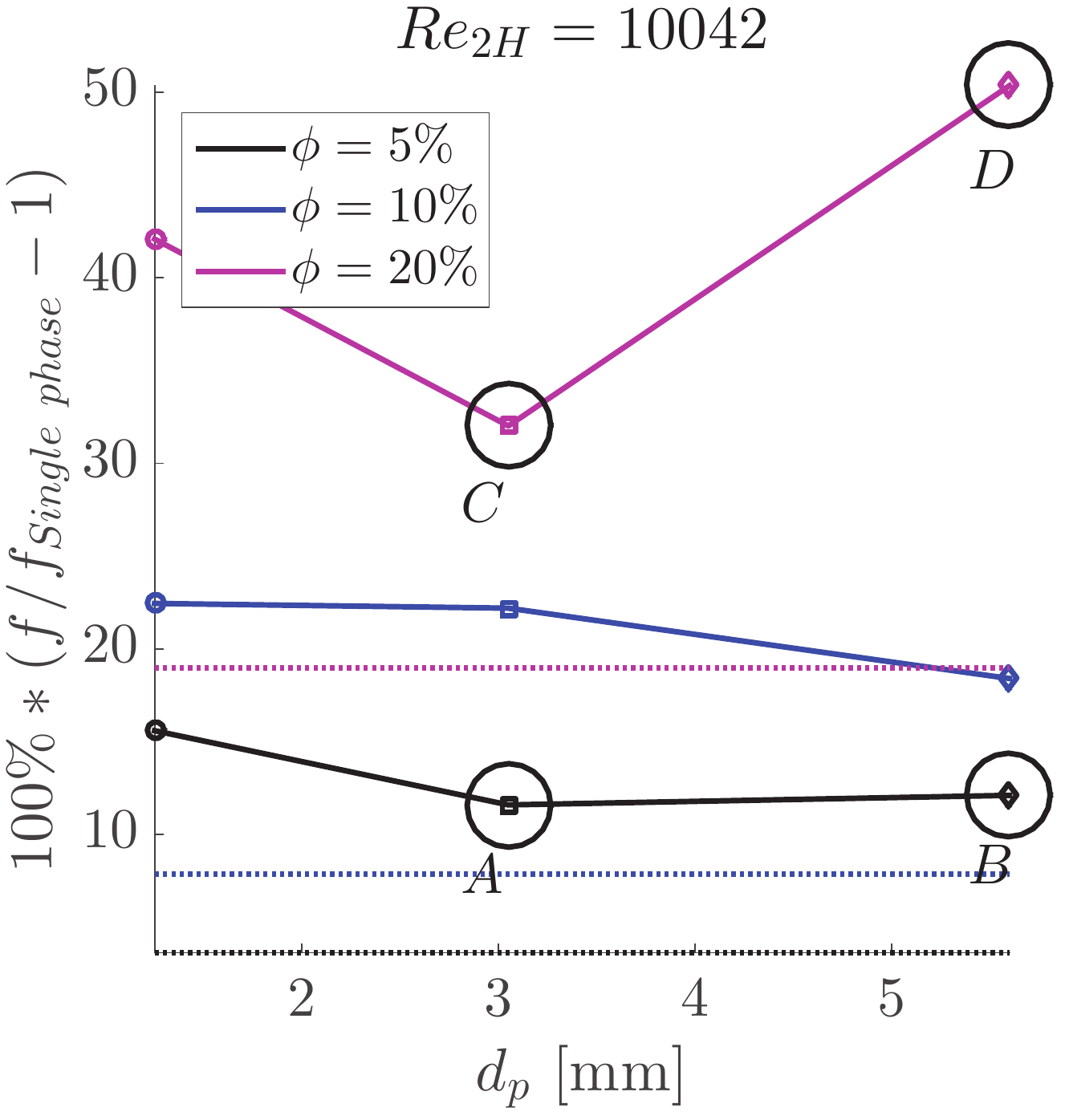}
  \caption{}
  \label{fig:df for Re10000}
\end{subfigure}%
\begin{subfigure}{.44\textwidth}
  \centering
  \includegraphics[height=0.95\linewidth]{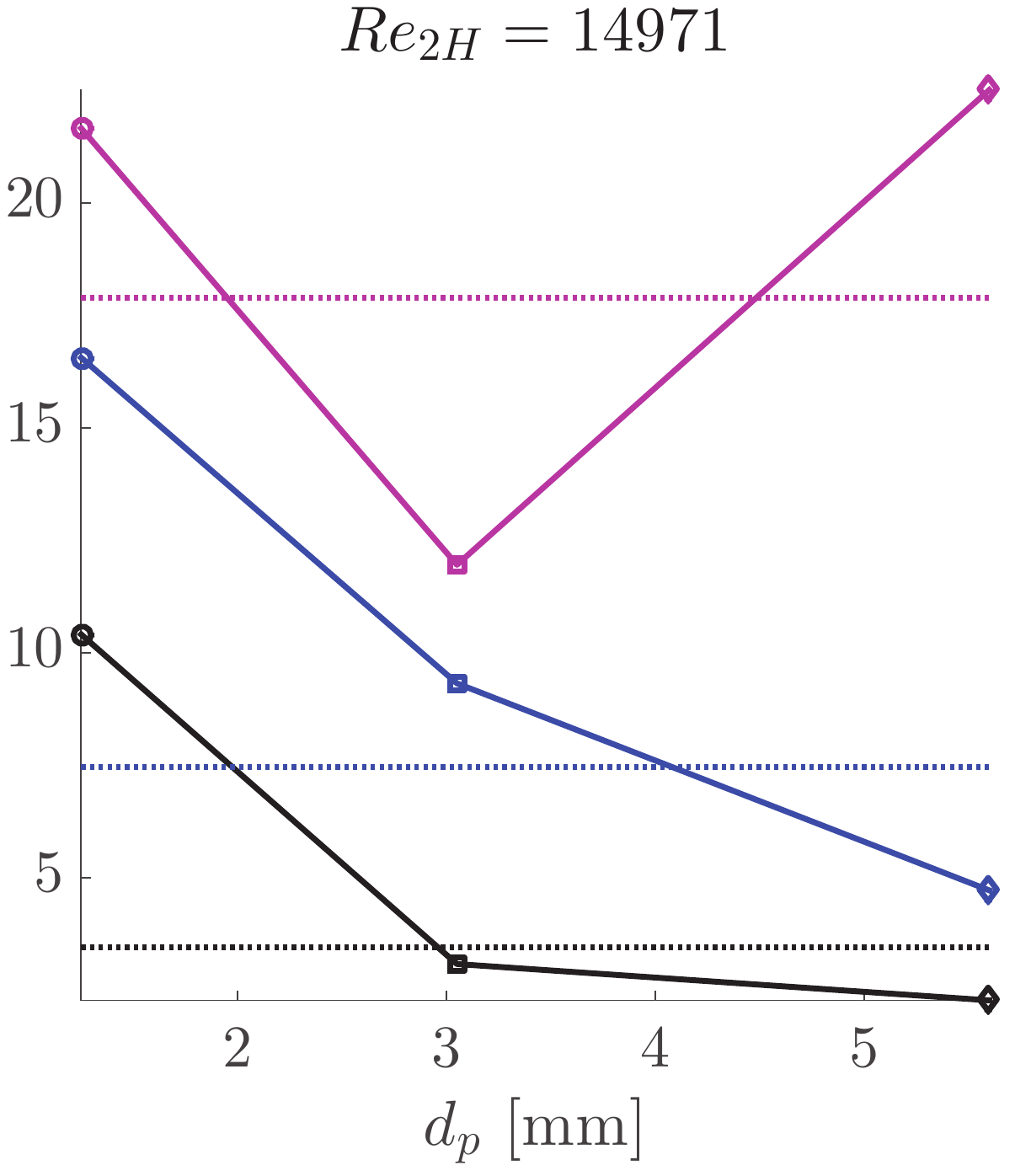}
  \caption{}
  \label{fig:df for Re15000}
\end{subfigure}
\begin{subfigure}{.44\textwidth}
  \centering
  \includegraphics[height=0.95\linewidth]{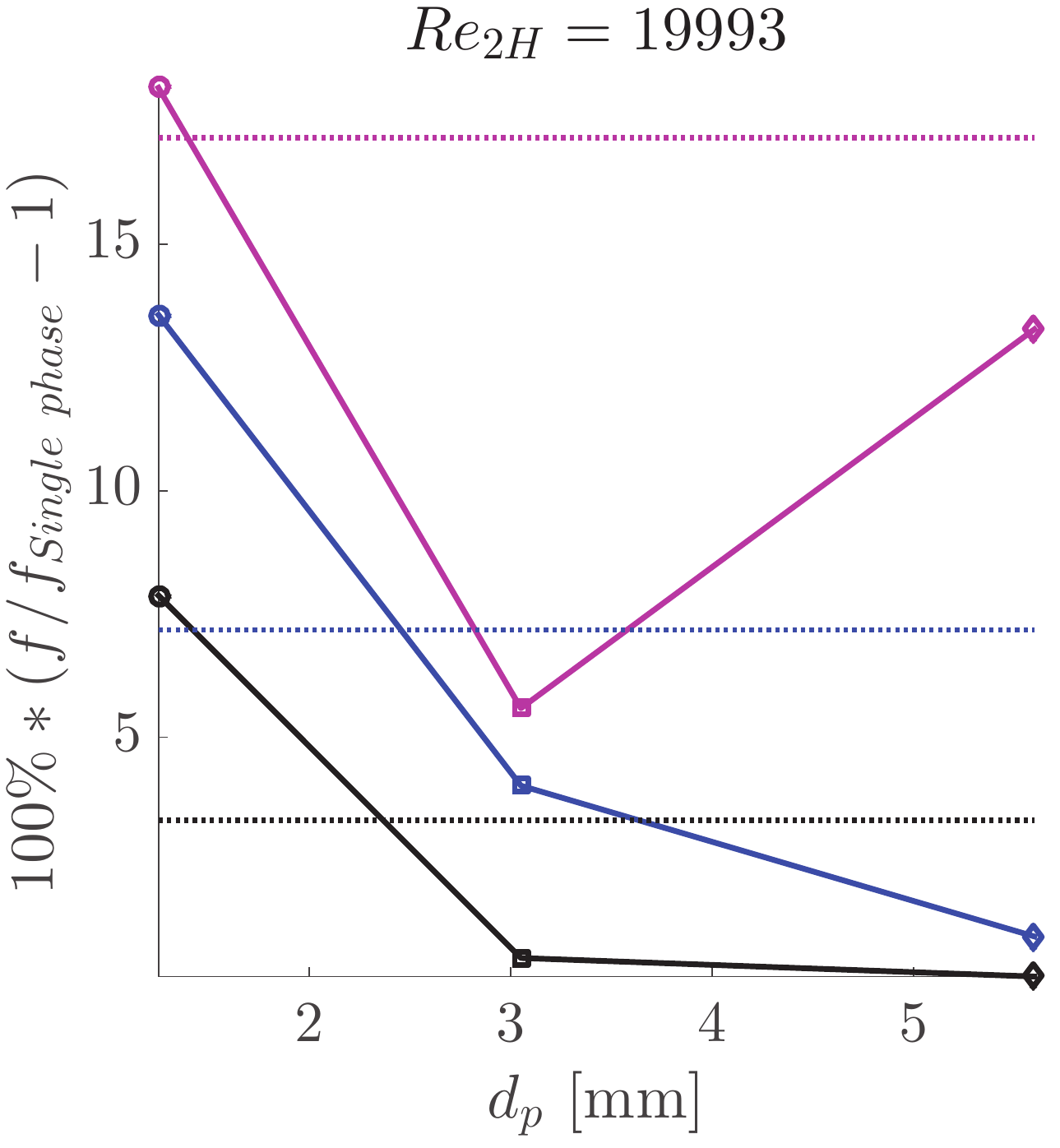}
  \caption{}
  \label{fig:df for Re20000}
\end{subfigure}%
\begin{subfigure}{.44\textwidth}
  \centering
  \includegraphics[height=0.95\linewidth]{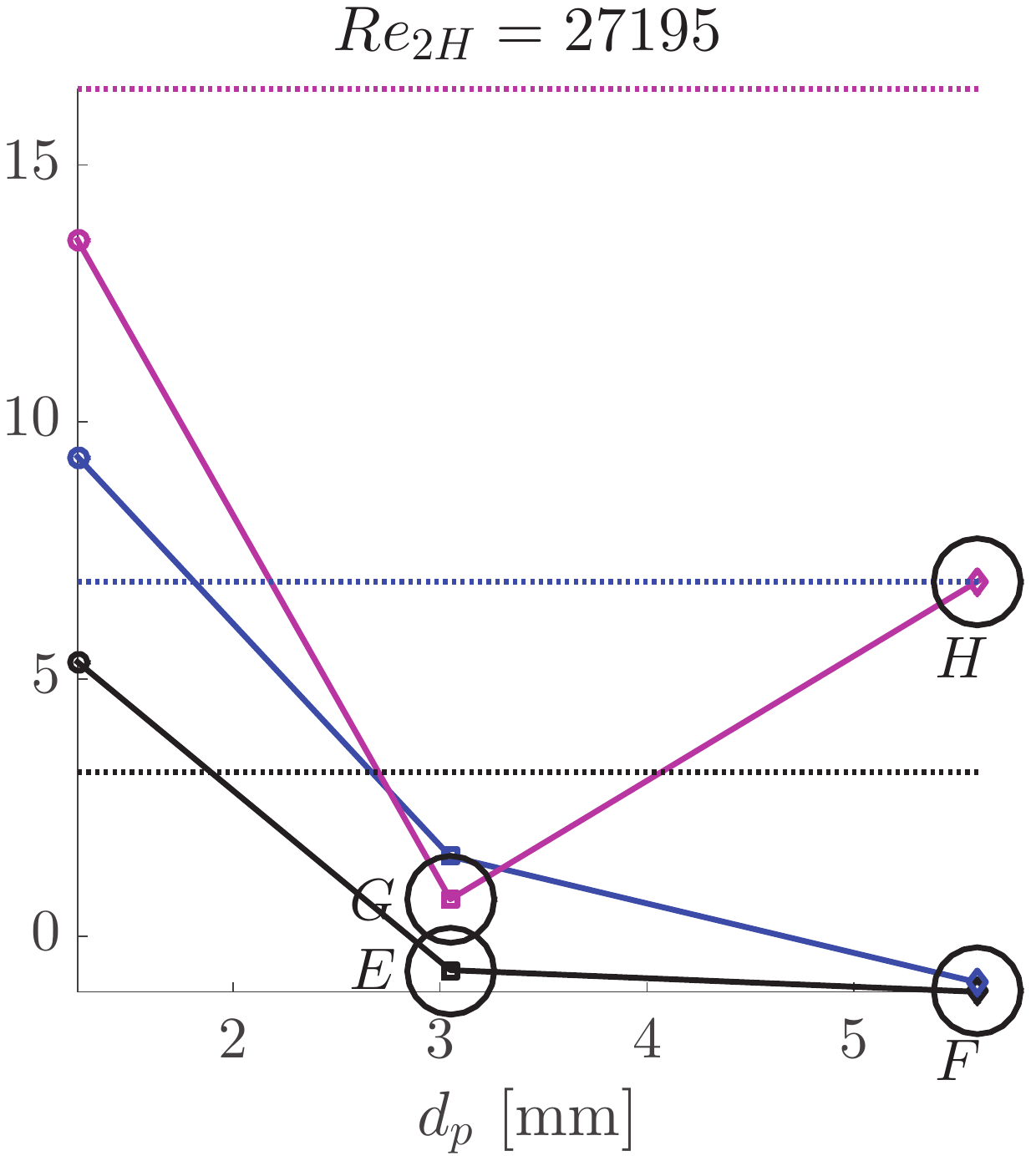}
  \caption{}
  \label{fig:df for Re27000}
\end{subfigure}
\caption{Percent change of the friction factor $f$ with respect to the friction factor of the single-phase flow versus the particle size $d_p$ and volume  fraction $\phi$ for 4 different $Re_{2H}$ (indicated on top of each sub-figure). 
The dashed lines (same color as for the data at the corresponding $\phi$) indicate the friction factor obtained with the values of the effective suspension viscosity from the Eilers fit (equation \ref{eqn:Eilers fit}). 
The large circles in figure (a) and (d) correspond to the cases where PIV and PTV data are reported. 
}
\label{fig:df plot}
\end{figure}

As seen in the figure, the drag increase  due to the particles strongly reduces as $Re_{2H}$ is increased. At the lowest $Re_{2H}\approx$ 10000, the pressure drop increases in the presence of particles by 10 to 50\% depending on the particle size and volume fraction. At the highest $Re_{2H}\approx$ 27000, it varies only from from -2 to 14\% of the single-phase value. 
These changes with $Re_{2H}$ will be related to the particle inertia and to the change in the particle dynamics: from sedimenting to fully suspended particles as seen in figure \ref{fig:Velocity vectors at 2 Re}. At the lowest $Re_{2H}\approx$ 10000, particles constitute a moving bed of higher concentration at the bottom of the duct. At this low $Re_{2H}$ the friction factor mainly depends on the particle volume concentration and it increases with  $\phi$. The particle size have a minor role on the pressure drop at the lowest concentrations considered, i.e. at $\phi$ = 5 and 10\%. At $\phi$ = 20\%, conversely, LP cause the largest pressure drop and MP cause the lowest pressure drop. So, we observe a dependence on the particle size at $\phi$ = 20\%. 

As the Reynolds number  is increased, most of the particles are now dispersed in suspension and hence, the pressure loss depends upon the mutual interaction between particles and turbulence in the bulk of the flow and close to the wall. 
An interesting observation is that at $Re_{2H}\approx$ 27000, the different concentrations of MP lead to nearly the same pressure drop as the single-phase flow, which we will explain as a result of two counteracting effects.

To better understand the changes in the friction factor at each $Re_{2H}$, the data points in figure \ref{fig:f_vs_Re plot} are displayed as relative change with respect to the single-phase flow in figure \ref{fig:df plot}. The dashed lines show the friction factor based on an effective suspension viscosity, obtained from the Eilers fit \citep{stickel2005fluid}; this empirical formula relates the effective viscosity to the nominal volume fraction $\phi$ in the limit of vanishing inertia,
\begin{equation}
  \frac{\eta_e}{\eta} = \Big(1 + \frac{5}{4}\frac{\phi}{1-\phi/0.65}\Big)^2.
  \label{eqn:Eilers fit}
\end{equation}
The effective viscosity $\eta_e$ is used to compute an effective Reynolds number $Re_e=U_{Bulk}2H/\eta_e$, in turn used to find the effective friction factor from equation \ref{eqn:Friction factor}. This simple approach predicts an increase in the friction factor with the particle concentration, although of different magnitude than that observed experimentally. We believe the difference is mainly due to the non-uniform particle distribution in a square duct,  although inertial effects are not negligible as will be discussed later. Modeling a turbulent flow using an effective suspension viscosity has been shown to be a good approximation for relatively small particles when the slip between fluid and particle velocity is small \citep{costa2016universal}. 

If we limit our attention to the cases with 5\% and 10\% particle volume fraction for increasing $Re_{2H}$, i.e. moving from \ref{fig:df for Re10000} to \ref{fig:df for Re27000}, the friction factor reduces as the particle size increases. The friction even becomes marginally smaller than for single-phase flows (implying a potential drag reduction) for the LP at $Re_{2H}\approx$ 27000 (see figure \ref{fig:df for Re27000}).  

If we consider the pressure drop for the highest volume fraction in this study ($\phi$ = 20\%), conversely, the drag reduces with increasing particle size from SP to MP but increases for LP. Thus, there is a consistent non-monotonic trend at $\phi$ = 20\%. This observation highlights the difficulties of {\it a priori} estimation of the pressure drop in suspension flows. For this high concentration, LP have the maximum relative pressure drop at the smallest $Re_{2H}$ whereas SP have the maximum relative pressure drop at the highest $Re_{2H}$. \sz{Interestingly, \cite{matas2003transition} observed in a pipe flow geometry a non-monotonic trend for the critical Reynolds number (for onset of turbulence) as a function of the particle size.}

\subsection{Particle distribution}

\begin{figure}
\centering
\begin{subfigure}{.4\textwidth}
  \centering
  \includegraphics[height=0.95\linewidth]{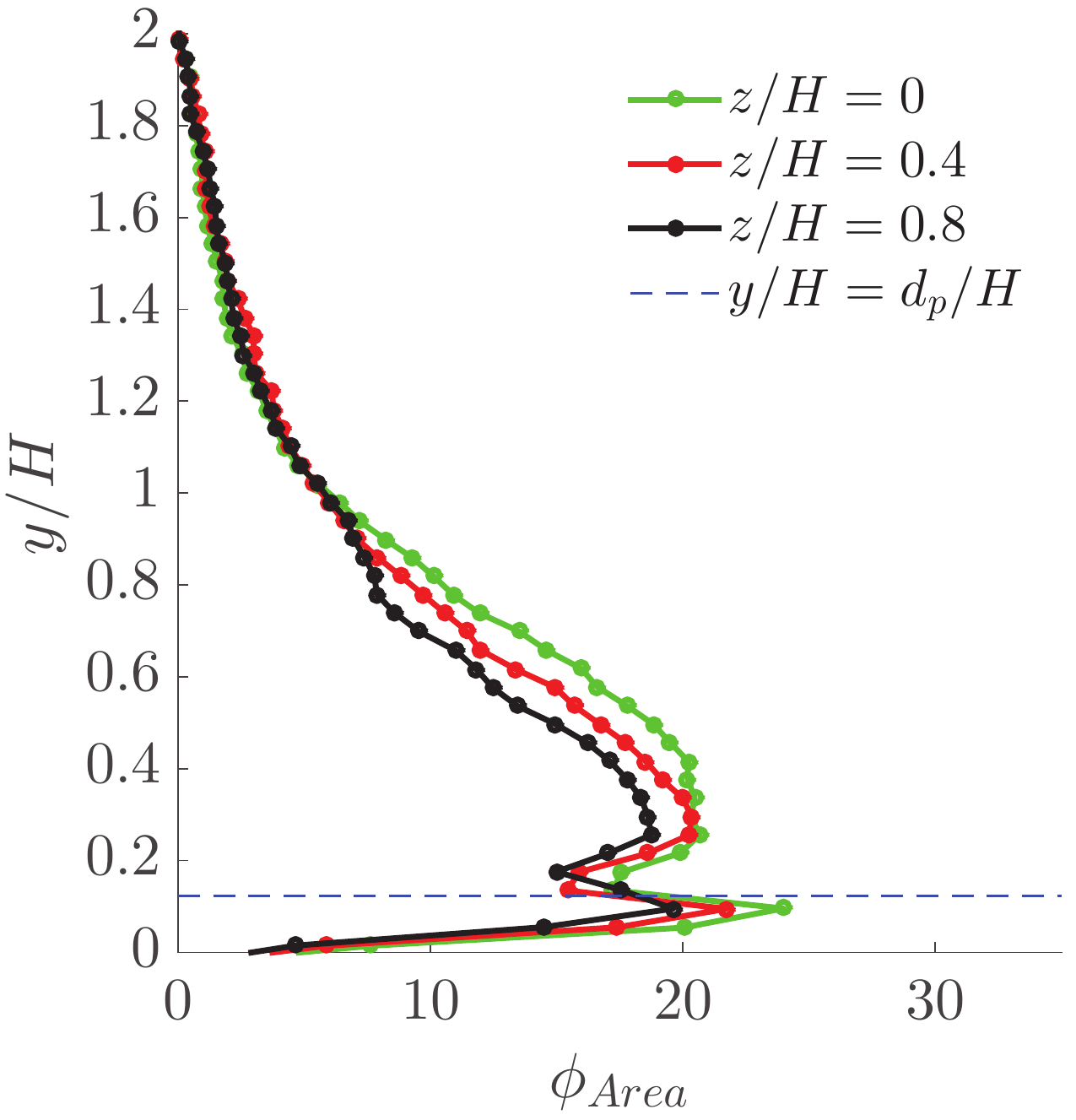}
  \caption{}
  \label{fig:3mm_5p_Re10000_phi}
\end{subfigure}%
\begin{subfigure}{.4\textwidth}
  \centering
  \includegraphics[height=0.95\linewidth]{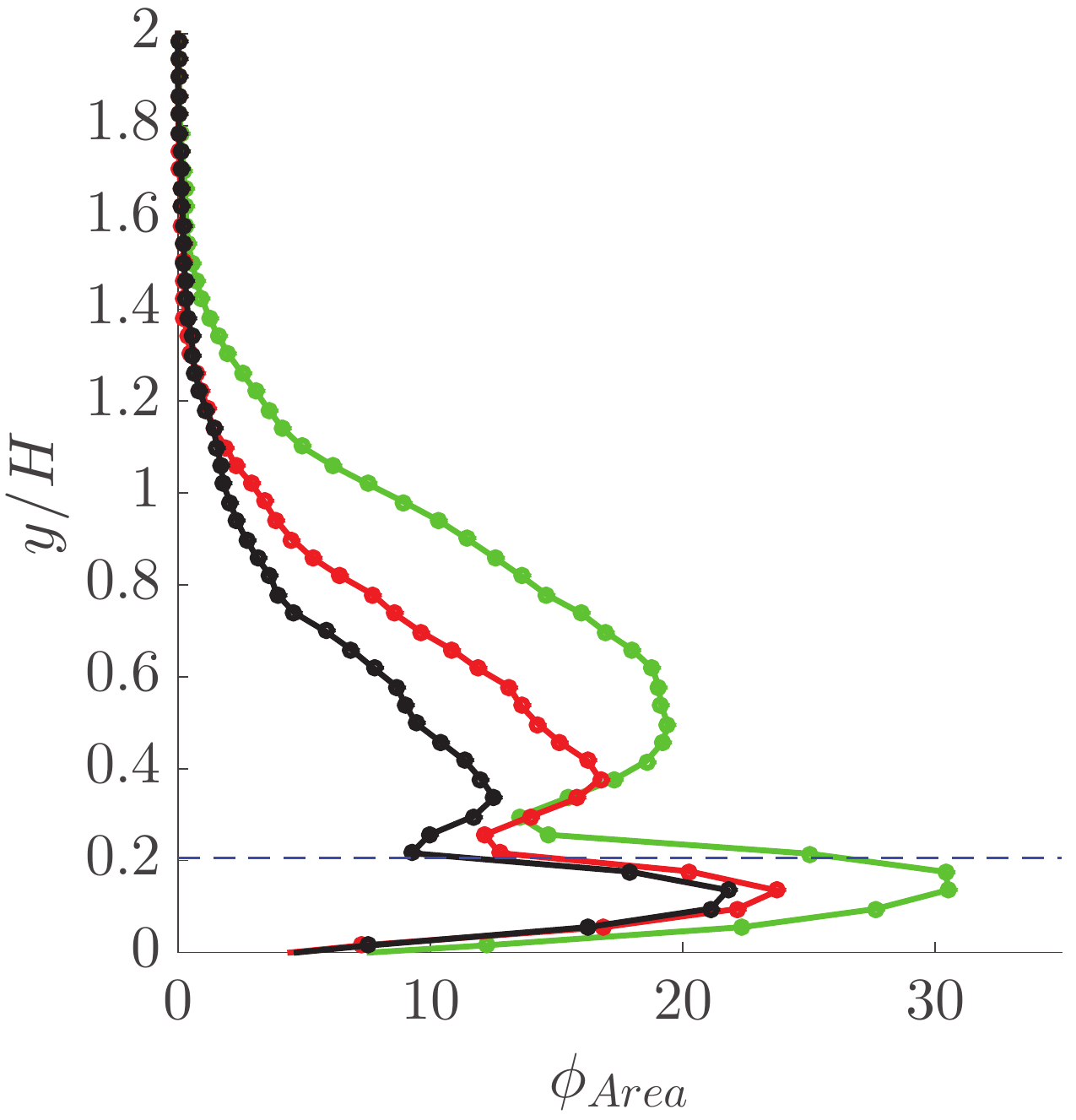}
  \caption{}
  \label{fig:6mm_5p_Re10000_phi}
\end{subfigure}
\begin{subfigure}{.4\textwidth}
  \centering
  \includegraphics[height=0.95\linewidth]{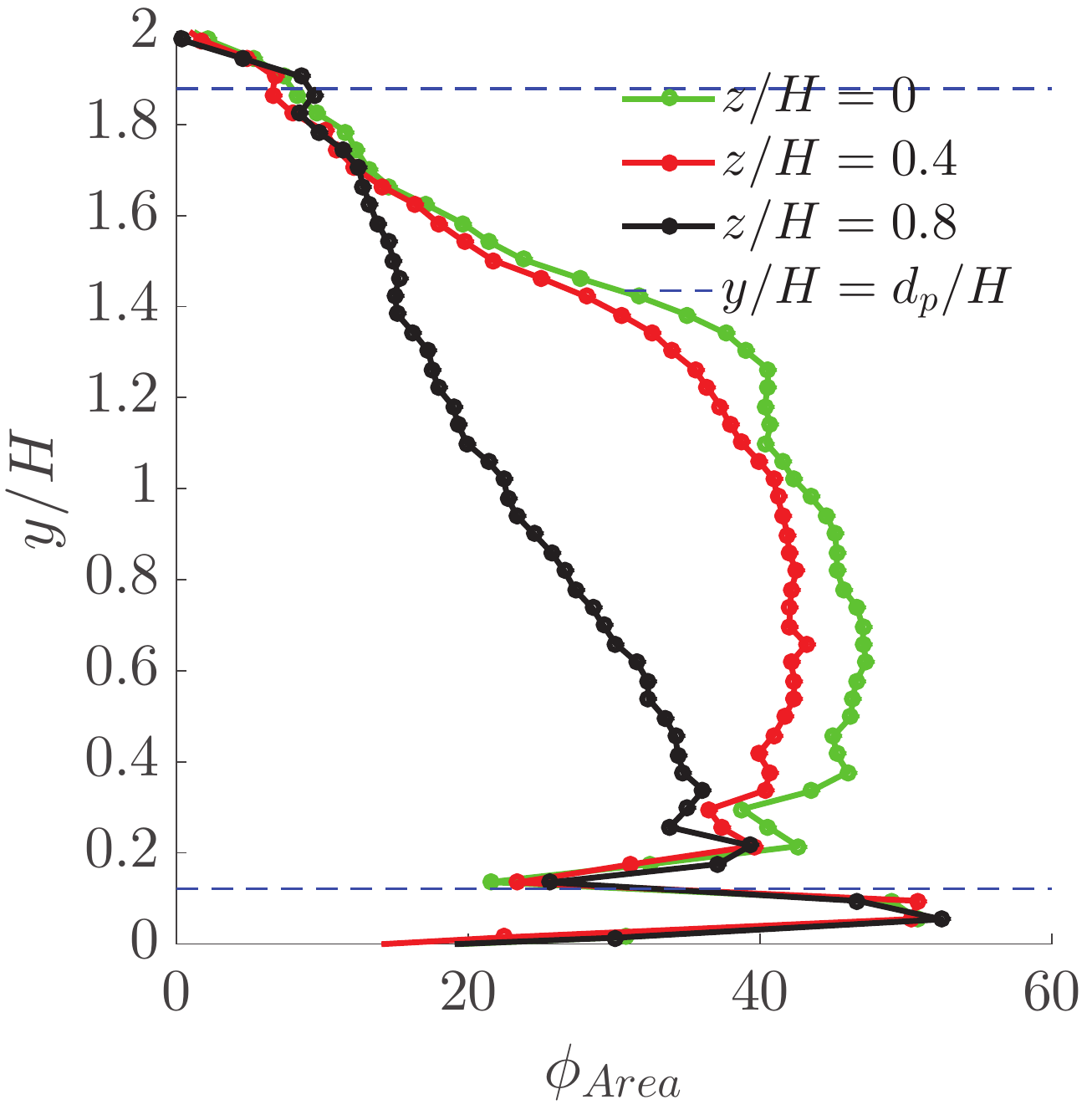}
  \caption{}
  \label{fig:3mm_20p_Re10000_phi}
\end{subfigure}%
\begin{subfigure}{.4\textwidth}
  \centering
  \includegraphics[height=0.95\linewidth]{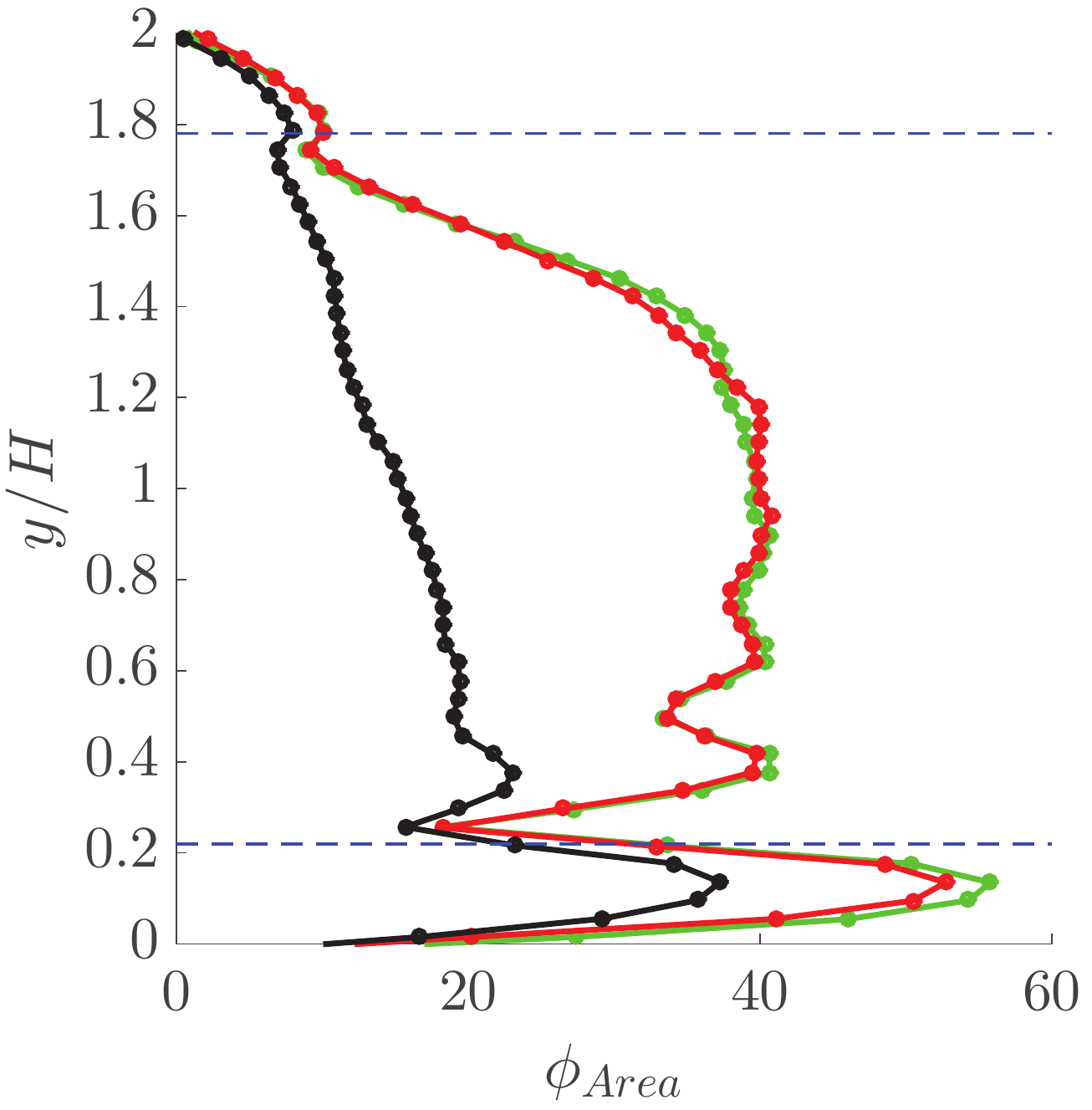}
  \caption{}
  \label{fig:6mm_20p_Re10000_phi}
\end{subfigure}
\caption{Area concentration profile at low $Re_{2H}\approx$ 10000 in 3 spanwise planes. (a) and (b) correspond to medium-sized (MP) and large-sized particles (LP) at $\phi$ = 5\% respectively. (c) and (d) correspond to MP and LP at $\phi$ = 20\%.}
\label{fig:Re10000_phi}
\end{figure}

\begin{figure}
\centering
\begin{subfigure}{.4\textwidth}
  \centering
  \includegraphics[height=0.95\linewidth]{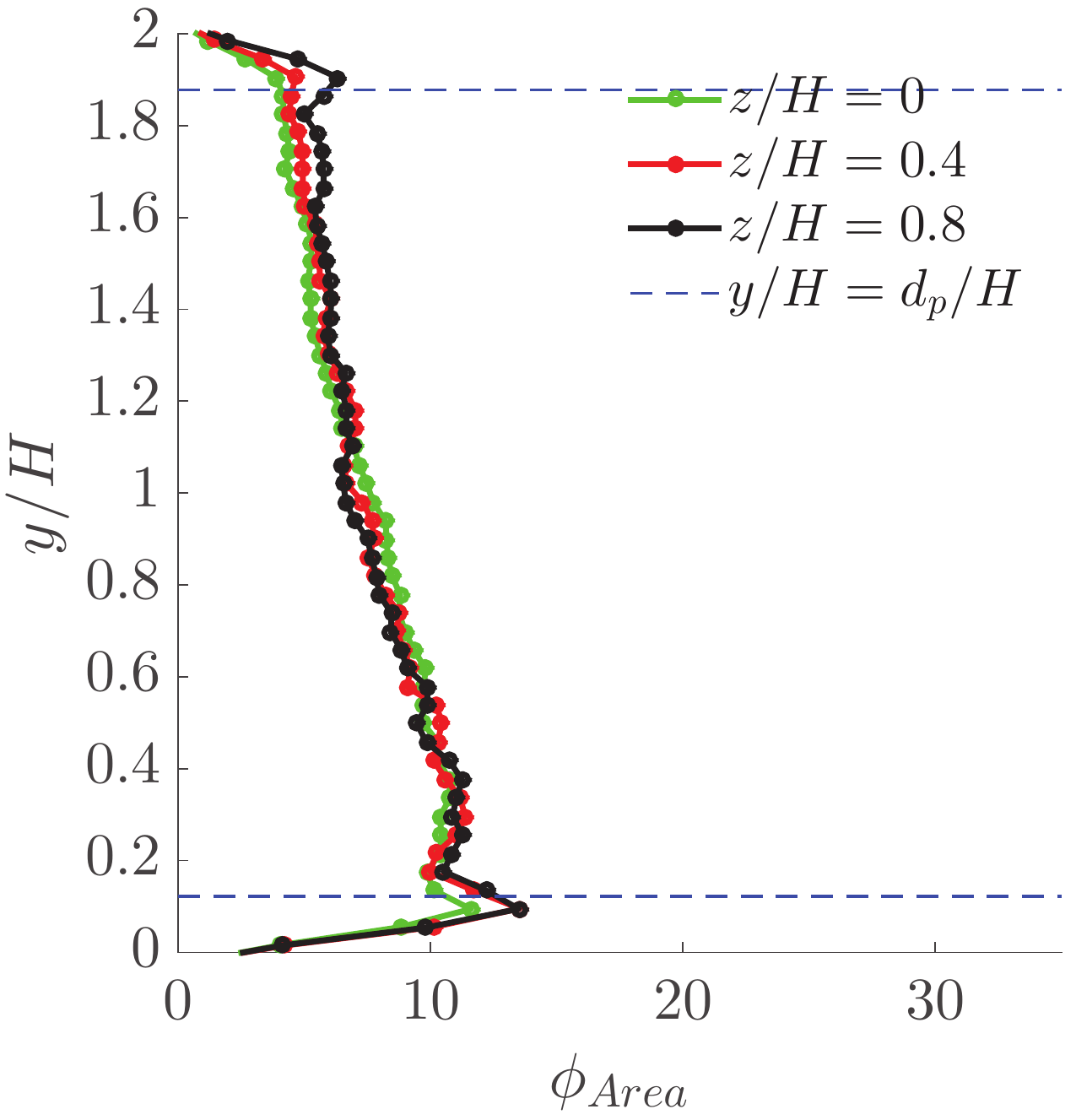}
  \caption{}
  \label{fig:3mm_5p_Re27000_phi}
\end{subfigure}%
\begin{subfigure}{.4\textwidth}
  \centering
  \includegraphics[height=0.95\linewidth]{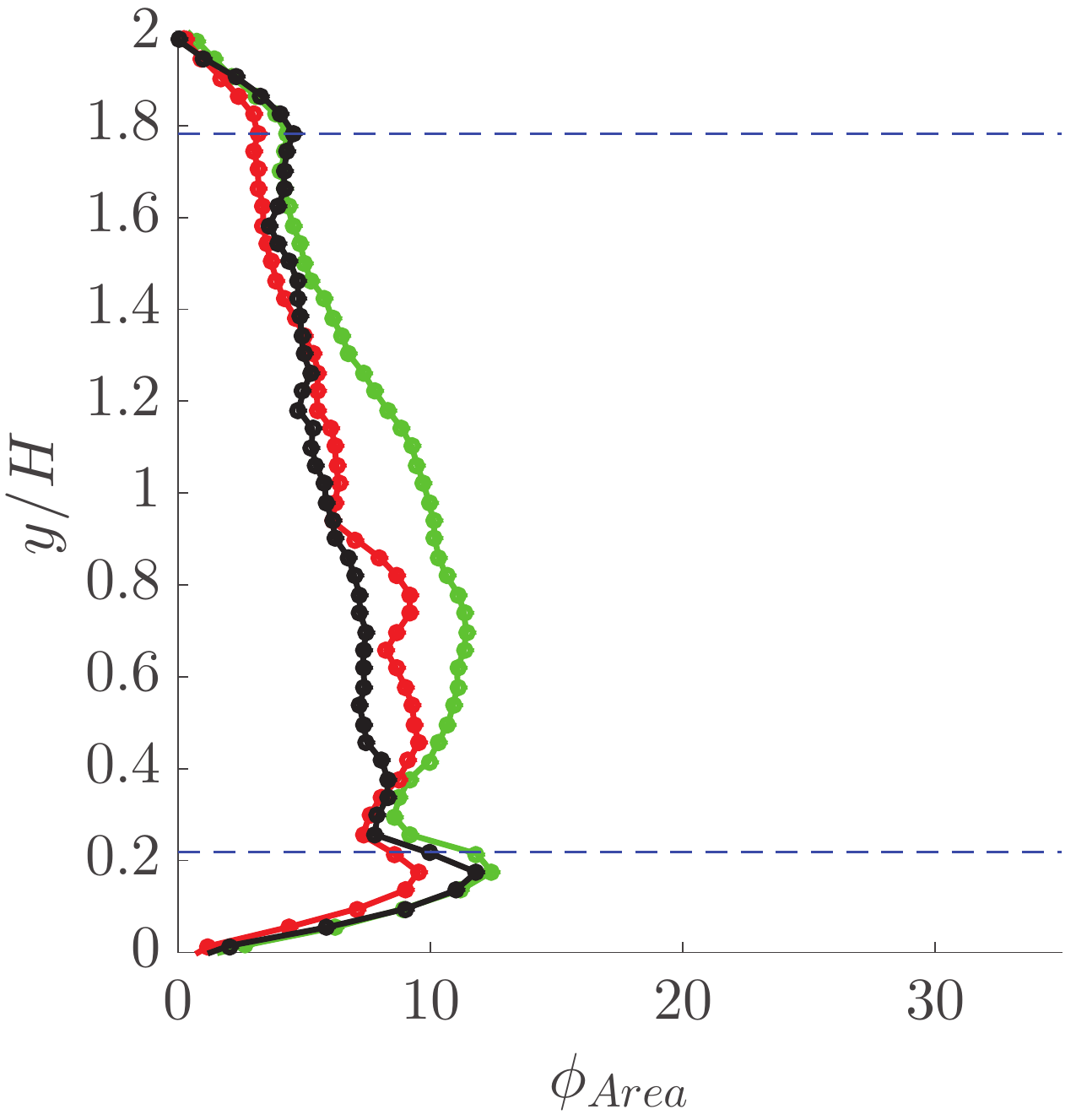}
  \caption{}
  \label{fig:6mm_5p_Re27000_phi}
\end{subfigure}
\begin{subfigure}{.4\textwidth}
  \centering
  \includegraphics[height=0.95\linewidth]{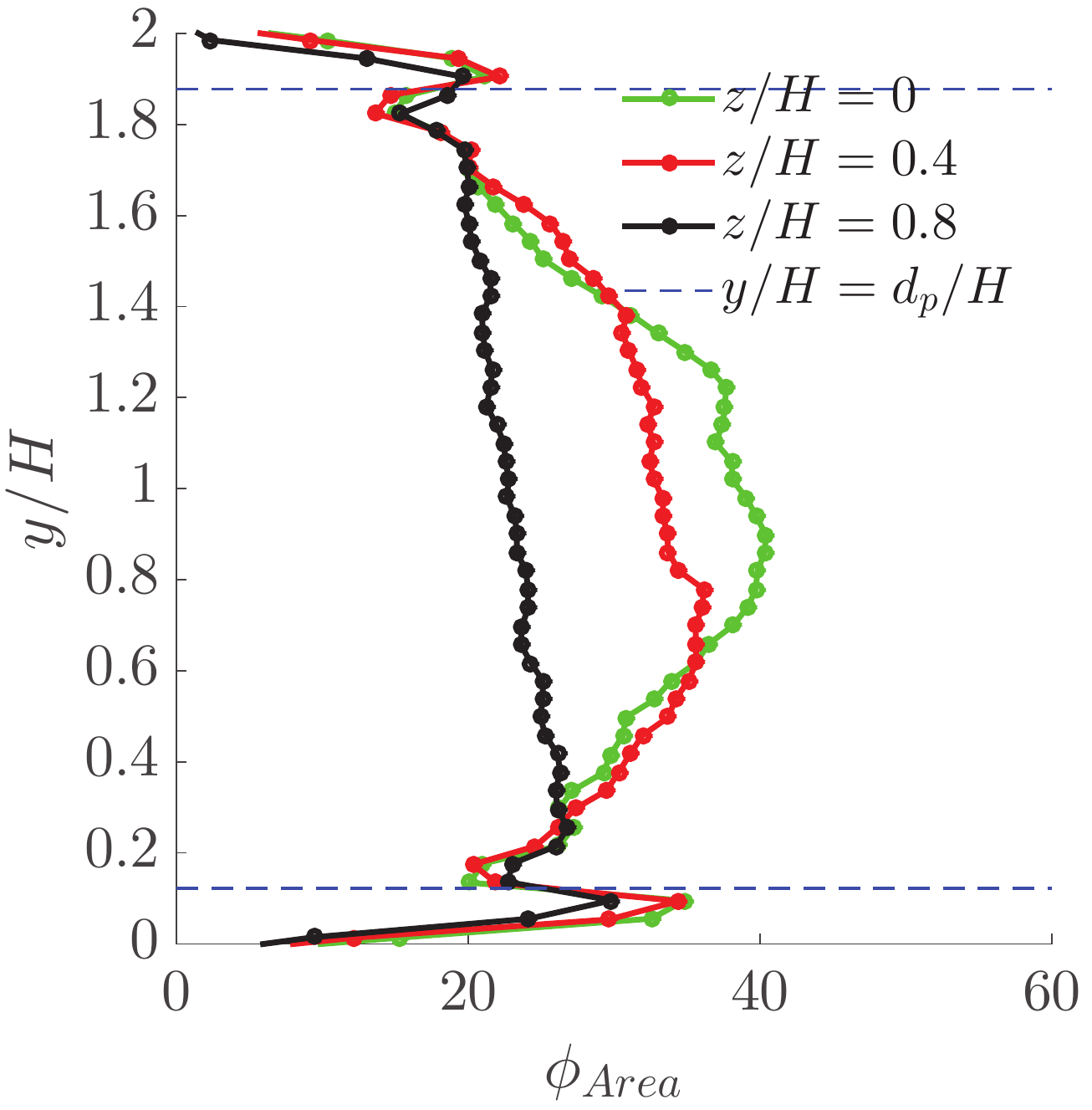}
  \caption{}
  \label{fig:3mm_20p_Re27000_phi}
\end{subfigure}%
\begin{subfigure}{.4\textwidth}
  \centering
  \includegraphics[height=0.95\linewidth]{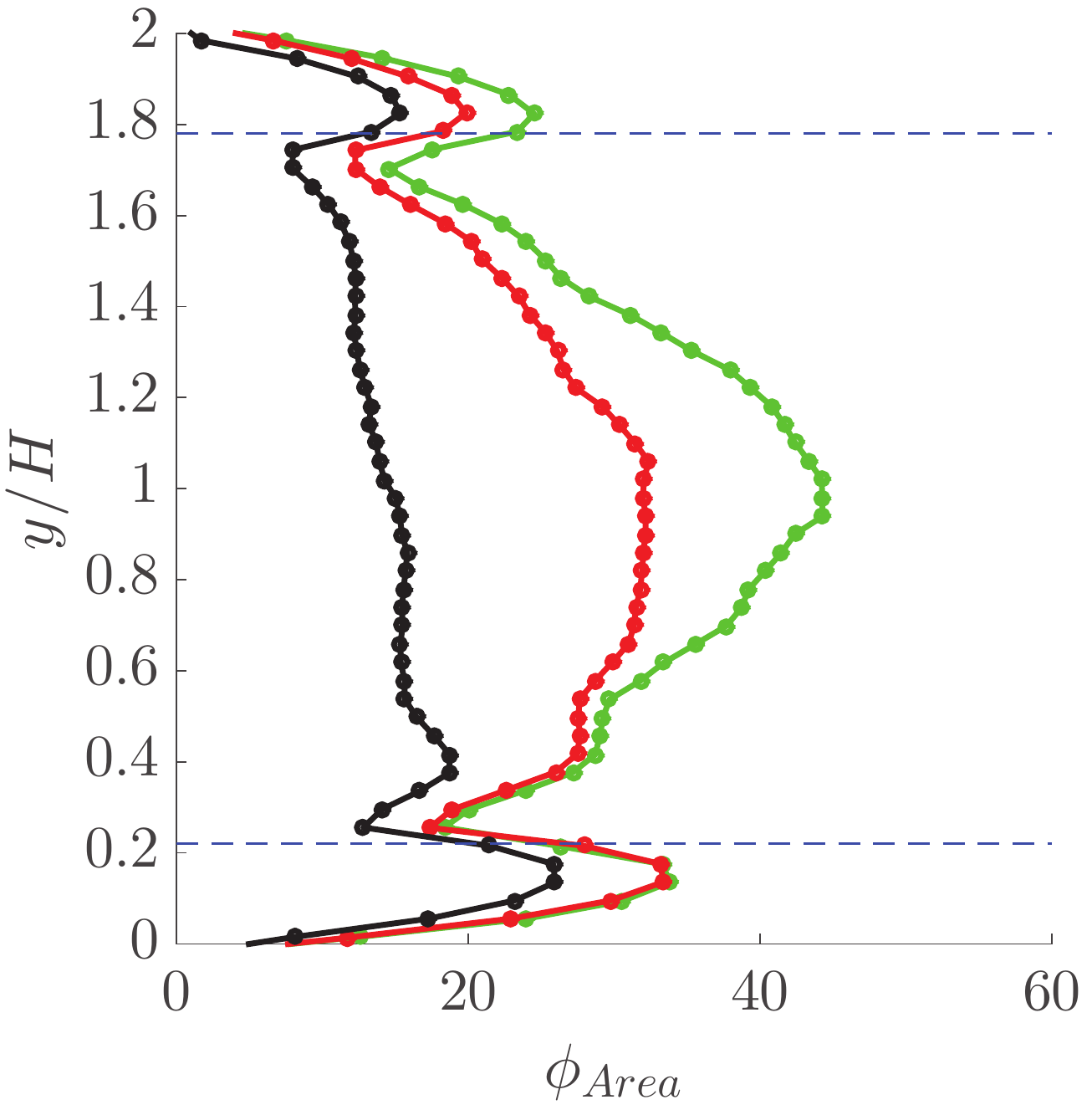}
  \caption{}
  \label{fig:6mm_20p_Re27000_phi}
\end{subfigure}
\caption{Area concentration profile at high $Re_{2H}\approx$ 27000 in 3 spanwise planes. (a) and (b) correspond to medium-sized (MP) and large-sized particles (LP) at $\phi$ = 5\%. (c) and (d) correspond to MP and LP at $\phi$ = 20\%.}
\label{fig:Re27000_phi}
\end{figure}

We will interpret our data focusing on the particle dynamics and therefore we start by reporting the particle concentration profiles. As seen in figure \ref{fig:Re10000_phi} for the lowest $Re_{2H}\approx$ 10000, particles sediment towards the bottom of the duct. The 3 colored lines in the figure correspond to 3 different spanwise planes: $z/H$ = 0, 0.4 and 0.8. At low volume fraction, $\phi$ = 5\%, (see figures \ref{fig:3mm_5p_Re10000_phi} and \ref{fig:6mm_5p_Re10000_phi}), the particle concentration is highest at the bottom wall, due to the presence of a particle-wall layer. Above this layer, the concentration reduces and goes to almost zero at the top wall. The concentration is higher on the plane of the wall bisector ($z/H$ = 0) and decreases while moving towards the side wall ($z/H$ = 1). For the highest volume fraction $\phi$ = 20\%, cf.\ figures \ref{fig:3mm_20p_Re10000_phi} and \ref{fig:6mm_20p_Re10000_phi} 
where we change the x--scale from 0--35\% to 0--60\%, particles tend to form a much denser bed with an area concentration close to 40\% above the particle wall layer.

On the other hand, for the highest $Re_{2H}\approx$ 27000, particles are dispersed throughout the volume of the duct since inertial forces overcome the influence of gravity (see figure \ref{fig:Re27000_phi}). Figures \ref{fig:3mm_5p_Re27000_phi} and  \ref{fig:6mm_5p_Re27000_phi} shows the particle concentration at $\phi$ = 5\% and compared to the case with $Re_{2H}\approx$ 10000, the particle distribution is more uniform. For $\phi$ = 20\%, the concentration distribution shows nearly equal peaks of high concentration at the top and the bottom walls along with a maximum at the center of the duct. More detailed features about the particle distribution profile will be discussed together with the velocity field in section \ref{subsec:Velocity Statistics}.

\subsection{Velocity statistics}
\label{subsec:Velocity Statistics}

Next, we report and discuss the measurement of fluid and particle velocity statistics. We recall that PIV and PTV were not possible for the SP, which were not transparent enough. We therefore report velocity statistics for MP and LP and focus on the extreme cases in terms of particle volume fraction $\phi$ (5\% and 20\%) and bulk Reynolds number $Re_{2H}$ (10000 and 27000). Thus, four different cases for each particle size.

\subsubsection{Fully suspended particles ($Re_{2H}\approx$ 27000)}
\label{sec: Velocity statistics Fully suspended}

As the Reynolds number is increased, more particles are drawn into suspension, see figure \ref{fig:Re27000_phi}. At the highest $Re_{2H}\approx$ 27000, the values of Rouse number $Ro=$ 0.6 for MP and 1.2 for LP lie in the `suspended-load' regime \citep{fredsoe1992mechanics}. 
We therefore start by examining the results at high $Re_{2H}$, initially assuming that effect of gravity is negligible.\\

{\it High volume fraction ($\phi$ = 20\%)}\\

\begin{figure}
\centering

\begin{subfigure}{.32\textwidth}
  \centering
  $z/H$ = 0
  \includegraphics[height=1\linewidth]{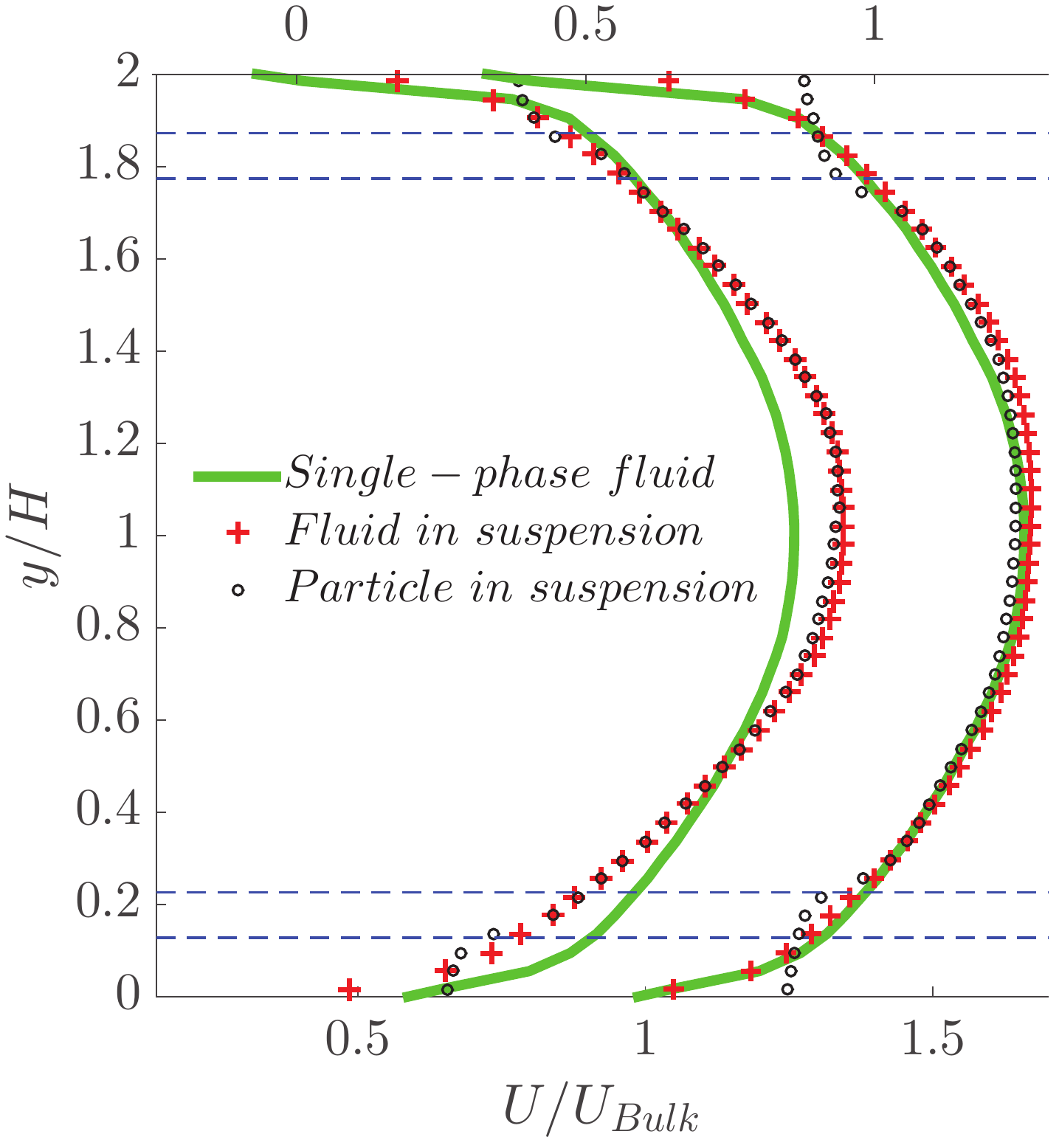}
  \caption{}
  \label{fig:3_6mm_20p_Re27000_sp00mm_Umean}
\end{subfigure}%
\begin{subfigure}{.32\textwidth}
  \centering
  $z/H$ = 0.4
  \includegraphics[height=1\linewidth]{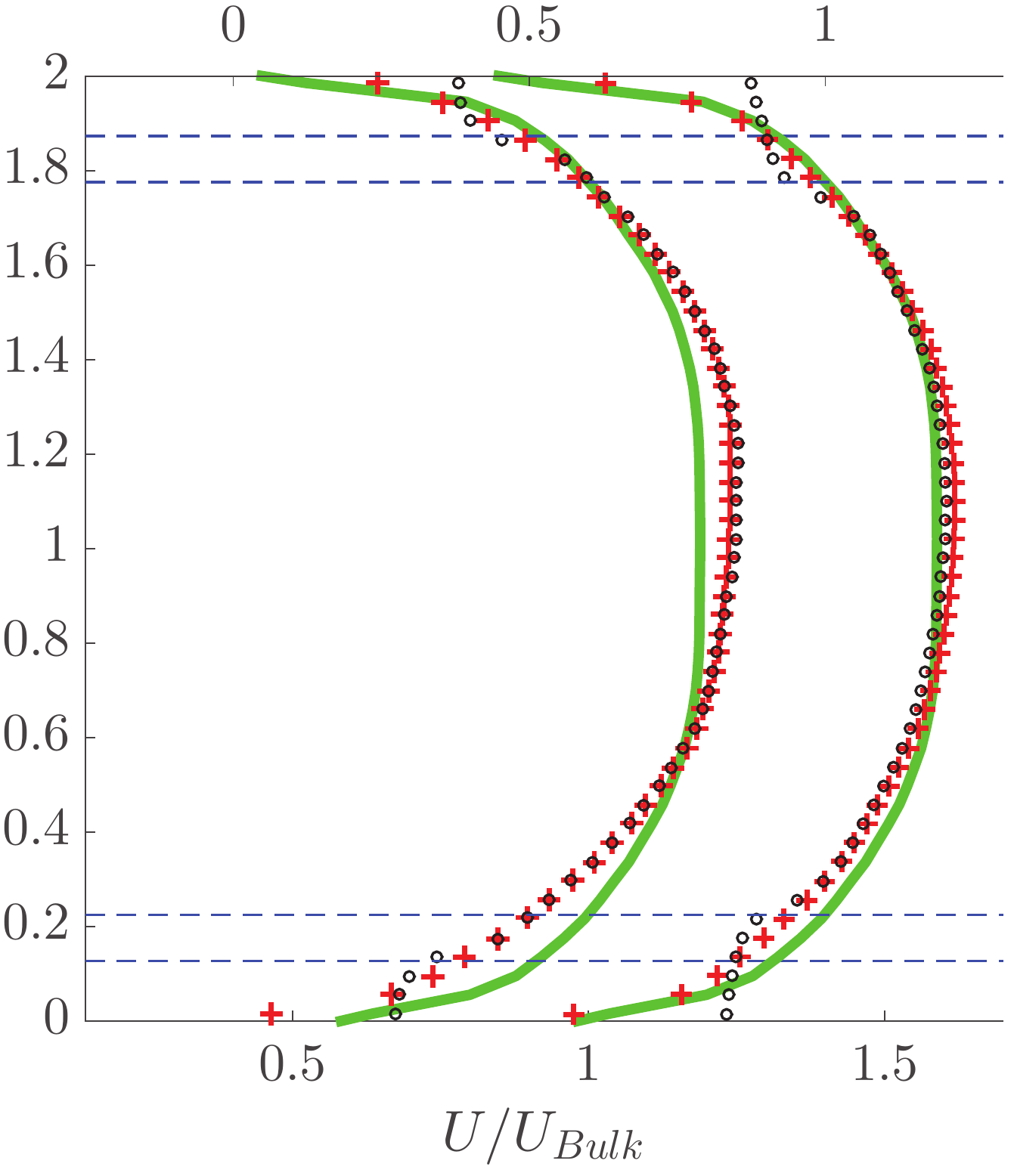}
  \caption{}
  \label{fig:3_6mm_20p_Re27000_sp10mm_Umean}
\end{subfigure}
\begin{subfigure}{.32\textwidth}
  \centering
  $z/H$ = 0.8
  \includegraphics[height=1\linewidth]{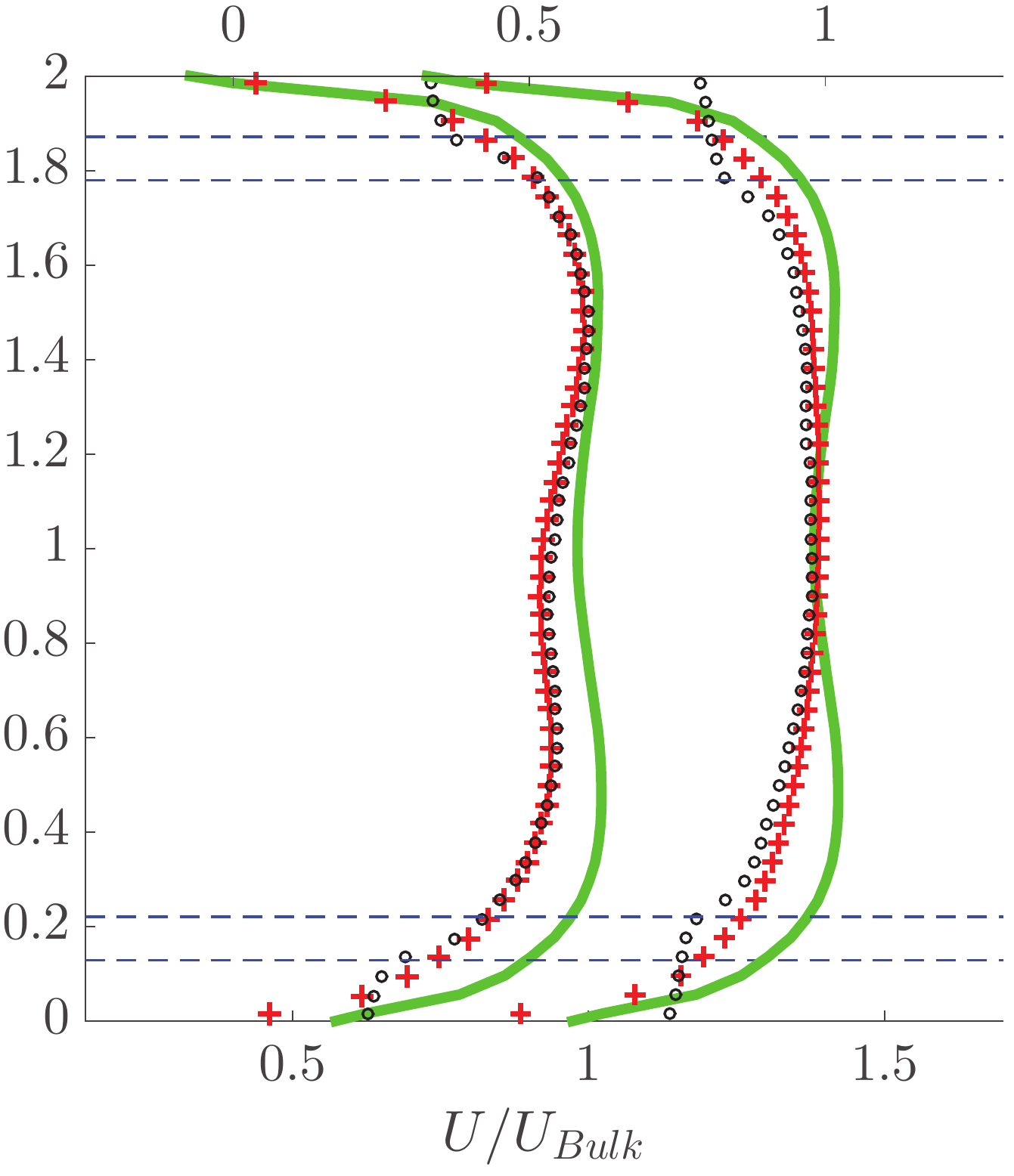}
  \caption{}
  \label{fig:3_6mm_20p_Re27000_sp20mm_Umean}
\end{subfigure}

\begin{subfigure}{.32\textwidth}
  \centering
  \includegraphics[height=1\linewidth]{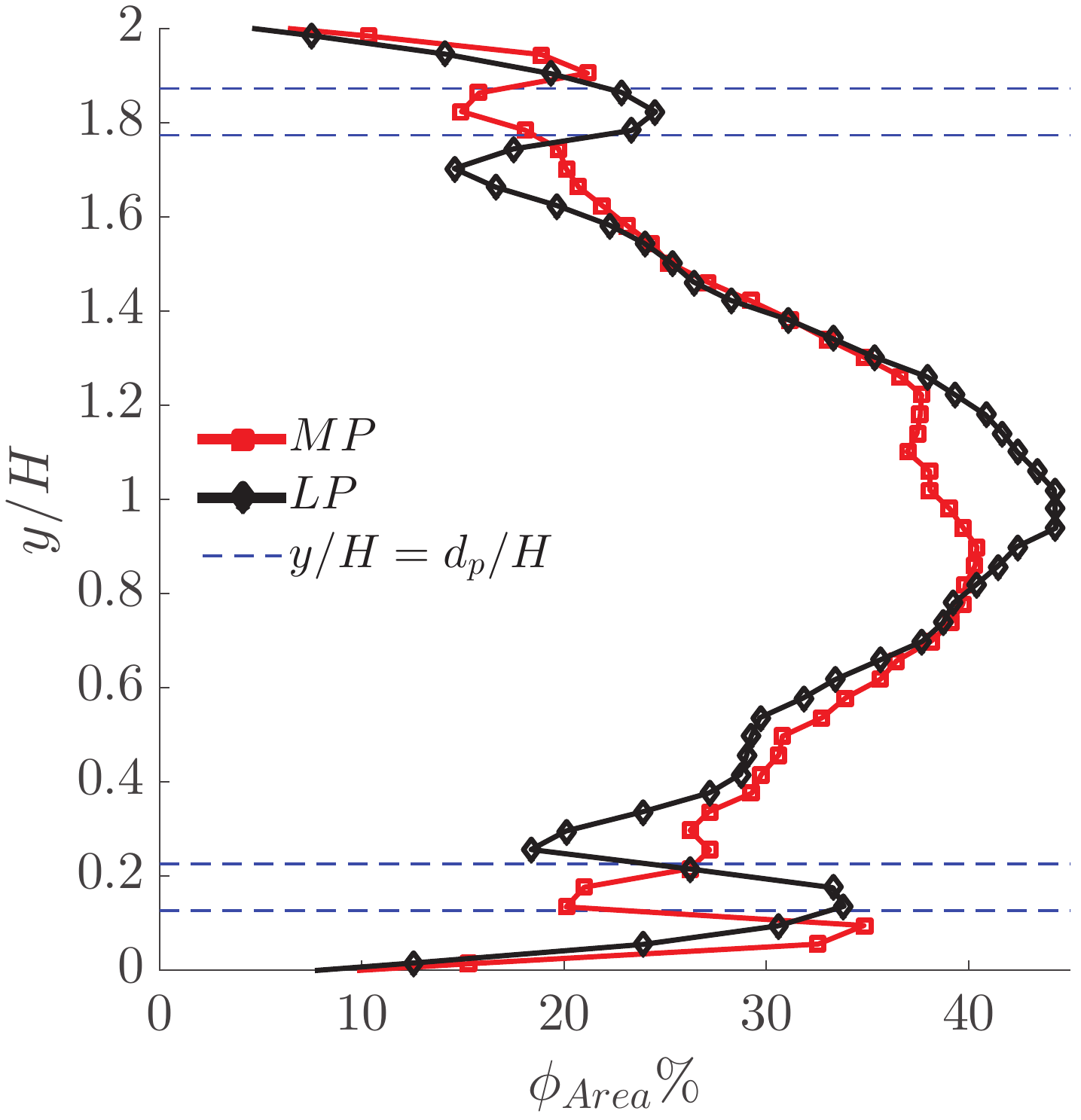}
  \caption{}
  \label{fig:3_6mm_20p_Re27000_sp00mm_phi}
\end{subfigure}%
\begin{subfigure}{.32\textwidth}
  \centering
  \includegraphics[height=1\linewidth]{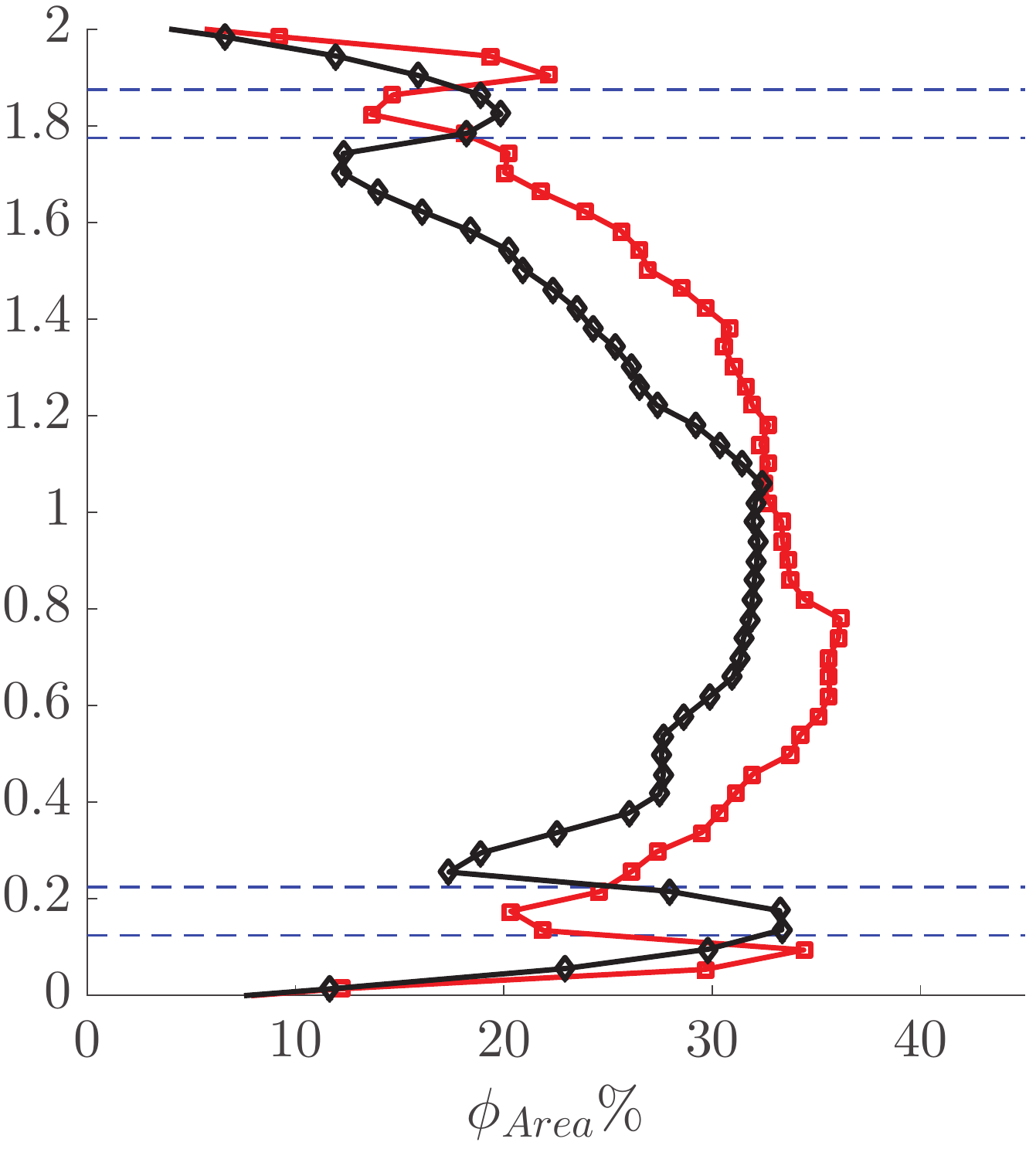}
  \caption{}
  \label{fig:3_6mm_30p_Re27000_sp10mm_phi}
\end{subfigure}
\begin{subfigure}{.32\textwidth}
  \centering
  \includegraphics[height=1\linewidth]{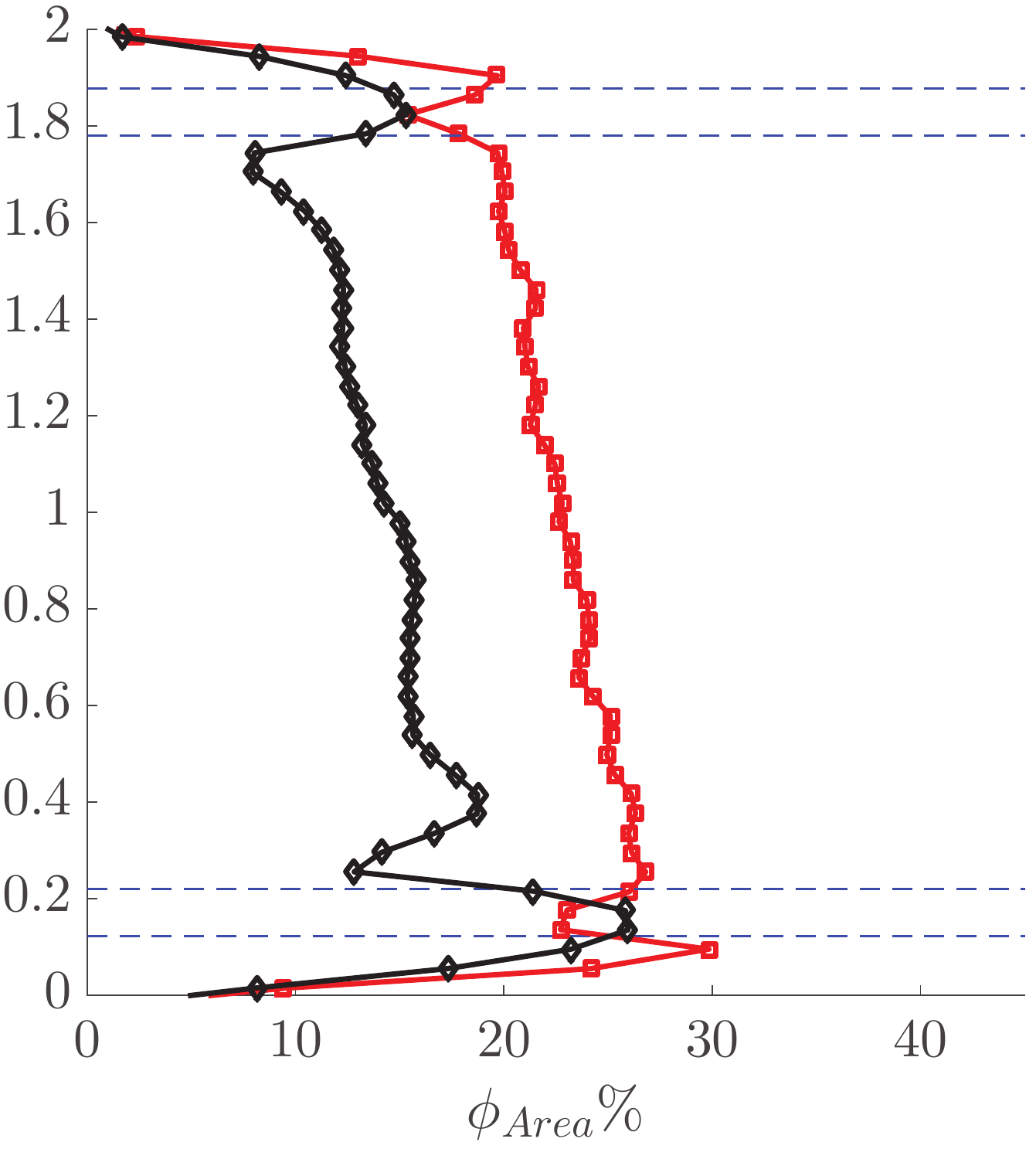}
  \caption{}
  \label{fig:3_6mm_20p_Re27000_sp20mm_phi}
\end{subfigure}

\begin{subfigure}{.32\textwidth}
  \centering
  $z/H$ = 0
  \includegraphics[height=1\linewidth]{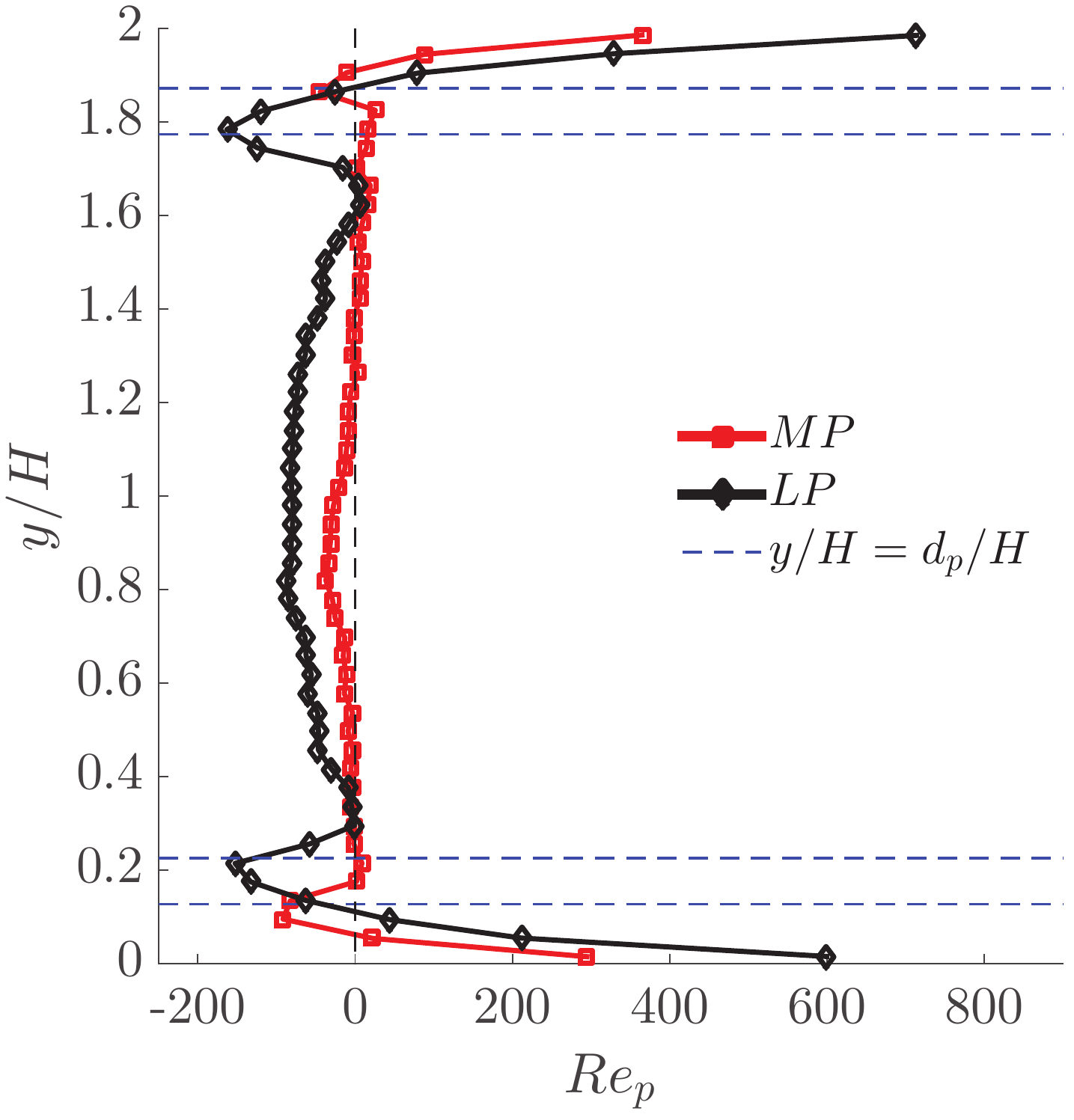}
  \caption{}
  \label{fig:3_6mm_20p_Re27000_sp00mm_Rep}
\end{subfigure}%
\begin{subfigure}{.32\textwidth}
  \centering
  $z/H$ = 0.4
  \includegraphics[height=1\linewidth]{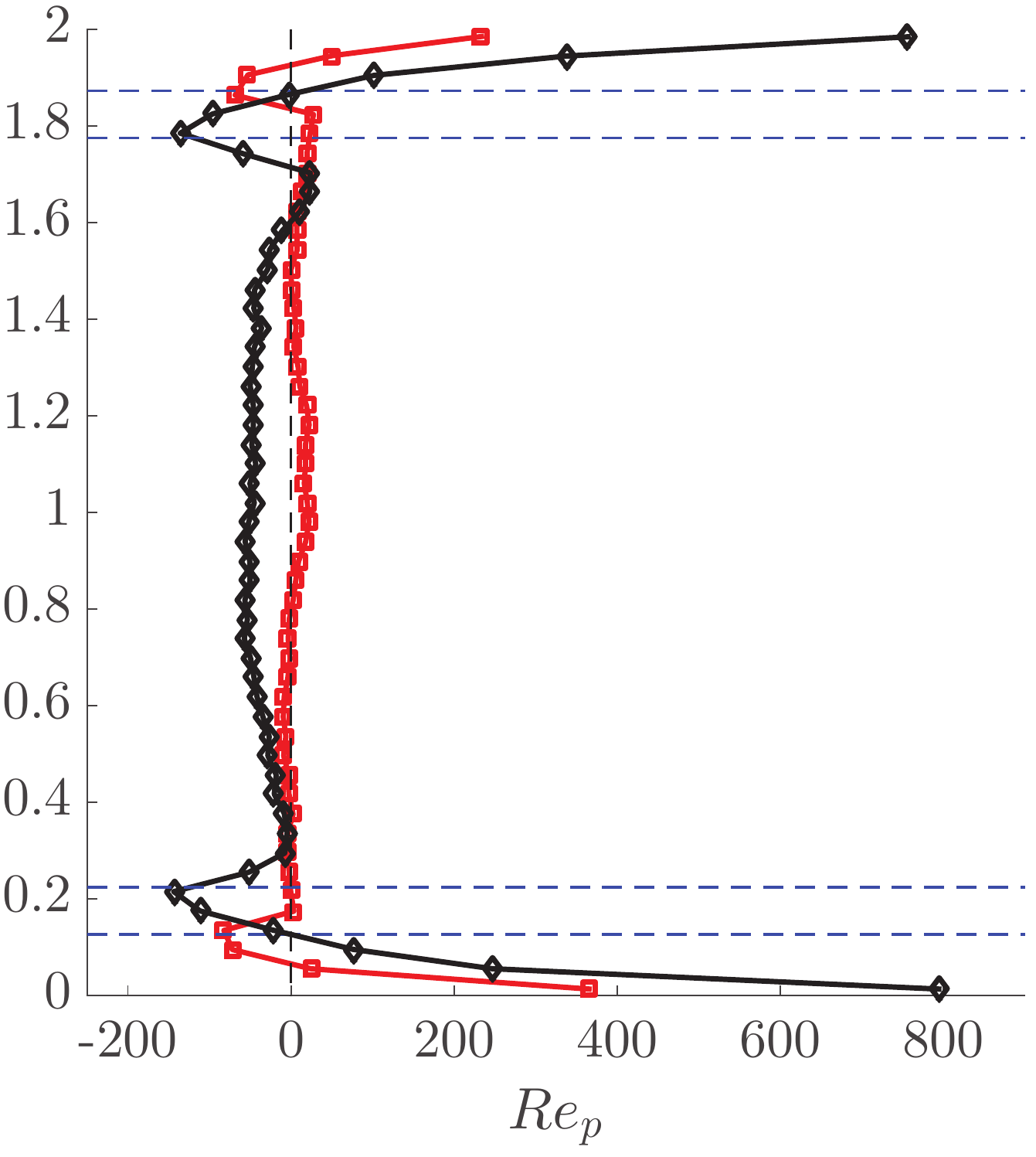}
  \caption{}
  \label{fig:3_6mm_20p_Re27000_sp10mm_Rep}
\end{subfigure}
\begin{subfigure}{.32\textwidth}
  \centering
  $z/H$ = 0.8
  \includegraphics[height=1\linewidth]{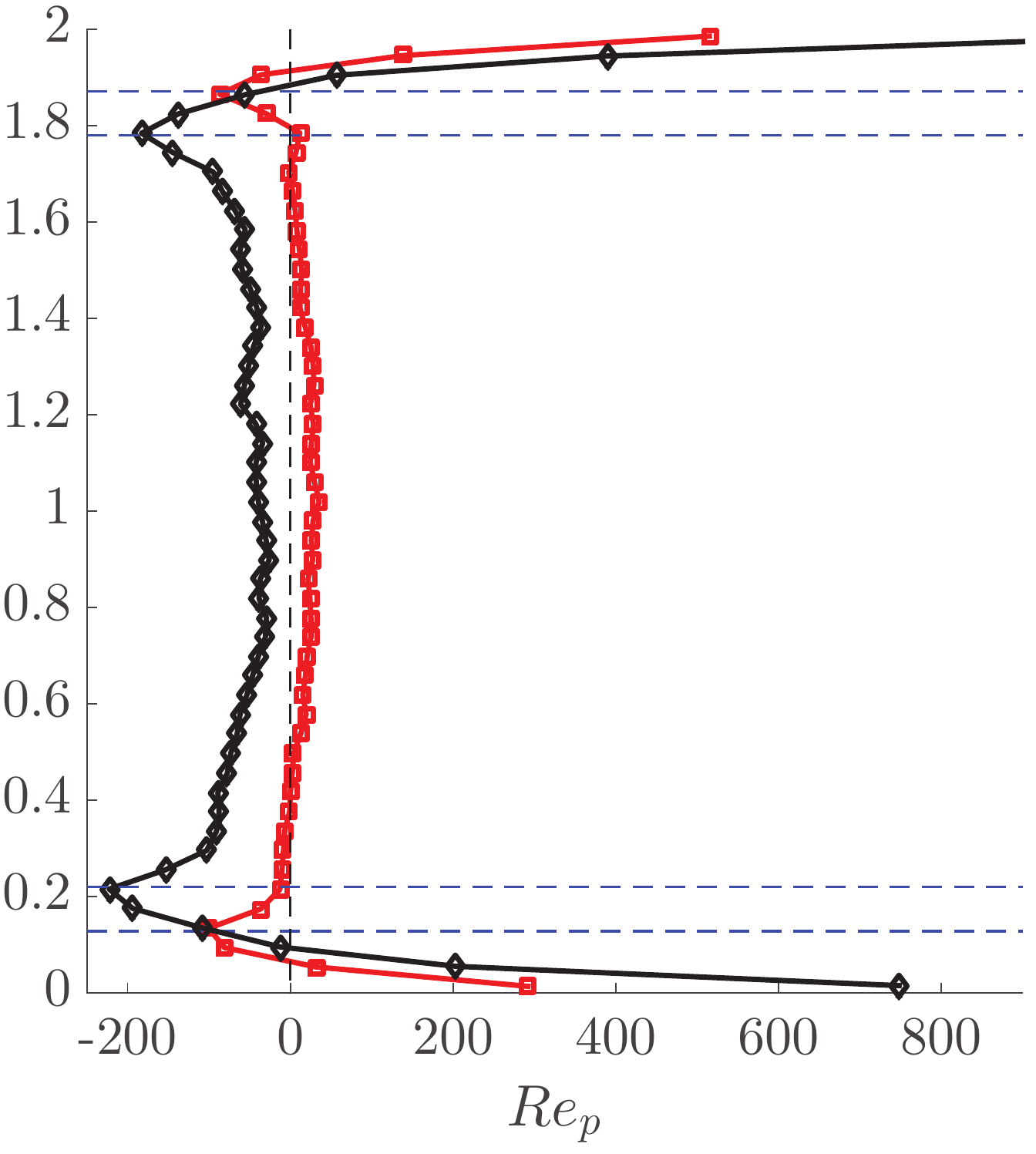}
  \caption{}
  \label{fig:3_6mm_20p_Re27000_sp20mm_Rep}
\end{subfigure}

\caption{Medium (MP, $2H/d_p$ = 16) and large-sized particles (LP, $2H/d_p$ = 9) in full suspension: $Re_{2H}\approx$ 27000, $\phi$ = 20\%, at three spanwise planes: $z/H$ = 0, 0.4 and 0.8. Shown above are profiles for the mean streamwise velocity (a)--(c), particle area concentration (d)--(f) and particle Reynolds number $Re_p$ (g)--(i). The blue dashed lines correspond to one particle diameter for each of the 2 particle sizes. The profiles in figures (a)--(c) that have been shifted to the right correspond to LP. The corresponding x-axis for these shifted profiles is on the top of the plots.}
\label{fig:3_6mm_20p_Re27000_Umean_phi_Rep}
\end{figure}

Let us first focus on the case with the highest volume fraction $\phi$ = 20\% where the flow statistics differ most from the corresponding single-phase case.
We compare the velocity profiles for both MP and LP in 3 different spanwise planes: $z/H$ = 0, 0.4 and 0.8. In the plots, the profiles for LP have been 
shifted to the right for clarity and the corresponding x-axis is on the top. Velocity profiles for the single-phase case at the same bulk $Re_{2H}$ are displayed with a green line for comparison purposes. 

Figure \ref{fig:3_6mm_20p_Re27000_sp00mm_Umean}--\ref{fig:3_6mm_20p_Re27000_sp20mm_Umean} show the mean-streamwise velocity profiles for both the fluid and particle phase. For MP, the velocity in the central region of the plane $z/H$ = 0 (see figure \ref{fig:3_6mm_20p_Re27000_sp00mm_Umean}) tends to be parabolic, characteristic of laminar flow, as compared to the flatter single-phase turbulent velocity profile; we note that a laminar-like velocity profile  has also been reported  in \citet{picano2015turbulent} from simulations of spheres in channel flow. 
On the other hand, the fluid velocity  does not change significantly when compared to the single-phase case for LP. A similar change is also seen in the plane at $z/H$ = 0.4, see figure \ref{fig:3_6mm_20p_Re27000_sp10mm_Umean}. We believe this to be a particle-size effect: the LP are around 0.1 times the full duct height ($2H/d_p$ = 9). Due to their relatively larger size, they transport momentum from high velocity regions to low velocity regions (i.e. from center towards the walls) more efficiently, which
 leads to a relatively flatter streamwise velocity profile. 

\begin{figure}
\centering

\begin{subfigure}{.32\textwidth}
  \centering
  $z/H$ = 0
  \includegraphics[height=1\linewidth]{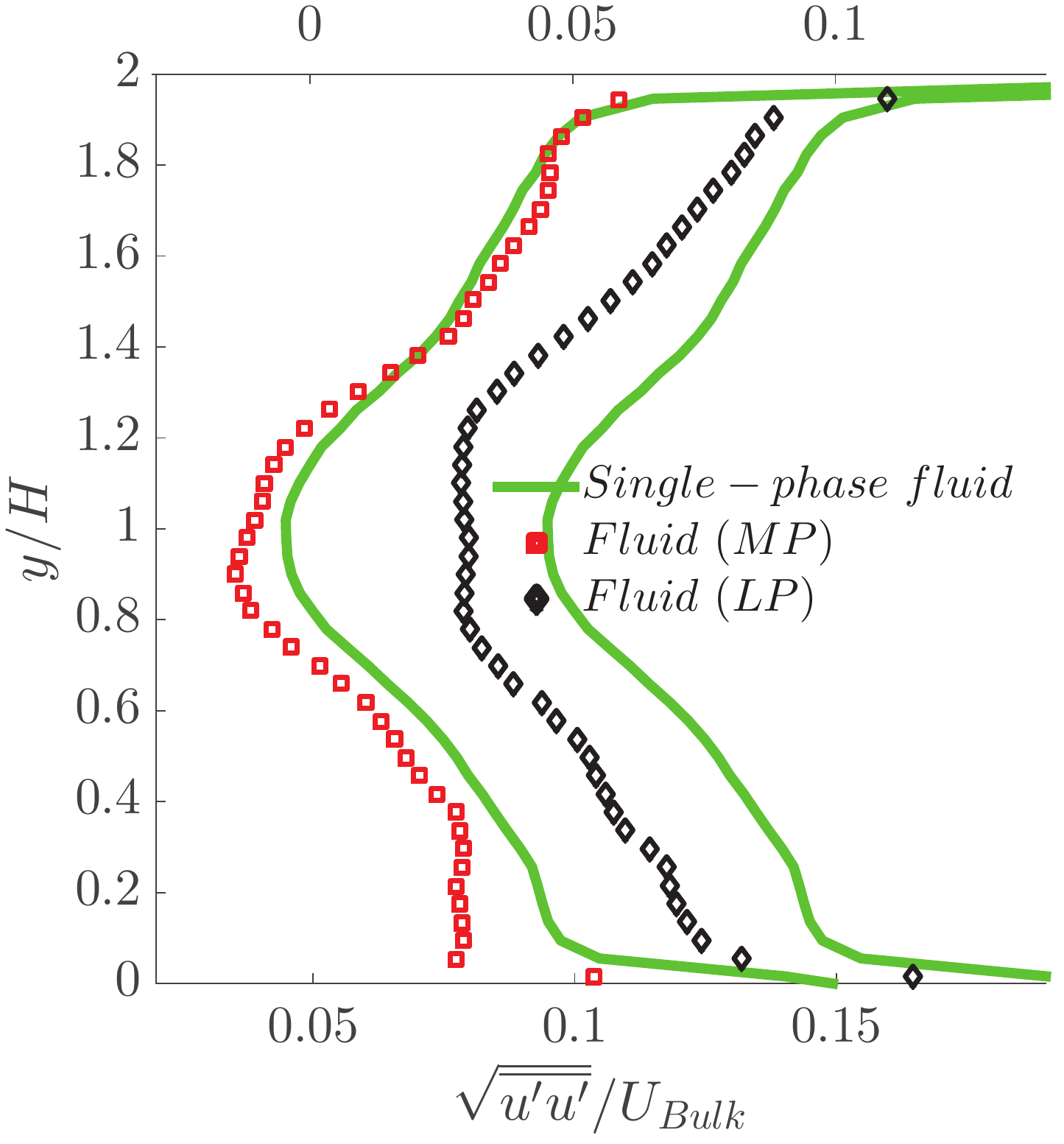}
  \caption{}
  \label{fig:3_6mm_20p_Re27000_sp00mm_urms}
\end{subfigure}%
\begin{subfigure}{.32\textwidth}
  \centering
  $z/H$ = 0.4
  \includegraphics[height=1\linewidth]{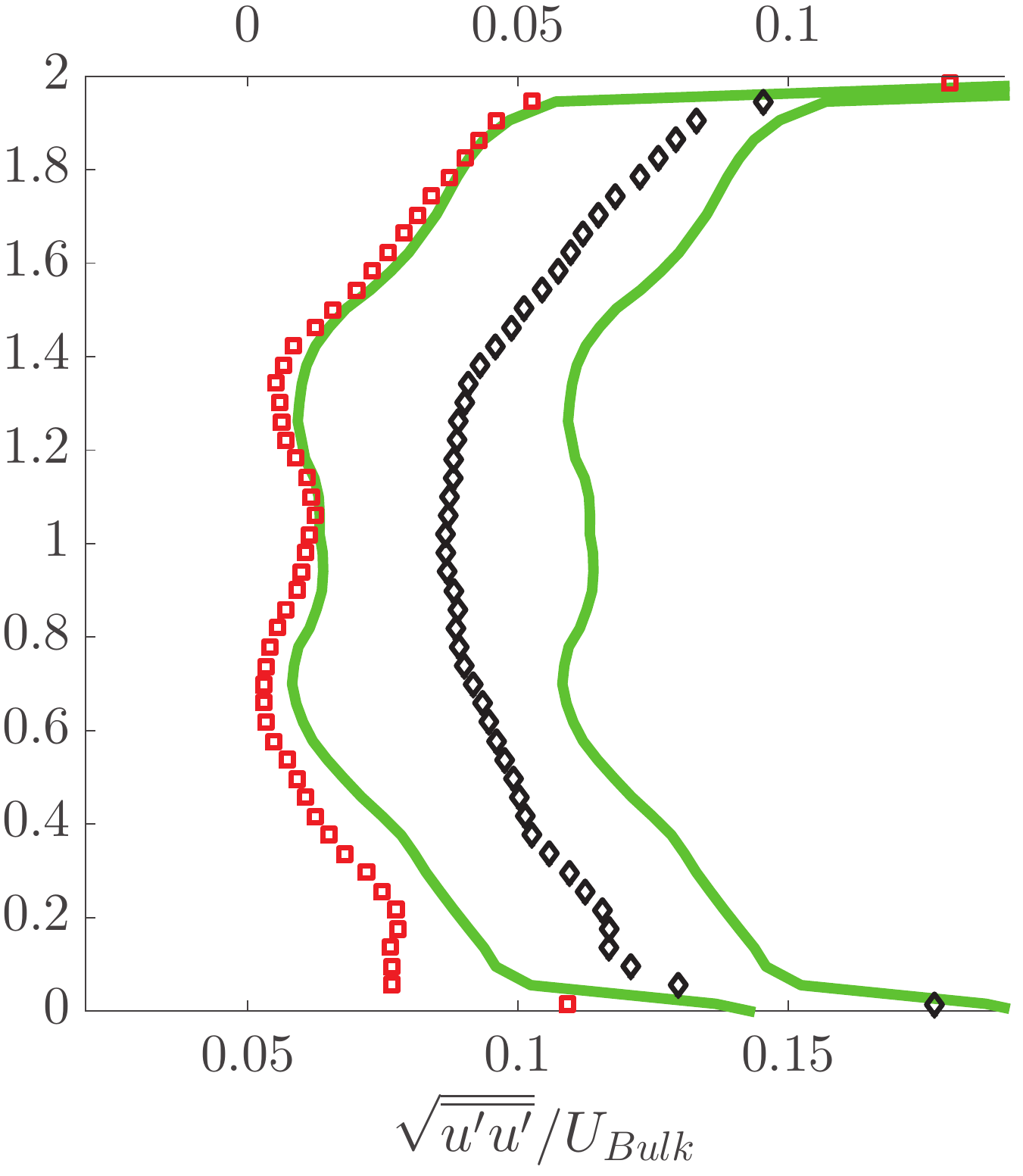}
  \caption{}
  \label{fig:3_6mm_20p_Re27000_sp10mm_urms}
\end{subfigure}
\begin{subfigure}{.32\textwidth}
  \centering
  $z/H$ = 0.8
  \includegraphics[height=1\linewidth]{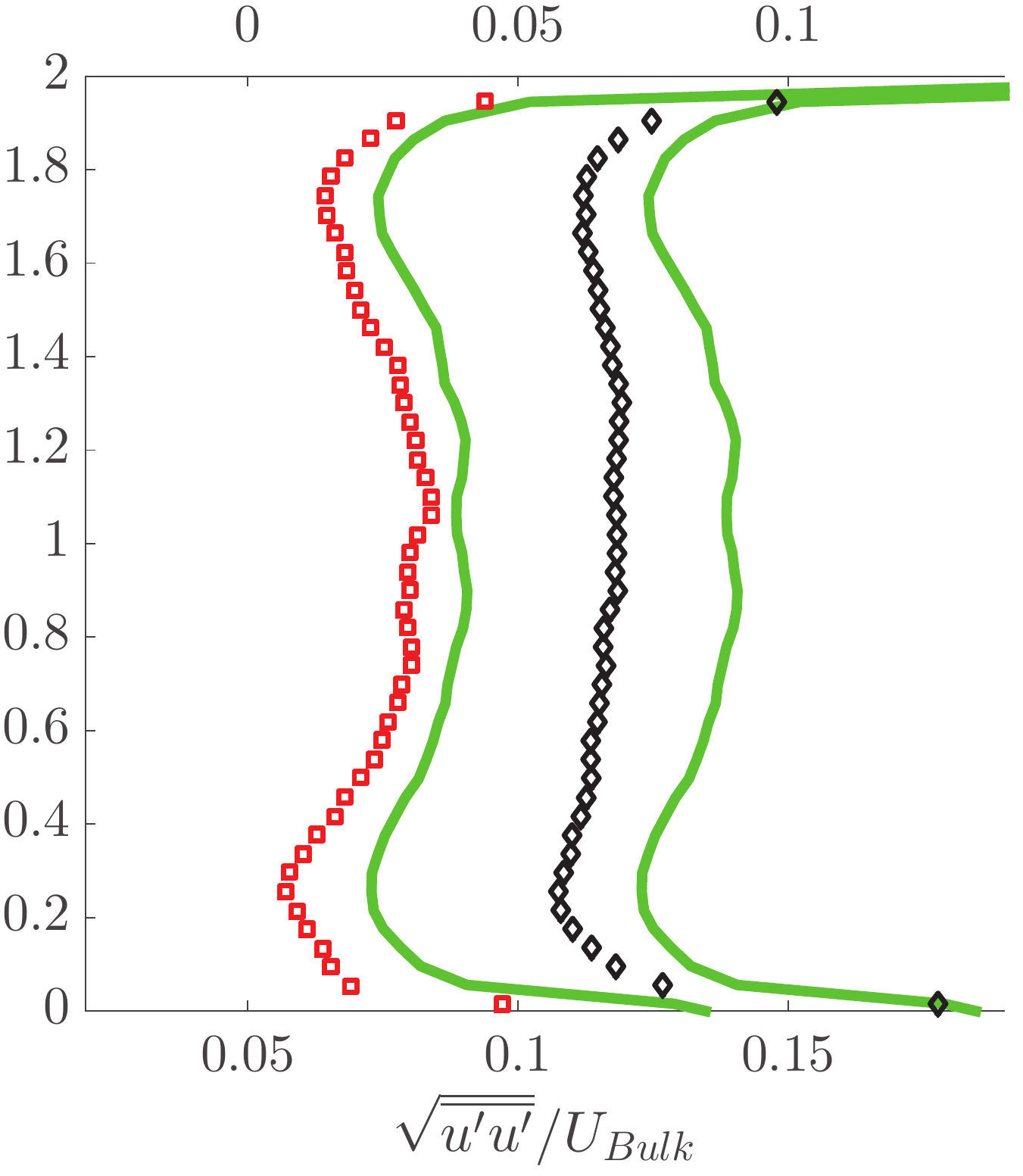}
  \caption{}
  \label{fig:3_6mm_20p_Re27000_sp20mm_urms}
\end{subfigure}

\begin{subfigure}{.32\textwidth}
  \centering
  \includegraphics[height=1\linewidth]{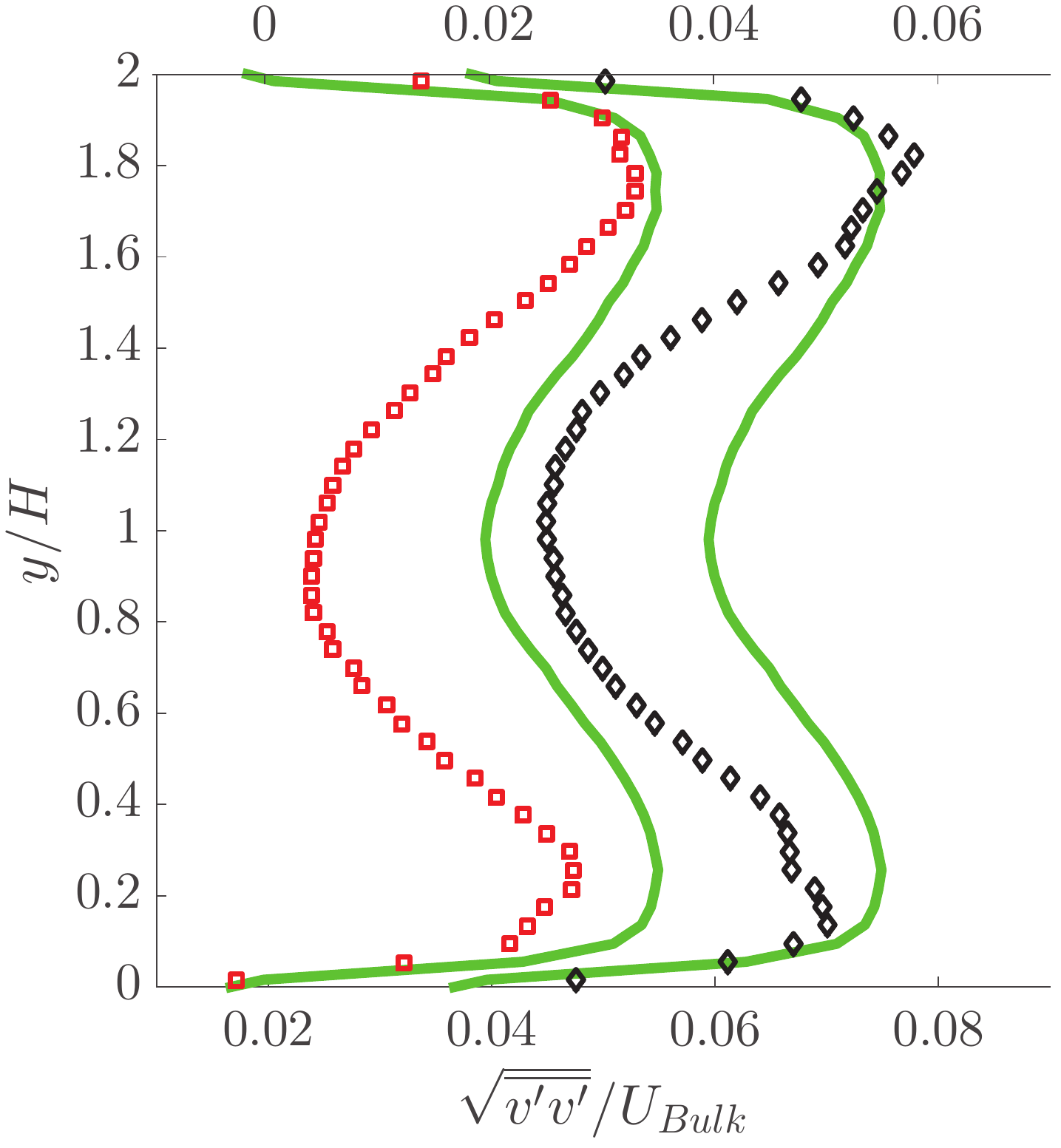}
  \caption{}
  \label{fig:3_6mm_20p_Re27000_sp00mm_vrms}
\end{subfigure}%
\begin{subfigure}{.32\textwidth}
  \centering
  \includegraphics[height=1\linewidth]{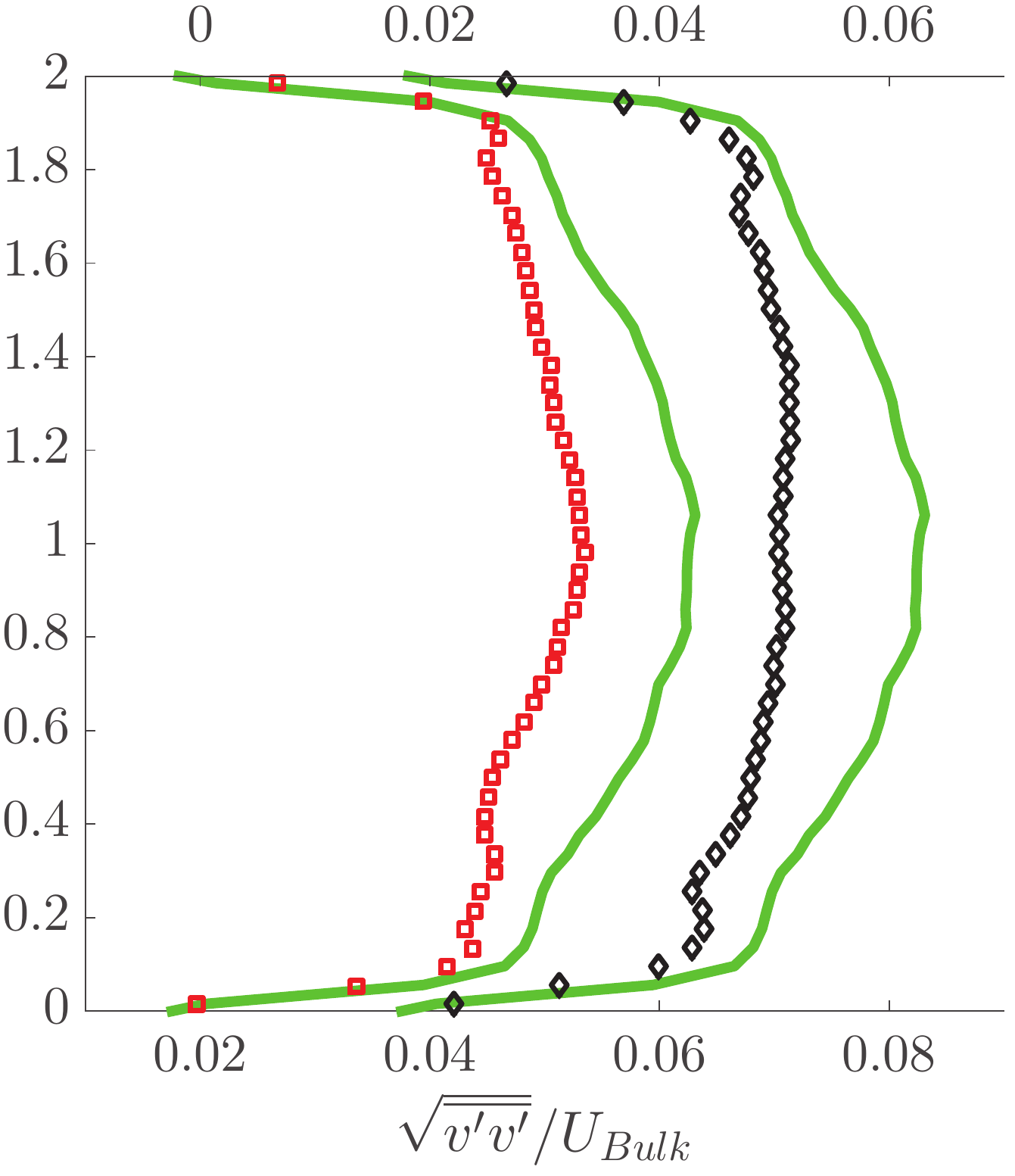}
  \caption{}
  \label{fig:3_6mm_20p_Re27000_sp10mm_vrms}
\end{subfigure}
\begin{subfigure}{.32\textwidth}
  \centering
  \includegraphics[height=1\linewidth]{pdf_figures/fig14f.pdf}
  \caption{}
  \label{fig:3_6mm_20p_Re27000_sp20mm_vrms}
\end{subfigure}

\begin{subfigure}{.32\textwidth}
  \centering
  \includegraphics[height=1\linewidth]{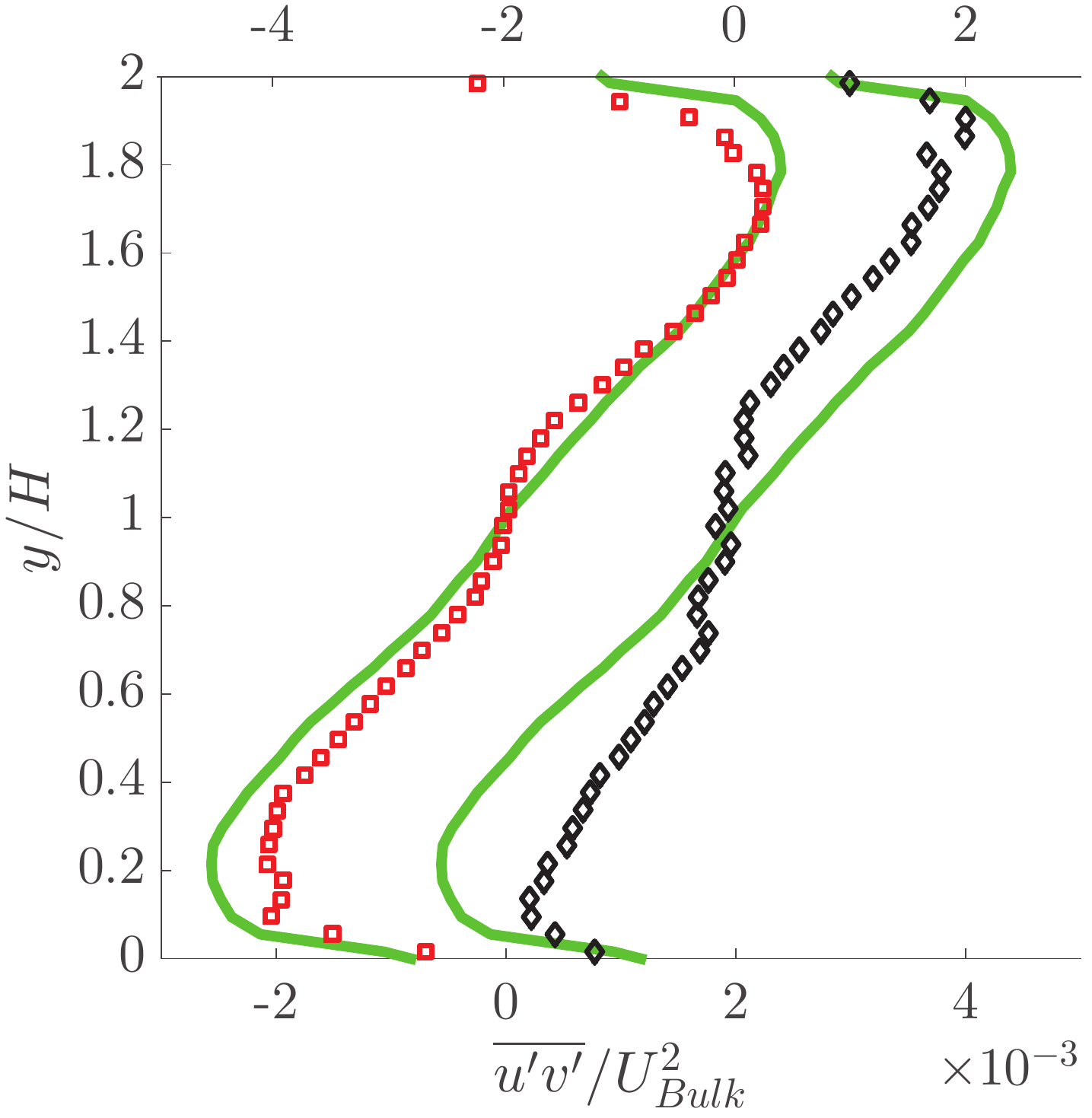}
  \caption{}
  \label{fig:3_6mm_20p_Re27000_sp00mm_uv}
\end{subfigure}%
\begin{subfigure}{.32\textwidth}
  \centering
  \includegraphics[height=1\linewidth]{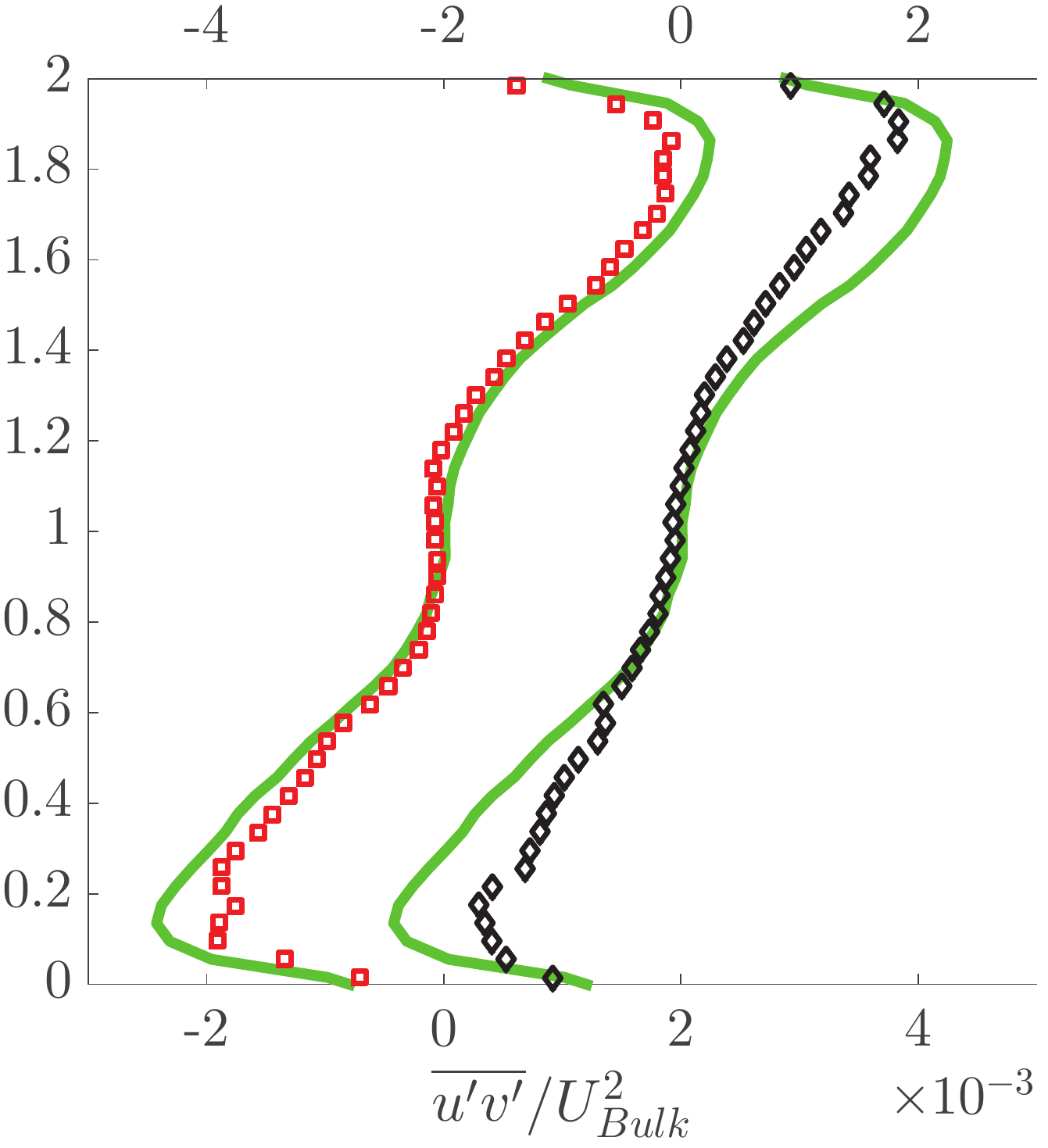}
  \caption{}
  \label{fig:3_6mm_20p_Re27000_sp10mm_uv}
\end{subfigure}
\begin{subfigure}{.32\textwidth}
  \centering
  \includegraphics[height=1\linewidth]{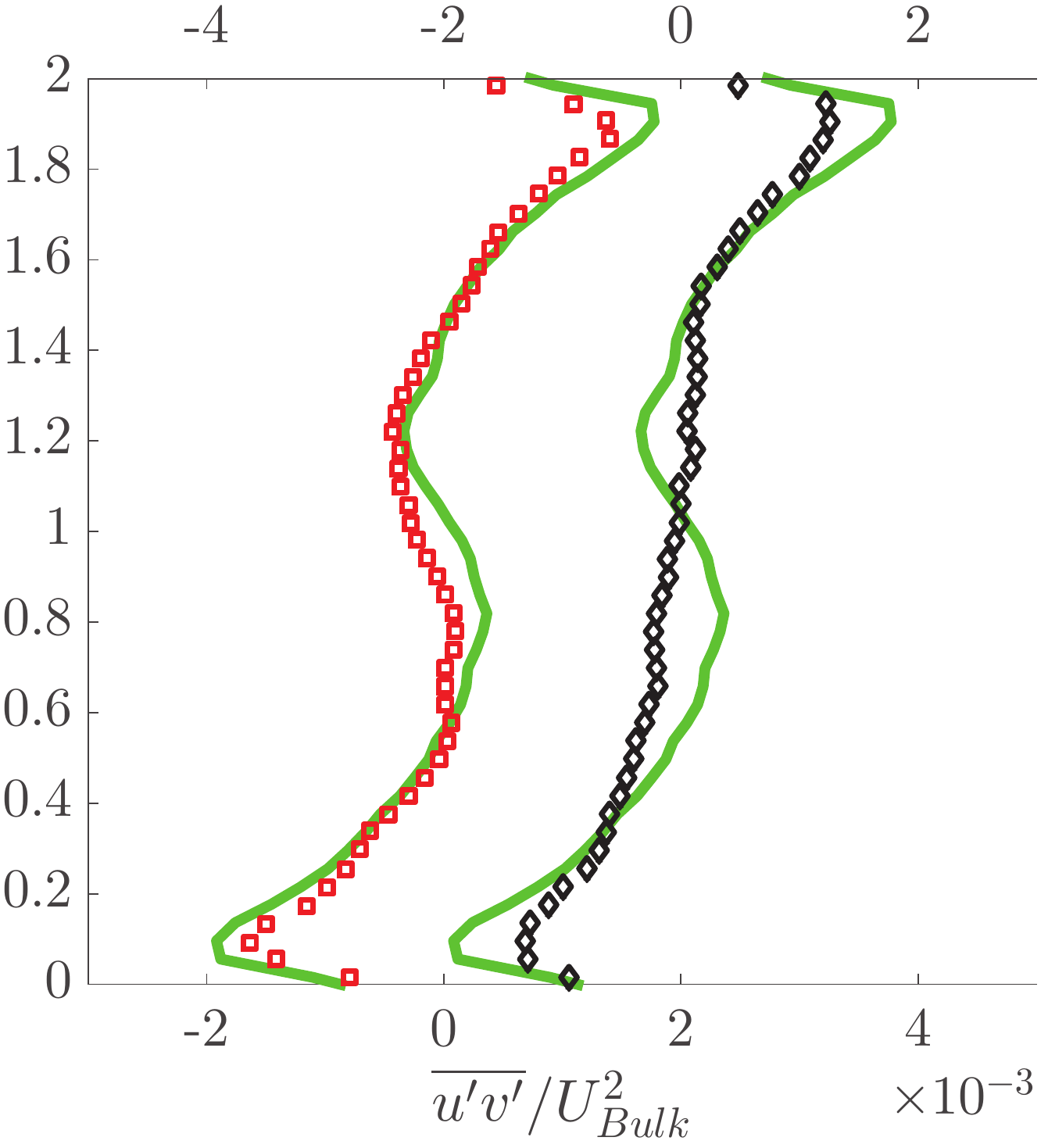}
  \caption{}
  \label{fig:3_6mm_20p_Re27000_sp20mm_uv}
\end{subfigure}

\caption{Medium (MP, $2H/d_p$ = 16) and large-sized particles (LP, $2H/d_p$ = 9) in full suspension: $Re_{2H}\approx$ 27000, $\phi$ = 20\%. Shown above are the profiles for the streamwise velocity fluctuations (a)-(c), wall-normal velocity fluctuations (d)-(f) and Reynolds stresses (g)-(i) of the fluid phase. Profiles shifted to the right correspond to LP and the corresponding x-axis is on the top.}
\label{fig:3_6mm_20p_Re27000_urms_vrms_uv}
\end{figure}

The velocity in the plane closest to the sidewall ($z/H$ = 0.8 in figure \ref{fig:3_6mm_20p_Re27000_sp20mm_Umean}) reduces at all elevations. This is expected because an increase of the flow rate in the duct core (close to $z/H$ = 0)
should be compensated by a decrease in flow rates close to the sidewalls if the volume flow rate is constant. 
For LP, 
the absence of velocity maxima of the fluid mean streamwise velocity profile at the top and bottom suggests that the secondary flow, i.e.\ corner vortices that carry streamwise momentum from the center towards the corners, has weakened. This may be explained by the reduction of  the turbulent activity, responsible for the generation of this secondary flow, which will be documented later.

The particle mean-streamwise velocity for both MP and LP is almost the same as that of the fluid except close to the top and bottom walls due to the slip of the particles at the walls \cite[also seen in the simulations by][]{picano2015turbulent,costa2016universal}. As mentioned above,  MP and LP have diameters of the order $d_p/\delta_{\nu}\approx$ 90 and 165, with $\delta_{\nu}$ the viscous length scale of the single-phase flow at the same $Re_{2H}$. This means that both particles extend well into the log layer of the reference single-phase flow. 

To understand the modification in the mean-streamwise velocity, we display
in figures \ref{fig:3_6mm_20p_Re27000_sp00mm_phi}--\ref{fig:3_6mm_20p_Re27000_sp20mm_phi} the concentration profiles for MP and LP side by side at the same spanwise position. 
The wall layering at the top and bottom walls is clear.
Particles in the duct mid-plane tend to migrate towards the center (see figure \ref{fig:3_6mm_20p_Re27000_sp00mm_phi}). This is due to inertial shear-induced migration as explained in \citet{fornari2016effect}. The concentration is more uniform close to the side-walls (see figure \ref{fig:3_6mm_20p_Re27000_sp20mm_phi}) due to the high intensity streaky turbulence in that region as observed in PIV movies. These vigorous velocity fluctuations tend to keep the particles in uniform suspension. However, the concentration of the LP at $z/H$ = 0.8 is lower than that of the MP because this plane nearly corresponds to the particle depleted layer, one diameter away from the sidewall. Note again that the concentration profiles appear a little skewed towards the bottom half due to sedimentation.

The difference between the particle velocity $U_p$ and the fluid velocity $U_f$, i.e.\ the apparent particle slip velocity, is depicted in figure \ref{fig:3_6mm_20p_Re27000_sp00mm_Rep}--\ref{fig:3_6mm_20p_Re27000_sp20mm_Rep} in terms of the particle Reynolds number $Re_p=(U_p-U_f)d_p/\nu_f$. 
As mentioned before, the slip velocity has its maximum value at the wall  ($Re_p\approx$ 500). \sz{This quantity is most likely overestimated near the wall, as the apparent particle motion is assumed to be only translational as measured using the PTV technique. In the near-wall region, particles are more likely to rotate according to the local fluid vorticity so that the local velocity of the solid phase is closer to the fluid phase yielding lower particle Reynolds number. The estimate of $Re_p$ becomes more accurate towards the core as the particle mean angular velocity becomes small.} Close to the center of the duct, the MP appear to travel with the mean fluid velocity resulting in a low $Re_p$. However, in the same region, the LP are slower than the fluid, with $Re_p$ of the order of 50. 
This could be due to the relatively higher relaxation time of the LP (see Stokes number in figure \ref{fig:Stokes number}) and to the higher concentration in the duct core. 

A finite particle Reynolds number $Re_p$ 
leads, in laminar flows, to inertial shear-thickening as described in \citet{picano2013shear}: finite inertia at the particle scale affects the particle relative motions and effectively increases the excluded volume around each particle. This increase in the effective excluded volume results in an enhanced effective viscosity.
For a fully turbulent flow, it is worthwhile to consider the theory of \citet{costa2016universal} who showed that a layer of particles, flowing near the wall with significant apparent slip velocity, is responsible for an increase in drag which cannot be modeled by an effective suspension viscosity. 
This theory assumes that the flow domain can be split into two regions: (i) a region close to the wall where the difference between the mean velocity of the two phases is 
substantial and (ii) a region away from the wall, where the mean flow is well represented by the continuum limit of a Newtonian fluid with an effective viscosity. 
This has been shown to predict the mean velocity and drag of turbulent plane channel flow. 
However, the particles studied here are larger than those in \cite{costa2016universal}  and a square duct has 2 inhomogeneous directions
and the two additional walls on the sides prevent the particle lateral motion. This confinement leads to a peak in the particle concentration profile at the duct core at high $\phi$, as shown in \citet{fornari2017suspensions}. \sz{The presence of weak secondary flow in the square duct would also influence the momentum balance, thus causing further deviations from a channel flow, where no secondary motion is seen.}

The root mean square (rms) of the fluid velocity fluctuations for both MP and LP is displayed in figure \ref{fig:3_6mm_20p_Re27000_urms_vrms_uv}. As in the previous figure, the statistics for LP are shifted towards the right for clarity. The fluctuations in the streamwise direction reduce, marginally for MP and more substantially for LP, in all the three planes considered (see figures \ref{fig:3_6mm_20p_Re27000_sp00mm_urms}--\ref{fig:3_6mm_20p_Re27000_sp20mm_urms}). 
This reduction is higher on the bottom half of the duct due to the asymmetry in the concentration distribution caused by gravity. 
The wall-normal velocity fluctuations (see figure \ref{fig:3_6mm_20p_Re27000_sp00mm_vrms}--\ref{fig:3_6mm_20p_Re27000_sp20mm_vrms}) reduce by nearly the same amount for both MP and LP in all the three spanwise planes examined. The reduction is highest close to the center, which could be related to the peak in particle concentration in this region. The  measured fluid Reynolds stress component (figure \ref{fig:3_6mm_20p_Re27000_sp00mm_uv}--\ref{fig:3_6mm_20p_Re27000_sp20mm_uv}) reduces more for LP than MP as compared to the single-phase case. Thus, it appears that LP are more effective in damping the turbulent fluctuations. \sz{This relatively larger damping of turbulence in LP might be the reason for the reduction of secondary motion and consequently the disappearance of the maxima in the mean streamwise velocity profile near the corners in the $z/H$ = 0.8 plane (cf figure \ref{fig:3_6mm_20p_Re27000_sp20mm_Umean}).}

Despite the larger reduction of the 3 measured components (out of 6) of the fluid Reynolds stress: $\overline{u'u'}$, $\overline{v'v'}$ and $\overline{u'v'}$ for LP over MP, the pressure drop for LP is larger than for MP. In particular, it increases by around 7\% compared to the single-phase case for LP whereas the increase is just 1\% for MP. \sz{At this highest $Re_{2H}$, the uncertainty in measuring the friction factor is of the order of $\pm$1\% (cf figure \ref{fig:f_vs_Re plot}).}
Recalling that the total momentum transfer 
arises from the contribution of viscous, turbulent and particle-induced stresses \citep{lashgari2014laminar},  it can be inferred that the reduction in turbulent stresses is compensated by an even higher increase in the particle-induced stress (assuming that the viscous stresses are smaller at such high $Re_{2H}$). As the particle-induced stress is  higher for LP than MP, the overall stress is higher for LP than MP despite the reduced turbulence activity in the flow seeded with LP.\\

\begin{figure}
\centering

\begin{subfigure}{.32\textwidth}
  \centering
  $z/H$ = 0
  \includegraphics[height=1\linewidth]{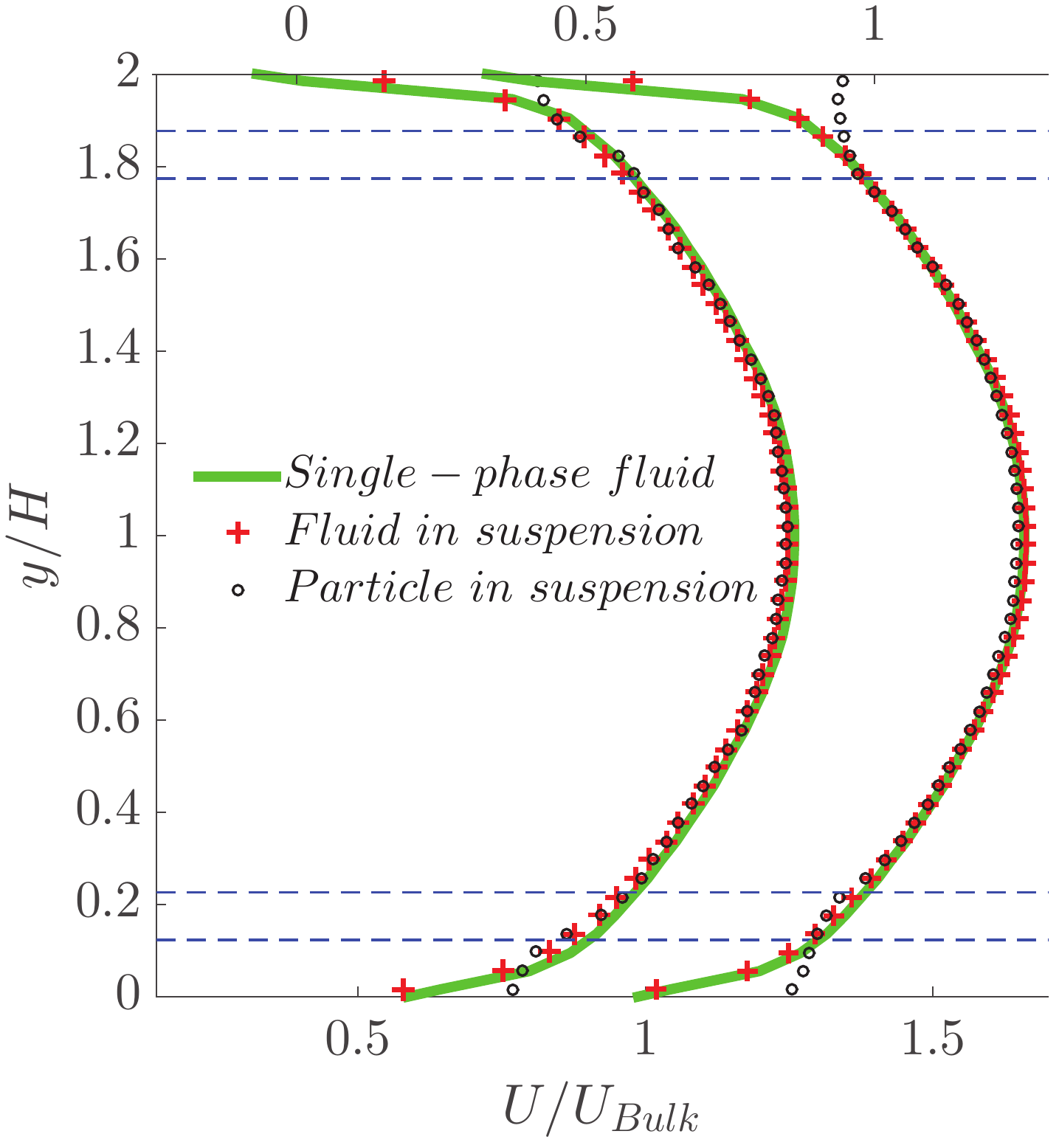}
  \caption{}
  \label{fig:3_6mm_5p_Re27000_sp00mm_Umean}
\end{subfigure}%
\begin{subfigure}{.32\textwidth}
  \centering
  $z/H$ = 0.4
  \includegraphics[height=1\linewidth]{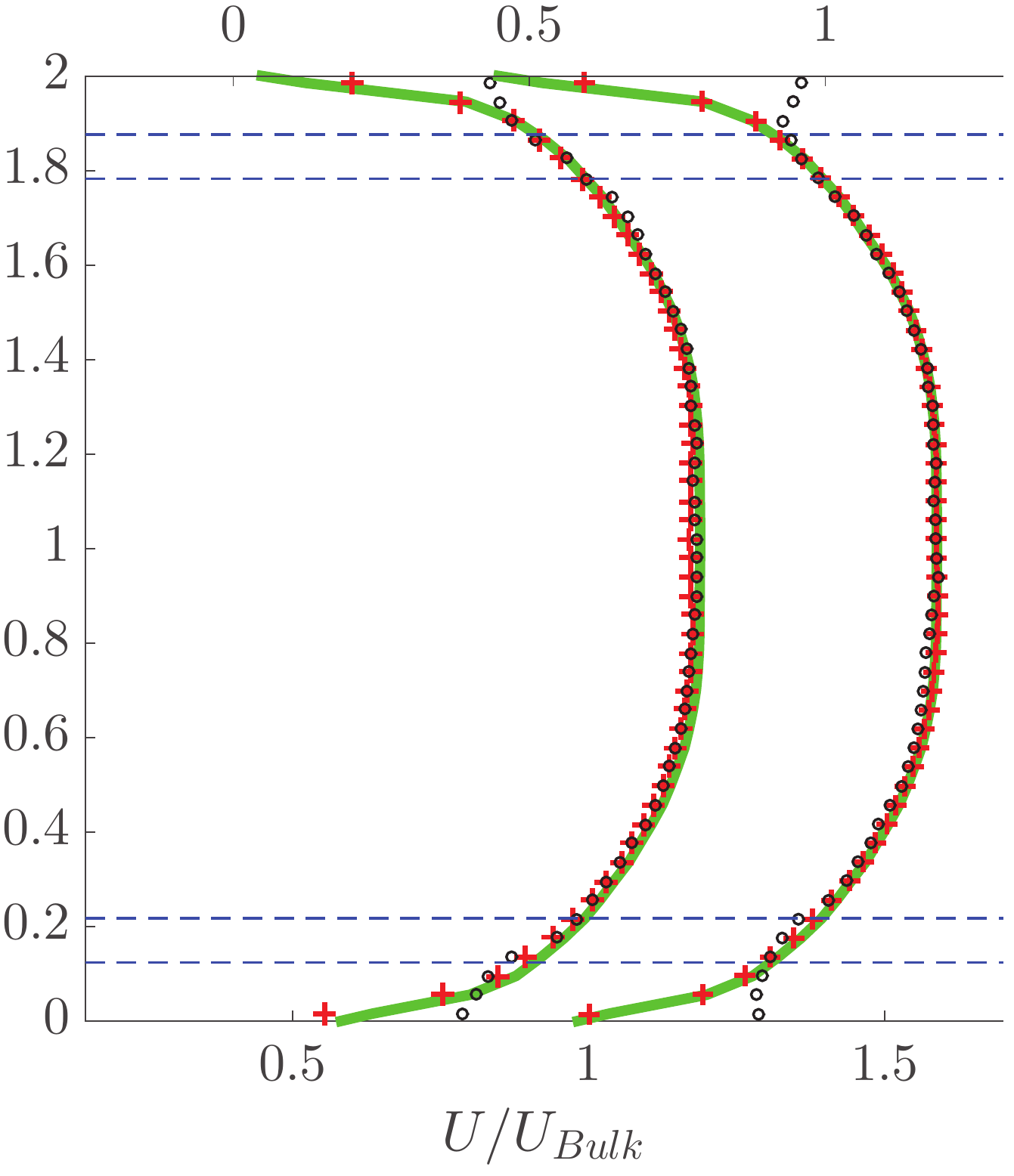}
  \caption{}
  \label{fig:3_6mm_5p_Re27000_sp10mm_Umean}
\end{subfigure}
\begin{subfigure}{.32\textwidth}
  \centering
  $z/H$ = 0.8
  \includegraphics[height=1\linewidth]{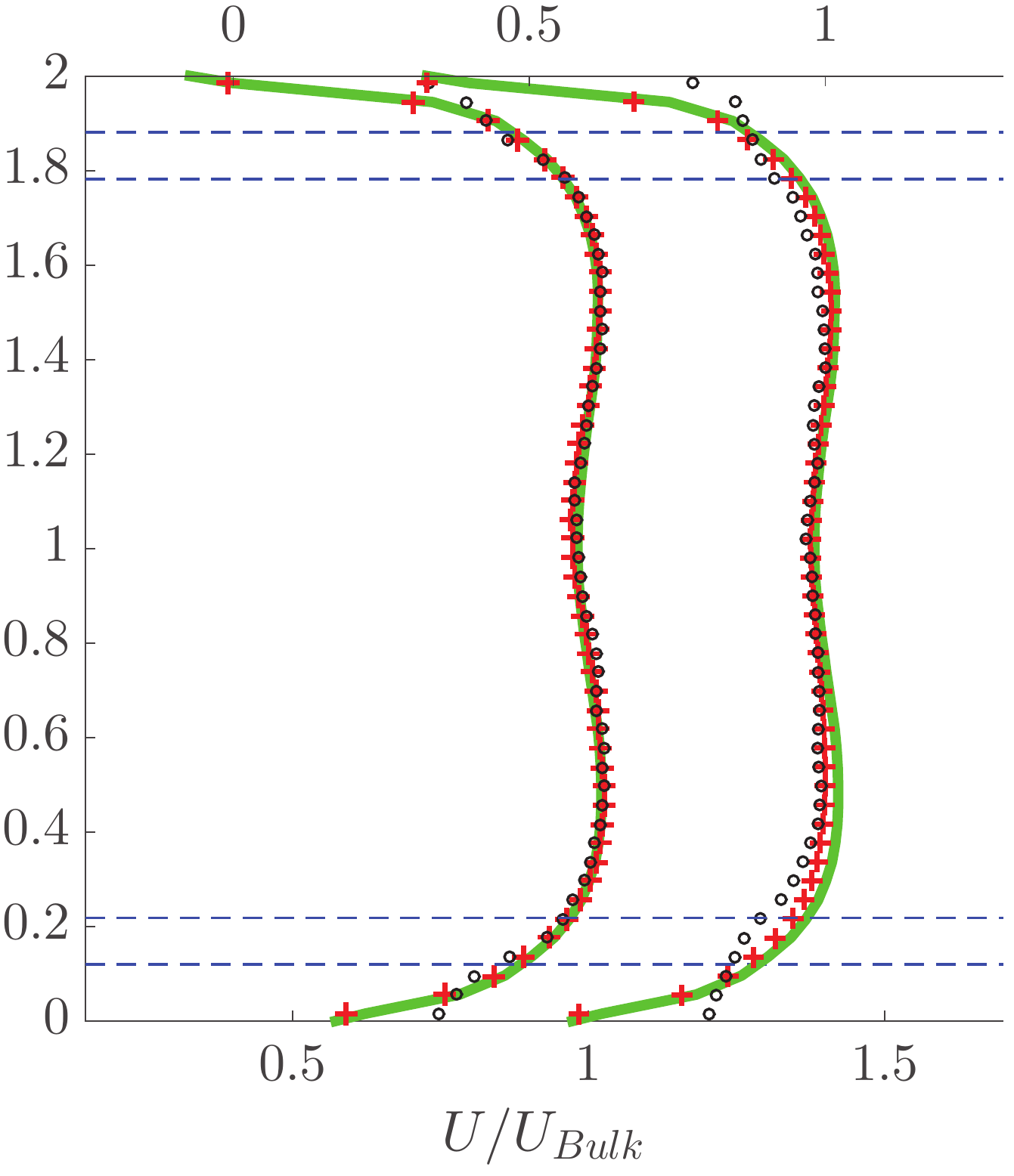}
  \caption{}
  \label{fig:3_6mm_5p_Re27000_sp20mm_Umean}
\end{subfigure}

\begin{subfigure}{.32\textwidth}
  \centering
  \includegraphics[height=1\linewidth]{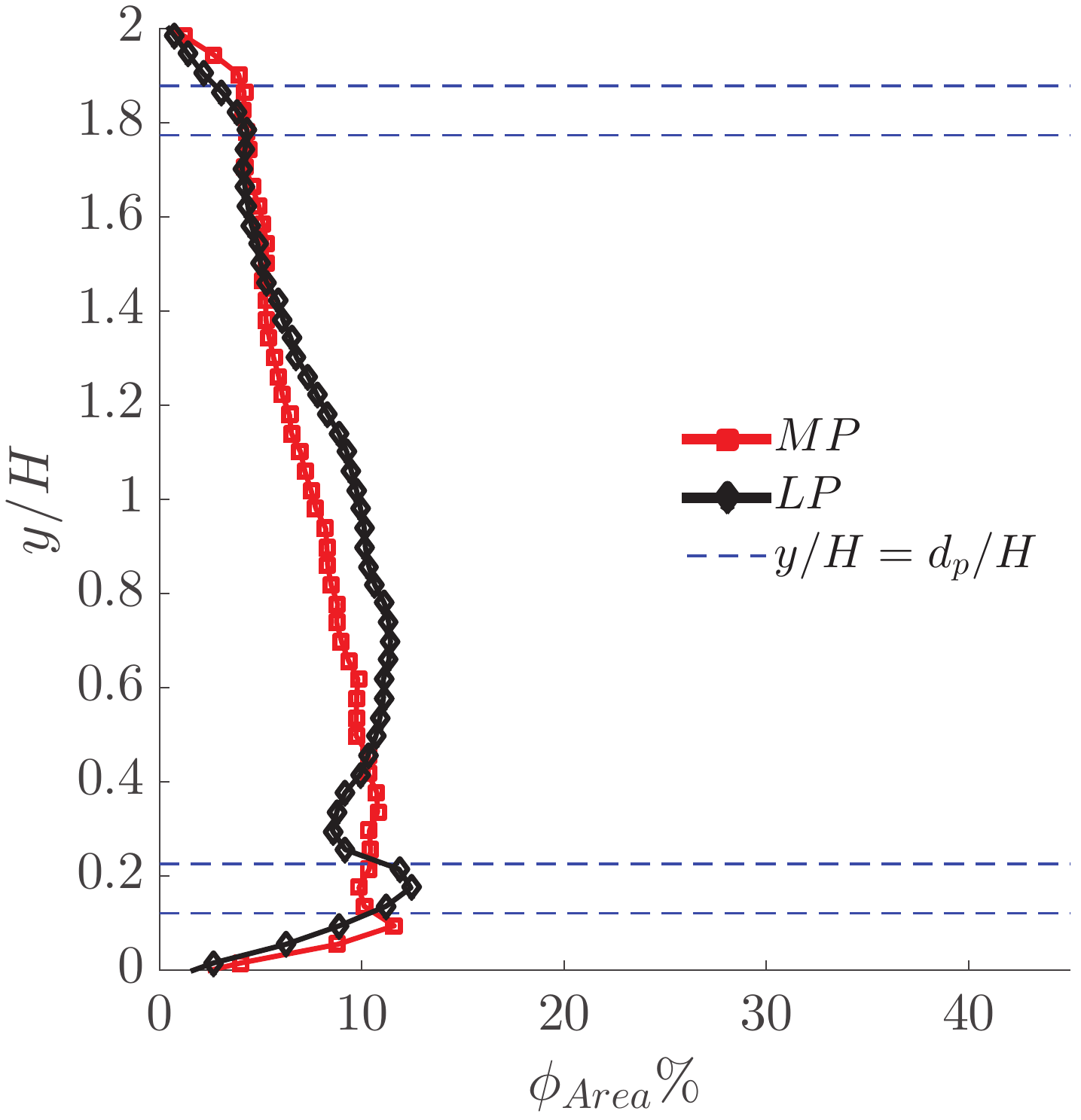}
  \caption{}
  \label{fig:3_6mm_5p_Re27000_sp00mm_phi}
\end{subfigure}%
\begin{subfigure}{.32\textwidth}
  \centering
  \includegraphics[height=1\linewidth]{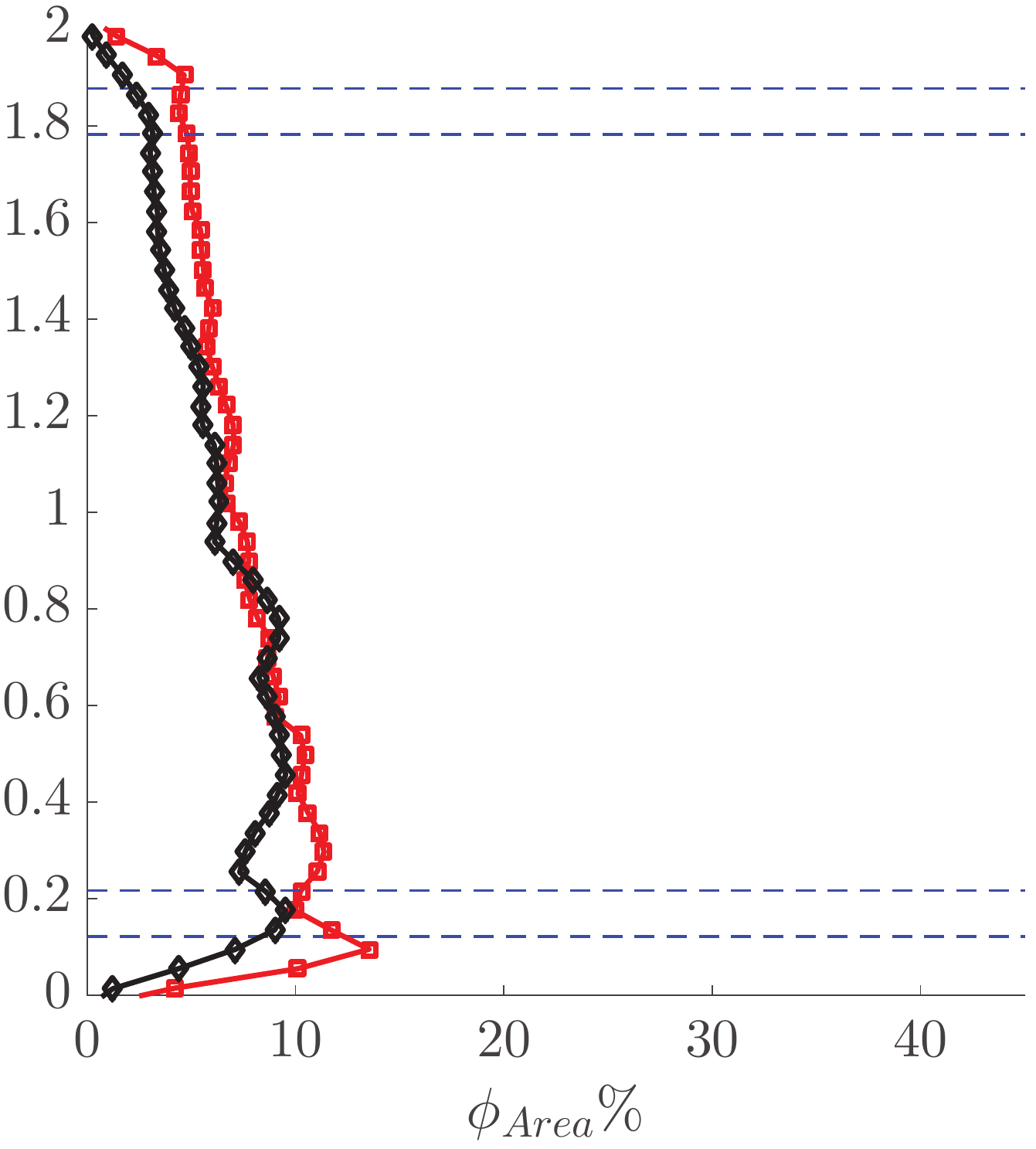}
  \caption{}
  \label{fig:3_6mm_5p_Re27000_sp10mm_phi}
\end{subfigure}
\begin{subfigure}{.32\textwidth}
  \centering
  \includegraphics[height=1\linewidth]{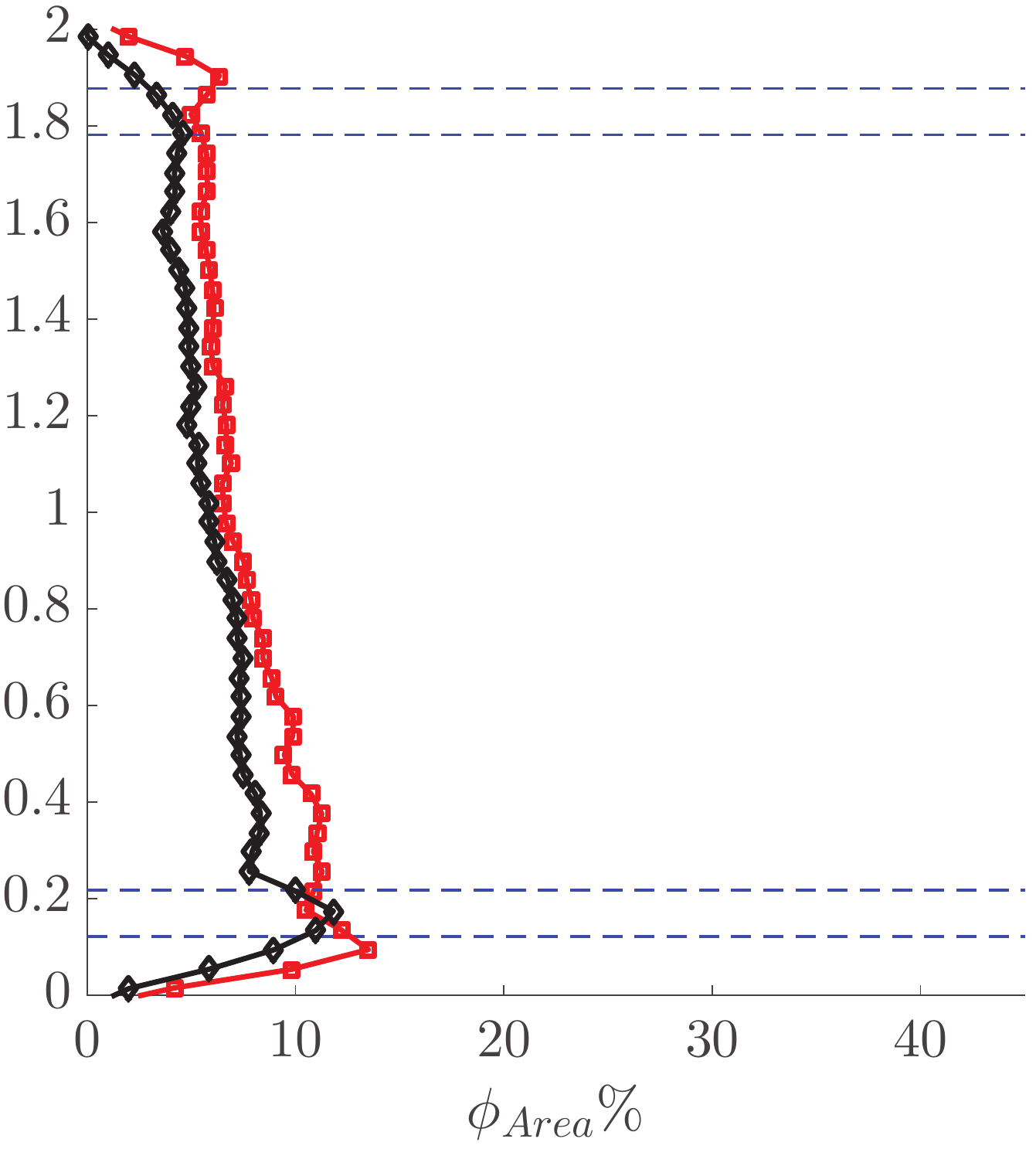}
  \caption{}
  \label{fig:3_6mm_5p_Re27000_sp20mm_phi}
\end{subfigure}

\begin{subfigure}{.32\textwidth}
  \centering
  \includegraphics[height=1\linewidth]{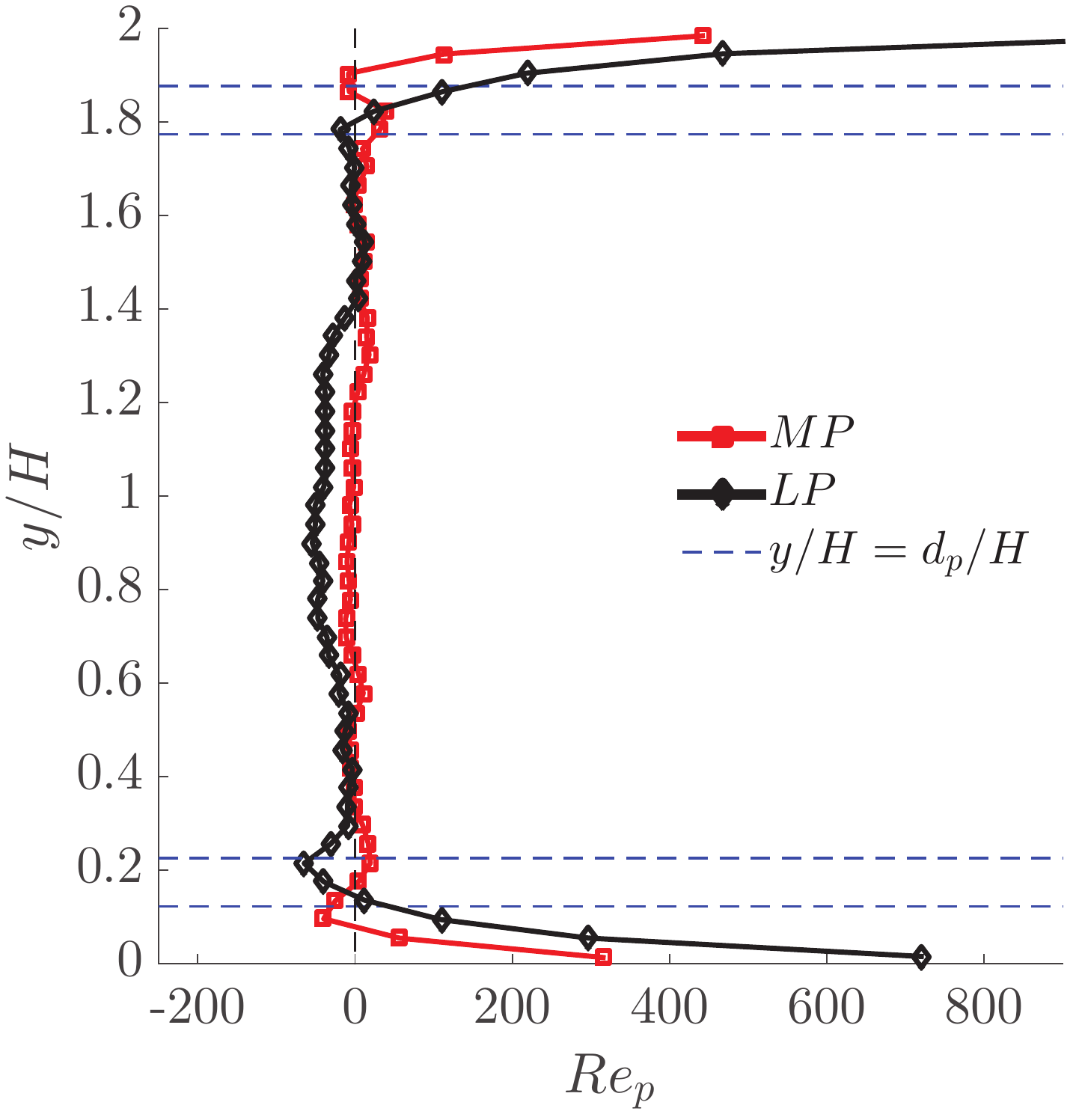}
  \caption{}
  \label{fig:3_6mm_5p_Re27000_sp00mm_Rep}
\end{subfigure}%
\begin{subfigure}{.32\textwidth}
  \centering
  \includegraphics[height=1\linewidth]{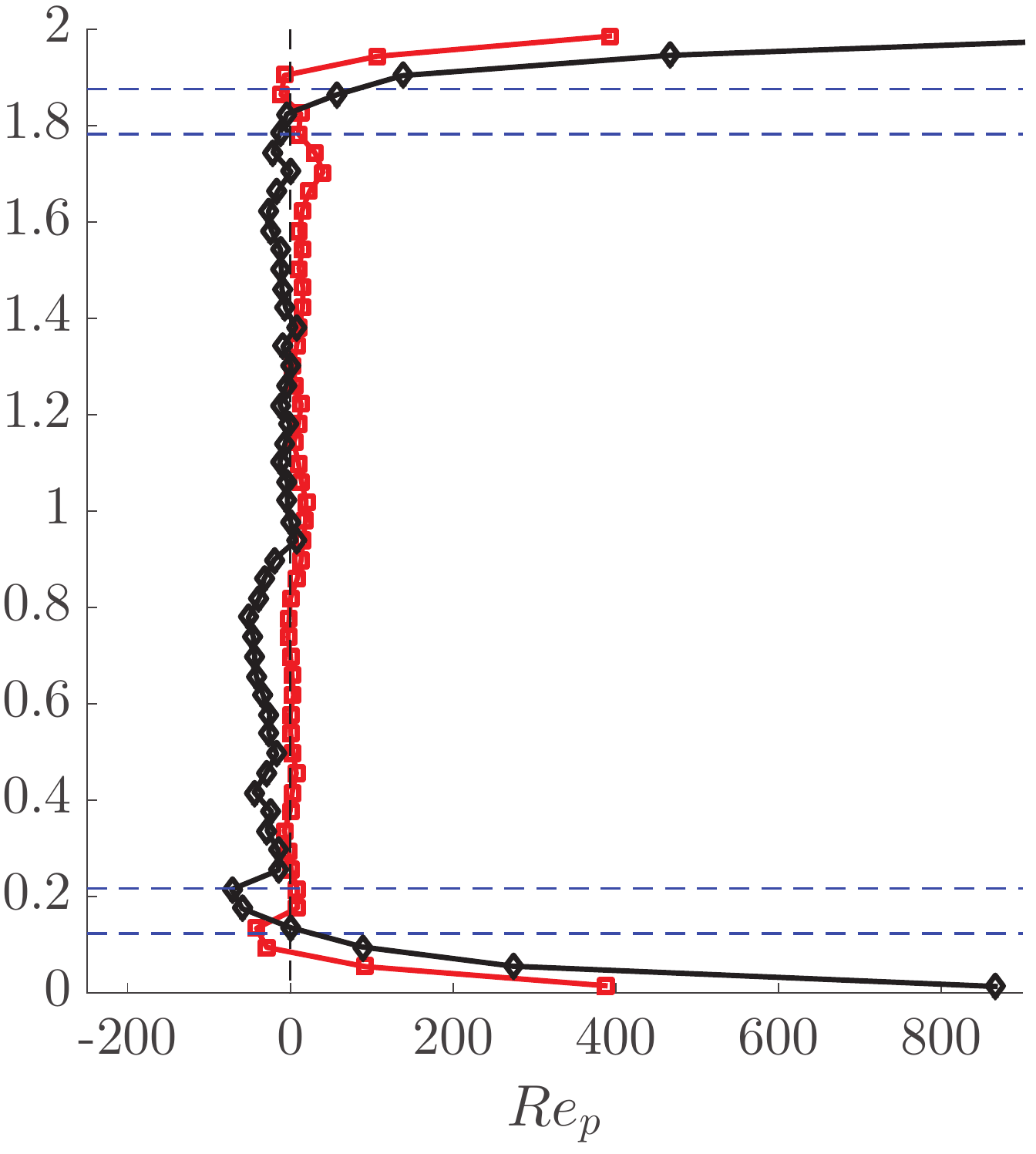}
  \caption{}
  \label{fig:3_6mm_5p_Re27000_sp10mm_Rep}
\end{subfigure}
\begin{subfigure}{.32\textwidth}
  \centering
  \includegraphics[height=1\linewidth]{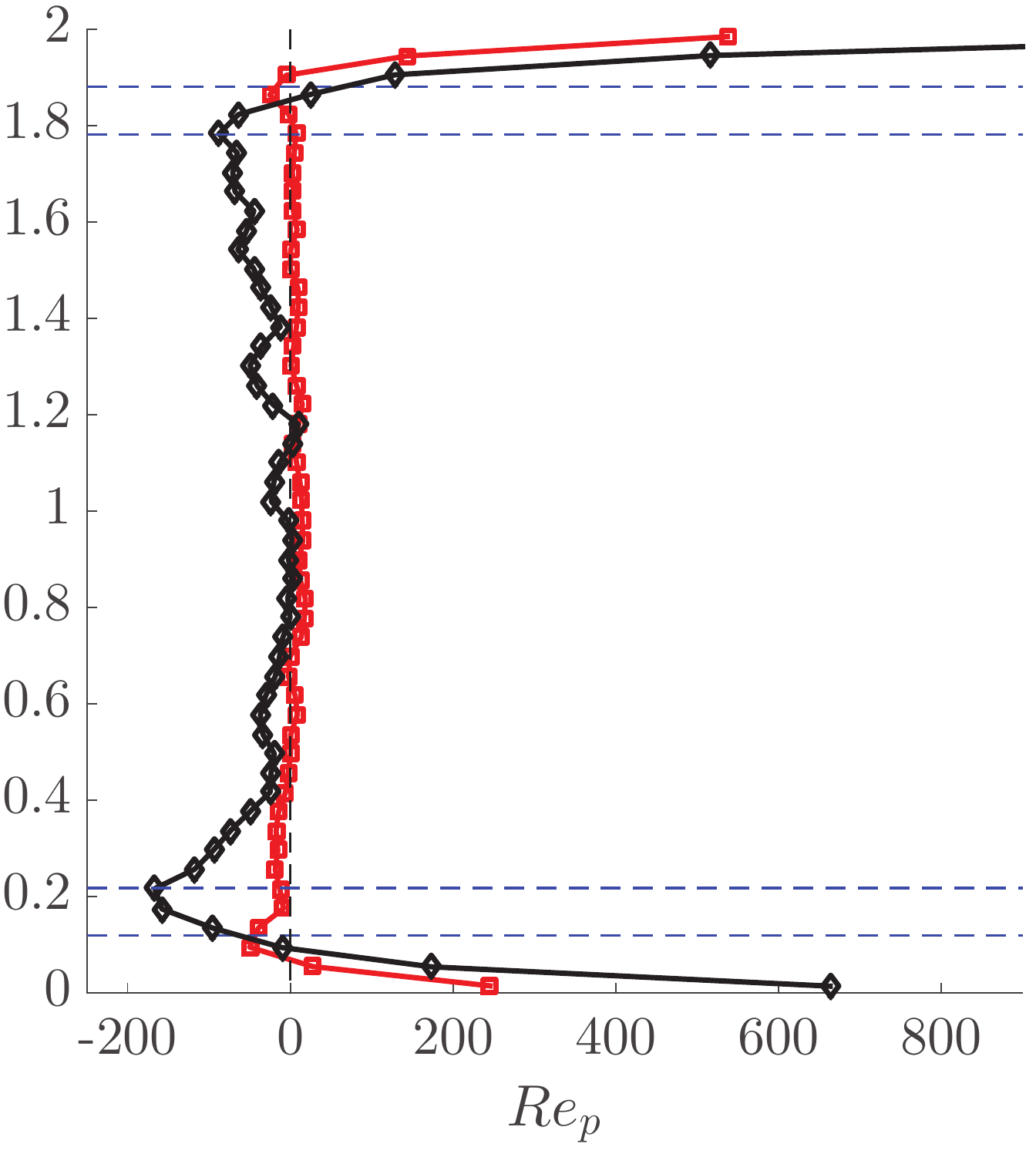}
  \caption{}
  \label{fig:3_6mm_5p_Re27000_sp20mm_Rep}
\end{subfigure}

\caption{MP ($2H/d_p$ = 16) and LP ($2H/d_p$ = 9) in full suspension: $Re_{2H}\approx$ 27000, $\phi$ = 5\%. Shown above are profiles for the mean streamwise velocity (a)--(c), particle area concentration (d)--(f) and particle Reynolds number $Re_p$ (g)--(i). The blue dashed lines correspond to one particle diameter for each of the 2 particle sizes. Profiles in figures (a)--(c), shifted to the right correspond to LP and the corresponding x-axis is on the top.}
\label{fig:3_6mm_5p_Re27000_Umean_phi_Rep}
\end{figure}

\begin{figure}
\centering

\begin{subfigure}{.32\textwidth}
  \centering
  $z/H$ = 0
  \includegraphics[height=1\linewidth]{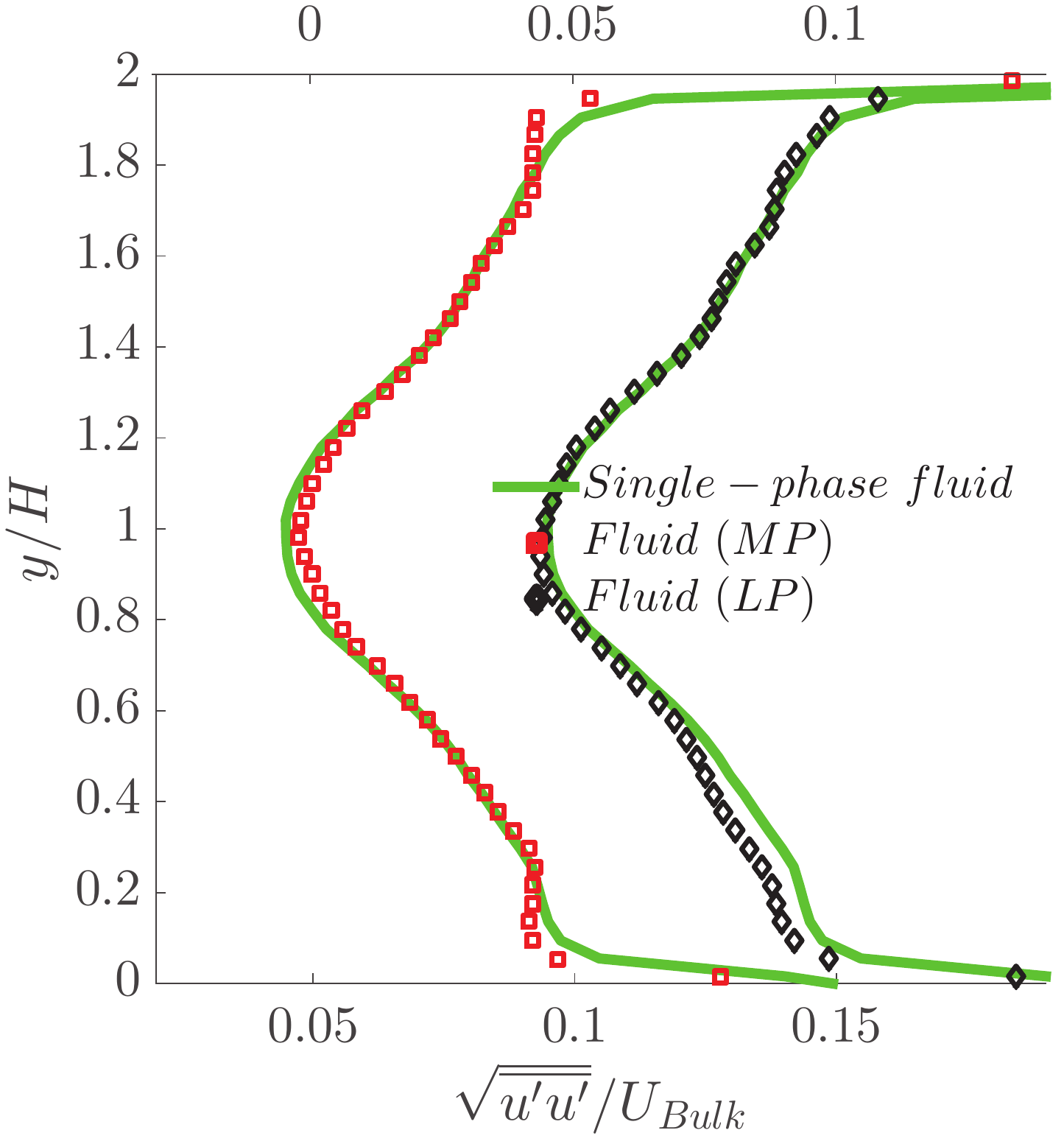}
  \caption{}
  \label{fig:3_6mm_5p_Re27000_sp00mm_urms}
\end{subfigure}%
\begin{subfigure}{.32\textwidth}
  \centering
  $z/H$ = 0.4
  \includegraphics[height=1\linewidth]{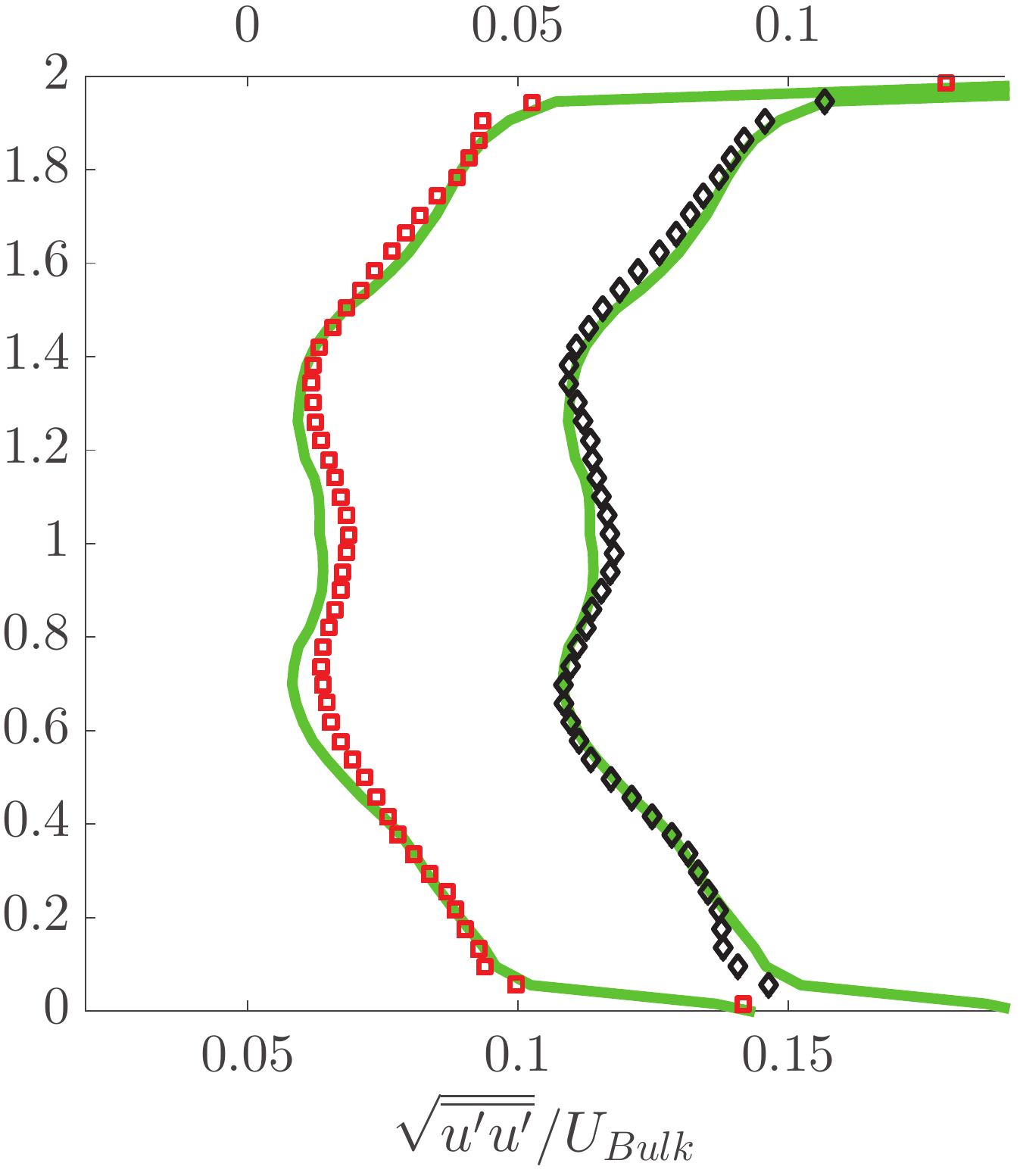}
  \caption{}
  \label{fig:3_6mm_5p_Re27000_sp10mm_urms}
\end{subfigure}
\begin{subfigure}{.32\textwidth}
  \centering
  $z/H$ = 0.8
  \includegraphics[height=1\linewidth]{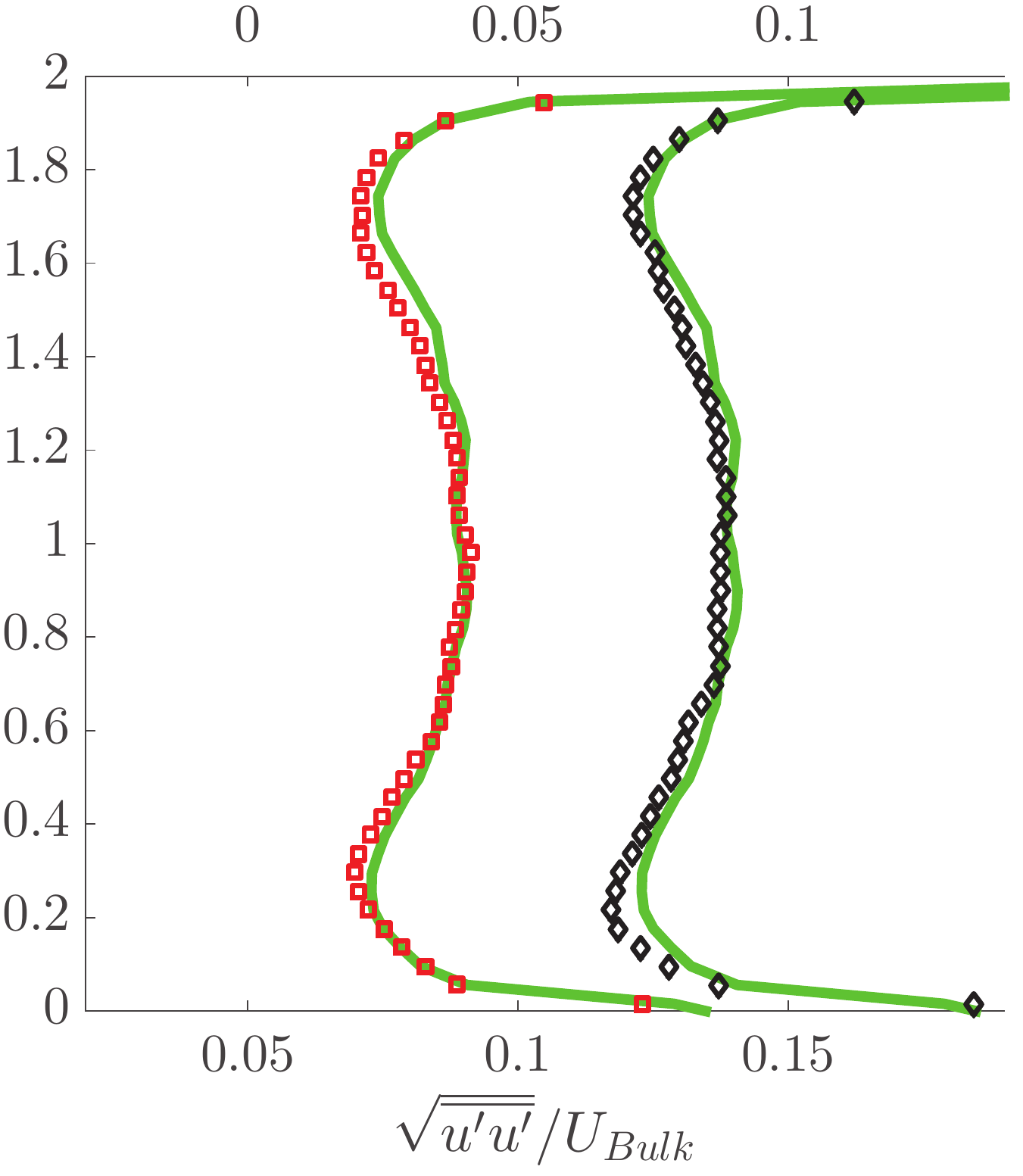}
  \caption{}
  \label{fig:3_6mm_5p_Re27000_sp20mm_urms}
\end{subfigure}

\begin{subfigure}{.32\textwidth}
  \centering
  \includegraphics[height=1\linewidth]{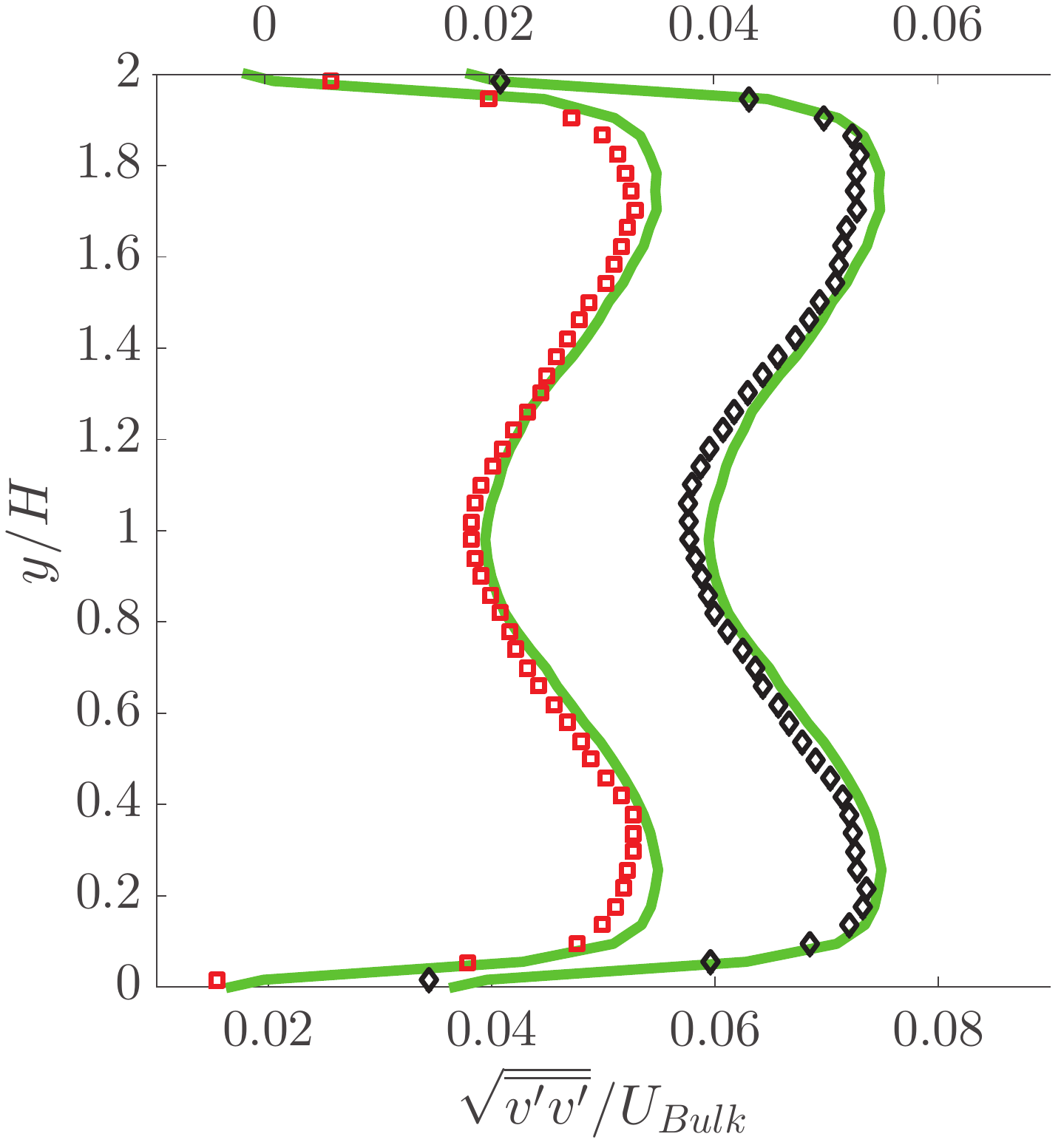}
  \caption{}
  \label{fig:3_6mm_5p_Re27000_sp00mm_vrms}
\end{subfigure}%
\begin{subfigure}{.32\textwidth}
  \centering
  \includegraphics[height=1\linewidth]{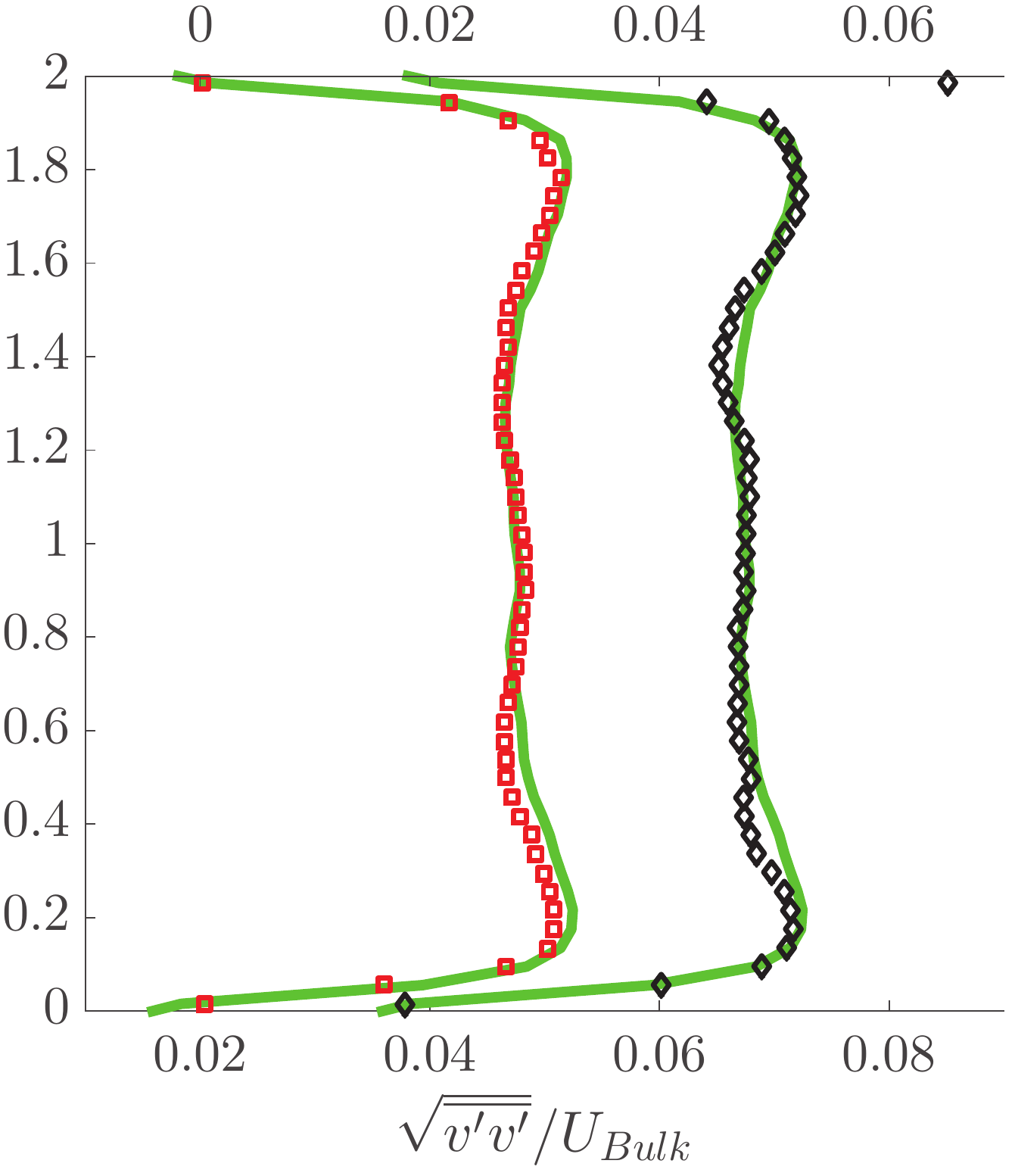}
  \caption{}
  \label{fig:3_6mm_5p_Re27000_sp10mm_vrms}
\end{subfigure}
\begin{subfigure}{.32\textwidth}
  \centering
  \includegraphics[height=1\linewidth]{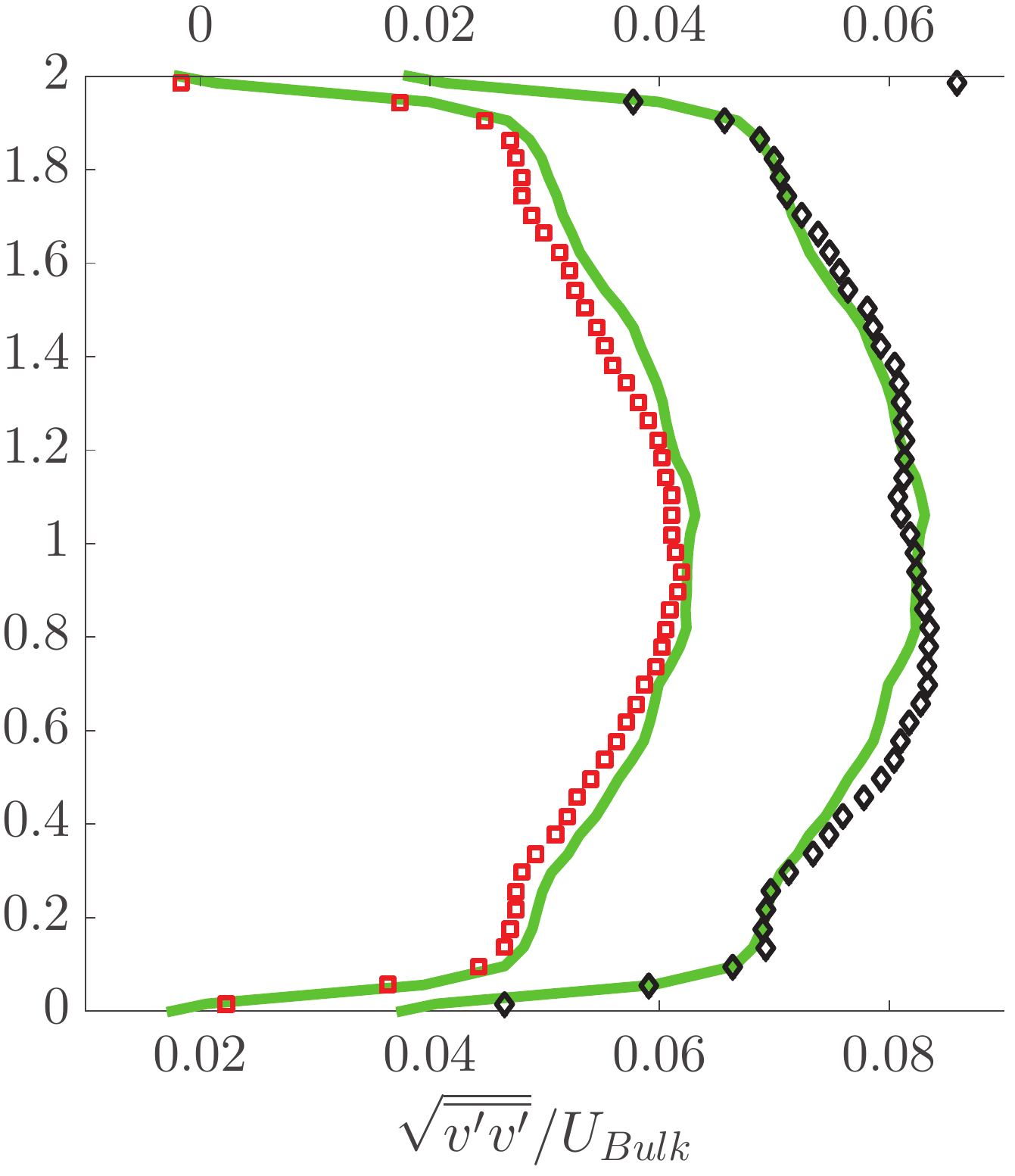}
  \caption{}
  \label{fig:3_6mm_5p_Re27000_sp20mm_vrms}
\end{subfigure}

\begin{subfigure}{.32\textwidth}
  \centering
  \includegraphics[height=1\linewidth]{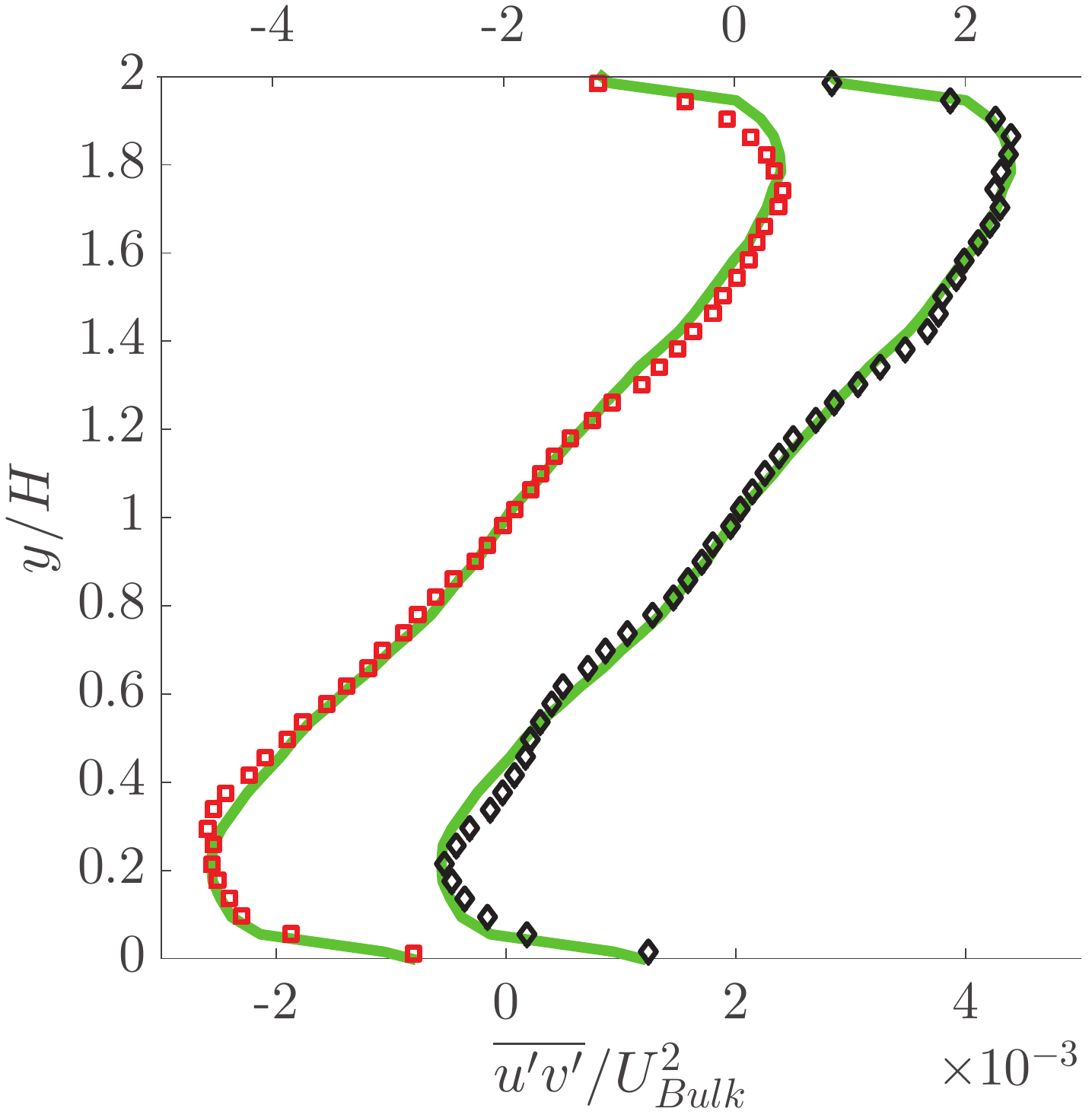}
  \caption{}
  \label{fig:3_6mm_5p_Re27000_sp00mm_uv}
\end{subfigure}%
\begin{subfigure}{.32\textwidth}
  \centering
  \includegraphics[height=1\linewidth]{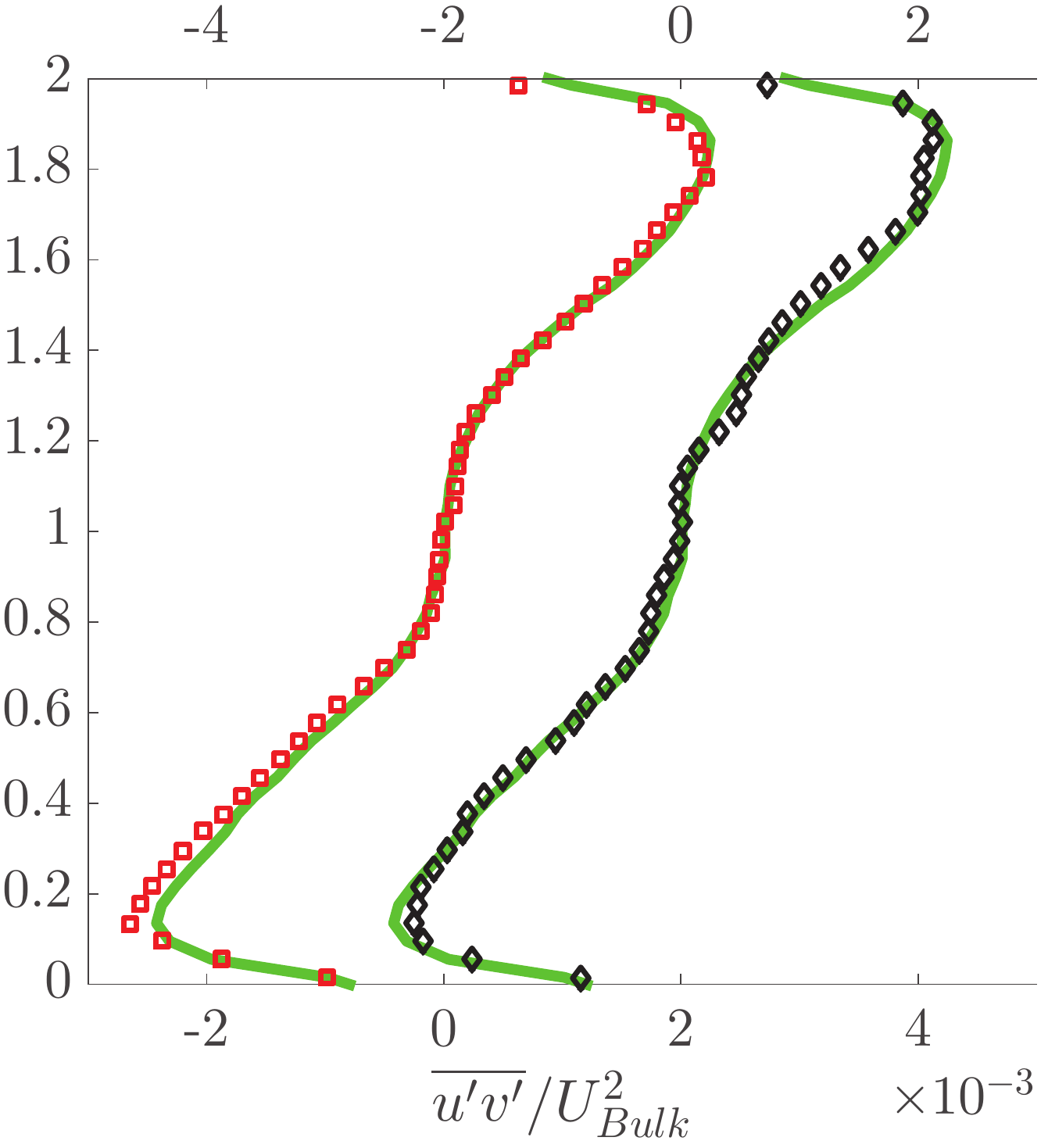}
  \caption{}
  \label{fig:3_6mm_5p_Re27000_sp10mm_uv}
\end{subfigure}
\begin{subfigure}{.32\textwidth}
  \centering
  \includegraphics[height=1\linewidth]{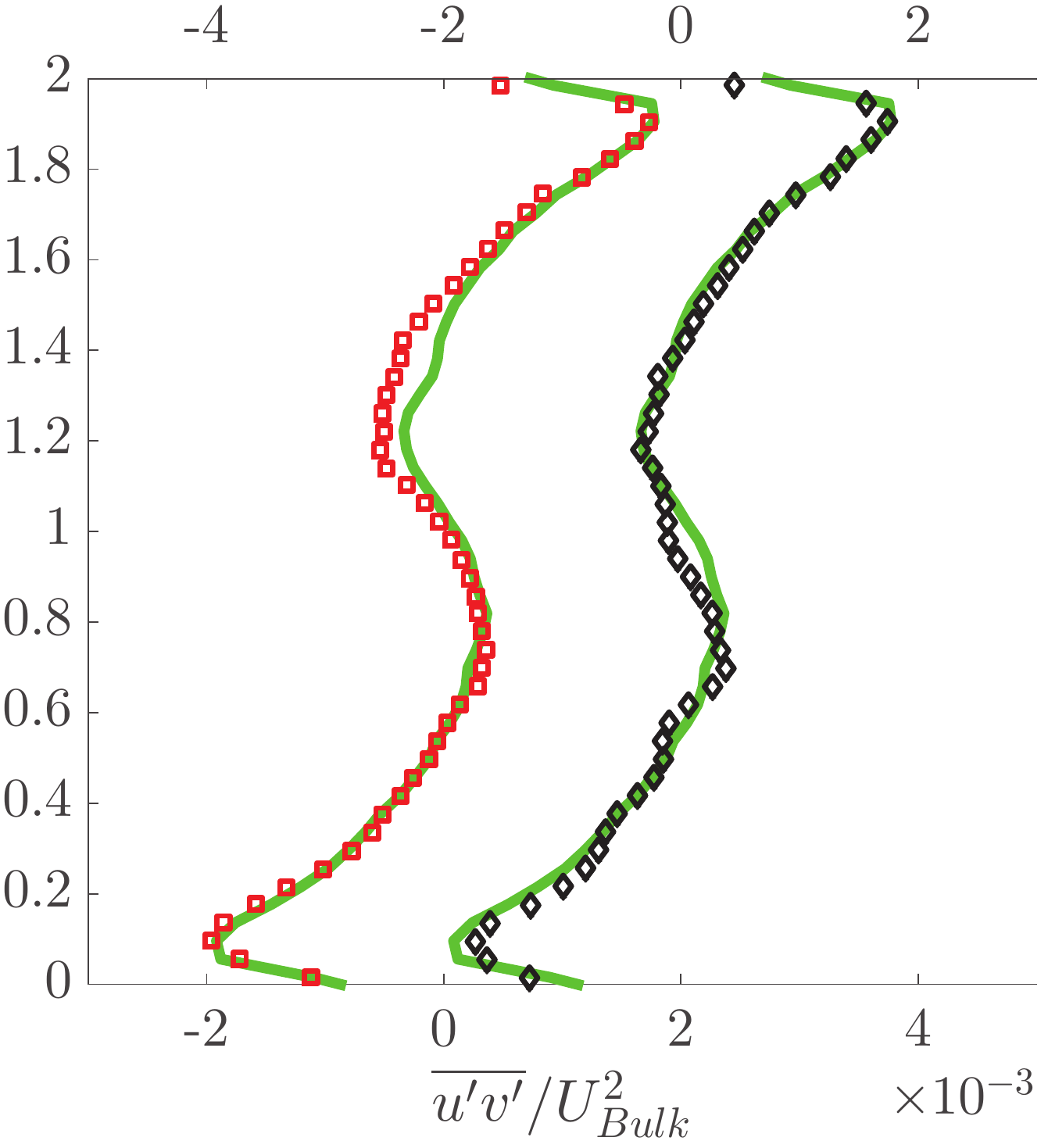}
  \caption{}
  \label{fig:3_6mm_5p_Re27000_sp20mm_uv}
\end{subfigure}

\caption{ Medium (MP, $2H/d_p$ = 16) and large-sized particles (LP, $2H/d_p$ = 9) in full suspension: $Re_{2H}\approx$ 27000, $\phi$ = 5\%.  Shown above are the profiles for the streamwise velocity fluctuations (a)-(c), wall-normal velocity fluctuations (d)-(f) and Reynolds stresses (g)-(i) of the fluid phase. Profiles shifted to the right correspond to LP and the corresponding x-axis is on the top.}
\label{fig:3_6mm_5p_Re27000_urms_vrms_uv}
\end{figure}

{\it Low volume fraction ($\phi$ = 5\%)}\\

Next we examine the velocity statistics for $\phi$ = 5\% and the largest Reynolds number considered, i.e.\ when the role of settling is assumed minor. Figures \ref{fig:3_6mm_5p_Re27000_sp00mm_Umean}--\ref{fig:3_6mm_5p_Re27000_sp20mm_Umean} show the mean-streamwise velocity for both MP and LP side by side. Unlike to the case with $\phi$ = 20\% discussed above, the overall changes with respect to the single phase flow are marginal. 
This is consistent with the friction factor in figures \ref{fig:f_vs_Re plot} and \ref{fig:df for Re27000}, which is very close to the values for single-phase flow at this volume fraction. 
The mean particle velocity is also nearly the same as that of the fluid, except at the walls. The distance from the walls over which the particle velocity deviates from the fluid velocity is larger for LP due to their larger size. One can also note that the magnitude of this slip velocity is higher for LP as the particle covers a larger distance from the wall and is exposed to a larger velocity difference.

Figures \ref{fig:3_6mm_5p_Re27000_sp00mm_phi}--\ref{fig:3_6mm_5p_Re27000_sp20mm_phi} compare the particle area-concentration profiles for MP and LP at the same spanwise location and nominal concentration of 5\%. It can be seen that the concentration is higher at the bottom than at the top wall indicating the presence of mild sedimentation even at these high $Re_{2H}$. The weak asymmetry in the concentration distribution is reflected in the mean-streamwise velocity profile. Nevertheless, the difference is marginal and one can assume that the effect of fluid inertia is much larger than the effect of gravity on the particle dynamics. 
The concentration profiles show a local peak in concentration at the bottom and top walls (stronger at the bottom wall), suggesting the presence of a particle wall-layer. Also, the peak of the concentration profile close to the wall does not occur at one particle radius, but slightly beyond that. This is due to the fact that the particles in the wall layer do not simply roll in contact with the wall at all times. \sz{Instead, they roll, slide and often bounce due to turbulent fluctuations and particle collisions. In addition, the (repulsive) wall-normal lubrication force also contributes to the shifted concentration peak.} 
In the mid-plane at $z/H$ = 0, the concentration of LP has a second local maxima at $y/H\approx$ 0.7: this is a signature of the increased inertial shear-induced migration.

We next consider the slip velocity between the two phases. Comparing to the case at $\phi$ = 20\%, the magnitude of the particle Reynolds number $Re_p$ is smaller at this lower volume fractions. This effect is even more pronounced for MP where $Re_p\approx$0 in the core of the flow (see figures \ref{fig:3_6mm_5p_Re27000_sp00mm_Rep}--\ref{fig:3_6mm_5p_Re27000_sp20mm_Rep}).
This smaller discrepancy between the velocity of the two phases is likely a consequence of a smaller frequency of particle-particle short-range interactions in more dilute conditions.

The rms of the streamwise and wall-normal velocity fluctuations in the 3 planes of measurement are displayed in 
figures \ref{fig:3_6mm_5p_Re27000_sp00mm_urms}--\ref{fig:3_6mm_5p_Re27000_sp20mm_urms} and \ref{fig:3_6mm_5p_Re27000_sp00mm_vrms}--\ref{fig:3_6mm_5p_Re27000_sp20mm_vrms}, whereas the  Reynolds shear stress $\overline{u'v'}$ is reported in figures \ref{fig:3_6mm_5p_Re27000_sp00mm_uv}--\ref{fig:3_6mm_5p_Re27000_sp20mm_uv}. 
A small local attenuation of the streamwise velocity fluctuations can be observed for MP close to the bottom and top wall (see figure \ref{fig:3_6mm_5p_Re27000_sp00mm_urms}), in correspondence with the particle wall layer, but, on the whole, it appears that for $\phi$ = 5\% the velocity fluctuations and their correlations do not deviate significantly from the single-phase case. Thus, the presence of particles do not cause significant differences in the second-order statistics. As shown in figure \ref{fig:df for Re27000}, the pressure drop for both MP and LP is $\approx$ 1\% lower than for the single phase flow. We can therefore infer that the stress induced by the particles do not have a major contribution towards the total stress for $\phi$ = 5\% at the highest $Re_{2H}\approx$ 27000 and turbulence is not significantly modified.

\subsubsection{Sedimenting particles ($Re_{2H}\approx$ 10000)}\label{sec:Velocity statistics: Sedimenting particles}

We now turn our attention to the case with the lowest $Re_{2H}\approx$ 10000, characterised by particles in sedimentation and resuspension as seen by the concentration distributions in figure \ref{fig:Re10000_phi}. The Rouse number $Ro$ is now 1.5 for MP and 3 for LP and ranges from the intermediate to the `bed-load' regime according to the classification in \cite{fredsoe1992mechanics}. 
\\

{\it Low volume fraction ($\phi$ = 5\%)}\\

\begin{figure}
\centering

\begin{subfigure}{.32\textwidth}
  \centering
  $z/H$ = 0
  \includegraphics[height=1\linewidth]{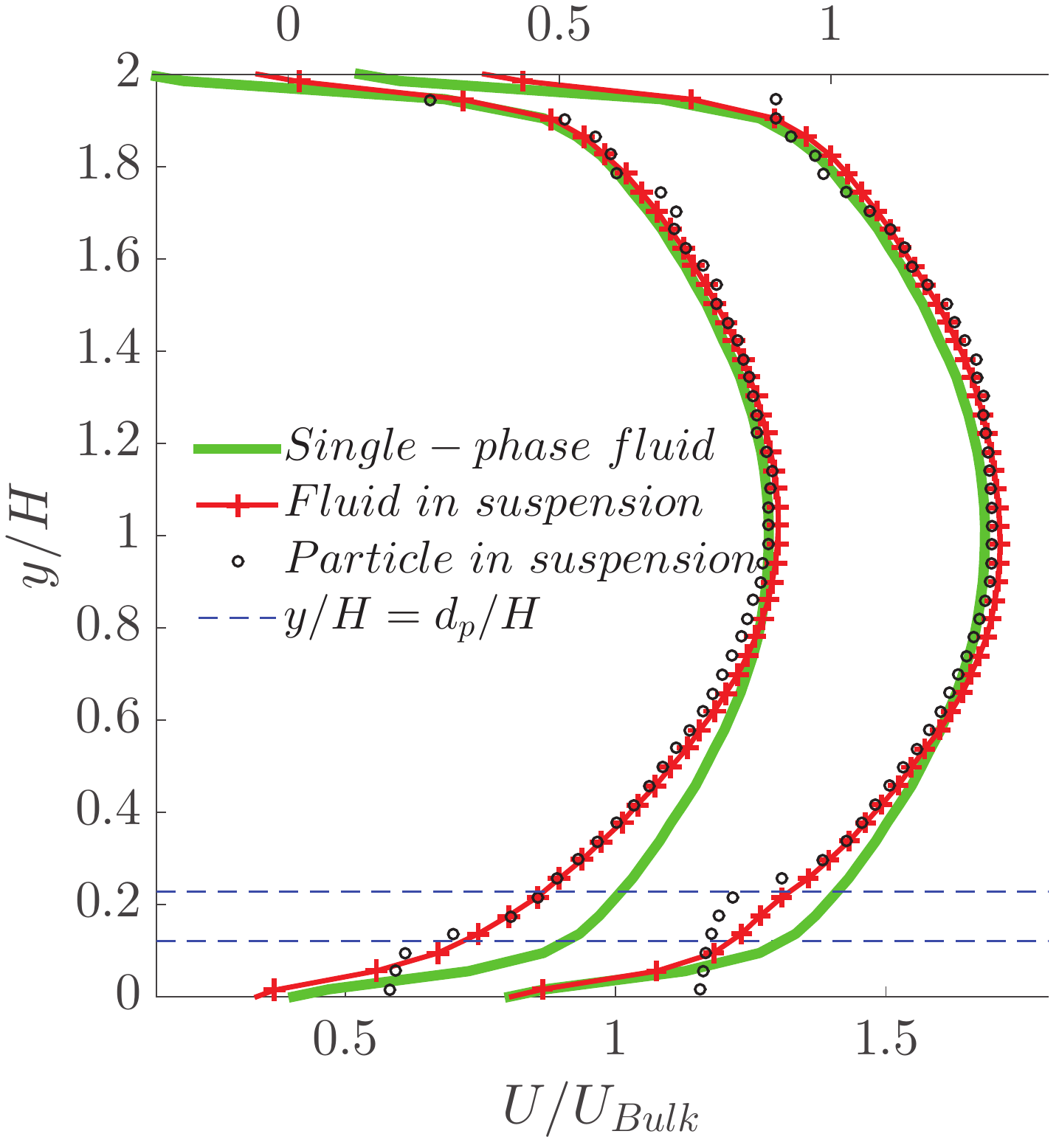}
  \caption{}
  \label{fig:3_6mm_5p_Re10000_sp00mm_Umean}
\end{subfigure}%
\begin{subfigure}{.32\textwidth}
  \centering
  $z/H$ = 0.4
  \includegraphics[height=1\linewidth]{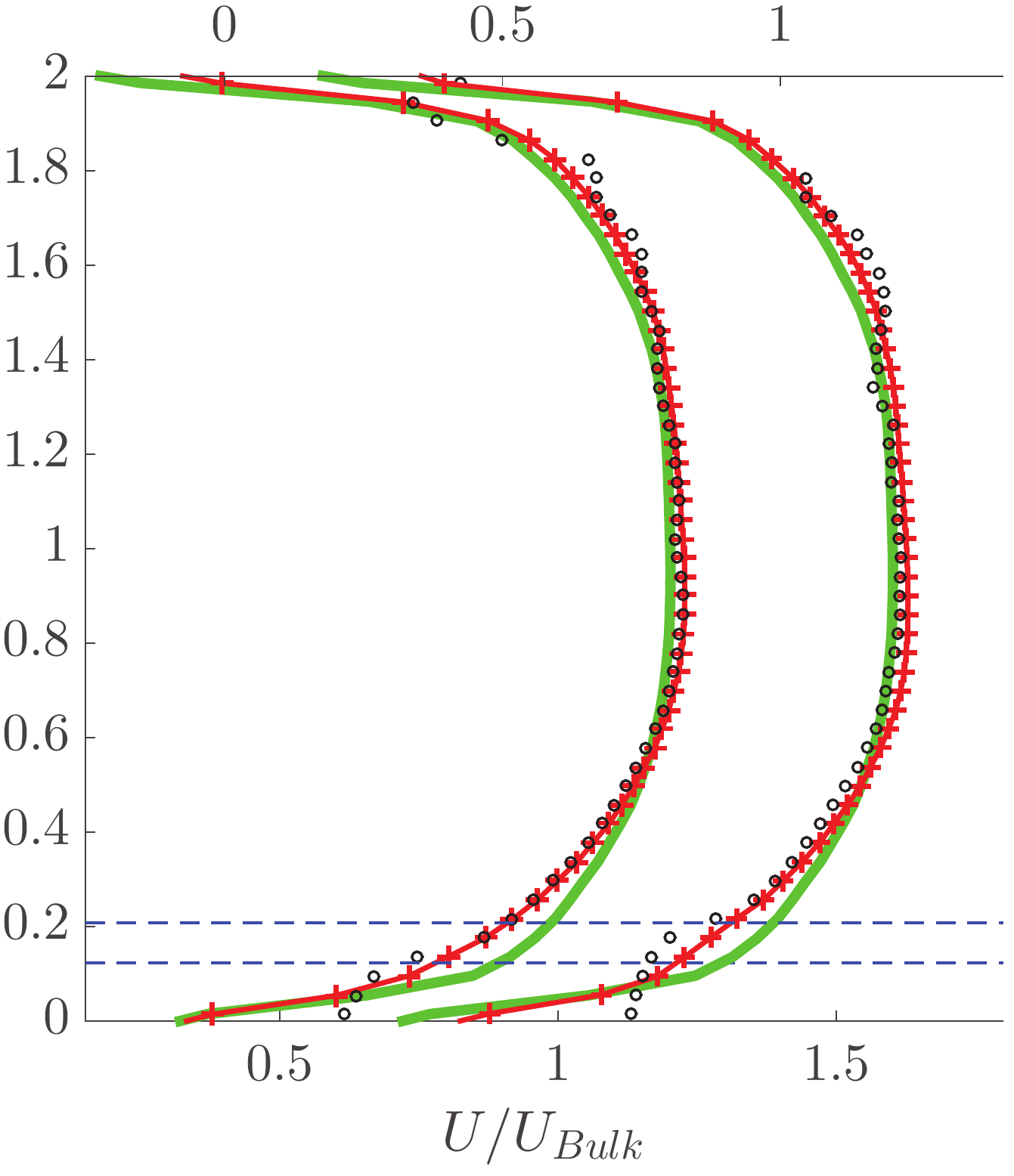}
  \caption{}
  \label{fig:3_6mm_5p_Re10000_sp10mm_Umean}
\end{subfigure}
\begin{subfigure}{.32\textwidth}
  \centering
  $z/H$ = 0.8
  \includegraphics[height=1\linewidth]{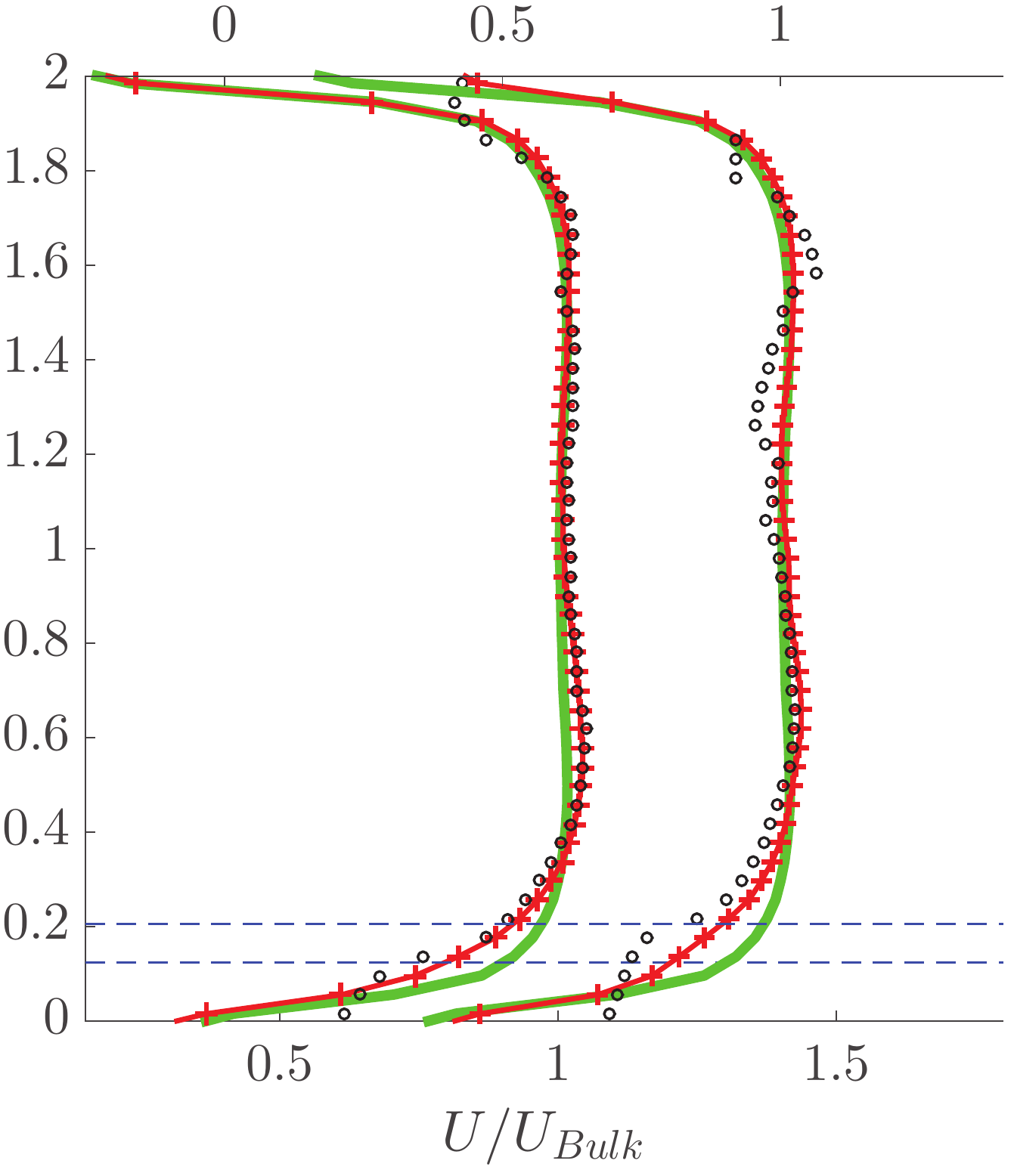}
  \caption{}
  \label{fig:3_6mm_5p_Re10000_sp20mm_Umean}
\end{subfigure}

\begin{subfigure}{.32\textwidth}
  \centering
  \includegraphics[height=1\linewidth]{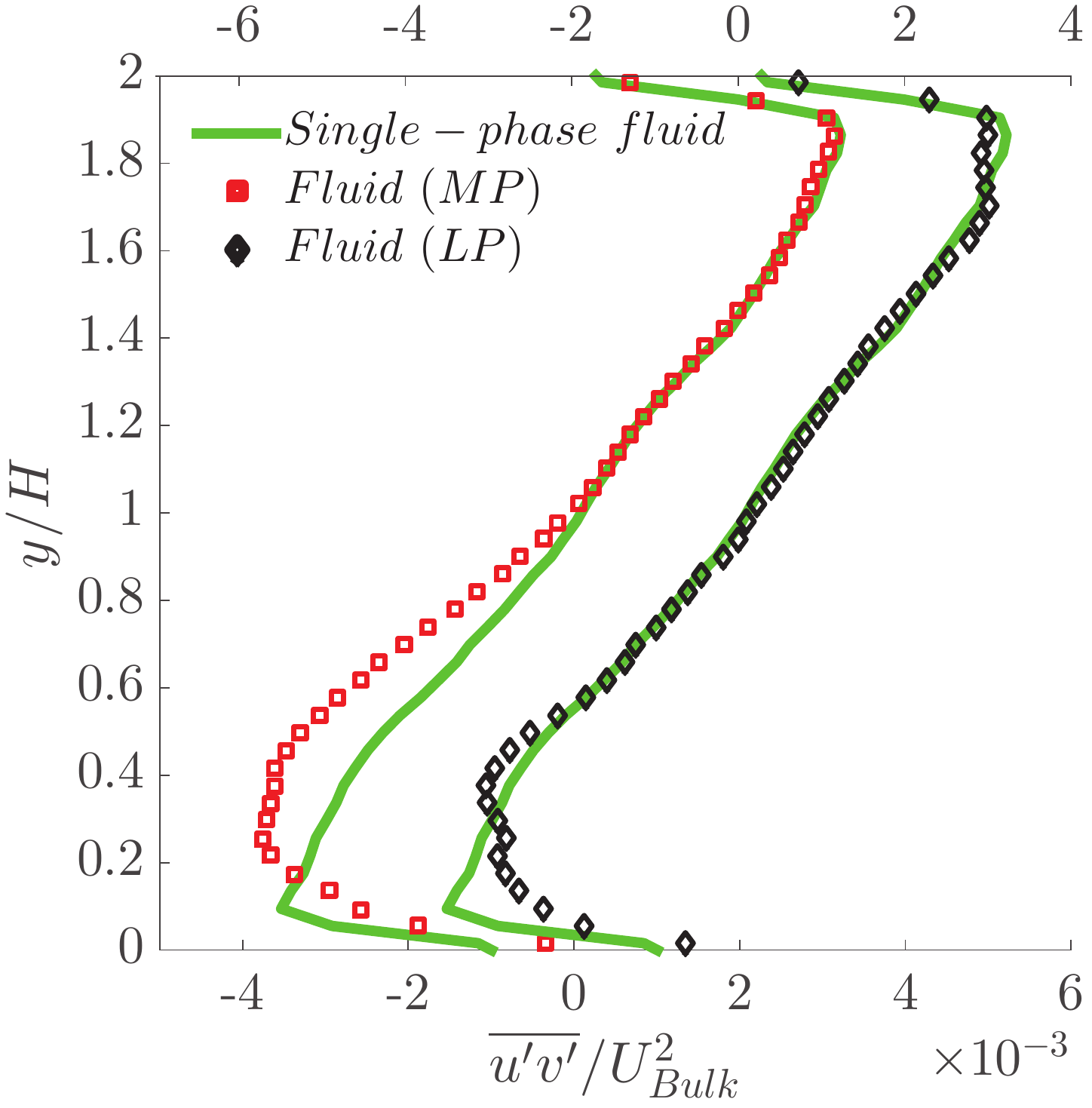}
  \caption{}
  \label{fig:3_6mm_5p_Re10000_sp00mm_uv}
\end{subfigure}%
\begin{subfigure}{.32\textwidth}
  \centering
  \includegraphics[height=1\linewidth]{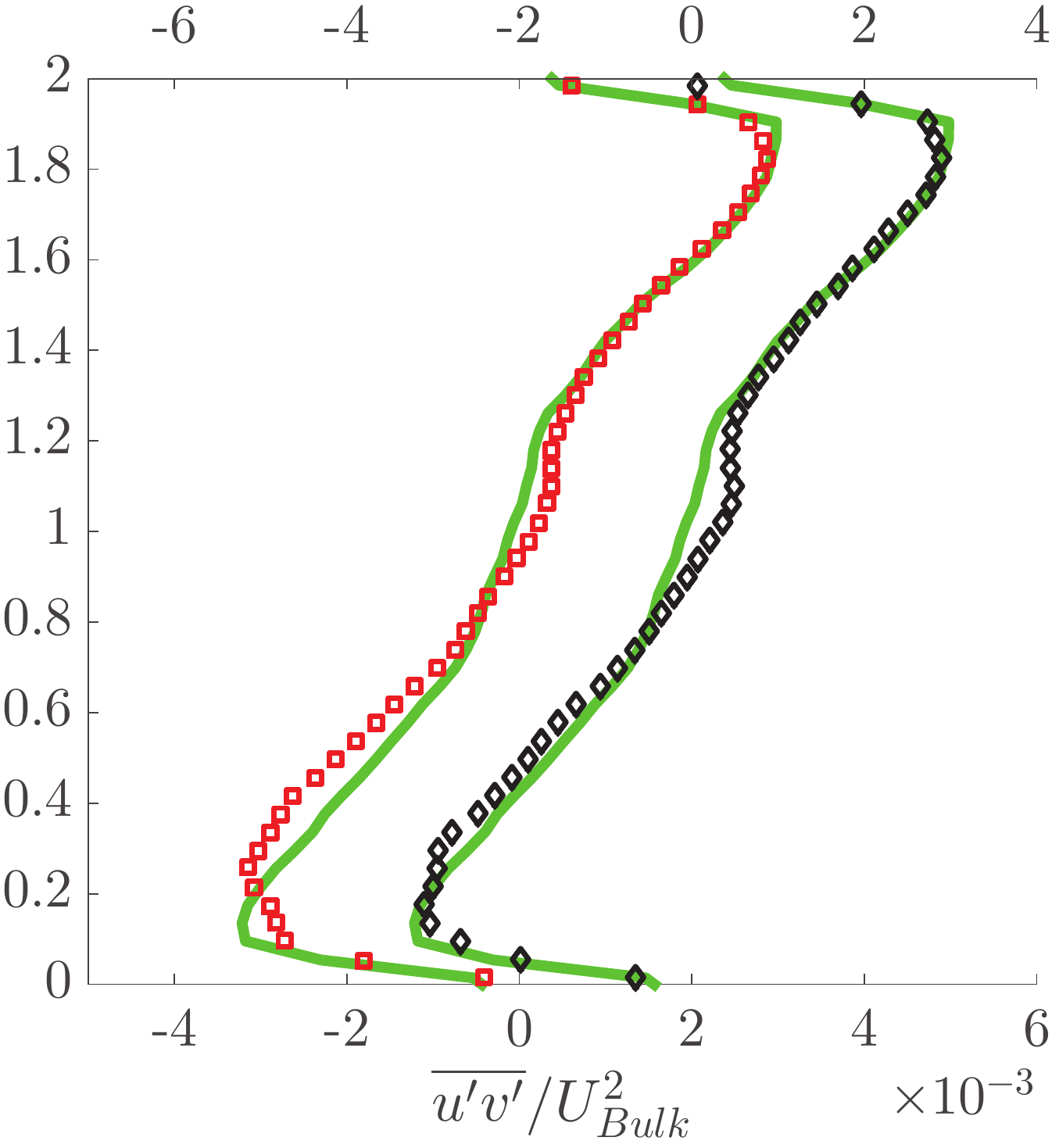}
  \caption{}
  \label{fig:3_6mm_5p_Re10000_sp10mm_uv}
\end{subfigure}
\begin{subfigure}{.32\textwidth}
  \centering
  \includegraphics[height=1\linewidth]{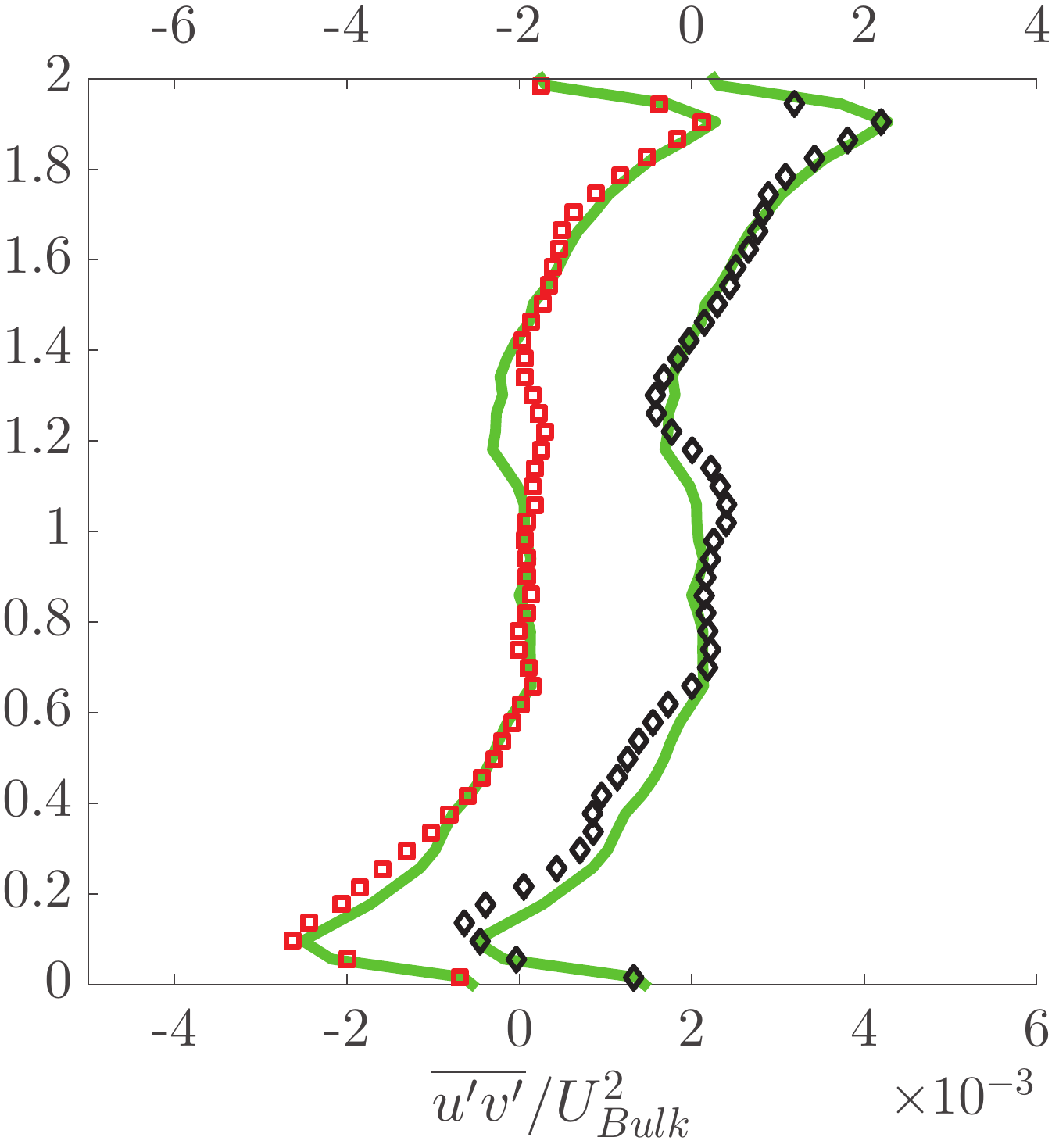}
  \caption{}
  \label{fig:3_6mm_5p_Re10000_sp20mm_uv}
\end{subfigure}

\caption{Comparsion of mean velocity and Reynolds stress for medium (MP) and large-sized particles (LP): $Re_{2H}\approx$ 10000, $\phi$ = 5\%. (a)-(c) Mean streamwise velocity and (d)-(f) Reynolds stress. The profiles in figures (a)--(f) that have been shifted to the right correspond to LP. The corresponding x-axis for these shifted profiles is on the top of the figures.}
\label{fig:3_6mm_5p_Re10000_Umean_uv}
\end{figure}

As shown in figures \ref{fig:3mm_5p_Re10000_phi} and \ref{fig:6mm_5p_Re10000_phi},
the particles reside mostly in the bottom half of the duct so that the velocity statistics be modified there the most. 
The fluid mean streamwise velocity, depicted in figures \ref{fig:3_6mm_5p_Re10000_sp00mm_Umean}--\ref{fig:3_6mm_5p_Re10000_sp20mm_Umean}, decreases in the region with higher particle concentration. The pressure drop for this low concentration is similar for both particle sizes ($\approx$ 12\% higher than the single-phase flow) and the mean flow is quite similar as well.

The Reynolds shear stress is shown in figure \ref{fig:3_6mm_5p_Re10000_sp00mm_uv}-\ref{fig:3_6mm_5p_Re10000_sp20mm_uv} for the same experiments;  it deviates the most from the single-phase case in the region of high particle concentration close to the bottom wall. It is interesting to note that the MP increase the fluctuations more than the LP in the region of high particle concentration. We believe that this is a particle number effect. The MP are $\approx$ 6 times more in number than the LP at the same volume fraction. At such relatively low $Re_{2H}$, particle induced effects increase as particle size reduces \citep{lin2017effects}. We also note that the peak in the Reynolds shear stress is displaced away from the bottom wall on the plane $z/H$ = 0 and that
 the streamwise fluid fluctuations are suppressed in the particle layer at the bottom wall for both MP and LP (see figure \ref{fig:3_6mm_5p_Re10000_urms_vrms} in the Appendix).\\

{\it High volume fraction ($\phi$ = 20\%)}\\

\begin{figure}
\centering

\begin{subfigure}{.32\textwidth}
  \centering
  $z/H$ = 0
  \includegraphics[height=1\linewidth]{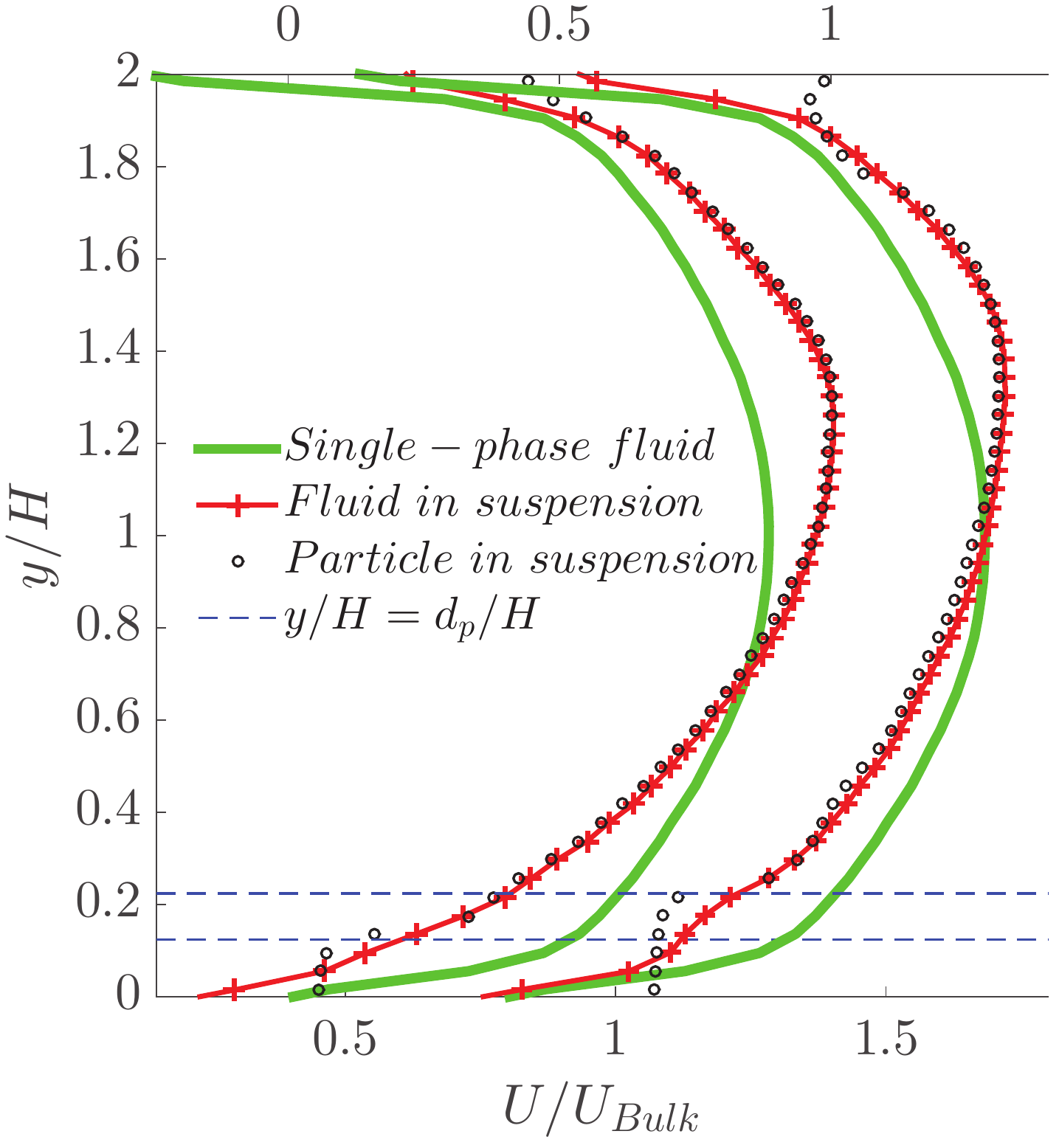}
  \caption{}
  \label{fig:3_6mm_20p_Re10000_sp00mm_Umean}
\end{subfigure}%
\begin{subfigure}{.32\textwidth}
  \centering
  $z/H$ = 0.4
  \includegraphics[height=1\linewidth]{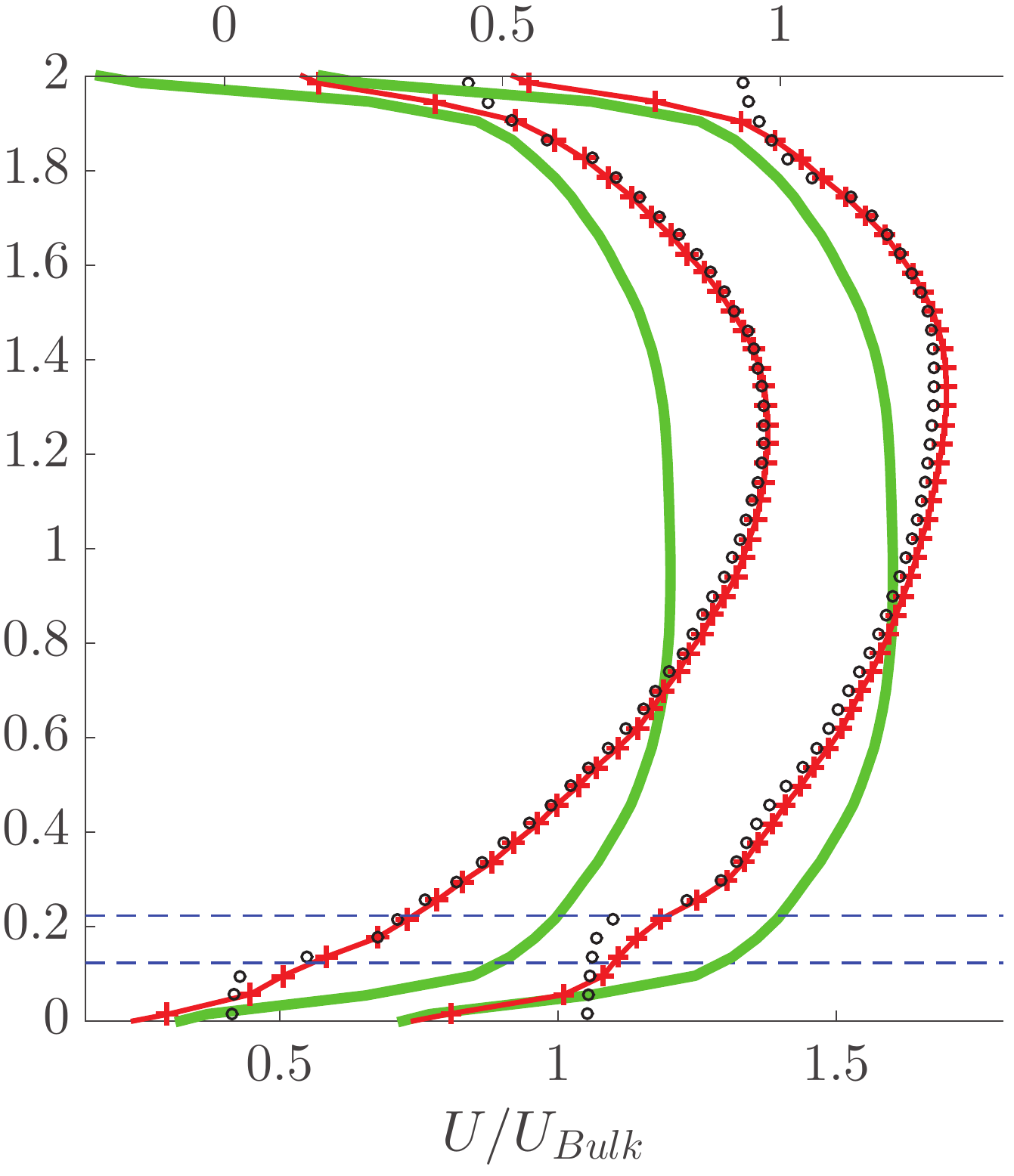}
  \caption{}
  \label{fig:3_6mm_20p_Re10000_sp10mm_Umean}
\end{subfigure}
\begin{subfigure}{.32\textwidth}
  \centering
  $z/H$ = 0.8
  \includegraphics[height=1\linewidth]{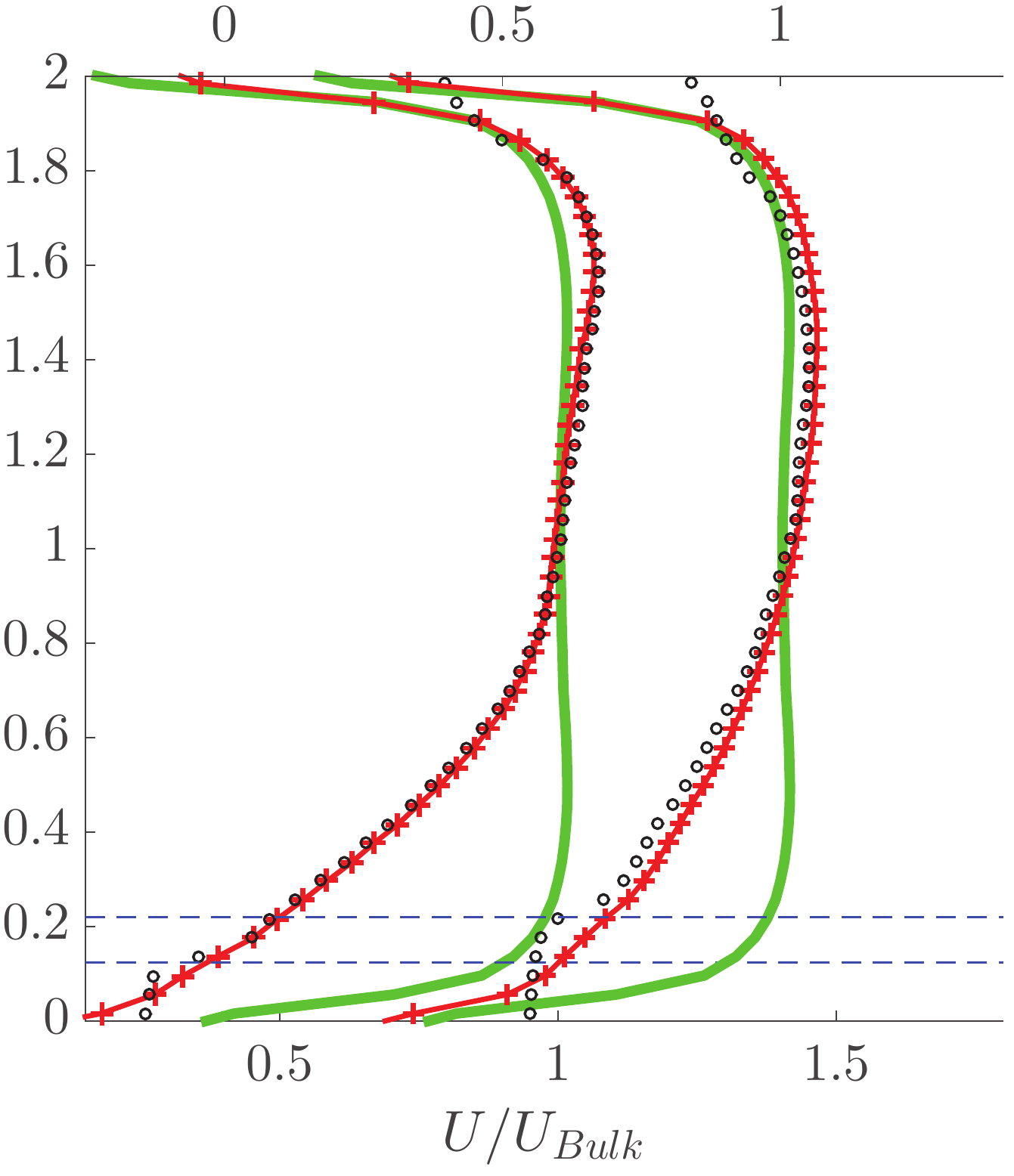}
  \caption{}
  \label{fig:3_6mm_20p_Re10000_sp20mm_Umean}
\end{subfigure}

\begin{subfigure}{.32\textwidth}
  \centering
  \includegraphics[height=1\linewidth]{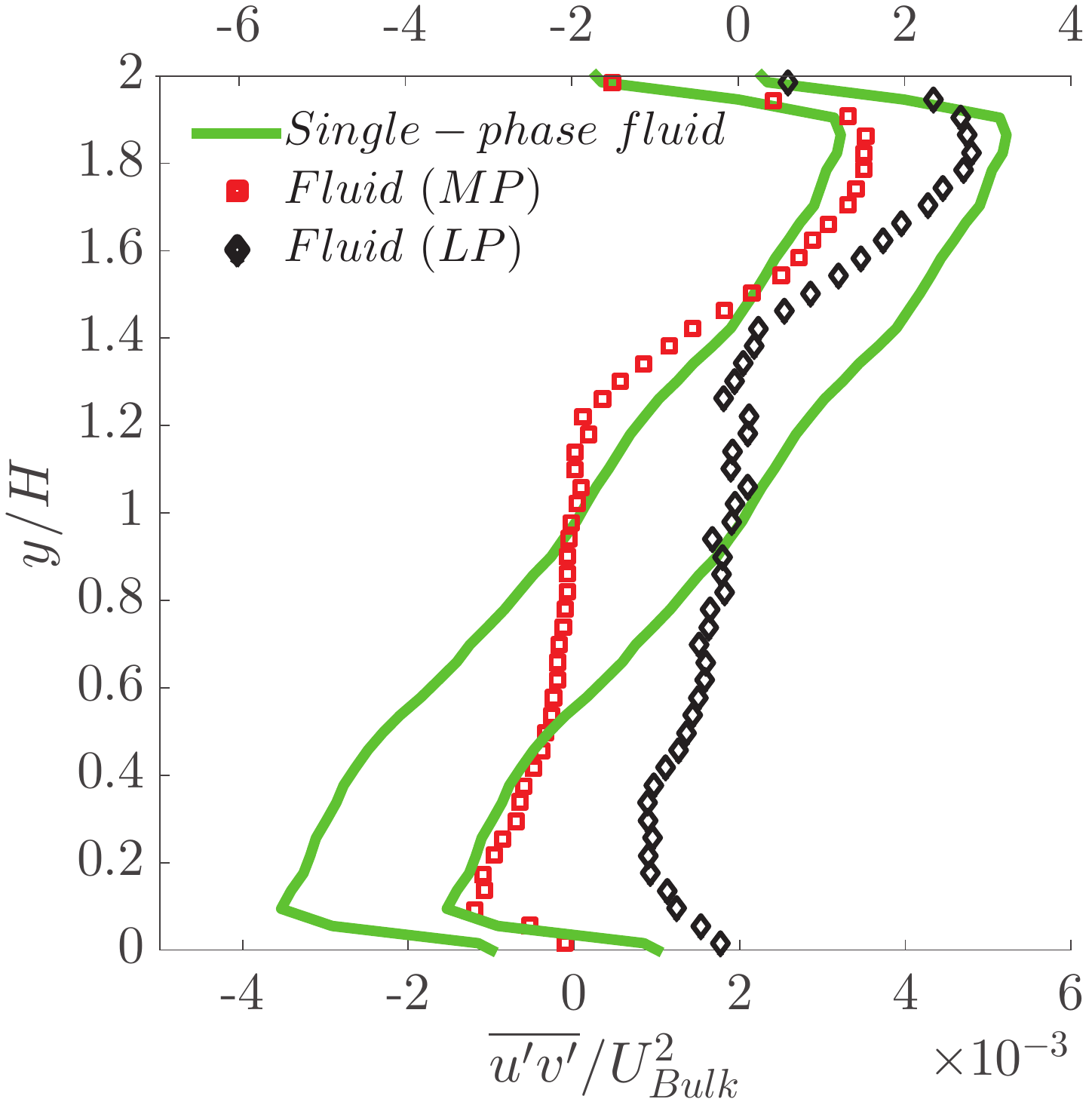}
  \caption{}
  \label{fig:3_6mm_20p_Re10000_sp00mm_uv}
\end{subfigure}%
\begin{subfigure}{.32\textwidth}
  \centering
  \includegraphics[height=1\linewidth]{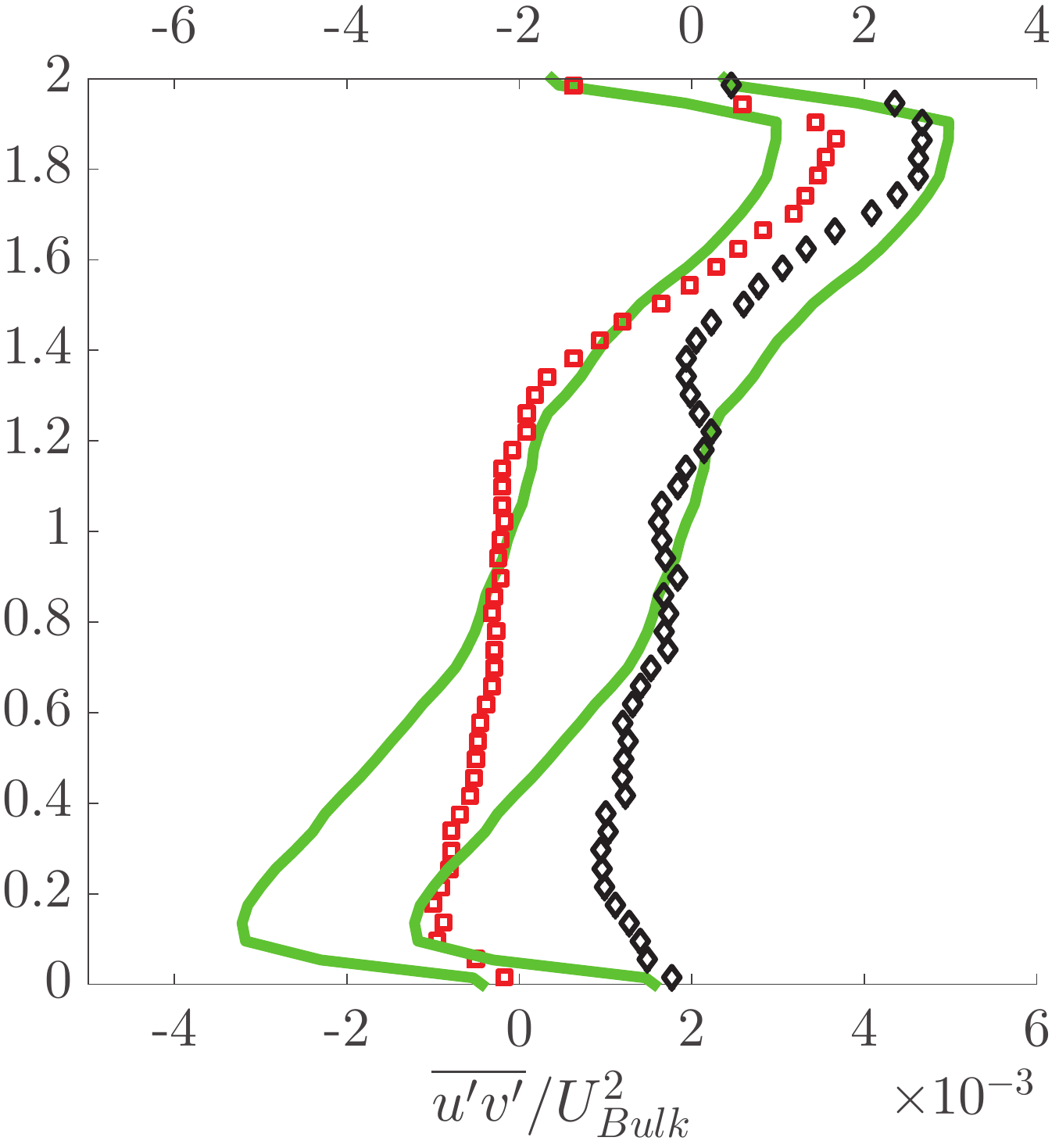}
  \caption{}
  \label{fig:3_6mm_20p_Re10000_sp10mm_uv}
\end{subfigure}
\begin{subfigure}{.32\textwidth}
  \centering
  \includegraphics[height=1\linewidth]{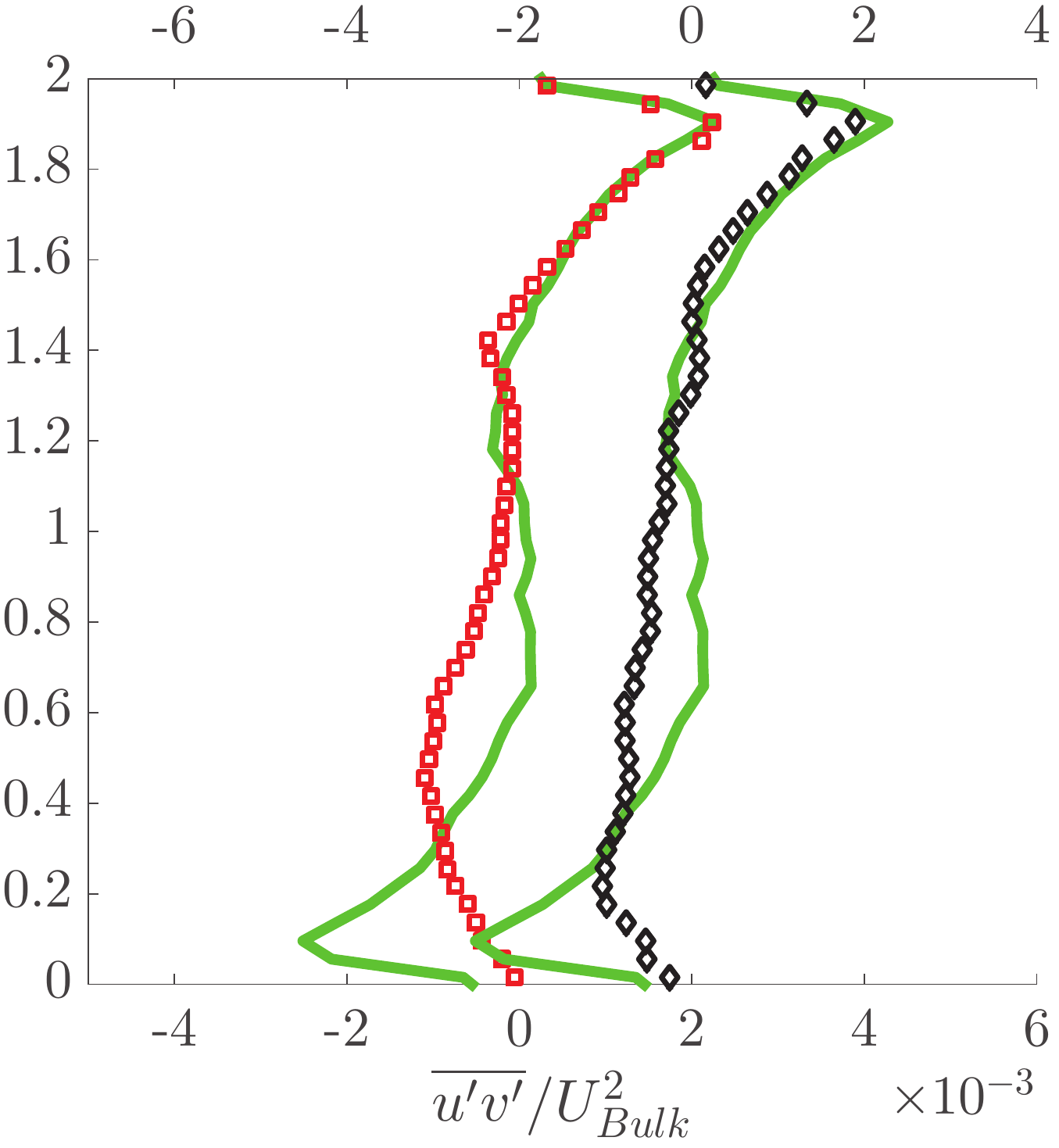}
  \caption{}
  \label{fig:3_6mm_20p_Re10000_sp20mm_uv}
\end{subfigure}

\caption{Comparsion of mean velocity and Reynolds stress for medium (MP) and large-sized particles (LP): $Re_{2H}\approx$ 10000, $\phi$ = 20\%. (a)-(c) Mean streamwise velocity and (d)-(f) Reynolds stress. Profiles shifted to the right correspond to LP and the corresponding x-axis is on the top.}
\label{fig:3_6mm_20p_Re10000_Umean_uv}
\end{figure}

The particle distribution, when increasing the volume fraction to 20\% at the same $Re_{2H}\approx$ 10000, was discussed in connection with
figures \ref{fig:3mm_20p_Re10000_phi} and \ref{fig:6mm_20p_Re10000_phi} for MP and LP. As noted above, the concentration is highest in the plane $z/H$ = 0 and decreases towards the sidewalls. Particle layering is clearly visible across the height of the channel, especially for the LP, which is indicative of a moving porous bed. The concentration peak corresponding to the particle layer at the top wall measured at higher Reynolds numbers
starts to appear already at this flow rate.

The fluid mean-streamwise velocity is substantially skewed (see figure \ref{fig:3_6mm_20p_Re10000_sp00mm_Umean}-\ref{fig:3_6mm_20p_Re10000_sp20mm_Umean}) with a large reduction in the lower half and an increase in the top half. The maximum streamwise velocity is higher for the case with MP, which can be explained 
 by the reduced porosity of the moving bed for MP. As MP are smaller, they are more closely packed, which
 lowers the bed porosity and causes a decrease of the fluid velocity in all the 3 spanwise planes. To compensate for the reduced fluid flow through the bed, the fluid accelerates in the region above the bed where the particle concentration is relatively low.

The apparent slip velocity, i.e. difference between fluid and particle velocity, is higher for LP than MP. The `bumpy' fluid velocity in the particle layer at the bottom wall can also be seen in all 3 planes suggesting that there is a stable particle layer slowing down the flow. This also confirms that our measurement procedures can indeed reveal flow fields under such high local-concentrations ($\phi\approx$ 50\%).

The Reynolds shear stress profiles are displayed in figures \ref{fig:3_6mm_20p_Re10000_sp00mm_uv}--\ref{fig:3_6mm_20p_Re10000_sp20mm_uv}. The reduced correlation of the streamwise and wall-normal component of the fluid velocity in regions of high particle concentration is clearly visible. Towards the top wall, where the particle concentration is lower, the Reynolds stresses tend to retain values similar to the single-phase case. Nevertheless, we note a small increase for the MP and a small decrease for the LP. Similar trends are observed in the intensity of the streamwise and wall-normal velocity fluctuations close to the top-wall (shown in figure 
\ref{fig:3_6mm_20p_Re10000_urms_vrms} in Appendix). This can be understood by recalling that the maximum streamwise velocity is higher for MP than in LP. Broadly speaking, the duct can be split in 2 regions: a region similar to a porous medium near the bottom wall and a region similar to a duct flow with low particle concentration near the top. The Reynolds number pertaining to this upper region is higher for MP than LP owing to the higher fluid velocity. Thus, the turbulence intensities are also higher for MP.

Finally, we have shown in \ref{fig:df for Re10000} that the pressure drop is higher with LP ($\approx$ 50\% higher than the single-phase flow) than MP ($\approx$ 32\% higher than the single-phase flow). LP are larger than MP and therefore form a bed of higher porosity, where the fluid flows through the gaps more effectively. Thus, these larger particles tend to be suspended more easily, despite their larger size.

\section{Discussion and conclusion}

We have presented measurements of pressure drop and fluid and particle velocity in a square duct laden with spherical particles for 3 particle volume fractions ($\phi$ = 5, 10 and 20\%) and 3 particle sizes ($2H/d_p\approx$ 40, 16 and 9).
Refractive index matched (RIM) hydrogel particles in water have enabled us to use PIV and PTV techniques to measure the fluid and particle velocities up to the relatively high volume fraction of $\phi$ = 20\%
for the two largest sizes considered. The particle to fluid density ratio ranges from 1.0035 to 1.01, so that especially at the lower Reynolds number considered settling is not negligible. In the analysis, we therefore consider the case at high $Re_{2H}$ as a fully-suspended regime, whereas at low $Re_{2H}$ we have a low-porosity moving particle bed.

The friction factor of the suspensions is found to be significantly larger than that of single-phase duct flow at the lower $Re_{2H}$ investigated; however, the difference decreases when increasing the flow rate and the total drag becomes close, in some cases slightly lower, than in single phase flow at the higher Reynolds number considered, $Re_{2H}=27000$.
With regards to the dependency on the particle size, the pressure drop is found to decrease with the particle diameter for volume fractions lower than $\phi$ = 10\% for nearly all $Re_{2H}$ investigated in this study. However, at the highest volume fraction $\phi$ = 20\%, we report a peculiar non-monotonic behavior:
the pressure drop first decreases and then increases with increasing particle size. 

\begin{figure}
\centering

\begin{subfigure}{.49\textwidth}
  \centering
  \includegraphics[height=1\linewidth]{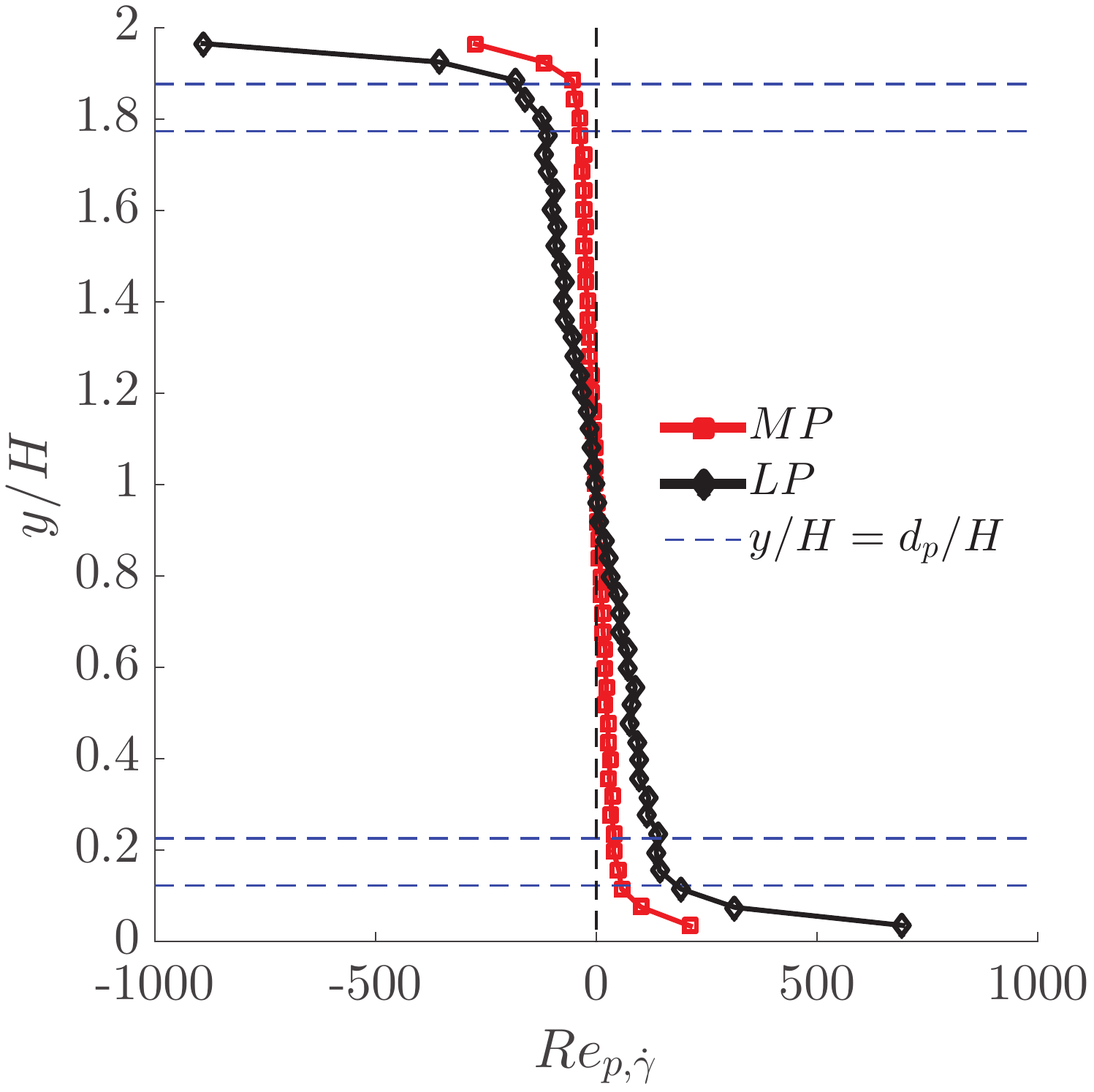}
  \caption{}
  \label{fig:Re_p_gamma_5p}
\end{subfigure}%
\begin{subfigure}{.49\textwidth}
  \centering
  \includegraphics[height=1\linewidth]{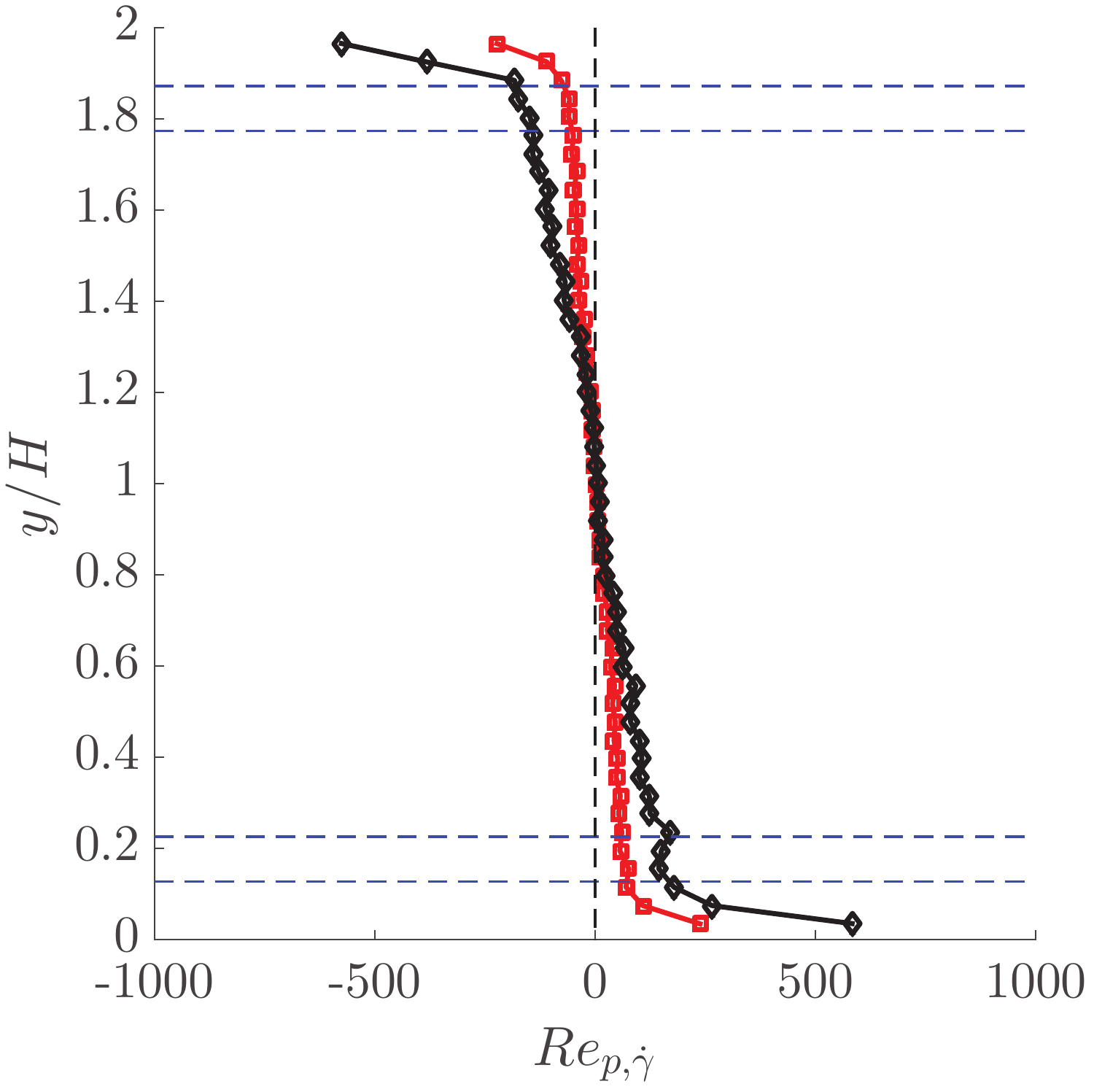}
  \caption{}
  \label{fig:Re_p_gamma_20p}
\end{subfigure}

\caption{Shear rate based particle Reynolds number $Re_{p,\dot{\gamma}}$ for (a) $\phi$ = 5\% and (b) $\phi$ = 20\% at $Re_{2H}\approx$ 27000 and $z/H$ = 0.}
\label{fig:Re_p_gamma}
\end{figure}

The decrease of the turbulent drag with particle size
at the lowest volume fractions is related to an attenuation of the turbulence. The larger particles more effectively break coherent eddies in the flow and reduce the turbulent stresses.
The drag increase from MP to LP at $\phi$ = 20\%, however, occurs despite a larger reduction of the turbulent stresses and it is due to significant particle induced stresses. 
Indeed, focusing on the highest $Re_{2H}$, when gravitational effects can be neglected, the pressure drop at a given volume fraction is defined by the balance between turbulent stresses and particle induced stresses, assuming viscous effects to be small and localized near the walls. As compared to MP, LP give a larger reduction in turbulent stresses. \sz{On the other hand, the particle-induced stresses appear to be larger for LP. This can be ascertained from the particle Reynolds number $Re_{p,\dot{\gamma}} = \dot{\gamma}(d_p/2)^2/\nu_f$, based on the local fluid shear rate $\dot{\gamma}$ and the particle radius as shown in figure \ref{fig:Re_p_gamma}. At the highest $Re_{2H}$, the $Re_{p,\dot{\gamma}}$ is larger for LP as compared to MP for all $\phi$. As shown in \cite{picano2013shear}, even at $Re_{p,\dot{\gamma}}$ = 10, significant inertial-shear-thickening is observed, especially for larger $\phi$. In the present case, we observe much higher values of $Re_{p,\dot{\gamma}}$ in the near-wall region.} This increase in particle induced stress exceeds the reduction in turbulent stresses at higher
 $\phi$ and lead to the overall increase of the pressure drop for LP as compared to MP at $\phi$ = 20\%.

\begin{figure}
\centering

\begin{subfigure}{.32\textwidth}
  \centering
  \includegraphics[height=1\linewidth]{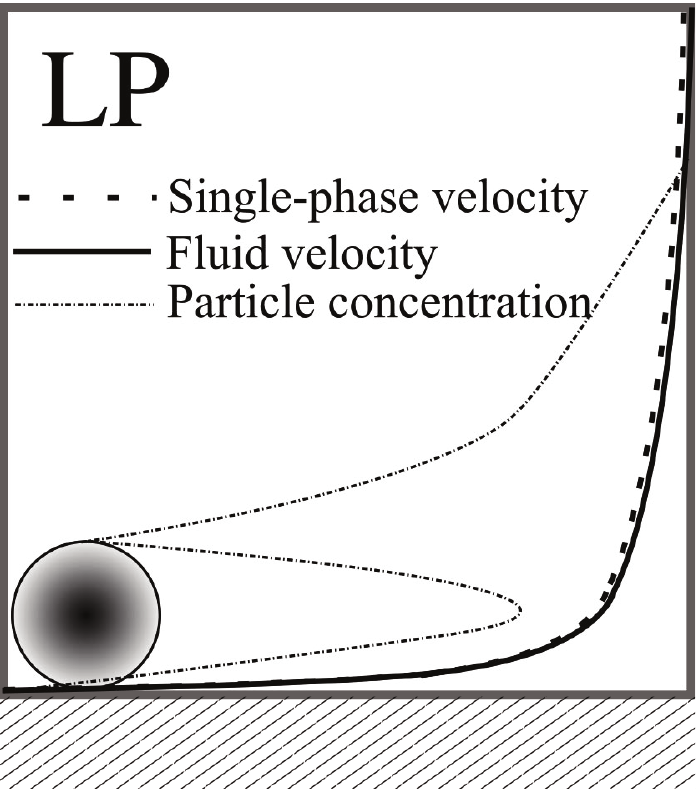}
  \caption{}
  \label{fig:LP}
\end{subfigure}%
\begin{subfigure}{.32\textwidth}
  \centering
  \includegraphics[height=1\linewidth]{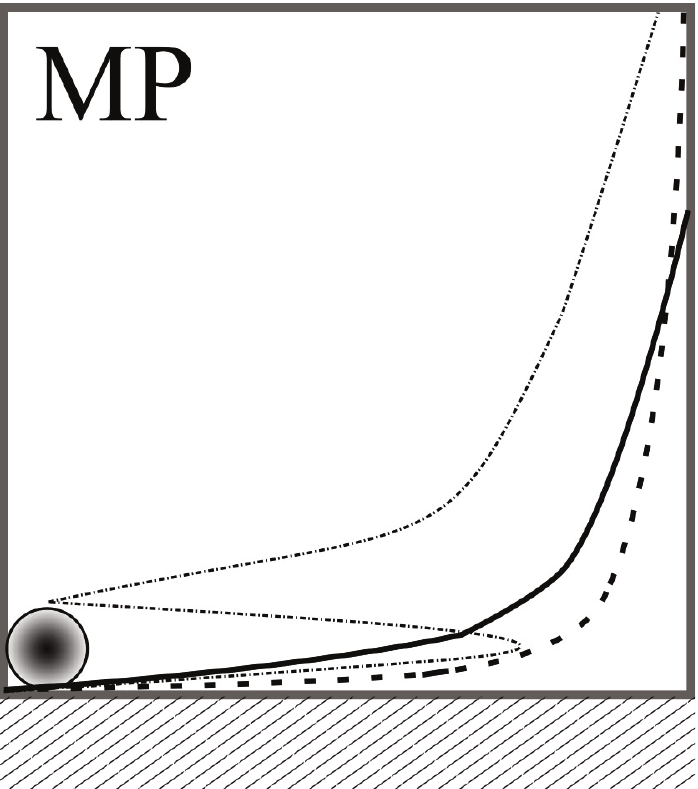}
  \caption{}
  \label{fig:MP}
\end{subfigure}
\begin{subfigure}{.32\textwidth}
  \centering
  \includegraphics[height=1\linewidth]{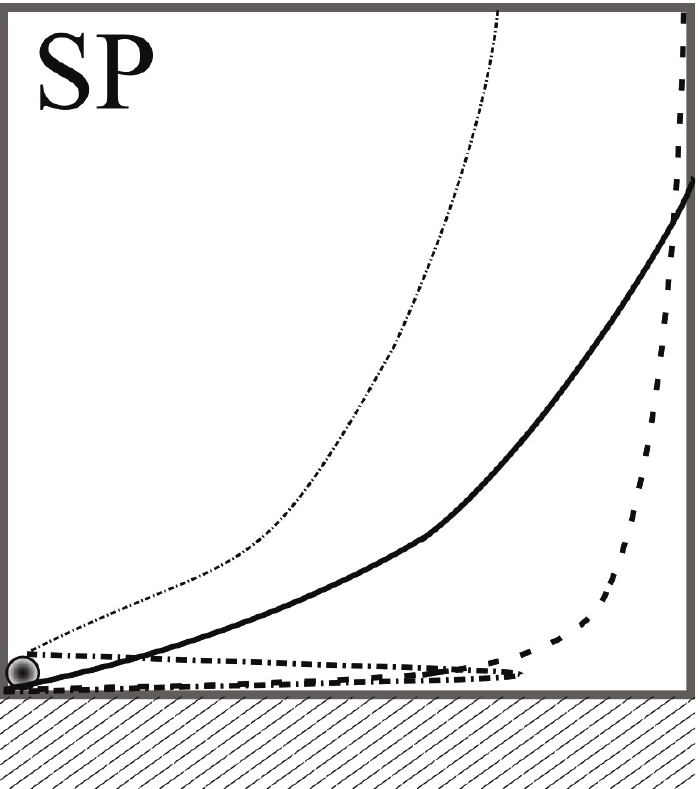}
  \caption{}
  \label{fig:SP}
\end{subfigure}

\caption{Schematic of change in mean-streamwise velocity and concentration profile \sz{in the plane of the wall-bisector} $z/H$ = 0 for (a) LP, (b) MP and (c) SP at $Re_{2H}\approx$ 27000 and $\phi$ = 20\%.}
\label{fig:SP_MP_LP}
\end{figure}

To explain the cause for the non-monotonic pressure drop at the highest volume fraction $\phi$ = 20\% and higher $Re_{2H}$ (where the influence of gravity on pressure drop is negligible) we  consider the variation of concentration and mean-streamwise velocity profile in the plane of the wall bisector, as sketched in figure \ref{fig:SP_MP_LP}. 
Velocity and concentration measurements have been performed for LP and MP only whereas a qualitative picture is proposed for SP based on the pressure drop measurements and the simulations in \cite{fornari2017suspensions} for neutrally buoyant particles as their size in inner units (20$\delta_\nu$) is in range of our experiments for SP (15--37$\delta_\nu$). The simulations are, however, performed at a lower $Re_{2H}$ = 5600 and the particle size in bulk units is $2H/d_{p}$ = 18, similar to MP in experiments. The simulations confirm the largest concentration at the core of the duct at $\phi$ = 20\% and that the Reynolds stresses are typically lower than those of the single phase flow, as also generally found in this study. \cite{fornari2017suspensions} suggest that turbulence is largely attenuated at this $\phi$: turbulence production is reduced, but this is compensated by a large increase in energy injection due to particle-induce stresses.

In light of our observations and as depicted in figure \ref{fig:SP_MP_LP}, the fluid velocity is very similar to the single phase case in the case of LP, despite the larger turbulence attenuation, whereas it becomes more laminar-like for MP and is believed to be even more so as the particle size reduces to SP.  
The mean near-wall gradient increases as the particle size increases due to the particle layer forming at the wall and to the associated increase of the momentum transport 
\citep{costa2016universal}.
\sz{At a constant volume fraction the number of particles increases as the particle size decreases. The increased number of particles means more contact points and collisions between the wall and the particles, potentially increasing friction even if the momentum of each particles decreases. If so, we have two competing mechanisms as particle size reduces.} This may explain the minimum in the pressure drop at MP and larger values for SP and LP. Note that SP are yet significantly larger than the small scales in the flow, of the order of about 30 viscous units and that the total drag is anyway above that of single-phase turbulence at $\phi=20\%$ but below that of a turbulent flow with the effective suspension viscosity at the highest Reynolds number considered.

At low $\phi$ = 5--10\%, \cite{fornari2017suspensions} found that the turbulence production is slightly larger than in single phase flow. Together with energy injection due to particle presence, the overall turbulence activity is found to increase by a few percent. In our experiments, at $\phi$ = 5\% and high $Re_{2H}$, the mean streamwise velocity and the fluid Reynolds stress profiles do not change substantially as compared to single phase flow between MP and LP. Nevertheless, a decreasing drag from SP to LP suggests that any increase in turbulence activity is a decreasing function of the particle size. As mentioned above, as larger particles have more inertia, they do not follow the fluid as nicely as smaller particles do and display larger resistance to deformation. Additional experimental investigations close to the wall and at $\phi$ = 10\% along with detailed numerical simulations for particles with size corresponding to SP and LP are needed to shed light on this.

At lower $Re_{2H}$ (e.g. 10000 and 15000) settling effects become important, especially at $\phi$ = 20\% when particles move as part of a relatively dense particle bed and particles close to the bottom wall are shielded from the fast moving fluid above them. 
At lower $\phi$ the thickness of the sediment bed is smaller and hence a larger fraction of particles can be resuspended by the moving fluid. 
In our experiments at constant flow rate, we observe the fluid flow above the MP bed to be faster as these smaller particles form a bed of lower permeability, and the drag to be large for LP as these contribute to the momentum transfer with larger particle-induced stresses.
As discussed above, the density of SP is higher than MP and LP (see table \ref{tab:Particle property}) and their Shields number lies in between the values for MP and LP. This can explain why we have an intermediate value of pressure drop for SP at these two lower $Re_{2H}$ and  $\phi$ = 20\%. 
As $Re_{2H}$ increases, more particles are suspended in the bulk of the flow and the total drag can be explained by the inertial effects and particle dynamics just discussed. 

It is also noteworthy that the total drag is above that of a continuum medium with an effective suspension viscosity at $Re_{2H} \approx 10000$ and is below that value when increasing the particle size and the flow Reynolds number, which may be explained by turbulence modulations and particle migrations inside the duct. This confirms that particle dynamics and inhomogeneities are fundamental for accurate models of suspensions flows in the turbulent regime, as shown for their rheological behaviour in simple laminar shear flows.

\section*{Acknowledgements}

This work was supported by the European Research Council Grant No. ERC-2013-CoG-616186, TRITOS, from the Swedish Research Council (VR), through the Outstanding Young Researcher Award to LB.


\bibliographystyle{jfm}
\bibliography{biblio.bib}

\section*{Appendix}
\label{sec:Appendix}

\begin{figure}
\centering

\begin{subfigure}{.24\textwidth}
  \centering
  $z/H$ = 0
  \includegraphics[height=1\linewidth]{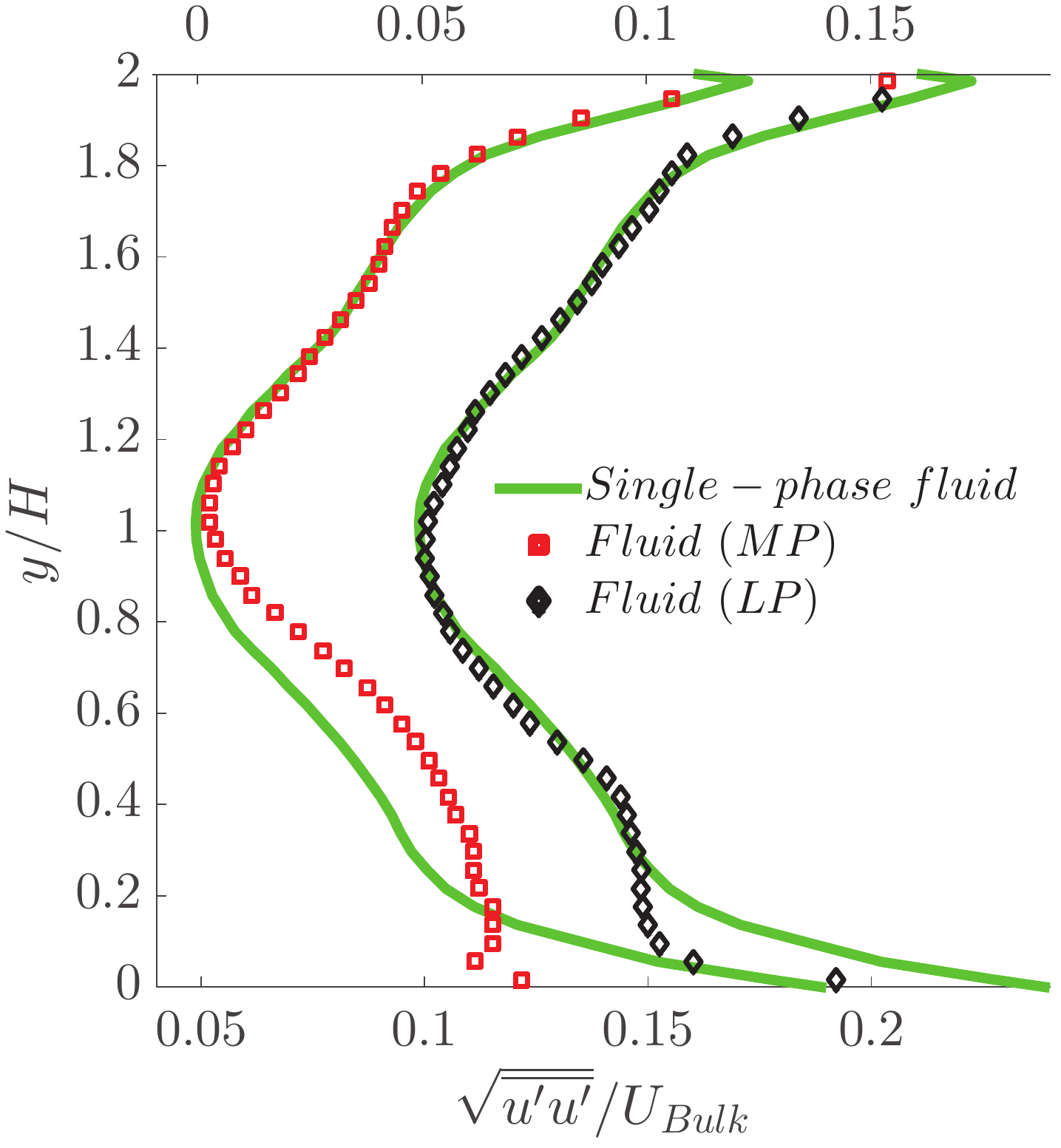}
  \caption{}
  \label{fig:3_6mm_5p_Re10000_sp00mm_urms}
\end{subfigure}%
\begin{subfigure}{.24\textwidth}
  \centering
  $z/H$ = 0.4
  \includegraphics[height=1\linewidth]{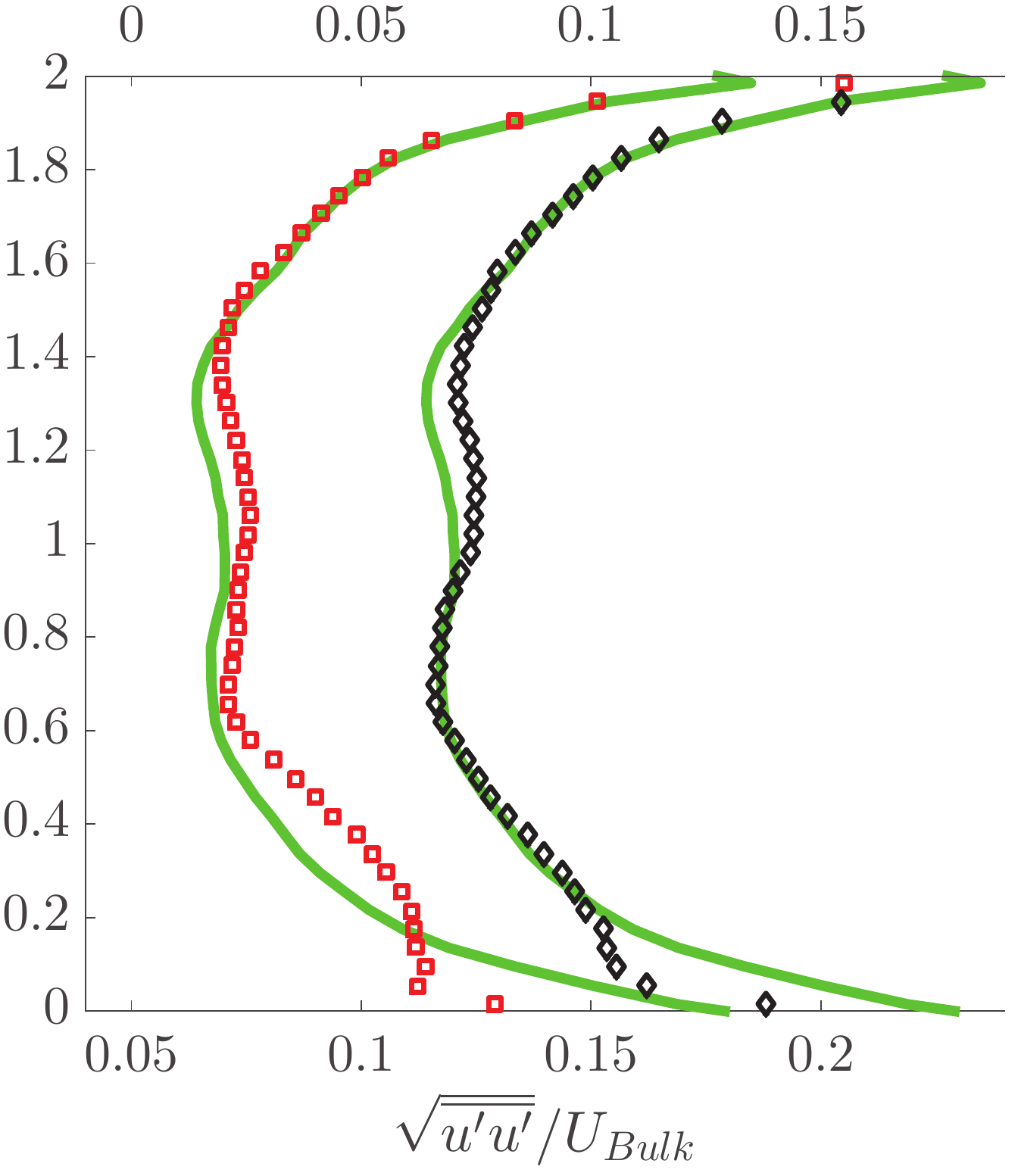}
  \caption{}
  \label{fig:3_6mm_5p_Re10000_sp10mm_urms}
\end{subfigure}
\begin{subfigure}{.24\textwidth}
  \centering
  $z/H$ = 0.8
  \includegraphics[height=1\linewidth]{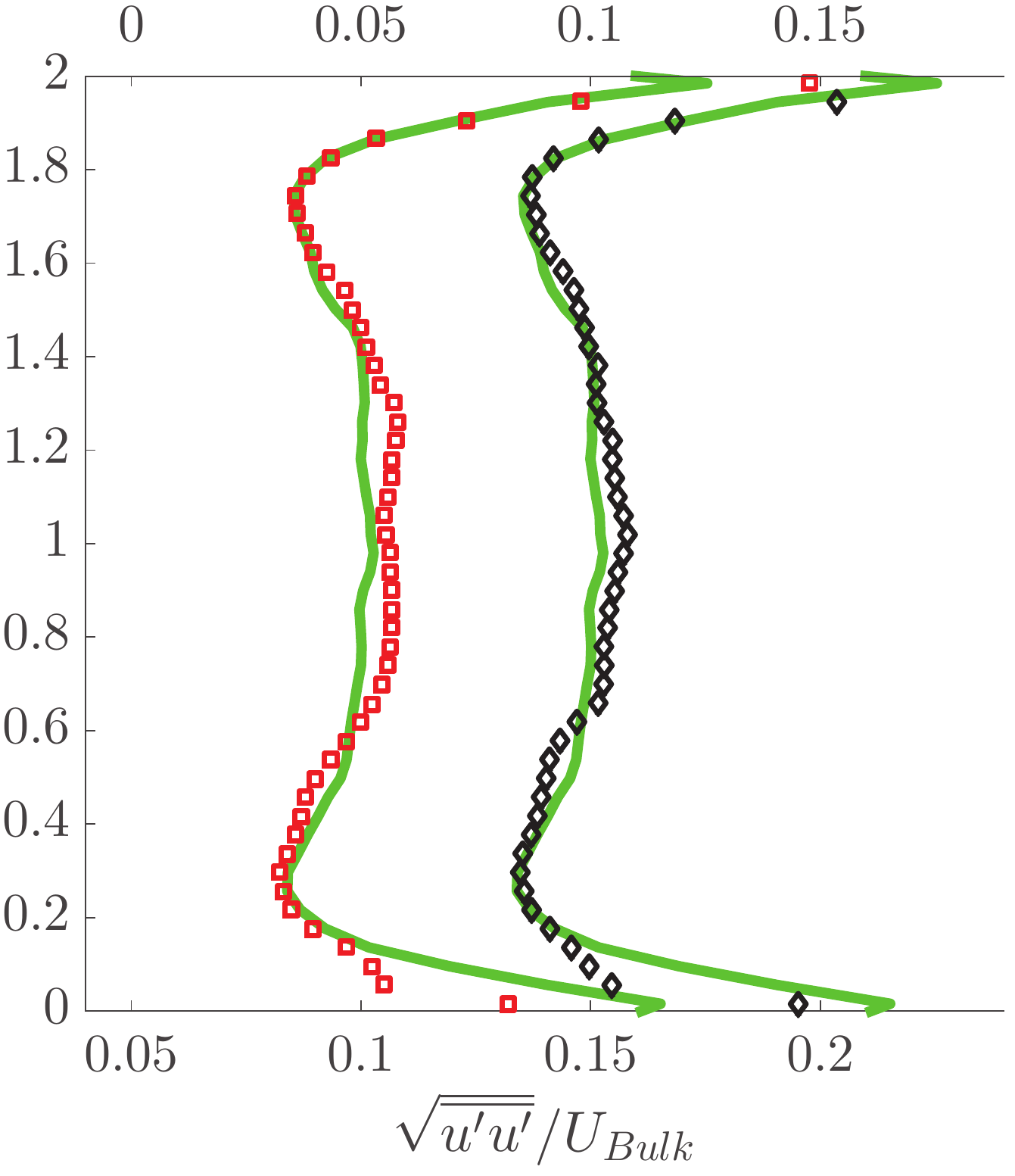}
  \caption{}
  \label{fig:3_6mm_5p_Re10000_sp20mm_urms}
\end{subfigure}

\begin{subfigure}{.24\textwidth}
  \centering
  \includegraphics[height=1\linewidth]{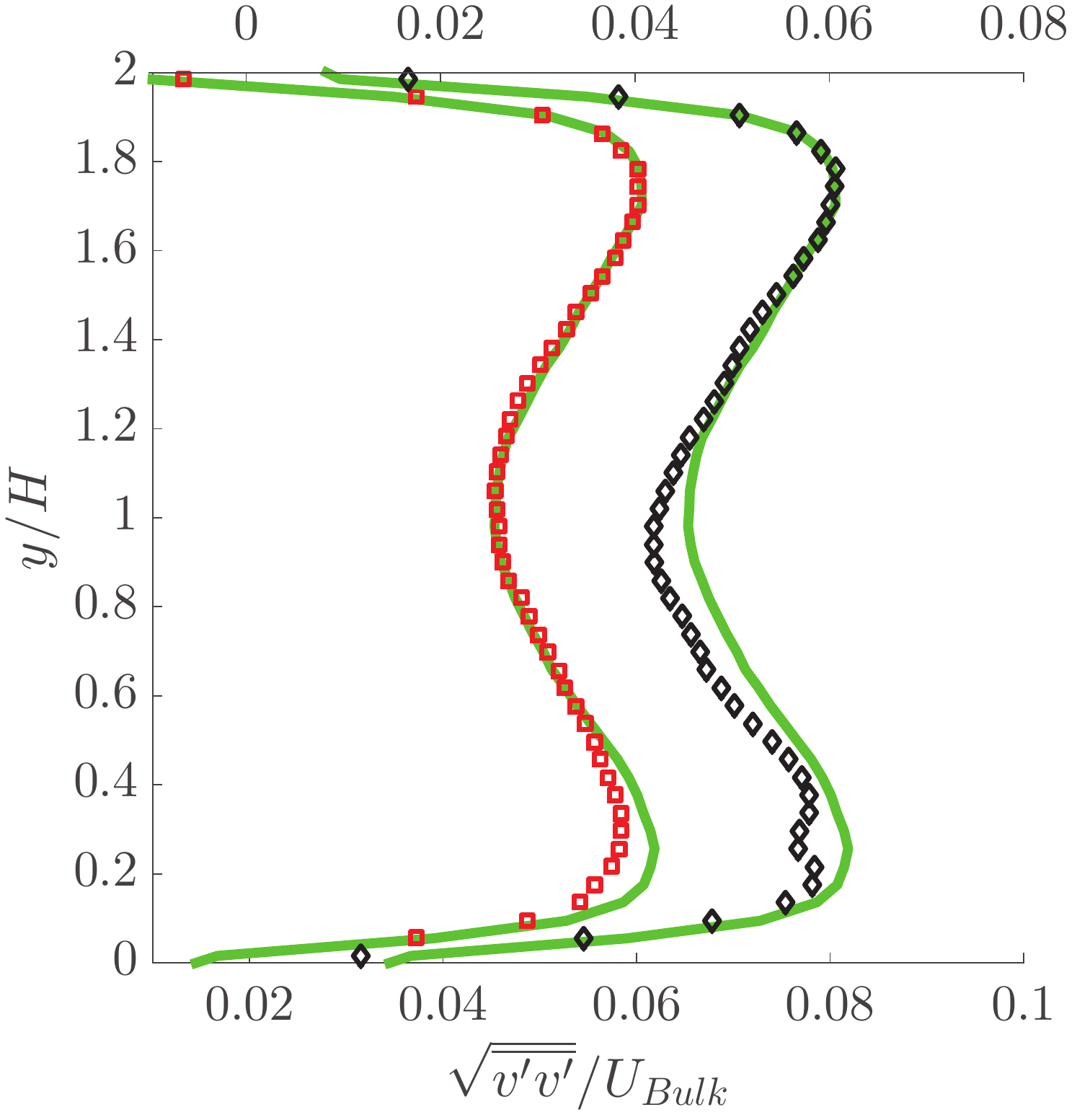}
  \caption{}
  \label{fig:3_6mm_5p_Re10000_sp00mm_vrms}
\end{subfigure}%
\begin{subfigure}{.24\textwidth}
  \centering
  \includegraphics[height=1\linewidth]{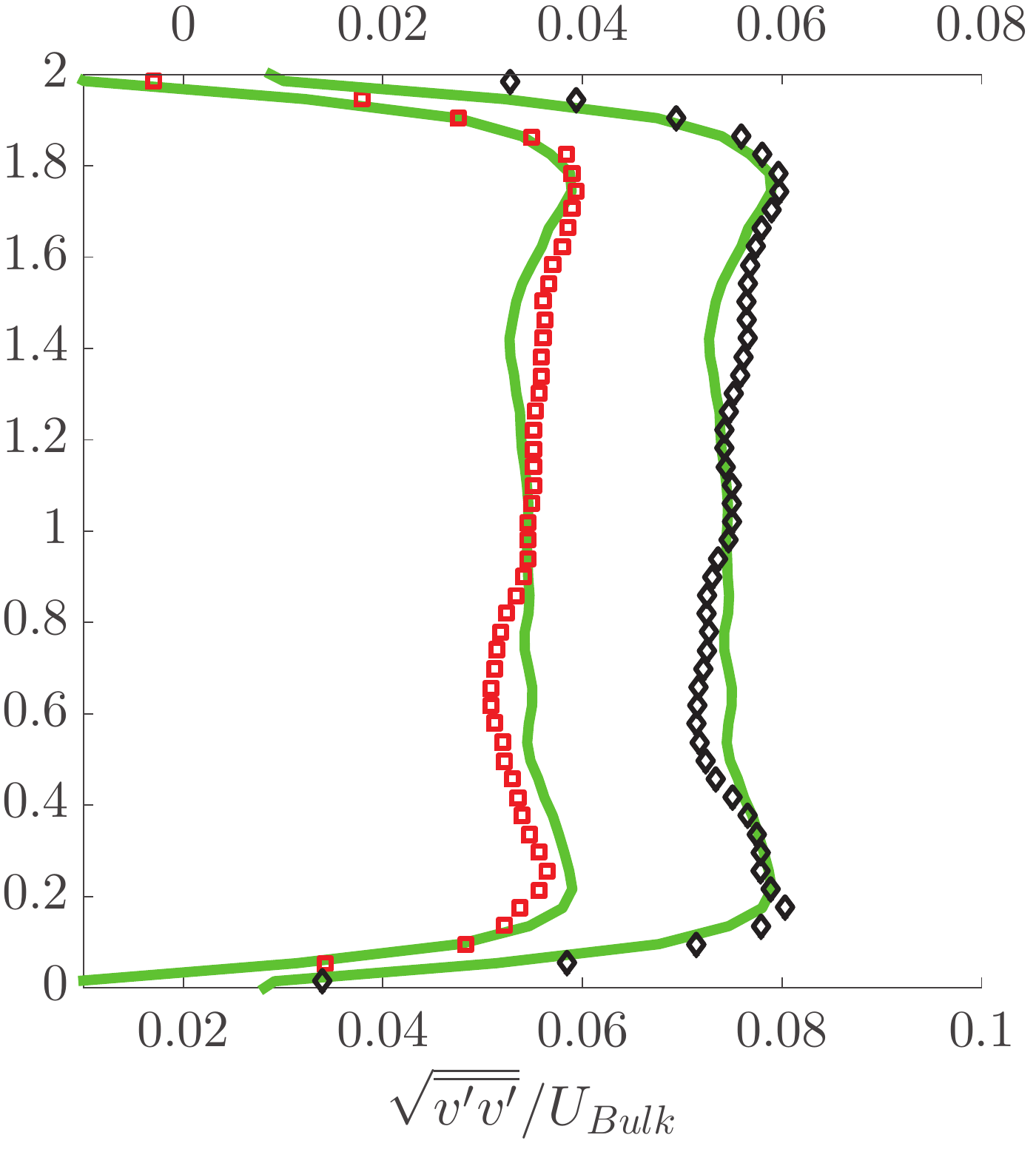}
  \caption{}
  \label{fig:3_6mm_5p_Re10000_sp10mm_vrms}
\end{subfigure}
\begin{subfigure}{.24\textwidth}
  \centering
  \includegraphics[height=1\linewidth]{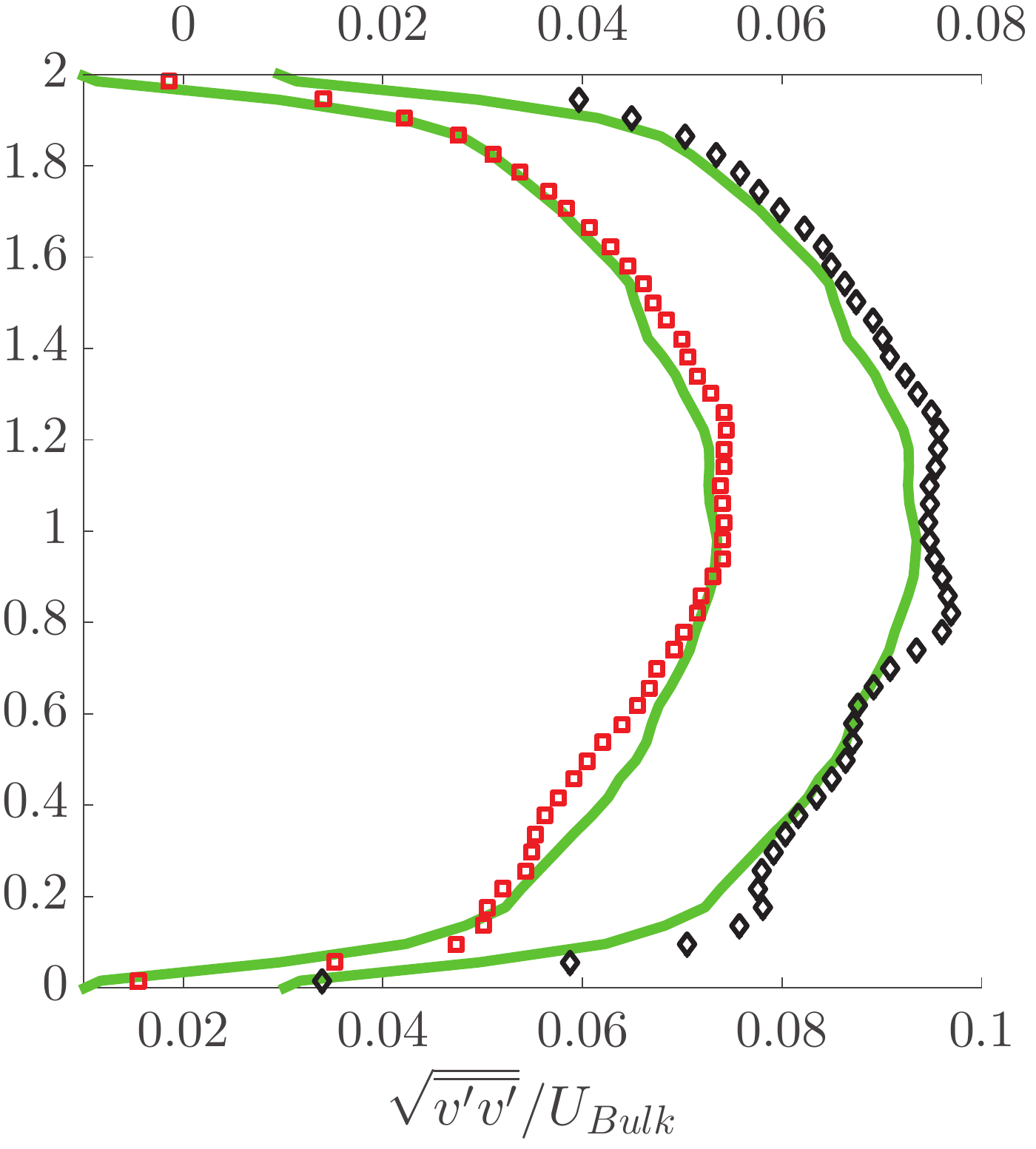}
  \caption{}
  \label{fig:3_6mm_5p_Re10000_sp20mm_vrms}
\end{subfigure}
\caption{Comparsion of streamwise (a)-(c) and wall-normal (d)-(f) rms velocities for medium (MP) and large-sized particles (LP): $Re_{2H}\approx$ 10000, $\phi$ = 5\%. Profiles shifted to the right correspond to LP and the corresponding x-axis is on the top.}
\label{fig:3_6mm_5p_Re10000_urms_vrms}
\end{figure}

\begin{figure}
\centering
\begin{subfigure}{.24\textwidth}
  \centering
  $z/H$ = 0
  \includegraphics[height=1\linewidth]{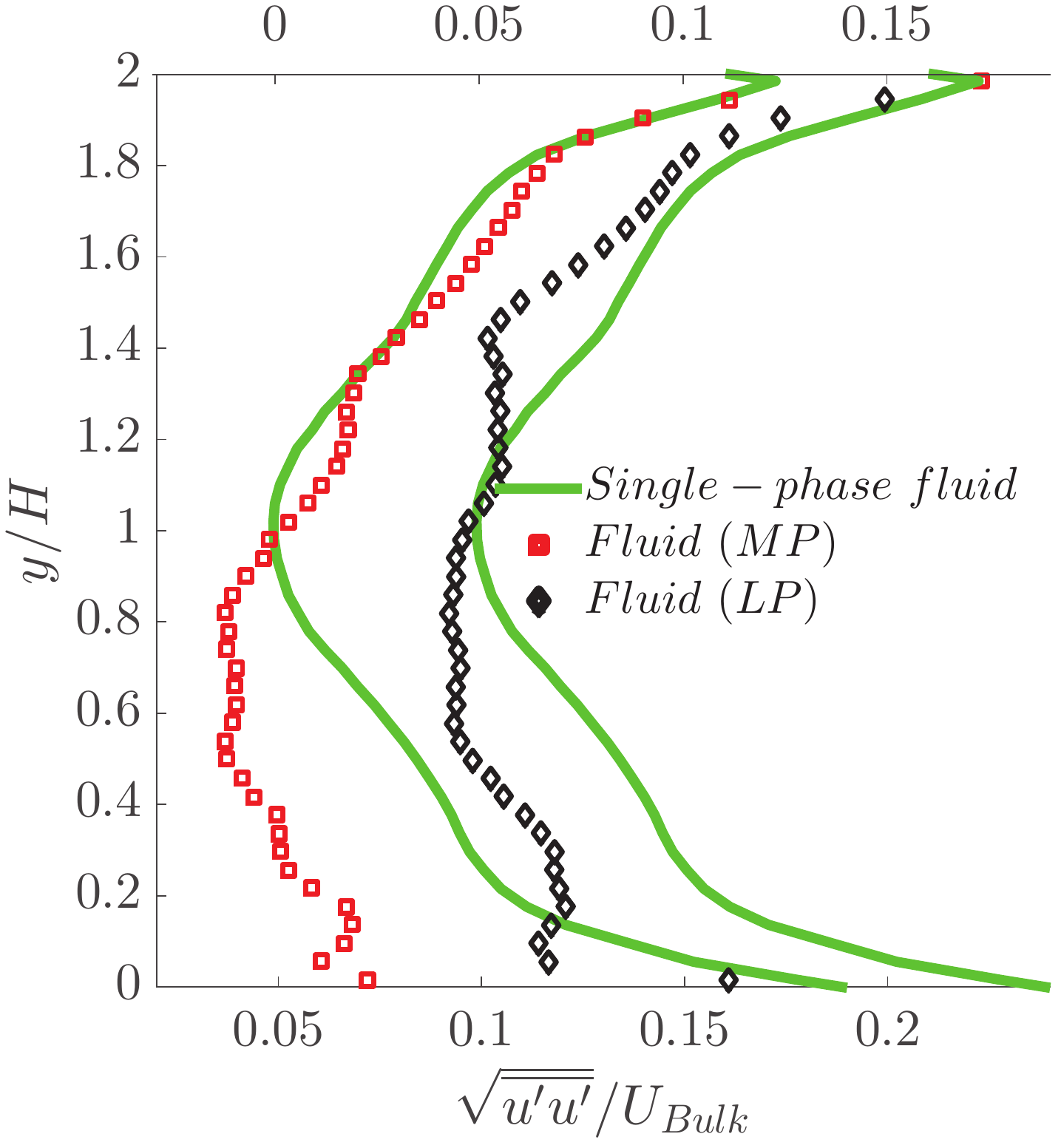}
  \caption{}
  \label{fig:3_6mm_20p_Re10000_sp00mm_urms}
\end{subfigure}%
\begin{subfigure}{.24\textwidth}
  \centering
  $z/H$ = 0.4
  \includegraphics[height=1\linewidth]{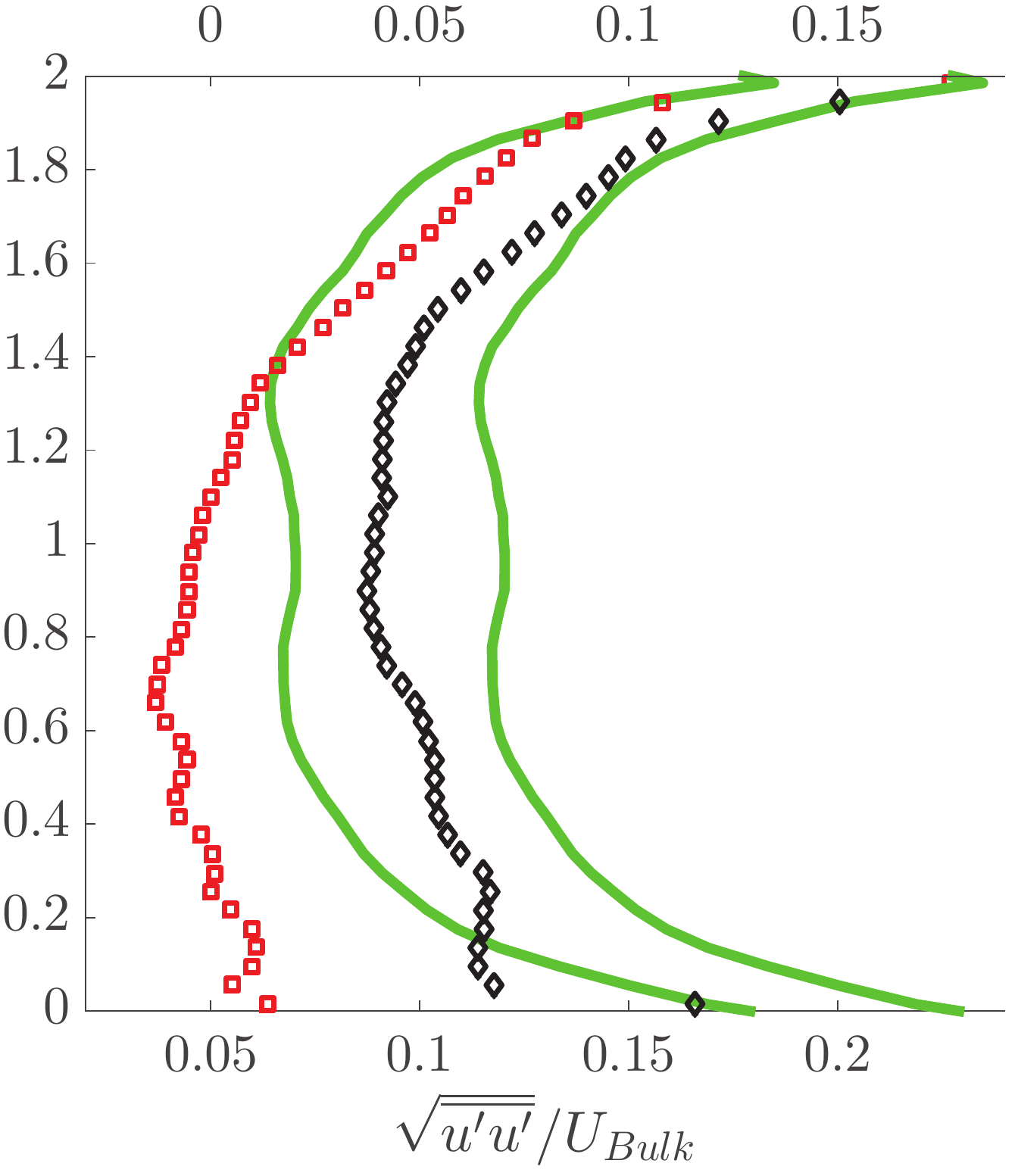}
  \caption{}
  \label{fig:3_6mm_20p_Re10000_sp10mm_urms}
\end{subfigure}
\begin{subfigure}{.24\textwidth}
  \centering
  $z/H$ = 0.8
  \includegraphics[height=1\linewidth]{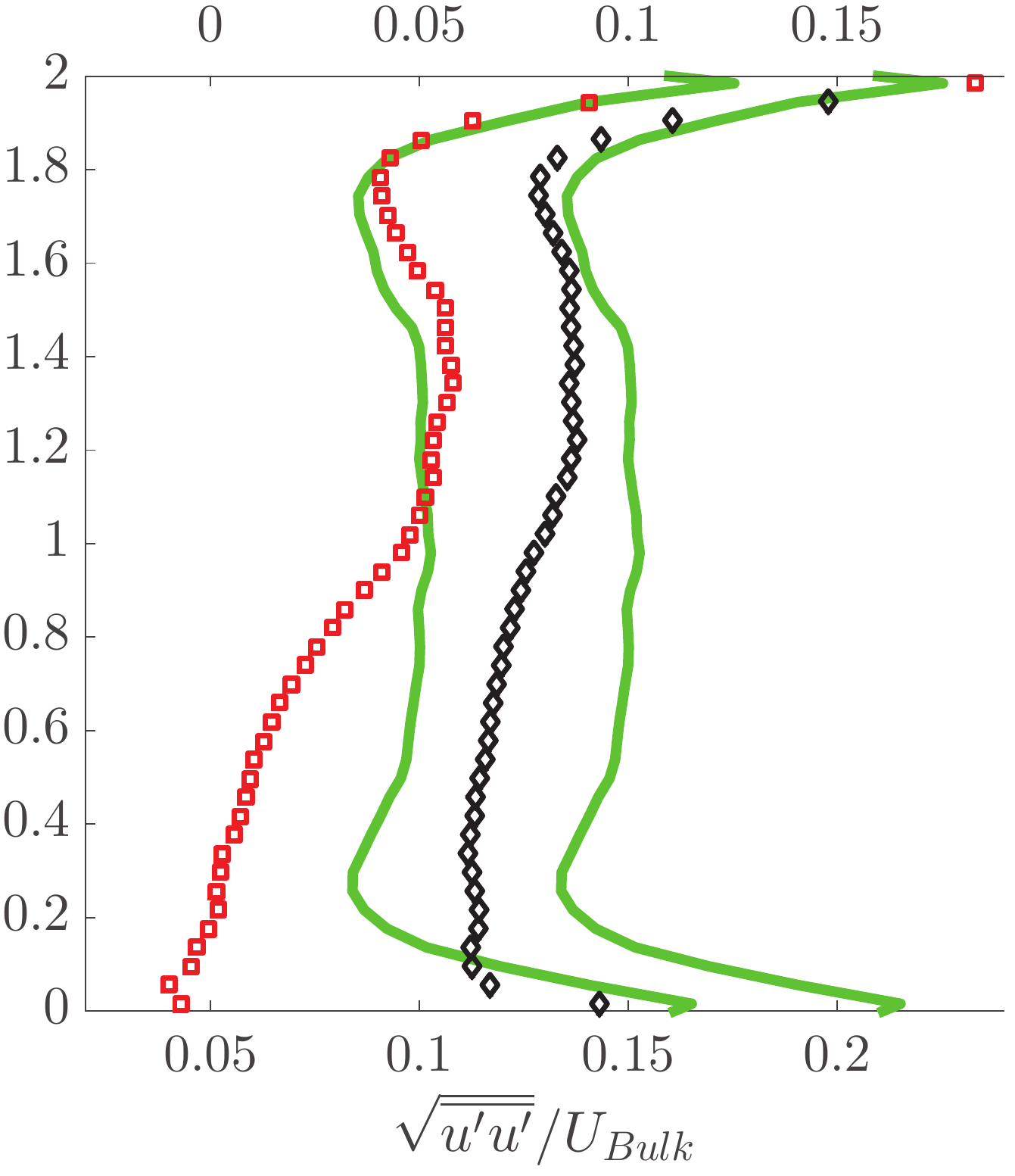}
  \caption{}
  \label{fig:3_6mm_20p_Re10000_sp20mm_urms}
\end{subfigure}

\begin{subfigure}{.24\textwidth}
  \centering
  \includegraphics[height=1\linewidth]{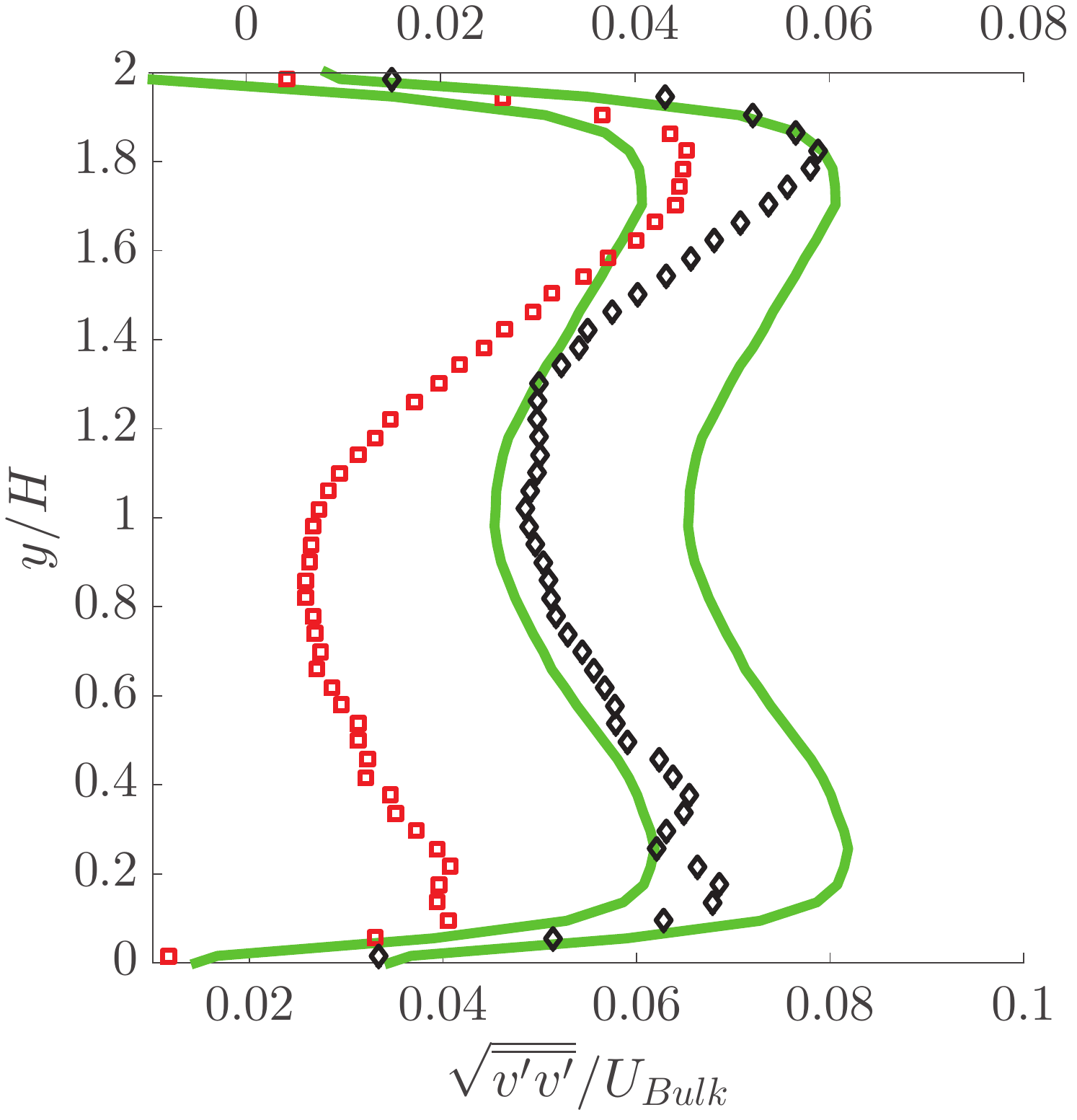}
  \caption{}
  \label{fig:3_6mm_20p_Re10000_sp00mm_vrms}
\end{subfigure}%
\begin{subfigure}{.24\textwidth}
  \centering
  \includegraphics[height=1\linewidth]{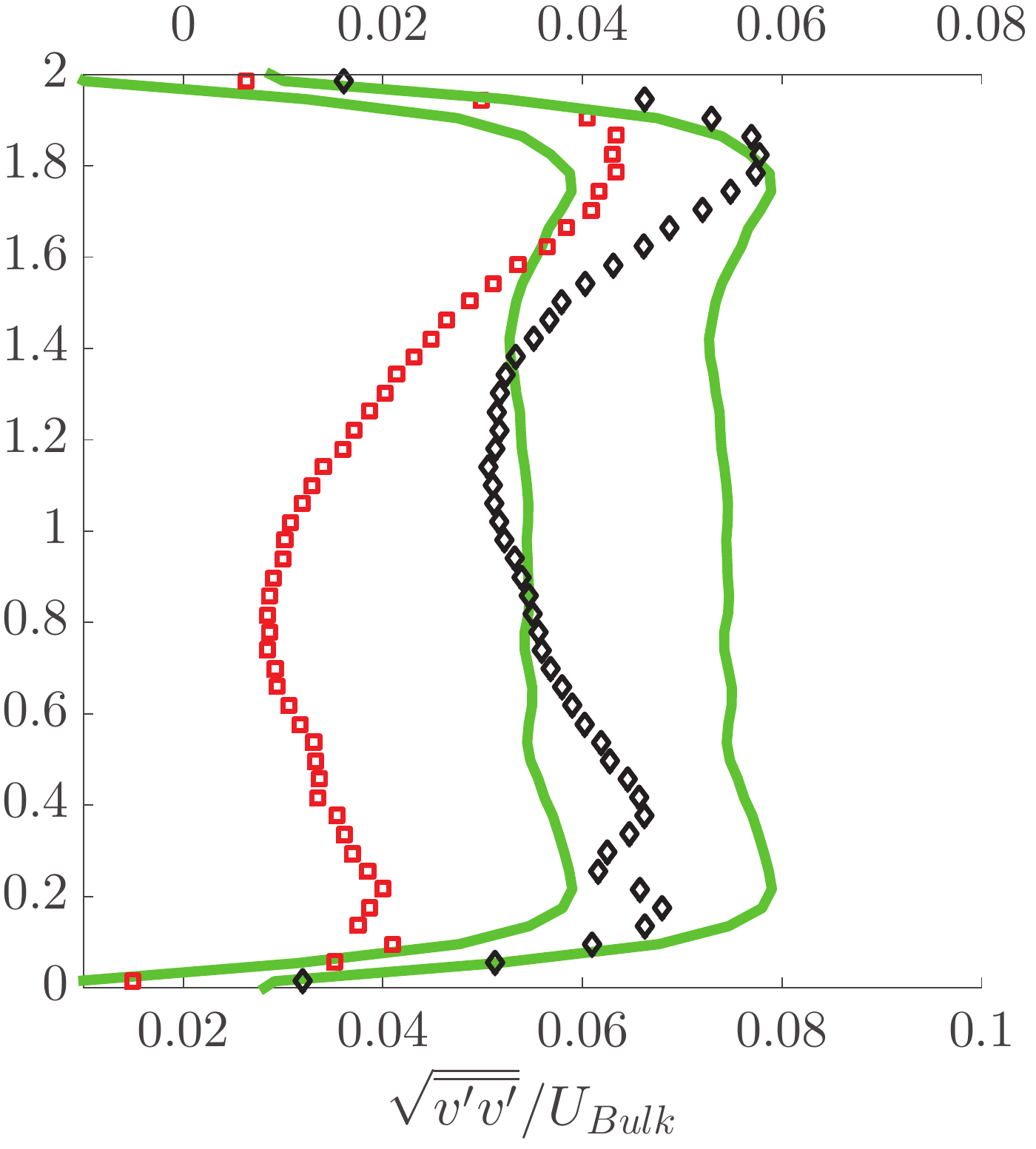}
  \caption{}
  \label{fig:3_6mm_20p_Re10000_sp10mm_vrms}
\end{subfigure}
\begin{subfigure}{.24\textwidth}
  \centering
  \includegraphics[height=1\linewidth]{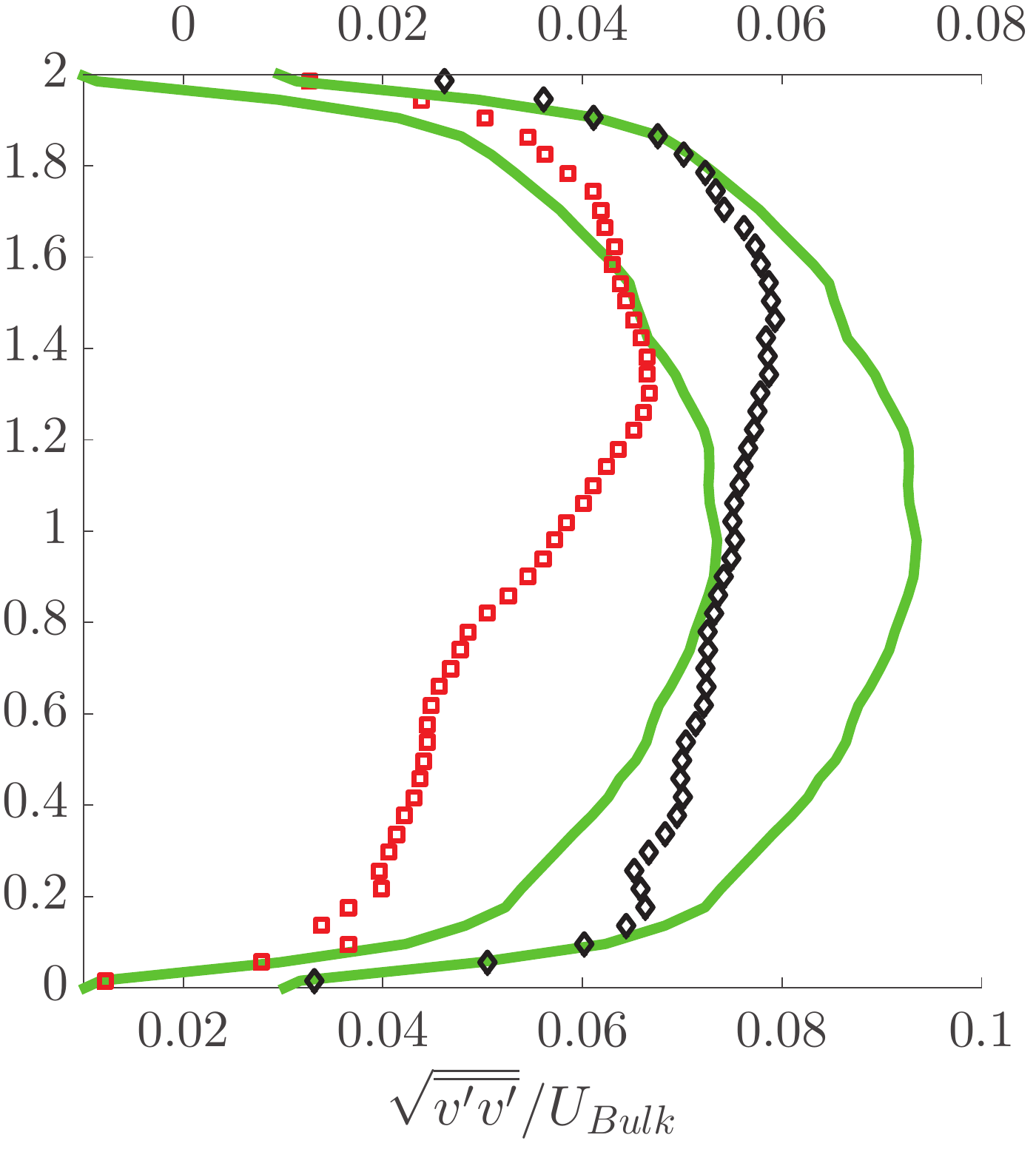}
  \caption{}
  \label{fig:3_6mm_20p_Re10000_sp20mm_vrms}
\end{subfigure}
\caption{Comparsion of streamwise (a)-(c) and wall-normal (d)-(f) rms velocities for medium (MP) and large-sized particles (LP): $Re_{2H}\approx$ 10000, $\phi$ = 20\%. Profiles shifted to the right correspond to LP and the corresponding x-axis is on the top.}
\label{fig:3_6mm_20p_Re10000_urms_vrms}
\end{figure}

For completeness, we report in figure \ref{fig:3_6mm_5p_Re10000_urms_vrms} the rms of the streamwise and wall-normal velocity fluctuations for MP and LP at $Re_{2H}\approx$ 10000 and $\phi$ = 5\%. The figure shows that the streamwise velocity fluctuations increase in case of MP in the lower part of the duct. Very close to the bottom wall the streamwise velocity fluctuations reduce both for MP and LP.
Similarly, figure \ref{fig:3_6mm_20p_Re10000_urms_vrms} displays the rms of the streamwise and wall-normal velocity fluctuations for MP and LP at $Re_{2H}\approx$ 10000 and $\phi$ = 20\%. 

\clearpage

\end{document}